\documentclass[11pt,makeidx,twoside]{phdthesis}

\usepackage{hyperref}
\usepackage{imakeidx}
\usepackage{makeidx}
\usepackage[utf8]{inputenc}
\usepackage{amsmath}
\usepackage{amssymb}
\usepackage{epsfig}
\usepackage{graphics}
\usepackage{float}
\usepackage{here}
\usepackage{rotating}
\usepackage{float}
\usepackage{multirow}
\usepackage{comment}
\usepackage{epigraph}
\usepackage{algorithm}
\usepackage{algorithmic}
\usepackage{amsfonts}
\usepackage{amstext}
\usepackage{graphicx}
\usepackage{url}
\usepackage{bm}
\usepackage{wrapfig}
\usepackage{cite}
\usepackage{dsfont}
\usepackage{epstopdf}
\usepackage{subfig}
\usepackage{avant}
\usepackage{makeidx}    
\usepackage{fancyhdr}
\usepackage{amsfonts}
\usepackage{amssymb}
\usepackage{amstext}
\usepackage{graphicx}
\usepackage{amsmath}
\usepackage{url}
\usepackage{epstopdf}
\usepackage{color}
\usepackage{xcolor}
\usepackage{bbm}
\usepackage{array}
\usepackage{amsfonts}
\usepackage{amssymb}
\usepackage{amstext}
\usepackage{pdfpages}
\usepackage{epigraph}
\usepackage{calligra}
\usepackage{yfonts,color}
\usepackage{lettrine}
\usepackage{avant}
\usepackage{fancyhdr}
\usepackage{epsfig}
\usepackage{float}
\usepackage{here}
\usepackage{rotating}
\usepackage{multirow}
\usepackage{comment}
\usepackage{epigraph}
\usepackage{amsfonts}
\usepackage{amstext}
\usepackage{amsthm}
\usepackage{url}
\usepackage{color}
\usepackage{xcolor}
\usepackage{bm}
\usepackage{dsfont}
\allowdisplaybreaks
\colorlet{mygreen}{green!40!gray}

\hypersetup{
colorlinks,
citecolor=mygreen,
linkcolor=black,
urlcolor=blue
}

\colorlet{linkequation}{blue}
\newcommand*{\refeq}[1]{%
  \begingroup
    \hypersetup{
      linkcolor=linkequation,
      linkbordercolor=linkequation,
    }%
    \ref{#1}%
  \endgroup
}

\colorlet{linkfigure}{red}
\newcommand*{\reffig}[1]{%
  \begingroup
    \hypersetup{
      linkcolor=linkfigure,
      linkbordercolor=linkfigure,
    }%
    \ref{#1}%
  \endgroup
}

\colorlet{linktable}{cyan}
\newcommand*{\reftab}[1]{%
  \begingroup
    \hypersetup{
      linkcolor=linktable,
      linkbordercolor=linktable,
    }%
    \ref{#1}%
  \endgroup
}

\makeindex

\newtheorem{definition}{Definition}

\graphicspath{{./figures/}}

\begin{document}


\includepdf[pages=-]{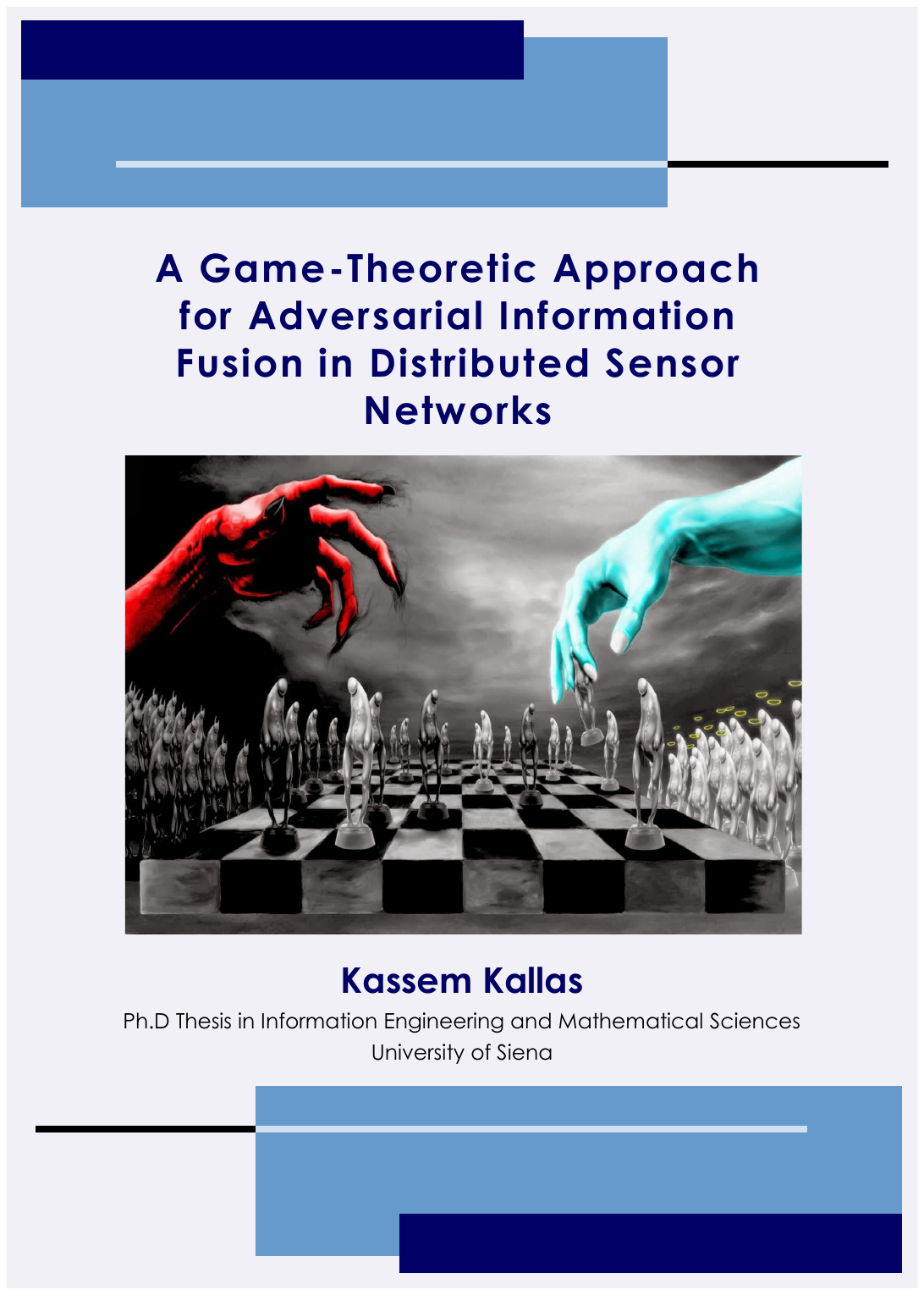} 

\pagestyle{fancyplain}
\begin{titlepage}

\begin{center}

\begin{large}
\textbf{UNIVERSIT\`A DEGLI STUDI DI SIENA} \\
\end{large}

\vspace{2pt}
\begin{large}
\textsc{Facolt\`a di Ingegneria}\\
\end{large}
\vspace{3pt}
\begin{normalsize}
\textsc{Dipartimento di Ingegneria dell'Informazione e Scienze Matematiche} \\
\end{normalsize}

\begin{figure}[!h]
\begin{center}
\includegraphics[width=3.5cm]{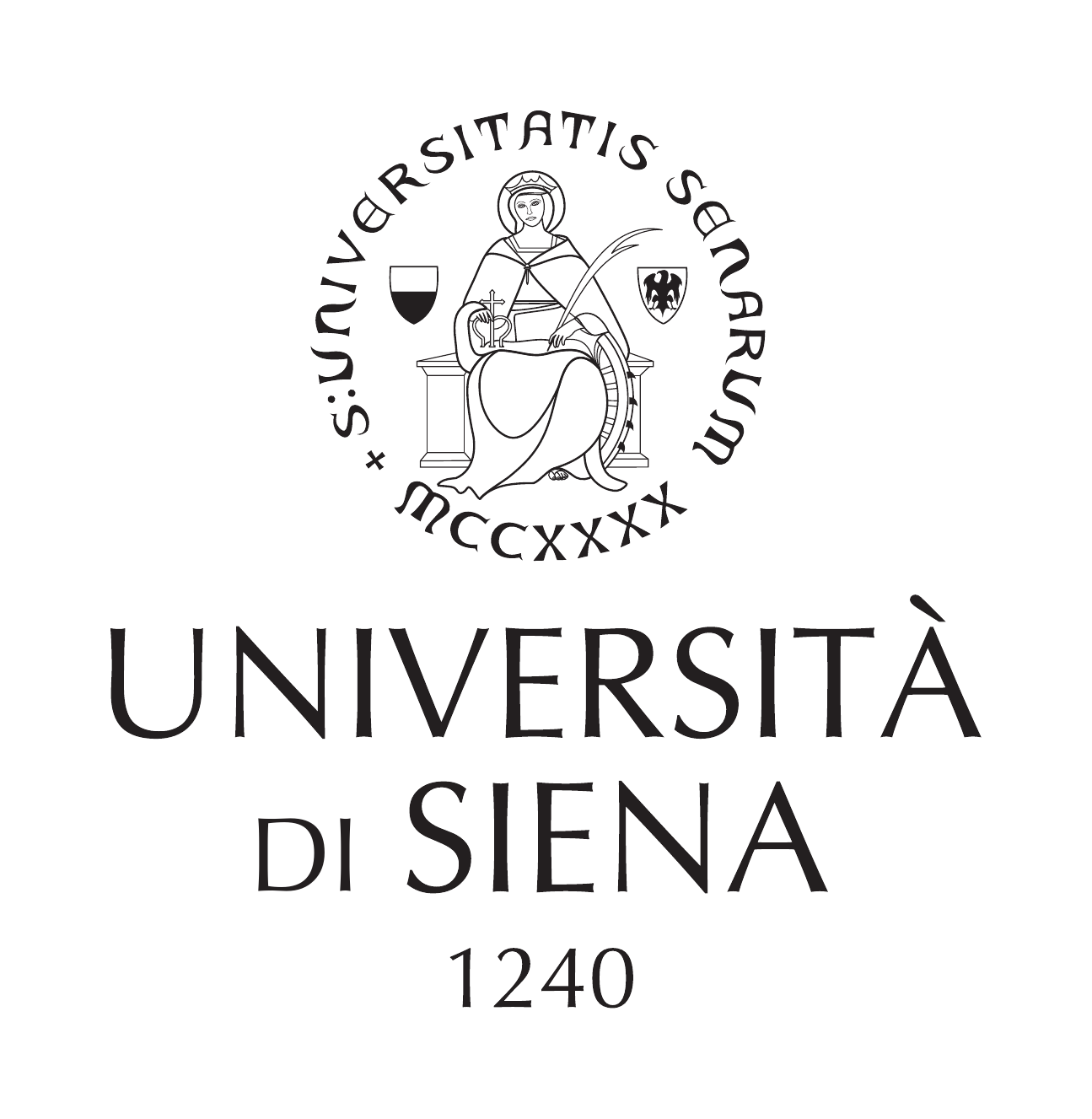}
\end{center}
\end{figure}

\vspace{-1.5cm}
\begin{LARGE}
\begin{center}
\textbf{A Game-Theoretic Approach for Adversarial Information Fusion in Distributed Sensor Networks}
\end{center}
\end{LARGE}

\begin{large}
\LARGE
\calligra{Kassem Kallas}
\end{large}

\vspace{10pt}
\begin{small}
\textsl{Ph.D Thesis in Information Engineering and Mathematical Sciences \\ \textit{XXIX Cycle}}
\end{small}

\vspace{12pt}

\begin{small}

\begin{center}
 \textit{Supervisor}
\\ \vspace{3pt} Prof. Mauro Barni
\end{center}
\begin{center}
 \textit{Co-Supervisor}
\\ \vspace{3pt} Prof. Andrea Abrardo
\end{center}

\vspace{10 pt}

\begin{minipage}{0.4\textwidth}
\begin{flushleft} 
\emph{External reviewers}\\
 \vspace{4pt} Prof. Vincenzo Matta
\\ \vspace{4pt} Prof. Vito Fragnelli\\
\end{flushleft}
\end{minipage}
~
\begin{minipage}{0.4\textwidth}
\begin{flushright} 
\emph{Examination Committee}
\\ \vspace{4pt} Prof. Vincenzo Matta
\\ \vspace{4pt} Prof. Vito Fragnelli
\\ \vspace{4pt} Prof. Andrea Garzelli
\end{flushright}
\end{minipage}

\end{small}


\vspace{1 pt}

\line(1, 0){338} \\
\begin{normalsize}
\textsc{Siena \\Month Day, Year}
\end{normalsize}

\end{center}
\end{titlepage}

\pagenumbering{roman}

\cleardoublepage \lhead[]{\fancyplain{}{\rightmark}}
\chead[\fancyplain{}{}]{\fancyplain{}{}}
\rhead[\fancyplain{}{\leftmark}]{\fancyplain{}{}}

\tableofcontents
\listoffigures
\listoftables

\cleardoublepage
\addcontentsline{toc}{chapter}{List of Symbols}

\chapter*{List of Symbols}

\begin{table}[h]
\renewcommand{\arraystretch}{1.2}
\centering
\begin{tabular}{|p{3.2cm}| p{8cm}|}
\hline
$H_0$ & null hypothesis\\ \hline
$H_1$ & alternative hypothesis \\ \hline
$n$ & number of nodes in the network\\ \hline
$\mathbf{x}_i$ & observation vectors available to sensor $i$ \\ \hline
$S_i$ & the system state under hypothesis $H_i, i \in \{0,1\}$  \\ \hline
$P(H_0)$ & a-prior probability that the system is in state $S_0$\\ \hline
$P(H_1)$ & a-prior probability that the system is in state $S_1$\\ \hline
$P(x|H_j)$ & the observation probability density conditioned to hypothesis $H_j$\\ \hline
$S^* \in \{0,1\}$ & the global decision at the fusion center regarding $S^*$\\ \hline
$C_{ij}$ & cost of deciding $H_i$ when $H_j$ is true \\ \hline
$\mathcal{C}$ & average cost or risk function for Bayesian detector \\ \hline
$\Lambda(x)$ & likelihood ratio test regarding the observation $x$ \\ \hline
$\lambda$ & decision threshold \\ \hline
\end{tabular}
\end{table}

\begin{table}[h]
\renewcommand{\arraystretch}{1.2}
\centering
\begin{tabular}{|p{3.2cm} | p{8cm}|}
\hline
$P_{FA}$ & probability of false alarm \\ \hline
$P_{MD}$ & probability of missed detection \\ \hline
$P_{D}$ & probability of detection \\ \hline
$P_{null}$ & probability to decide $H_0$ when it is true \\ \hline
$P_e$ & probability of error \\ \hline
$\lambda_{NP}$ & local Neyman-Pearson likelihood decision threshold \\ \hline
$\alpha_{NP}$ & acceptable false alarm for Neyman-Pearson detector \\ \hline
$\mathcal{F}$ & Lagrange function for Neyman-Pearson detector optimization \\ \hline
$\lambda_i, i \in \{0,1\}$ & decision threshold for hypothesis $H_i$ for local SPRT detector \\ \hline
$\alpha_{ST}$ & local SPRT detector constraint on false alarm probability\\ \hline
$\beta_{ST}$ & local SPRT detector constraint on missed detection probability\\ \hline
$u_i$ & information sent by sensor $i$ to the FC \\ \hline
$P_{d_i}$ & local probability of detection at node $i$ \\ \hline
$P_{{fa}_i}$ & local probability of false alarm at node $i$ \\ \hline
$P_{{md}_i}$ & local probability of missed detection at node $i$ \\ \hline
$Q_D$ & global probability of detection at the FC \\ \hline
$Q_{FA}$ & global probability of false alarm at the FC \\ \hline
$Q_{D_{AND}}$ & global probability of detection for the AND rule \\ \hline
$Q_{FA_{AND}}$ & global probability of false alarm for the AND rule\\ \hline
$Q_{D_{OR}}$ & global probability of detection for the OR rule \\ \hline
$Q_{FA_{OR}}$ & global probability of false alarm for the OR rule\\ \hline
\end{tabular}
\end{table}

\begin{table}[h]
\renewcommand{\arraystretch}{1.2}
\centering
\begin{tabular}{|p{3.2cm} | p{8cm}|}
\hline
$Q_{D_{kn}}$ & global probability of detection for the $k$-out-of-$n$ rule \\ \hline
$Q_{FA_{kn}}$ & global probability of false alarm for the $k$-out-of-$n$ rule\\ \hline
$U_{SLC}$ & Square Law Combining information fusion result  \\ \hline
$U_{MRC}$ & Maximum Ratio Combining information fusion result  \\ \hline
$U_{SC}$ & Selection Combining information fusion result  \\ \hline
$\zeta$  & decision threshold of the soft combination rules \\ \hline
$\Upsilon_i, i \in \{0,1\}$ & decision threshold for hypothesis $H_i$ for global SPRT detector \\ \hline
$\alpha_{FC}$ & global SPRT detector constraint on false alarm probability\\ \hline
$\beta_{FC}$ & global SPRT detector constraint on missed detection probability\\ \hline
$\mathcal{G} = (\mathcal{N}, \mathcal{E})$ & graph $\mathcal{G}$ with the set of vertices $\mathcal{N} = \{n_1, ...,n_n\}$ \\ 
& and the set of edges $\mathcal{E}$\\ \hline
$\mathcal{N}_i$ & neighborhood list of node $i$ \\ \hline
$x_i(k)$ &  state value of node $i$ at iteration $k$ of the consensus algorithm \\ \hline
$w_i$ & weight assigned to node $i$ state update value \\ \hline
$\epsilon$ & consensus update parameter \\ \hline
$\bar{x}$ & consensus value or the average value of all initial state values \\ \hline
$\mathcal{X}_{M}^2$  & chi-square distribution with $M$ degrees of freedom \\ \hline
$\Gamma(.)$ & the incomplete gamma function \\ \hline
$Q(.)$ & the generalized Marcum \emph{Q-function} \\ \hline
$\gamma_S$ & the SNR exprienced by the Energy Detector \\ \hline
\end{tabular}
\end{table}

\begin{table}[h]
\renewcommand{\arraystretch}{1.2}
\centering
\begin{tabular}{|p{3.2cm} | p{8cm}|}
\hline
$\mathcal{S}_i$ & strategy set available to player $i$ \\ \hline
$v_l$ & payoff (or utility) of player $l$\\ \hline
$G(N,\mathcal{S},\mathbf{v})$ & Game with $N$ players, strategy set $\mathcal{S}$ and payoff vector $\mathbf{v}$   \\ \hline
$\mathbf{ss}$ & strategy profile vector for a game with $N$ players  \\ \hline
$\Pi(\mathcal{Z})$ & set of all probability distributions over the set $\mathcal{Z}$  \\ \hline
$r_i$  & report sent by node $i$ to the FC \\ \hline
$\alpha$ & fraction of nodes (or links) under attack \\
& or the probability that a node (or link) is under attack\\ \hline
$\mathbf{\tilde{x}}_i$ & attacked observation seen by node $i$\\ \hline
$\tilde{x}_i(0)$ & falsified initial data of node $i$ in consensus network\\ \hline
$\tilde{w}_i$ & tampered weight for node $i$'s state value in consensus network \\ \hline
$\Delta_i$ & attack value in falsification attack on consensus network \\ \hline
$u_i(k)$ &  attack value of consensus disruption attack at node $i$ for the $k+1$ iteration step \\ \hline
$w_i$ & weight assigned to node $i$ state update value \\ \hline
$r_{ij}$ & report by node $i$ at instant $j$  \\ \hline
$m$ & observation window size \\ \hline
$P_{mal}$  & node malicious probability or crossover probability of the attacked links \\ \hline
$u_{ij}$ & decision by node $i$ at instant $j$ \\ \hline
$\Gamma_{i}$ & hard reputation score of node $i$ \\ \hline
$d_{int}(j)$ & intermediate decision at instant $j$ at the FC \\ \hline
\end{tabular}
\end{table}

\begin{table}[h]
\renewcommand{\arraystretch}{1.2}
\centering
\begin{tabular}{|p{3.2cm} | p{8cm}|}
\hline
$\eta$ & isolation threshold \\ \hline
$R_{ij}$ & soft reputation score of node $i$ at instant $j$\\ \hline
$DF(\mathcal{S}_{FC},\mathcal{S}_{FC},v)$ & decision fusion game with $\mathcal{S}_{FC}$ the strategy set for the FC, \\
& $\mathcal{S}_{B}$ the strategy set for Byzantines, and payoff $v$   \\ \hline
$\mathbf{ss}$ & strategy profile vector for a game with $N$ players  \\ \hline
$P_{e,ar}$ & probability of error after removal of Byzantines  \\ \hline
$P_{ISO}^B$  & probability of correct isolation of Byzantines \\ \hline
$P_{ISO}^H$ & probability of erroneous isolation of honest nodes \\ \hline
$P_{mal}^{FC}$ & the FC guess of $P_{mal}$\\ \hline
$P_X(x)$ & probability mass function of the random variable $x$\\ \hline
$S^m$ & sequence of system states random variable with instantiation $s^m$  \\ \hline
$P_{S_j}(i), i \in \{0,1\}$ & probability that a system is at state $S_j$ at time $i$ \\ \hline
$U_{ij}$ &  random variable for the local decision of node $i$ at instant $j$ with instantiation $u_{ij}$ \\ \hline
$A^n = (A_1,\dots,A_n)$ & binary random sequence for Byzantines positions with $a^n$ its instantiation \\ \hline
$\mathbf{R}= \{R_{ij}\}$ & random matrix of all received reports by FC \\
& with $\mathbf{r}= \{r_{ij}\}$ as its instantiation \\ \hline
$P(a^n)$ & probability of Byzantines sequence \\ \hline
$\varepsilon$  & local decision error at the nodes \\ \hline
$\delta$ & the probability that the FC will receive a wrong report \\ \hline
$\Gamma_{i}$ & hard reputation score of node $i$ \\ \hline
$m_{eq}^{(i)}$ & the number of instants of which the report is equal to the system state for node $i$ \\ \hline
\end{tabular}
\end{table}

\begin{table}[h]
\renewcommand{\arraystretch}{1.2}
\centering
\begin{tabular}{|p{3.2cm} | p{8cm}|}
\hline
$E[N_B]$ & expected number of Byzantines \\ \hline
$\mu_{A_i}$ & expected value of $A_i$\\ \hline
$H(A^n)$ & entropy distribution of Byzantines \\\hline
$h(\mu_{A_i})$ & binary entropy function for the expected value of $A_i$  \\ \hline
$h$ & the FC expected maximum number of Byzantines   \\ \hline
$\mathcal{I} = \{1,\dots,n\}$  & indexing set of size $n$ \\ \hline
$\mathcal{I}_k$ & set of all $k$-subsets of $\mathcal{I}$ \\ \hline
$I$ & random variable with indexes of byzantine nodes\\ \hline
$P(I)$ & equivalent to $P(a^n)$\\ \hline
$n_B$ & fixed number of Byzantines in the network knwon to the FC  \\ \hline
$DF_{Byz}(\mathcal{S}_B,\mathcal{S}_{FC},v)$ &decision fusion game with $\mathcal{S}_B$ the strategy set of Byzantines, \\
& $\mathcal{S}_{FC}$ the strategy set of the FC, and $v$ the payoff \\ \hline
$P_{mal}^{B}$ & malicious probability strategy of the Byzantines \\ \hline
$\mathcal{S}_B^q$ & quantized Byzantines's strategy set \\ \hline
$\mathcal{S}_{FC}^q$ & quantized FC's strategy set \\
& with $\mathbf{r}= \{r_{ij}\}$ as its instantiation \\ \hline
$\mathbf{V}$ & payoff matrix for each pair of strategies \\ \hline
$P_e^*$  & probability of error at the equilibrium \\ \hline
$P(P_{mal}^{B})$ & probability assigned by Byzantines to a strategy in mixed strategy Nash equilibrium \\ \hline
$P(P_{mal}^{FC})$ & probability assigned by FC to a strategy in mixed strategy Nash equilibrium \\ \hline
$\rho$ & state transition probability in a two-state markov model \\ \hline
\end{tabular}
\end{table}

\begin{table}[h]
\renewcommand{\arraystretch}{1.2}
\centering
\begin{tabular}{|p{3.2cm} | p{8cm}|}
\hline
$m_{vf}^{(l)}$ & variable-to-function message for factor $l$ \\ \hline
$m_{fv}^{(l)}$ & function-to-variable message for factor $l$ \\ \hline
$X_i$ & random variable for the measurement at node $n_i$ with $x_i$ its instantiation  \\ \hline
$\mathcal{N}(\mu,\sigma)$ & normal distribution with mean $\mu$ and standard deviation $\sigma$ \\ \hline
$\mathcal{N}_H$ & set of nodes of uncorrupted measurements \\ \hline
$\tilde{x}_i$ & fake state update value for node $i$ \\ \hline
$n_A$ & number of attacked links \\ \hline
$\Delta$ & measurement falsification attack value  \\ \hline
$p$ & success probability of measurement falsification attack \\ \hline
$\mathcal{R}$ & set of remaining nodes after removal  \\ \hline
$CDD(\mathcal{S}_A,\mathcal{S}_D,v)$ & consensus-based distributed detection game with $\mathcal{S}_A$ the strategy set of attacker, \\
& $\mathcal{S}_D$ the strategy set of the defender, and $v$ the payoff \\ \hline
$SNR$ & signal to noise ratio \\ \hline
\end{tabular}
\end{table}



\cleardoublepage
\addcontentsline{toc}{chapter}{Abstract}


\chapter*{Abstract}

\bigskip

\PARstart{\textcolor{red}E}very{} \textit{day we share our personal information through digital systems which are constantly exposed to threats. For this reason, security-oriented disciplines of signal processing have received increasing attention in the last decades: multimedia forensics, digital watermarking, biometrics, network monitoring, steganography and steganalysis are just a few examples. Even though each of these fields has its own peculiarities, they all have to deal with a common problem: the presence of one or more adversaries aiming at making the system fail. Adversarial Signal Processing lays the basis of a general theory that takes into account the impact that the presence of an adversary has on the design of effective signal processing tools.}

\textit{By focusing on the application side of Adversarial Signal Processing, namely adversarial information fusion in distributed sensor networks, and adopting a game-theoretic approach, this thesis contributes to the above mission by addressing four issues. First, we address decision fusion in distributed sensor networks by developing a novel soft isolation defense scheme that protect the network from adversaries, specifically, Byzantines. Second, we develop an optimum decision fusion strategy in the presence of Byzantines. In the next step, we propose a technique to reduce the complexity of the optimum fusion by relying on a novel near-optimum message passing algorithm based on factor graphs. Finally, we introduce a defense mechanism to protect decentralized networks running consensus algorithm against data falsification attacks.}




\newcommand{\publ}{}

\pagestyle{fancyplain}
\renewcommand{\headrulewidth}{0.3pt}
\renewcommand{\footrulewidth}{0.0pt}
\renewcommand{\plainfootrulewidth}{0.0pt}
\renewcommand{\plainheadrulewidth}{0pt}
\renewcommand{\sectionmark}[1]{\markright{\it \thesection.\ #1}}
\renewcommand{\chaptermark}[1]{\markboth{
       \it \thechapter.\ #1}{}}
\lhead[\thepage]{\fancyplain{\publ}{\rightmark}}
\chead[\fancyplain{}{}]{\fancyplain{}{}}
\rhead[\fancyplain{}{\leftmark}]{\fancyplain{}{\thepage}}
\lfoot[]{}
\cfoot[]{}
\rfoot[]{}

\pagenumbering{arabic}

\part{Introduction, Basic Notions, and State of Art}


\chapter{Introduction}
\label{chapter:intro}

\begin{flushright}
\emph{"Security against defeat implies defensive tactics; ability to defeat the enemy means taking the offensive."}
\\
Sun Tzu, "The Art of War"

\emph{"When you play the game of thrones, you win or you die. There is no middle ground."}
\\
Cersei Lannister, "Game of Thrones"


\end{flushright}
\bigskip
\section{Motivation}
\PARstart{\color{red}I}n{}  the era of digital revolution, intelligent and digital systems are invading our lives. This evolution has a fundamental impact on social, political and economical domains both at personal and society level.

While this digital world is of extreme importance and contributes to the health of our society, its ultra-fast growth creates new opportunities to perpetrate digital crimes, that is cybercrimes, all the more, that this new kind of criminal activity does not need anymore the physical presence of criminals on the crime scene. Criminals and victims are no more limited to territorial borders since crimes are perpetrated in a virtual cyberspace. These crimes can target economy and finance, public health and national security.

Cybercrimes\index{cybercrimes|textbf} encompass a spectrum of activities ranging from violating personal privacy to illegal retrieval of digital information about a firm, a person and so on. Crimes like fraud, child pornography, violation of digital privacy, money laundering, and counterfeiting stand on the middle of the cybercrimes spectrum. Due to the anonymity provided by the internet, criminals are concealed over the cyberspace to attack\index{attack|textbf} their specific victims\index{victims|textbf}. Another part of these crimes aims at altering the data of individuals within corporations or government bureaucracies for either profit or political objectives. The other part of the spectrum is occupied by crimes that aim at disrupting internet functionality. These range from spam, hacking, and Denial of Service (DoS)\index{Denial of Service|textbf} attacks, to cyberterrorism\index{cyberterrorism|textbf} that, according to the U.S. Federal Bureau of Investigation (FBI)\index{Federal Bureau of Investigation (FBI)|textbf}, is any \textit{"premeditated, politically motivated attack against information, computer systems, computer programs, and data which results in violence against non-combatant targets by sub-national groups or clandestine agents"}.
The public awareness about the danger of cyberterrorism has grown dramatically since 11 September 2001 attacks. Especially nowadays, due to the existence of numerous terrorist groups that are interested in attacking a vast list of targets. These groups are benefiting from the cyberspace to recruit personnel to get involved into terrorist activities. As an evidence about the economical impact of cybercrimes, McAfee\index{McAfee|textbf} reported in 2014 that the estimated cost of cybercrime on the global economy is more than 400 billion dollars \cite{McAfeeReport2014}.

For all these reasons, the fight against cybercrime occupies a top position in the priorities list of many governments around the globe. For instance, in the United States, within the department of justice, the FBI's Cyber Division is the agency responsible to combat cybercrime \cite{FBICyberCrime}. Other agencies like the U.S. Secret Service (USSS) and U.S. Immigration and Customs Enforcement (ICE) have specific branches committed to fight cybercrimes \cite{HomeLandSec}. Moreover, the USSS runs the National Computer Forensic Institute (NCFI) that offers training courses in computer forensics to help the state and local law officers, prosecutors, and judges to conduct basis electronic crimes investigations, respond to network intrusion incidents, conduct computer forensics examination, and strengthen their prosecution and adjudication \cite{CyberForensicsBook}. 
In addition, the Internet Crime Complaint Center (IC3) serves as a partnership between the FBI and the National White Collar Crime Center -known as NW3C-. IC3 aims to provide the public with a suitable and reliable reporting mechanism to submit information to FBI. For this sake, IC3 accepts online complaints from victims of internet crimes or any interested third party \cite{IC3online}.

Attention to and fight against cybercrimes\index{cybercrimes|textbf} is not limited to governmental institutions, it also involves scientific researchers\index{scientific researchers|textbf} with various backgrounds from all around the globe. Researchers devote their effort to develop effective solutions to these security problems\index{security problems|textbf} and take effective steps toward secure defense solutions and algorithms to combate cybercrime. Signal processing\index{Signal processing|textbf} researchers are on the top of the list of scientists engaged in such an effort. Increasing attention has been devoted to disciplines like multimedia forensics\index{multimedia forensics|textbf} \cite{BarniTondiSourceIdentificationGame},\cite{Boh12}, digital watermarking\index{digital watermarking|textbf} \cite{perez2006watermarking}, steganography and steganalysis\index{steganography and steganalysis|textbf} \cite{kessler2011overview}, biometrics\index{biometrics|textbf} \cite{jain2005biometric}, network intrusion detection, spam filtering \cite{barreno2010security},\cite{lowd2005adversarial}, traffic monitoring \cite{deng2005countermeasures}, videosurveillance \cite{dufaux2008scrambling} and many others. Despite enormous differences, all these fields are characterized by a unifying feature: the presence of one or more adversary\index{adversary|textbf} aiming at system failure. So far, the problem of coping with an adversary has been addressed by different communities with very limited interaction among them. It is not surprising, then, that similar solutions are re-invented several times, and that the same problems are faced again and again by ignoring that satisfactory solutions have already been discovered in contiguous fields. The lack of a unifying view makes difficult to grasp the essence of the addressed problems and work out effective solutions. While each adversarial scenario has its own peculiarities, there are some common and fundamental problems whose solution under a unified framework would speed up the understanding of the associated security problems and the development of effective and general solutions. The absence of a unifying framework raises the need for a general theory that takes into account the presence of an adversary and its effect on the design of effective signal processing tools, i.e. a theory of Adversarial Signal Processing (Adv-SP)\index{Adversarial Signal Processing|textbf}, a.k.a. Adversary-aware Signal Processing \cite{AdvSP}.

Adv-SP is an emerging discipline that aims at studying signal processing techniques explicitly thought to withstand the intentional attacks of one or more adversaries aiming at system failure. Its final aim is modeling the interplay between a Defender, wishing to carry out a certain processing task, and an Attacker, aiming at impeding it. A natural framework to model this interplay relies on Game-Theory\index{Game Theory|textbf} since it provides a powerful mathematical model of conflict and cooperation between rational decision-makers\index{rational decision-makers|textbf}. This framework helps to overcome the so called "cat $\&$ mouse" loop in which researchers and system designers continuously develop new attacks and countermeasures in a never-ending loop. By adopting the game-theoretical formalization\index{Game-Theory|textbf}, a tremendous step toward the development of the theoretical foundation of Adv-SP has been already achieved, \cite{BarniTondiSourceIdentificationGame},\cite{BarniTondiHTwithTrainingData}, \cite{BarniTondiSecurityMargin} and \cite{BarniTondiSourceDistinguishability} are just a few examples. While these works aim at developing a general Adv-SP theory, in this thesis, we apply some general ideas from the Adv-SP field to the problem of Adversarial Information Fusion\index{adversarial information fusion|textbf} in Distributed Sensor Networks\index{distributed sensor networks|textbf}. In these networks, some distributed sensors, for instance autonomous sensors, actuators, mobile devices, must provide some information about the state of an observed system. In the centralized approach, the information collected by the sensors is sent to a \textit{"Fusion Center"}\index{Fusion Center|textbf} (FC). By using all the information\index{information|textbf} received from the nodes, the FC is responsible of making final global decision\index{global decision|textbf} about the state of the system of interest. The actual process of integrating the information submitted by several sources into a coherent understanding of the system state is called \textit{"Information Fusion"}. Therefore, Information Fusion\index{information fusion|textbf}, in general, refers to particular mathematical functions, algorithms, methods and procedures for combining information. This term is very flexible and the classification of various techniques depends on different perspectives i.e type of information, type of data representation, level of information abstraction, and others \cite{BookModellingInfoFusion}. Information fusion techniques are extensively used in several fields. In economics for instance, information fusion is used to compute the Retail Price Index (RPI) which is a measure of the change of the average prices over a certain amount of time, or the Human Development Index (HDI) that is a metric to assess the social and ecomic development levels of countries, and many others. In addition, in biology, fusing DNA and RNA sequences is another form of information fusion. Moreover, in Computer Science and Artificial Intelligence, information fusion is used widely, i.e sensors data fusion in robotics, fusion of images in computer vision, ensemble methods for data mining, decision making systems, multi-agents systems and many others \cite{BookModellingInfoFusion},\cite{chao1987evidential},\cite{xiong2002multiSenSorInfoFusion}.

In this thesis, we focus on Information Fusion in Distributed Sensor Networks in the presence of adversaries. Specifically, we address a setup wherein some of the sensors might be interested in corrupting the information fusion process to pursue an exclusive benefit from the system under inspection, forge the knowledge about the state of the system or corrupt the whole network functionality. These malicious\index{malicious|textbf} and misbehaving nodes\index{misbehaving nodes|textbf}\footnote{Throughout the thesis, we will alternatively use the words sensor and node to refer to a distributed sensor network entity.} are known as adversaries and due to their presence, the problem at hand is called "Adversarial Information Fusion". Following this setting, the defender or the network designer is asked to modify the fusion process to take into account that a part of the network is under the dominance of the adversaries. The modification of the fusion process is to be implemented at the FC if the network is centralized, while, on the other hand, is to be implemented locally at the nodes when the network is fully decentralized.

Adversarial Information Fusion\index{adversarial information fusion|textbf} in Distributed Networks is of great importance in many applications. Cognitive Radio Networks\index{Cognitive Radio Networks|textbf} (CRN) offer a first example. The electromagnetic spectrum is a naturally limited resource that is ever-demanded due to the explosion of wireless technology \cite{haykinspectrumsensing},\cite{xiaoCRN}. The use of a fixed access policy is ineffective because it assigns spectrum portions exclusively to licensed users. A report by Federal Communication Commission\index{Federal Communication Commission|textbf} (FCC) in 2002 revealed that the licensed spectrum is heavily underutilized by the owners \cite{FCC2002}; this raises the need for a more flexibile and efficient spectrum allocation policy. Dynamic Spectrum Access\index{Dynamic Spectrum Access|textbf} (DSA) is a promising solution to underutilization of the spectrum \cite{liangCRDSA}. 
In DSA there are two types of users: licensed users known as Primary Users\index{Primary Users|textbf} (PU) and second priority users known as Secondary User\index{Secondary User|textbf} (SU). A PU has the highest priority to access the spectrum resource since he is the license holder; SUs on the other hand, are allowed to access the spectrum when it is free or in a shared manner providing no harmful interference to PUs \cite{liangCRDSA}. DSA requires that SUs are able to sense, monitor and access the spectrum in an efficient, intelligent and dynamic way. An SU device with cognitive capabilities is known as Cognitive Radio (CR). This concept was introduced by Joseph Mitola III in his PhD thesis in year 2000 \cite{Mitola1999CR}, \cite{Mitola2000CR}. Cognitive Radio is Software Defined Radio (SDR) device that is a fully programmable wireless device that can sense its surrounding environment and adapt its transmission and reception parameters accordingly. The intelligent and dynamic adaptation capabilities of the CR are the reason behind calling it a "Brain Empowered Wireless Communication" \cite{HaykinBrainEmpowered}. 

Cognitive Radios must sense the spectrum by monitoring the PU activity in order to decide if the spectrum is occupied or not. This decision can be made either locally by exchanging the "measurements" between CRs or remoltely by sending the measurements to a FC that "fuses" the received information and broadcasts back the final global decision. The adversaries in CRN can modify their measurements to cause a wrong decision about the spectrum occupancy. This wrong decision can have many effects, for instance, cause harmful interference to PU's transmission, exclusive use of the spectrum by the adversaries or even just confusing the network.

Wireless Sensor Networks\index{Wireless Sensor Networks|textbf} (WSN) offer another important example. A WSN is a group of spatially distributed sensors that are responsible to monitor a physical phenomenon, health and environmental conditions like temperature, sound, pressure etc... In WSN, the sensors are responsibile to measure the physical phenomenon of interest and then pass the information gathered to the nearby nodes or to a FC which fuses all the data received and comes out with a global decision. If the adversary can control some of the sensors, based on its objective and knowledge of the physical system, it can perform arbitrary attacks by flooding the network with random information or devise a strategic attack by sending specific wrong information to force a precise false global decision. These behaviors can disrupt severely the network's functionality and operation and consequently, corrupt the whole WSN.

The emerging field of multimedia forensics offers an additional example. Due to nowdays powerful and user-friendly softwares, editing digital media such as images, video or audio no longer requires professional skills. Typically, editing is used to enhance the media quality, e.g. by enhancing image contrast, denoising an audio track or re-encoding a video to reduce its size. However, altering a digital media can serve less 'innocent' purposes. For instance, to remove or implant evidence or to distribute fake content so to create a deceiving forgery. This makes the truthfulness of the message conveyed by media contents doubtfull and to be questioned since "seeing is not believing"  anymore and a photographic image cannot be considered as an evidence to support any fact. Multimedia Forensics tackles with this problem based on the observation that any processing tends to leave traces that can be exploited to expose the occurrence of manipulations \cite{Bar13},\cite{BarniTondiSourceIdentificationGame}. Very often, the creation of a forgery involves the application of more than one single processing tool, thus leaving a number of traces that can be used by the forensic analyst; this consideration suggests to analyze the "authenticity" of digital medias by using more than one tool. Furthermore, these tools are far from ideal and often give uncertain or even wrong answers. Especially because after each improvement in forensic tools,  the adversaries impose an opposite effort to devise more powerful techniques that leave minor evidence into the forged content in the attempt to impair the forensic detection tools.
Therefore, whenever possible, it may be wise to employ more than one tool searching for the same trace. By taking into account the presence of the adversaries, fusing all the local decisions to make a final decision about document's authenticity can improve forgery detection \cite{fontanidecisionfusion}. 

Online reputation systems\index{Online reputation systems|textbf} are an additional example.  
A reputation system gathers evidence from agents about objects like products, good, services, business, users or digital contents in order to come out with reputation scores. Most online commercial systems collect user feedbacks as evidence to compute scores for the objects of interest. These scores have a major influence on new online agent's decision.  Thus, they provide enough incentive for attackers to manipulate them by providing false/forged feedbacks. By doing so, the attackers try to increase or decrease the reputation of an object and hence, manipulate the decisions of possibile new agents \cite{onlinereputationsystemssec}.

\section{Goals and Contributions}
Various types of adversaries exist for distributed sensor networks and they can be classified depending on many factors: their objectives, their behavior, the amount of information they have about the system under control as well as the network, and so on \cite{padmavathi2009survey}, \cite{wood2002denial}, \cite{wang2006surveyWSN}, \cite{chen2009sensorsurvey}. Attacks in which adversaries have full control of a number of nodes and behave arbitrarily to disrupt the network are referred to as \textit{Byzantines}\index{Byzantine Attacks|textbf}. The term Byzantine Attack is originated from the Byzantines general problem stated by Lamport et al. in \cite{ByzantineGeneralProblem} as follows: \textit{"a group of generals of the Byzantine army camped with their troops around an enemy city. Communicating only by messenger, the generals must agree upon a common battle plan. However, one or more of them may be traitors who will try to confuse the others. The problem is to find an algorithm to ensure that the loyal generals will reach agreement"}. The relation between this problem and distributed sensor networks is straightforward as Byzantine generals play the role of internal adversaries. A Byzantine adversary may have various behaviors, such as lying about network connectivity, flooding network with false traffic, forging control information, modifying the information sent by nearby sensors (e.g., peer to peer networks), or dominating the control of a strategic set of sensors in the network and colluding \cite{Vemp13}. As an example, in cooperative spectrum sensing in cognitive radio networks, Byzantine attacks are known as Spectrum Sensing Data Falsification Attack"\index{Spectrum Sensing Data Falsification Attack|textbf} (SSFD) \cite{tolerant_scheme}, \cite{SecureCSSinCRN2008},\cite{yu2009defense}. In addition, various Byzantine attacks in wireless sensor networks can severely degrade the WSN performance and corrupt its functionality \cite{ByzantinesinWSN2011}, etc.. \cite{Vemp13}.

\subsection{Goal of the Thesis}

With the above ideas in mind, in this thesis we consider the problem of Byzantine attacks in distributed sensor network in a binary decision setup \cite{kayIIbookdetectiontheory}, \cite{scharf1991statistical}. In the literature binary decision\index{binary decision|textbf} is also referred to as binary detection\index{binary detection|textbf} (or Binary Hypothesis Test\index{Binary Hypothesis Test|textbf}), since, in many applications, the decision problem refers to the detection of the presence or absence of a certain phenomenon or signal.
For instance, in source identification in multimedia forensics, the analyst aims to distinguish which between two sources (e.g. a photo camera and a scanner) generated a specific digital content or to identify the specific device used to acquire the content\cite{BarniTondiSourceIdentificationGame}. In Cognitive Radio Networks, the FC or the network wants to distinguish between two cases: the presence or absence of PU signal in the spectrum of interest \cite{Mitola2000CR}, \cite{HaykinBrainEmpowered}. In addition, in WSN, the network wants to detect the presence of a certain physical phenomenon \cite{DDinWSNVarhs}. As a last example, machine learning binary detection and classification for various applications (e.g. spam filtering) to differentiate to which class a specific data belongs \cite{MachineLeaningEX1}, \cite{MachineLeaningEX2}.

The goal of this thesis is to study the problem of adversarial information fusion in distributed sensor networks, analyzing the effects of the Byzantines on the binary detection problem and developing possible defense strategies\index{defense strategies|textbf} to mitigate the effect of the attacks on the information fusion process. Later, by assuming that Byzantines are not na\"{i}ve and because \textit{"to every action, there is always a reaction"}\cite{newtonprincipiabook}, we formalize the interplay betwen the Byzantines (the Attackers) and the Defender (the network designer) in a Game-Theoretic fashion following the Adv-SP setup. We adopt such a formalization  to study the optimal behavior of the attacker and defender. We analyze the existence of equilibrium points\index{equilibrium point|textbf} for the game and we then evaluate the performance achieved by the players at the equilibrium so to understand who is going to "win" the game.  

\subsection{Contribution of the Thesis}
We start by considering an adversarial decision fusion\index{adversarial decision fusion|textbf} in which the nodes send to the FC a vector of binary decisions about the state of a system over an observation window. Considering this setup, as a first contribution, we develop a novel soft identification and isolation\index{soft identification and isolation|textbf} scheme to exclude the reports sent by the Byzantines from the decision fusion process. By \textit{isolation}\index{isolation|textbf} we mean the process whereby the FC removes Byzantines's  decisions from the fusion process after identifying them among the nodes in the distributed network. By adopting this soft scheme, the FC can assign a reliability value to each node. Moreover, as an additional contribution, we formalize the competition between the Byzantines and the FC in a game-theoretic sense and we study the existence of an equilibrium point for the game. Then, we derive the payoff in terms of the decision error probability when the players "play" at the equilibrium. 

As a second contribution, we derive the optimum decision fusion rule\index{optimum fusion rule|textbf} in the presence of Byzantines in a centralized setup. By observing the system over an observation window, we adopt the Maximum A Posteriori Probability\index{Maximum A Posteriori Probability|textbf} (MAP) rule while assuming that the FC knows the attack strategy of the Byzantines and their distribution across the network. With regard to the knowledge that the FC has about the distribution of Byzantines over the network, we consider several cases. First, we examine an unconstrained maximum entropy scenario in which the uncertainty about the distribution of Byzantines is maximum, which means that the a-prior infromation available at the FC about the Byzantines's distribution is minimum. Then, we consider a more favorable scenario to the FC in which the maximum entropy case is subject to a constraint. In this scenario, the FC has more a-priori information about Byzantines's distribution i.e the average or the maximum number of Byzantines in the network. Finally, we consider the most favorable situation in which the FC knows the exact number of Byzantines present in the network. Concerning the complexity of the optimal fusion rule, we develop an efficient implementation based on dynamic programming. Thereafter, we introduce a game-theoretic framework to cope with the lack of knowledge regarding the Byzantines strategy. In such a framework, the FC makes a "guess" by slecting arbitrarily a Byzantine's attacking strategy within the optimum fusion rule. By considering the decision error probability as the payoff, we study the performance of the Byzantines as well as the FC at the game equilibrium for several setups when the players adopt their best possibile strategies. 

By revisiting the complexity of the optimum fusion rule, as an additional contribution, we propose a novel message passing approach based on factor graph. We consider a more general model for the observed system in which we examine both independent and Markovian sequences. Then, we show that the message passing algorithm can give a near-optimal performance while reducing the complexity from exponential to linear as a function of the observation window size. 

In the last part of the thesis, we consider a decentralized version of the data fusion process. In this setup, the nodes detect the state of the system by iteratively exchanging their observations with each other in order to reach an agreement about the status of the system. The decentralized fusion algorithm is known as \textit{consensus algorithm} \cite{degroot1974consensus}, \cite{olfati2007consensuscooperation}. In this scenario, we focus on a case in which the adveraries attack the links between the system being monitored and the sensors. To make the network more robust, we propose a primary isolation step to be carried at the node level to filter out the falsified information injected by the attacker.
In turn, as a reaction, the adversary may adjust the strength of the falsification attack to avoid that the forged measurements are discarded. So, we employ game-theory to model the competition between the adversary and the network. Then, we use numerical simulations to derive and study the equilibrium points of the game and the performance at the equilibrium.

\section{Thesis Overview}

We start by briefly introducing some basic notions of detection theory in Chapter \ref{chapter:DF} specifying some information fusion techniques. Chapter \ref{chapter:GoT} presents a short introduction to Game Theory. Then, in Chapter \ref{chapter:SecurityThreats} we review some of the most common security threats and defenses in distributed detection systems.

In Chapter \ref{chapter:CDC} we develop a novel soft identification and isolation scheme for Byzantines and we model the competition between them and the FC through game-theory.

In Chapter \ref{chapter:TIFS_SPL} we derive the optimum fusion rule in the presence of Byzantines and we consider a game-theoretic framework to model the interplay between them and the FC. Then, we study the optimum behavior of the players at the equilibrium of the game considering several Byzantines distributions in the network.

In Chapter \ref{chapter:InfoFusion}, we explain a near-optimal message passing approach based on factor graph is explained that reduces the complexity of the optimum Fusion rule.

Chapter \ref{chapter:GameSec} deals with the decentralized binary detection in distributed networks using the consensus algorithm under falsification attack. We make the algorithm more robust by proposing a primary isolation step to filter out the measurements coming from the links under the control of the adversaries. Then, by formalizing the game between the network and the adversary in a game-theoretic setup and derive the best performance for both players.

The thesis ends Chapter \ref{chapter:Conclusion} where we give some guide lines for future research.

\section{Publication List}
The research activity of this thesis resulted in the following publications:

\textit{Chapter 5} \\
A. Abrardo, M. Barni, \textbf{K. Kallas}, and B. Tondi, "Decision fusion with corrupted reports in multi-sensor networks: a game-theoretic approach," in Proceedings of CDC'15, \emph{IEEE Conference on Decision and Control}, Los Angeles, California, December 2014.

\textit{Chapter 6} \\
Andrea Abrardo, Mauro Barni, \textbf{Kassem Kallas}, and Benedetta Tondi, "A Game-Theoretic Framework for Optimum Decision Fusion In the Presence of Byzantines", in \emph{IEEE Transactions on Information Forensics and Security}, vol. 11, no. 6, pp. 1333-1345, June 2016.

\textit{Chapter 7} \\
Andrea Abrardo, Mauro Barni, \textbf{Kassem Kallas}, and B. Tondi, "A Message Passing Approach for Decision Fusion in Adversarial Multi-Sensor Networks", Submitted to Elsevier Journal on Information Fusion.

\textit{Chapter 8} \\
\textbf{Kassem Kallas}, Benedetta Tondi, Riccardo Lazzeretti and Mauro Barni, "Consensus Algorithm with Censored Data for Distributed Detection with Corrupted Measurements: A Game-Theoretic Approach," \emph{Proceedings of the 7th Conference on Decision and Game Theory for Security}, New York, NY, November, 2016. 


\chapter{Distributed Detection and Information Fusion in Sensor Networks}
\label{chapter:DF}

\begin{flushright}
\emph{"We are to admit no more causes of natural things than such as are both true and sufficient to explain their appearances." }
\\
Isaac Newton, "Philosophi$\ae$  Naturalis Principia Mathematica" \\

\emph{"Hypotheses should be subservient only in explaining the properties of things but not assumed in determining them, unless so far as they may furnish experiments." }
\\
Isaac Newton
\end{flushright}

\bigskip
\section{Introduction}
\PARstart{\textcolor{red}D}istributed{} sensor networks consist of a set of spatially distributed sensors that operate as data collectors\index{data collectors|textbf} or decision makers\index{decision makers|textbf} to monitor a shared phenomenon. This is a common case in many real world situations like air-traffic control, economic and finance, medical diagnosis, electric power networks, wireless sensor networks, cognitive radio networks, online reputation systems, and many others.  Usually, in centralized networks, if there are no power, channel, communication or privacy constraints, the sensors can send the full raw information they collect to a FC. However, real life situations are different and several constraints must be considered e.g. when sensors are spatially distributed over a large territorial area, when the channel bandwidth is limited, or even when the sensors are supplied with short life power sources. To address these limitations, sensors must perform some local processing before sending a compressed version of the collected information to the FC. The abstraction level of the information summary can vary a lot. For instance, it can be a quantized set of the raw information, a soft summary statistic like an average or a likelihood value, or even a single information bit.

By means of a fusion rule, the FC integrates the information received from the sensors to make a global decision regarding the system or phenomena under probation. The definition of the fusion rule depends on the a-prior information available about the sensors, the transmission channel and the phenomena as well as the information type provided by the sensors. Fusion rules can be as simple as voting rules or advanced sophisticated statistical rules. 

In the rest of this chapter, first, we briefly introduce some basic notions of detection theory and outline some detection techniques used locally at the sensors as well as the corresponding decision strategy. Then, we list some common information fusion techniques that can be employed by the FC in the centralized setup. Regarding the decentralized case, we introduce very shortly the emerging field of signal processing over graphs and we explain decentralized fusion by means of consensus algorithm as one of its application. Finally, we give an overview of cognitive radio networks and cooperative spectrum sensing as an example of distributed detection and information fusion.

\section{Detection Theory}
\label{Sec:Detection_Theory}

Detection theory is a methodology to model the decision making process in order to decide among various classes of items and how to react to that decision. Making decisions and detecting events is not restricted to human being and other living creatures but it also includes intelligent devices and machines. Detection theory is fundamental in the design of electronic and signal processing systems \cite{kayIIbookdetectiontheory} as it has been applied widely in information systems i.e. in radar systems, digital communications, sensor networks, image processing, bio-medicine, control systems, seismology, cognitive radio networks and many others. The main common objective of detection theory is being able to decide about the existence and the status of a phenomenon or event. For example, a radar system must decide about the presence or absence of an aircraft or any other target, a sensor network has to detect the presence or absence of a natural incident, medical test must detect if a certain disease is present or not, and image processing may aim at detecting the presence of a specific object or feature in an image.

\begin{figure}[h]
\centering
\includegraphics[scale=0.3]{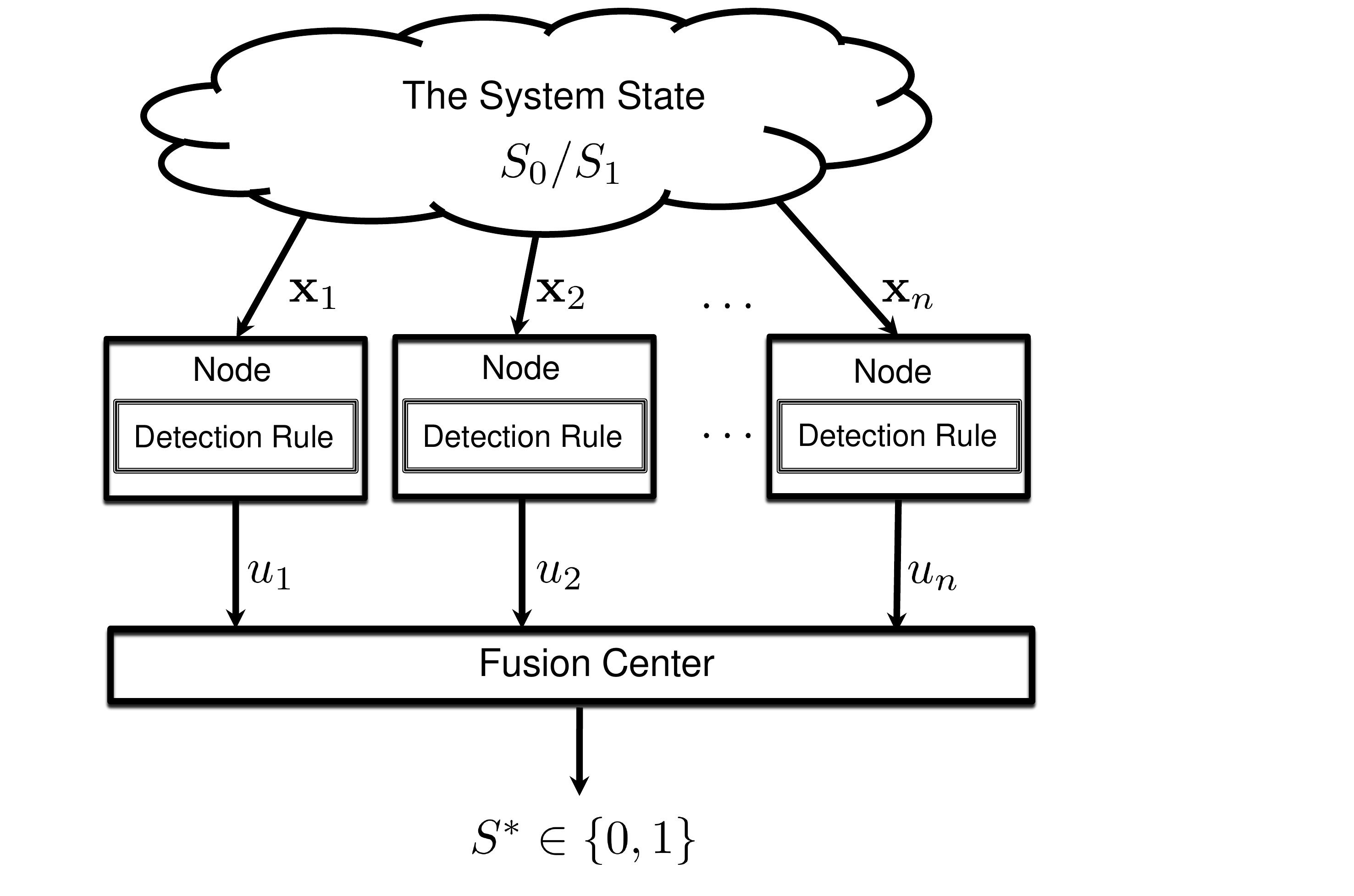}
\caption{\textit{Parallel Topology}}
\label{fig:Parallel_Topology}
\end{figure}

The most common case is when the sensors are faced with a binary detection problem i.e. they must decide about the presence or absence of a phenomenon\index{phenomenon|textbf}. In the literature, a widely used synonym for the same problem is binary hypothesis testing. Usually, the two hypotheses are denoted by $H_0$ and $H_1$ where, $H_0$ is called the null hypothesis and represents the absence of the phenomenon of interest\index{system state|textbf}, whereas $H_1$ is called the alternative hypothesis and represents the presence of the phenomenon. A more general situation is when the sensors have to decide between a set of $M$ hypotheses. However, this case is out of the scope of this thesis. By focusing on the binary detection problem\index{binary detection|textbf}, in the setup considered in this chapter, the system state\index{system state|textbf}\footnote{from now on, we interchange the use of the phenomenon, the event and the system state\index{system state|textbf} to refer to the system under probation.} is observed by a network of $n$ sensors through vectors $\mathbf{x}_1, \mathbf{x}_2,\dots,\mathbf{x}_n$. The sensors can decide about the system state\index{system state|textbf} by producing the information $u_1, u_2,\dots,u_n$ that, depending on the information abstraction\index{information abstraction|textbf}, can be an information value or a decision bit. The system state\index{system state|textbf} $S_i, i \in \{0,1\}$ can be in $S_0$ under hypothesis $H_0$ and in $S_1$ under hypothesis $H_1$. $P(H_0)$ and $P(H_1)$ are the a-priori probabilities that the system is under hypothesis $H_0$ and $H_1$, respectively. The sensors are not assumed to communicate with each other and compute their local information independently and send it the FC, which in turn, has to come out with a global decision $S^* \in \{0,1\}$ regarding the state $S_i$. The above setup is illustrated in Figure \reffig{fig:Parallel_Topology}.

In the following subsection, we consider the case of a single sensor observing the system through a variable $x$. The sensor employs a certain detection technique\index{detection technique|textbf} in order to make a decision about the state of the system. We will present the most common techniques to perform binary detection and decision locally at the sensor.
\subsection{Bayesian Detection}
In Bayesian detection\index{Bayesian detection|textbf}, two fundamental pieces of information must be available to the sensors: the a-prior probabilities about the system states\index{system state|textbf} $P(H_0)$ and $P(H_1)$, and the observation probability densities conditioned to the hypotheses, namely, $p(x|H_0)$ and $p(x|H_1)$. After the decision in favor of $H_i, i \in \{0,1\}$, four situations are possible; among them two are correct decisions and the others are erroneous. These cases are shown in Table \reftab{tab.binaryDD}.
\begin{table}[]
\centering
\caption{\textit{Decision cases in binary detection}\label{tab.binaryDD}}
\begin{tabular}{l|l|l|}
\cline{2-3}
                                              & \multicolumn{2}{l|}{System State $H_j$} \\ \hline
\multicolumn{1}{|l|}{\multirow{2}{*}{Decision $H_i$}} & $H_0|H_0$       & $H_0|H_1$        \\ \cline{2-3} 
\multicolumn{1}{|l|}{}                        &  $H_1|H_0$       & $H_1|H_1$           \\ \hline
\end{tabular}
\end{table}

Each decision is taken by the sensor at a cost $C_{ij}$ referring to the case of deciding $H_i$ while $H_j$ is true. Clearly, the erroneous decision costs $C_{01}$ and $C_{10}$ cost the sensor more than the correct decisions ($C_{00}$ and $C_{11}$). Following this formulation, a sensor prefers to employ a decision rule which minimizes the average cost or risk function $\mathcal{C}$ given by
\begin{equation}
\mathcal{C} = \sum\limits_{i=0}^1 \sum\limits_{j=0}^1 C_{ij} P(H_j) P(H_i|H_j)
\label{eq.BayesRisk}
\end{equation}
In \cite{Var97} it is shown that the decision rule that minimizes $\mathcal{R}$ is given by the Likelihood Ratio Test (LRT) as
\begin{equation}
\Lambda(x) = \frac{P(x|H_1)}{P(x|H_0)} \underset{H_0}{\overset{H_1}{\gtrless}} \frac{P(H_0)(C_{10}-C_{00})}{P(H_1)(C_{01}-C_{11})},
\label{eq.LRTDD1}
\end{equation}
where, the left hand side of the equation $\Lambda(x)$ is the likelihood ratio\index{likelihood ratio|textbf}, while the right hand side is the decision threshold $\lambda$ given by 
\begin{equation}
\lambda = \frac{P(H_0)(C_{10}-C_{00})}{P(H_1)(C_{01}-C_{11})}
\end{equation}
So, the LRT test can be written as
\begin{equation}
\Lambda(x) \underset{H_0}{\overset{H_1}{\gtrless}} \lambda
\label{eq.LRTDD2}
\end{equation}
Consequently, the sensor decides in favor of $H_1$ when the $\Lambda(x)$ is greater than $\lambda$ and in favor of $H_0$ otherwise.

The Log-likelihood ratio test\index{likelihood ratio|textbf} (LLRT) is obtained by applying the logarithm to both sides of Equation (\refeq{eq.LRTDD2}):
\begin{equation}
\log \Lambda(x) \underset{H_0}{\overset{H_1}{\gtrless}} \log \lambda
\label{eq.LRTDD3}
\end{equation}

\subsection{Detection Performance Metrics}
For a sensor, the performance of the adopted detection rule is evaluated based on correct and wrong decision probabilities. The wrong decision probabilities are the probability of false alarm\index{probability of false alarm|textbf}, $P_{FA}$, and the probability of missed detection\index{probability of missed detection|textbf}, $P_{MD}$, and are given by

\begin{equation}
\begin{split}
& P_{FA} =  P(H_1|H_0) \\
& P_{MD} =  P(H_0|H_1)
\end{split}
\label{eq.PFAPMD}
\end{equation}

These terminologies originate from radar theory to indicate the cases of missing an existing target and raising an alarm when the target is absent. A false alarm refers to a case in which the sensor mistakenly decides for $H_1$ while the true system state\index{system state|textbf} is $H_0$, whereas a missed detection occurs when the sensor decides for $H_0$ and the true system state\index{system state|textbf} is $H_1$. In statistics, the probabilities $P_{FA}$ and $P_{MD}$ are known as Type I and Type II error probabilities as well as false positive and false negative, respectively. 
Consequently, the correct detection probability $P_D$ and the null probability $P_{null}$ (usually called as true negative) are given by
\begin{equation}
\begin{split}
& P_{D} =  P(H_1|H_1) = 1-P_{MD} \\
& P_{null} = P(H_0|H_0) = 1-P_{FA}
\end{split}
\end{equation}
and hence, the overall decision error probability is given as 
\begin{equation}
P_e = P(H_0) P_{FA} + P(H_1) P_{MD}
\end{equation}
In order to evaluate the performance of a detector, a common graphical representation known as the Receiver Operating Characteristics (ROC)\index{Receiver Operating Characteristics|textbf}is used \cite{ROC1982meaning}. The use of this curve is not restricted to radar and detection theory but also extended to medicine, radiology, bio-metrics, machine learning and data mining research. Usually, the ROC curve shows the performance of the detector by plotting the $P_D$ versus $P_{FA}$ by varying the decision threshold. Other forms of the curve are constructed by using other detection probabilities.
\begin{figure}[h]
\centering
\includegraphics[scale=0.35]{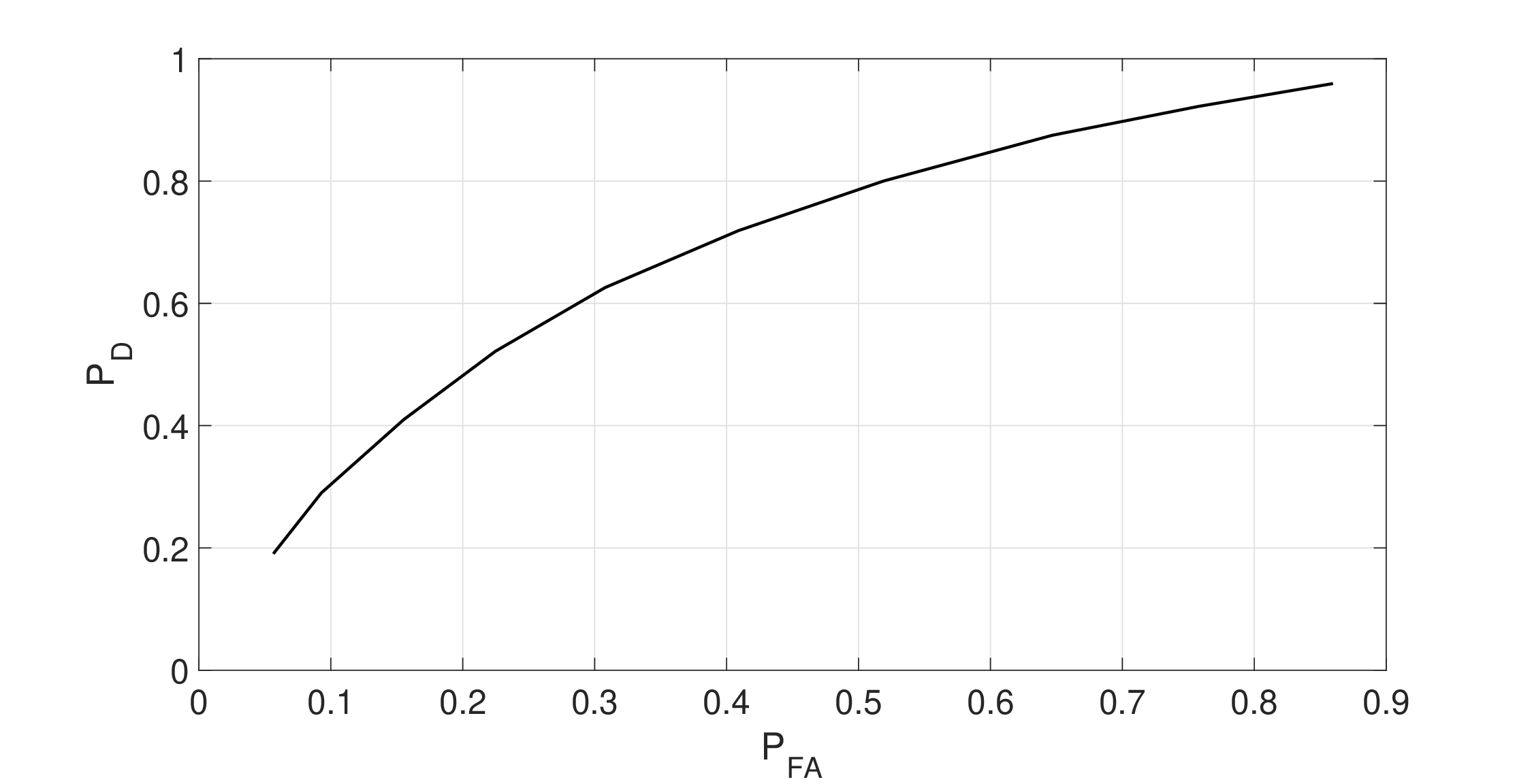}
\caption{\textit{ROC curve example}}
\label{fig:ROC_EX}
\end{figure}
An example of a ROC curve is depicted in Figure 
\reffig{fig:ROC_EX}. From this Figure, it can be seen that the worst case performance of a detector is on the straight line where $P_{FA} = P_D$ since, in this case, the detector is completely "blind" and decides by just flipping a coin. Below this line, the detector can flip its decision to return back to correct decision region. The optimal operation point for any detector is the point that maximizes $P_D$ and minimizes $P_{FA}$. From the curve, it is the nearest point on the top left corner of the graph. An ideal detector is a detector with an operating point exactly at that corner with $P_D = 1$ and $P_{FA} = 0$.

\subsection{Neyman-Pearson Detection}

\begin{figure}[h]
\centering
\includegraphics[scale=0.5]{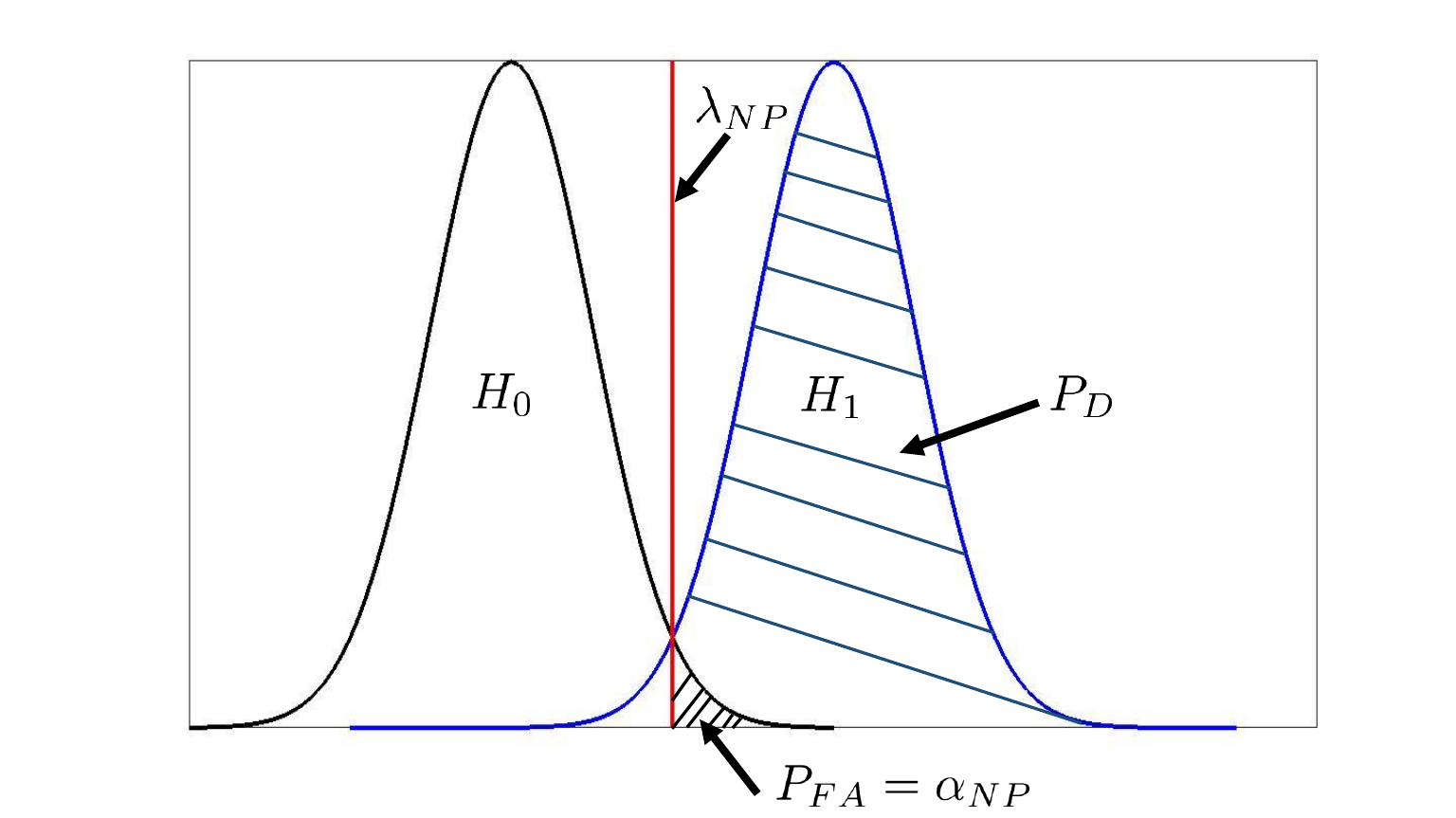}
\caption{\textit{Neyman-Pearson Setup}}
\label{fig:NP_Test}
\end{figure}

In practice, the a-priori probabilities required to implement a Bayesian detector are difficult to be known. In addition, assigning the costs $C_{ij}$ is difficult or even impossible. Thus, selecting an "optimal" threshold for the LRT test cannot be guaranteed. Neyman-Pearson\index{Neyman-Pearson|textbf} (NP) detection is a design criteria that overcomes these limitations. By constraining one type of decision error (usually the $P_{FA}$), the NP detector minimizes the other type. By doing so, the NP detector does not need the a-priori probabilities but instead, it needs to specify a maximum tolerable error for one error type among the two. This detection technique can achieve two opposite objectives: it minimizes the false alarm probability $P_{FA}$ at one hand, and maximizes the probability of detection $P_D$ on the other hand. Formally speaking, the NP detector constraints $P_{FA}$ to an acceptable value $\alpha_{NP}$ and maximizes the detection probability $P_D$ as illustrated in Figure \reffig{fig:NP_Test}. 
Hence, the LRT for the NP setup becomes\begin{equation}
\log \Lambda(x) \underset{H_1}{\overset{H_0}{\gtrless}} \log \lambda_{NP} .
\label{eq.NP_LRT}
\end{equation}
The decision threshold $\lambda_{NP}$ is computed by letting $P_{FA}=\alpha_{NP}$. A common method to compute it is using Lagrange Multipliers\index{Lagrange Multipliers|textbf}. Consider a function $\mathcal{F} = P_{MD} + \lambda_{NP}\{ P_{FA} -\alpha_{NP}\}$ with $\lambda_{NP} \geq 0$, then, the Lagrange multiplier $\lambda_{NP}$ is selected to satisfy the following condition
\begin{equation}
P_{FA} = \int_{\lambda_{NP}}^{\infty} p(\Lambda / H_0) d\Lambda = \alpha_{NP}
\label{eq.NP_Alfa}
\end{equation}
The threshold $\lambda_{NP}$ obtained by solving this integral, is optimal for the LRT test.
\subsection{Sequential Detection}
In some situations, the sensor decision is based on a vector of observations instead of a single observation and the number of observations needed to make a decision is not fixed. In this situation, the information is gathered sequentially by the sensor over an observation window. To minimize the delay, the sensor make its decision as soon as the collected information is sufficient to make an acceptable "accurate" decision. With sequential detection, a new information is collected only when the available observations are not sufficient to make a decision. For this reason, the sequential detector uses two thresholds for the LRT test, so to collect a new observation only when the LRT value falls between the two thresholds.

\begin{figure}[h]
\centering
\includegraphics[scale=0.35]{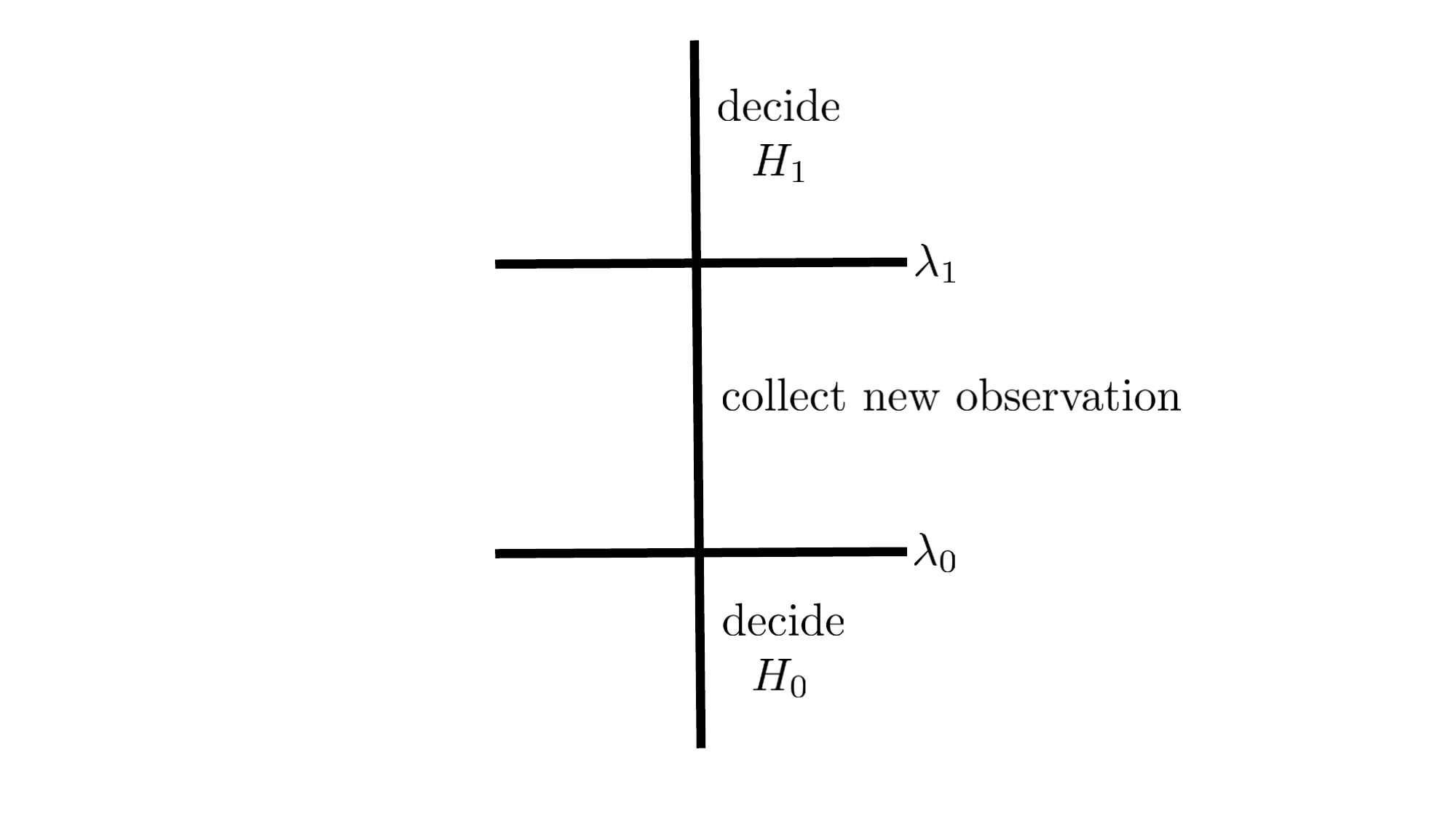}
\caption{\textit{SPRT detector}}
\label{fig:SPRT_detector}
\end{figure}

The Neyman-Pearson approach to sequential detection has been developed by Abraham Wald and known as the Wald's Sequential Probability Ratio Test (SPRT) \cite{WaldSPRT1},\cite{WaldSPRT2}. In each step of SPRT, the sensor compares the LRT value to two thresholds $\lambda_0$ and $\lambda_1$ determined based on pre-defined values for $P_{FA}$ and $P_{MD}$. If the value falls between $\lambda_0$ and $\lambda_1$, the sensor takes a new observation. On the other hand, the sensor decides for $H_1$ if the LRT result is greater than $\lambda_1$ whereas, it decides for $H_0$ when the value is smaller than $\lambda_0$. The decision scenario of the SPRT detector is depicted in Figure \reffig{fig:SPRT_detector}.

The vector $\mathbf{x_K} = [x_1, x_2, \dots , x_K]$ represents the observations gathered by a sensor at the $K^{th}$ time instant. Then, the LRT of the SPRT is constructed as follows
\begin{equation}
\Lambda(x_K) = \frac{p(\mathbf{x_K}| H_1)}{p(\mathbf{x_K} | H_0)} = \prod\limits_{k=1}^K \frac{p(x_k | H_1)}{p(x_k | H_0)}.
\label{eq.LRT_SPRT}
\end{equation}

The thresholds $\lambda_0$ and $\lambda_1$ are computed by constraining $P_{FA}$ and $P_{MD}$ to $\alpha_{ST}$ and $\beta_{ST}$, respectively, and are computed as follows
\begin{equation}
\begin{split}
& \lambda_0 =  \frac{\beta_{ST}}{1-\alpha_{ST}} \\
& \lambda_1 =  \frac{1- \beta_{ST}}{\alpha_{ST}}.
\end{split}
\end{equation}
The main drawback of this detection method is the time required to make a decision. In addition, the detector can get stuck in a never-ending loop between the thresholds $\lambda_0$ and $\lambda_1$ due to the low quality of the observations. On the contrary, the advantage of the SPRT is that it takes on average fewer observations than a fixed size observation test \cite{Var97}.

\section{Information Fusion Rules}
In this section, we consider the parallel topology\index{parallel topology|textbf} depicted in Figure \reffig{fig:Parallel_Topology}. This is, the most common topology to model distributed sensor networks. In the following, we give an overview of the commonly used fusion rules\index{fusion rules|textbf} that can be implemented in both centralized and decentralized network setups. The choice of a fusion rule depends on many factors; for instance, the processing capability of the sensors, the available information about the system and the network, the channel bandwidth and quality, the energy consumption, the presence of attacks and adversaries and many others.
\subsection{Centralized Fusion}
In centralized networks\index{centralized networks|textbf}, after local processing\index{local processing|textbf} each sensor $i$ sends its information $u_i$ to the FC. Each sensor is assumed to have local detection and false alarm probabilities $P_{d_i}$, $P_{{fa}_i}$. The performance of the fusion rule deployed at the FC is measured using global detection and false alarm probabilities denoted as $Q_D$ and $Q_{FA}$, respectively. Two classes of fusion rules are considered: simple and advanced rules. In the former, the processing burden is low and the amount of a-priori information required at the FC is small; while in the latter, more processing is required and more information must be known in advance since most of these rules are statistically based.

\subsubsection{Simple Fusion Rules}
The term "simple" refers to the fact that the operations performed at the FC  are computationally cheap. These rules can be used in the absence of a-priori information about the system and the network. We start by considering the case of binary reports, a case which is suitable for bandwidth limited applications. By receiving a pool of binary bits, the FC can apply a "hard" or "voting" fusion rule, namely, AND, OR and $k$-out-of-$n$ rule\cite{asymptoticdetectionsystem}. The decision bit is computed locally at the sensor by performing an LRT detection as in Section \ref{Sec:Detection_Theory}, then, the node sends a bit $1$ when the LRT decides for $H_1$ and $0$ otherwise. Using the AND rule\index{AND rule|textbf}, the FC decides for $S^* = 1$ only when all the sensors decide in favor of $H_1$,

\begin{equation}
\begin{split}
& S^* = 1: \textrm{if} \quad \sum\limits_{i=1}^n u_i = n   \\
& S^* = 0:  \textrm{otherwise}
\end{split}
\label{eq:AND_rule}
\end{equation}
while, by applying the OR rule\index{OR rule|textbf} \cite{ghasemicollaborativeSS}, it decides for $S^* = 1$ if any of the nodes decide for $H_1$. The global decision of the OR rule is given by
\begin{equation}
\begin{split}
& S^* = 1: \textrm{if} \quad \sum\limits_{i=1}^n u_i \geq 1   \\
& S^* = 0: \text{otherwise}
\end{split}
\label{eq:OR_rule}
\end{equation}
The most general voting rule is the $k$-out-of-$n$ rule\index{$k$-out-of-$n$ rule|textbf} wherein the FC decides $S^* = 1$ when at least $k$ nodes out of $n$ decide for $H_1$. A special case is the majority rule\index{majority rule|textbf} where $k = \frac{n}{2}$. The $k$-out-of-$n$ rule is formalized by
\begin{equation}
\begin{split}
& S^* = 1: \textrm{if} \quad \sum\limits_{i=1}^n u_i \geq k \\
& S^* = 0: \text{otherwise}
\end{split}
\label{eq:KN_rule}
\end{equation}
The performance of the fusion rule are evaluated by $Q_D = P( S^* = 1 | H_1)$  and $Q_{FA} = P(S^* = 1 | H_0)$  as the global probabilities of detection and false alarm. By assuming that the each sensor makes its decision independently of the others, the expressions for the performance are given below. $Q_{D_{AND}}$ and $Q_{FA_{AND}}$ obtained by applying the AND rule are given by
\begin{equation}
\begin{split}
& Q_{D_{AND}} = \prod\limits_{i=1}^n  P_{d_i}, \\
& Q_{FA_{AND}} = \prod\limits_{i=1}^n P_{{fa}_i},
\end{split}
\label{eq:AND_rule_performances}
\end{equation}
while, for the OR rule, $Q_{D_{OR}}$ and $Q_{FA_{OR}}$ are given by
\begin{equation}
\begin{split}
& Q_{D_{OR}} = 1- \prod\limits_{i=1}^n (1 - P_{d_i}), \\
& Q_{FA_{OR}} = 1 - \prod\limits_{i=1}^n (1 - P_{{fa}_i}),
\end{split}
\label{eq:OR_rule_performances}
\end{equation}
and finally, for the $k$-out-of-$n$ rule
\begin{equation}
\begin{split}
& Q_{D_{kn}} = \sum\limits_{i=k}^n   \binom ni P_{d_i}^i  (1-P_{d_i})^{(n-i)} \\
& Q_{FA_{kn}} = \sum\limits_{i=k}^n  \binom ni P_{{fa}_i}^i  (1-P_{{fa}_i})^{(n-i)}.
\end{split}
\label{eq:KN_rule_performances}
\end{equation}
A comparative performance evaluation of the three voting rules under different settings is conducted in \cite{EnergyDetectioninCRN}.

When there are no limitations on the bandwidth, the overall performance of the fusion technique can be improved by sending more detailed information to the FC \cite{DataFusionCSSCRN}. This information can be a statistics or the LRT value about the system (known as "soft decision"\index{soft decision|textbf}). We present three simple and common information fusion rules: Square Law Combining\index{Square Law Combining|textbf} (SLC) \cite{SoftSLCcombination}, Maximal Ratio Combining\index{Maximal Ratio Combining|textbf} (MRC) and Selection Combining\index{Selection Combining|textbf} (SC)\cite{CSSPhDSoftCombining}. MRC is an optimal combination scheme when the Channel Side Information (CSI) is known at the FC \cite{simondigitalcomm}. The CSI is the Signal to Noise ratio (SNR) between the sensor $i$ and the system and it is denoted by $\gamma_i$. This information is not required by SC and SLC. By applying the SLC, the FC sums all the received data as follows
\begin{equation}
U_{SLC} = \sum\limits_{i=1}^n u_i ,
\label{eq:SLC}
\end{equation}
MRC is a modified version of SLC wherein a weight $w_i$ proportional to SNR is assigned to each information provided by the sensors as follows
\begin{equation}
U_{MRC} = \sum\limits_{i=1}^n w_i u_i.
\label{eq:MRC}
\end{equation}
On the other hand, the SC selects the information of the sensor experiencing the maximum SNR
\begin{equation}
U_{SC} =  \max_{\gamma_i} (u_1, u_2,\dots,u_n).
\label{eq:SC}
\end{equation}
Using soft combination rules\index{soft combination rules|textbf}, the decision for the value of $S^*$ is made by comparing the combined information to a threshold $\zeta$. In \cite{DDMultipleSensorIAdvanced}, \cite{DiversitySoftbetterHard},\cite{SoftCombinationCRN},\cite{CollaborativeTVdetectionDSA}, it is shown that at the cost of higher overhead and channel quality requirement, the soft fusion rules provide better performance than hard fusion rules. 

\subsubsection{Advanced Fusion Rules}
Many forms of advanced information fusion schemes exist. They depend on many factors ranging from the a-priori available information to the application scenarios wherein these schemes are applied. Examples of advanced information fusion techniques include evidential belief reasoning, fusion and fuzzy reasoning, rough set fusion for imperfect data, random set theoretic fusion and others \cite{fusion2013stateofart}. Here we consider statistical information fusion due to its perfect match with distributed sensor network applications. In addition, statistical information fusion has a very rich background and in the literature it is the most common approach applied to distributed sensor networks \cite{Var97}, \cite{detectionsensortutorial}. By adopting such an approach, the LRT has to be performed at the FC after receiving the information from the sensors. Given the vector $\mathbf{u} = {u_1, u_2, \dots, u_n}$ with the information sent to the FC, the Bayesian information fusion that minimizes the average cost of the global decision is given by
\begin{equation}
\Lambda(\mathbf{u}) = \frac{P(u_1,u_2,\dots,u_n|H_1)}{P(u_1,u_2,\dots,u_n|H_0)} 
\underset{S^*=0}{\overset{S^*=1}{\gtrless}}
\frac{P_0(C_{10}-C_{00})}{P_1(C_{01}-C_{11})} =\lambda
\label{eq.BayesianFC1}
\end{equation}
where, $C_{ij}$ is the cost of the global decision of the FC in favor of $H_i$ when the system is in state $H_j$. Hereafter, given the conditional independence assumption of local \emph{decisions} provided by the sensors, the Bayesian fusion becomes
\begin{equation}
\begin{aligned}
& \Lambda(\mathbf{u}) = \frac{P(u_1,u_2,\dots,u_n|H_1)}{P(u_1,u_2,\dots,u_n|H_0)} \\
&  = \prod_{i=1}^n \frac{P(u_i|H_1)}{P(u_i|H_0)} \\
& = \prod_{S_1} \frac{P(u_i=1|H_1)}{P(u_i=1|H_0)}\prod_{S_0} \frac{P(u_i=0|H_1)}{P(u_i=0|H_0)} \\
& = \prod_{S_1} \frac{1-P_{{md}_i}}{P_{{fa}_i}}\prod_{S_0} \frac{P_{{md}_i}}{1-P_{{fa}_i}}.
\end{aligned}
\label{eq.BayesianFC2}
\end{equation}
where, $S_j$ is the set of the sensors for which $u_i=j$. By applying the logarithm to the last part of Equation (\refeq{eq.BayesianFC2}), we obtain
\begin{equation}
\sum_{S_1} log \Big(\frac{1-P_{{md}_i}}{P_{{fa}_i}} \Big) + \sum_{S_0} log\Big(\frac{P_{{md}_i}}{1-P_{{fa}_i}}\Big) 
\underset{S^*=0}{\overset{S^*=1}{\gtrless}} log(\lambda)
\label{eq.ChairVarshney1}
\end{equation}
which also can be expressed as
\begin{equation}
\sum_{i=1}^n \big[ u_i log\Big(\frac{1-P_{{md}_i}}{P_{{fa}_i}}\Big) + (1-u_i) log\Big(\frac{P_{{md}_i}}{1-P_{{fa}_i}}\Big)\big] 
\underset{S^*=0}{\overset{S^*=1}{\gtrless}} log(\lambda)
\label{eq.ChairVarshney2}
\end{equation}
This is an optimal fusion rule\index{optimal fusion rule|textbf} and it is known as the Chair-Varshney rule \cite{OptFusion}. It can be seen as performing a weighted sum over the local decisions provided by the sensors. The Chair-Varshney rule\index{Chair-Varshney rule|textbf} requires the knowledge of the local performances of the sensors , the a-prior probabilities about the system state, and the costs.

These limitations can be overcome by using the Neyman-Pearson rule \cite{OptimalNPfusion} maximizing the detection probability at the FC ($Q_D$) while constraining the false alarm probability $Q_{FA}$.
\begin{equation}
\Lambda(\mathbf{u}) = \frac{P(u_1,u_2,\dots,u_n|H_1)}{P(u_1,u_2,\dots,u_n|H_0)} \underset{S^*=0}{\overset{S^*=1}{\gtrless}} \lambda.
\label{eq.NPFC1}
\end{equation}
The threshold $\lambda$ is set in such a way to satisfy the constraint on the global false alarm probability $Q_{FA}$, in the following
\begin{equation}
\sum_{\Lambda(\mathbf{u}) > \lambda} P(\Lambda(\mathbf{u})|H_0) = Q_{FA}.
\label{eq.NPFCth}
\end{equation}
If the FC must take a decision as soon as the information sent by the sensors is enough, the sequential probability ratio test\index{sequential probability ratio test|textbf} (SPRT) can be applied globally. In this case, if $M$ information samples are already enough, the FC can make the global decision. If not, the FC can collect more information from the sensors. The SPRT can be formalized as follows
\begin{equation}
\prod_{i=1}^M \frac{P(u_i|H_1)}{P(u_i|H_0)} = \Lambda(\mathbf{u}),
\label{eq.SPRT_FC}
\end{equation}
then, the decision is made according to the rule:
\begin{equation}
  \left \{
  \begin{aligned}
     & \Lambda(\mathbf{u}) \geq \Upsilon_1, &&  S^* = 1\\
     & \Upsilon_0 < \Lambda(\mathbf{u}) < \Upsilon_1 && \text{take new value} \\
     & \Lambda(\mathbf{u}) \leq \Upsilon_0, && S^* = 0
  \end{aligned} \right .
\label{eq.SPRT_FC1}
\end{equation} 
The thresholds $\Upsilon_0$ and $\Upsilon_1$ are computed by constraining both $Q_{FA} = \alpha_{FC}$ and $Q_{MD} = \beta_{FC}$ as in the following equation:
\begin{equation}
\begin{split}
& \Upsilon_1 =  \frac{1- \beta_{FC}}{\alpha_{FC}} \\
& \Upsilon_0 =  \frac{\beta_{FC}}{1-\alpha_{FC}}.
\end{split}
\end{equation}

\subsection{Decentralized Fusion}
In decentralized fusion\index{decentralized fusion|textbf}, the network by itself in the absence of the FC is responsible to decide about the system state\index{system state|textbf}. To do so, the information fusion techniques are implemented locally at the sensors. The global decision about the system state is reached by iterating a peer-to-peer information exchange among the sensors. Decentralized information fusion is desirable in many situations, for instance, when the sensors do not want to exchange information with a remote party due to privacy reasons or power constraints. Even more, in large networks, the FC can become a bottleneck or a single point of failure. Finally, the nature of the future communication networks is fully distributed and self-adaptive thus, it is preferable to avoid to rely on a central unit to control the data exchange and decision in the network. Recently, consensus algorithms\index{consensus algorithm|textbf} \cite{olfati2007consensuscooperation} have been used in distributed sensor networks and several other fields. In a consensus algorithm, each sensor communicates with its neighbors and updates its local information about the system by applying a local fusion rule that is a weighted combination of its own information and the information received from its neighbors. This process is repeated until the whole network reaches a steady state value which is used by all the sensors to decide about the state of the system. In the sequel, we give a short introduction to the emerging field of signal processing over graphs and then, as an application of the field, we introduce the consensus algorithm.

\subsubsection{Signal Processing over Graphs}
\begin{figure}[h]
\centering
\includegraphics[scale=0.3]{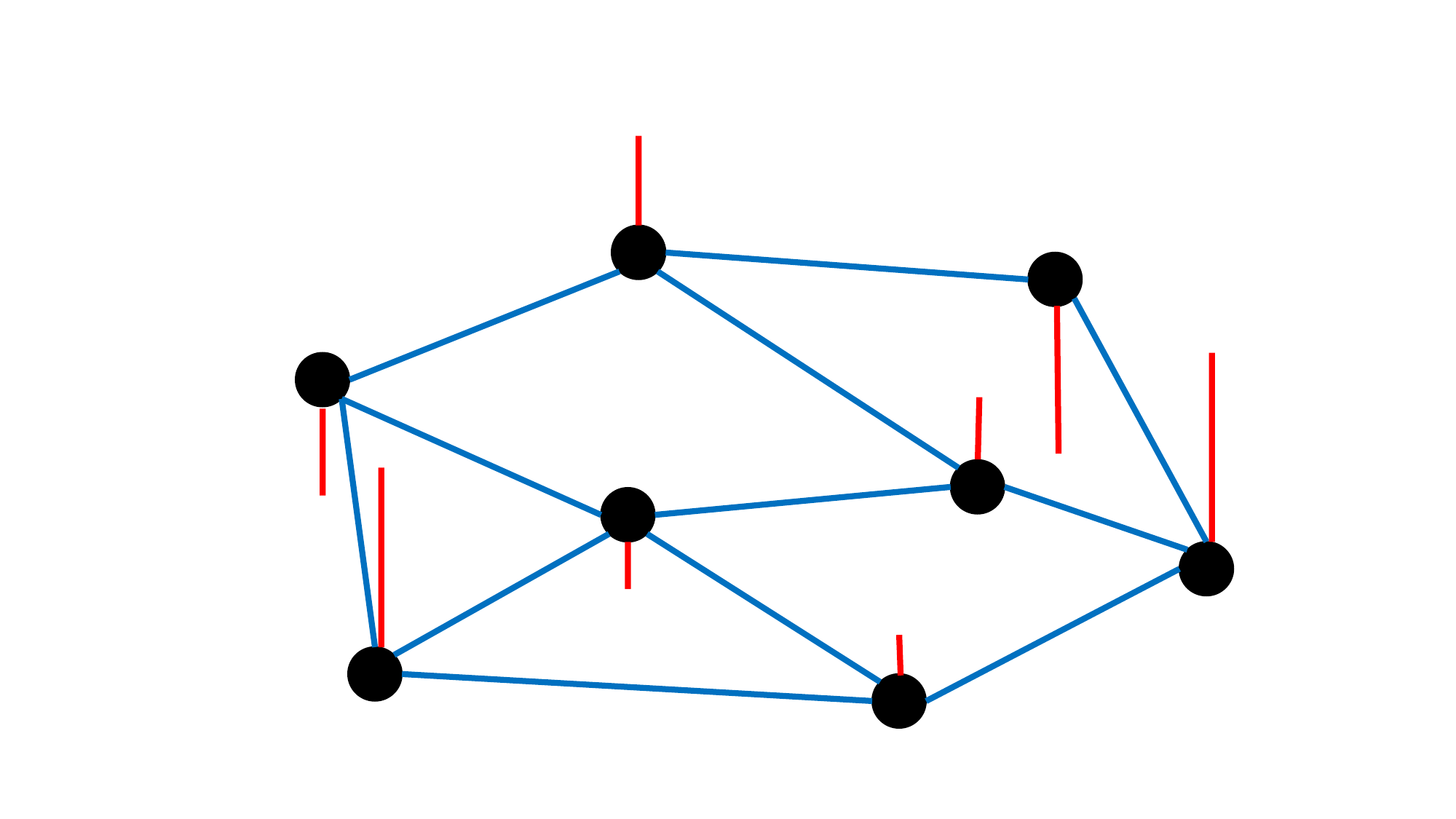}
\caption{\textit{Signals over a Graph. The length of the red bar represents the values of the signal and its direction indicates its value sign.}}
\label{fig:Signals_over_Graph}
\end{figure}

Signal processing over graphs\index{Signal processing over graphs|textbf} is an emerging field that combines algebraic and spectral graph theoretic concepts to process signals on graphs \cite{SPoverGraphs}. A graph is a generic data representation to describe the geometric structure of the data. Data forms can be found in many applications, including social, energy, transportation, sensor, information, and neuron networks. Each entity in these networks is represented by a vertex on the graph and each couple of related vertices are connected by an edge. The edge is usually associated to a weight that reflects the similarity between the vertices it connects, for instance, it can be concluded from the physical phenomenon between the vertices or can be inferred from the data itself. As an example, the edge weight may be inversely proportional to the physical distance between nodes in a network. Each vertex on the graph carries a data sample and signals on graphs can be seen as the set of the samples. A representative example of signals over graph is illustrated in Figure \reffig{fig:Signals_over_Graph}.

Many real life examples can be represented by signals over graphs. For example, spread of disease, human migration patterns, trading goods in transportation systems, sensing results in WSN or CRN and many others. For instance, the anatomical connectivity of functional regions of the cerebral cortex used in brain imaging applications \cite{BrainImagingex}. Additional example offered image processing where a graph-based filtering methods are applied for better edges and textures recognition \cite{imagegraphex1},\cite{imagegraphex2}. Cooperative spectrum sensing in cognitive radio networks offers another example. In this application, the vertices are the SUs performing the spectrum sensing, the edges are the communication links between them, and the signals are the measurements collected about the PU activity in the spectrum. The connection between signal processing over graphs and distributed sensor networks is straightforward, as sensors are represented by the vertices and the link connecting them as the edges. The signals carried by the sensors are the information gathered about the system of interest.

\subsubsection{Consensus Algorithm}
\textit{The Decentralized Distributed Sensor Netowrk Model} \\

\begin{figure}[h]
\centering
\includegraphics[scale=0.5]{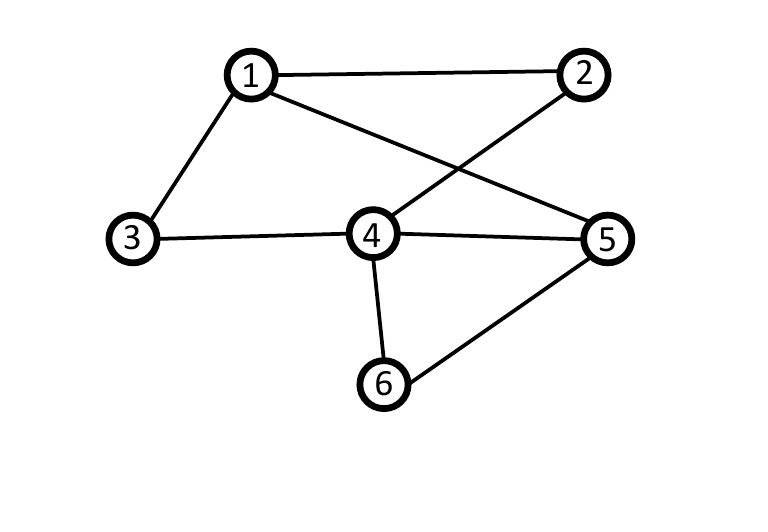}
\caption{\textit{A consensus network example.}}
\label{fig:Consensus_Net_Ex}
\end{figure}

A distributed sensor network can be modeled as an non-directed graph $\mathcal{G}$ where the information can be exchanged in both directions between sensors. A graph\index{graph|textbf} $\mathcal{G} = (\mathcal{N}, \mathcal{E})$ consists of the set of vertices $\mathcal{N} = \{n_1, ...,n_n\}$ and the set of edges $\mathcal{E}$  where $(n_i,n_j) \in \mathcal{E}$ if and only if there is a common communication link between $n_i$ and $n_j$,  i.e., they are neighbors.
The neighborhood of a node $n_i$ is indicated as $\mathcal{N}_i = \{n_j \in \mathcal{N} : (n_i,n_j) \in \mathcal{E}$ \}.  A graph $\mathcal{G}$ can be represented by its adjacency matrix $A = \{a_{ij}\}$ where $a_{ij} = 1$, if $(n_i,n_j) \in \mathcal{E}$, $0$ otherwise. An example of consensus network\index{consensus algorithm|textbf} is depicted in Figure \reffig{fig:Consensus_Net_Ex} and its corresponding matrix is given in Equation (\refeq{eq.AdjMtrxEX}).

\begin{equation}
A = 
 \begin{pmatrix}
0 & 1 & 1 & 0 & 1 & 0 \\
1 & 0 & 0 & 1 & 0 & 0 \\
1 & 0 & 0 & 1 & 0 & 0 \\
0 & 1 & 1 & 0 & 1 & 1 \\
1 & 0 & 0 & 1 & 0 & 1 \\
0 & 0 & 0 & 1 & 1 & 0 
 \end{pmatrix}  
\label{eq.AdjMtrxEX}
\end{equation}

The degree matrix $D$ of $\mathcal{G}$ is a diagonal matrix with $d_{ii} = a_{i1} + a_{i2}  + ... + a_{in}$, $d_{ij} = 0$, $\forall i$, $j \neq i$  \cite{godsilgraphtheory}.

\textit{The Consensus Algorithm} \\
Consensus algorithm for distributed detection is a protocol where the sensors locally exchange information with their neighbors in order to converge to an agreement about the state of a system \cite{olfati2007consensuscooperation}. It consists of three phases: the initial phase, the state update phase and the decision phase.
\begin{enumerate}
\item Initial phase: the sensors collect their initial information $x_i(0)$ about the system state, and exchange the information with their neighbors.
\item State update phase: at each time step $k$, each sensor updates its state based on the information received from its neighbors. Then, at step $k+1$ we have:
\begin{equation}
x_i(k+1) = x_i(k) + \frac{\epsilon}{w_i} \sum_{j \in \mathcal{N}_i} (x_j(k)-x_i(k))
\end{equation}
where, $0 < \epsilon < ( \max\limits_i \mathcal{N}_i)^{-1}$ is the update step parameter and $w_i$ is the weight assigned to neighbor's information value. A special case is the equal weight combining rule for which $w_i=1,  \forall i$. This phase is iterated until the sensors reach the consensus value $\bar{x} = \frac{1}{n} \sum_{i \in \mathcal{N}} x_i(0)$, which corresponds to the mean of the initial measurements. This is when the state update at the sensors is the same at consecutive iterations.
\item The final decision phase: this is the last phase in which all sensors compare the consensus value $\bar{x}$ to a threshold $\lambda$ to make the final decision $S^*$:
\begin{equation}
  S^* =\begin{cases}
    1, & \text{ if }  \bar{x} > \lambda,\\
    0, & \text{otherwise}.
  \end{cases}
\end{equation}
\end{enumerate}

\section{Cognitive Radio Networks: application of Distributed Detection and Information Fusion}

\begin{figure}[h]
\centering
\includegraphics[scale=0.35]{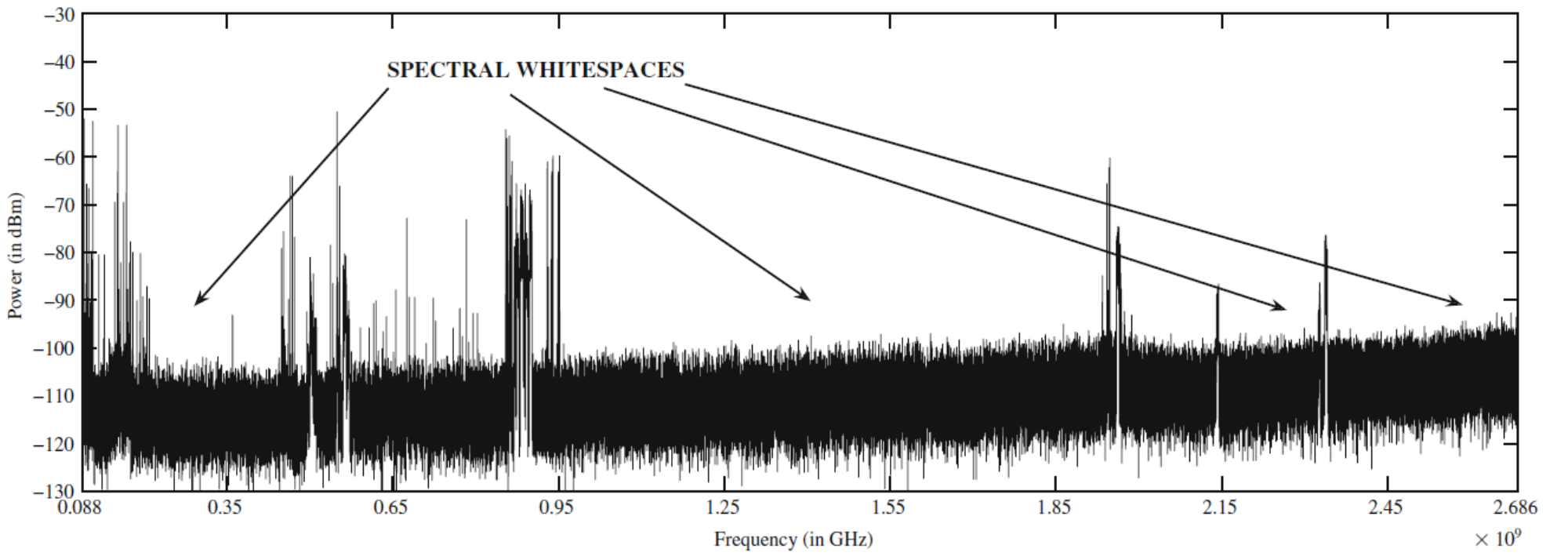}
\caption{\textit{A snapshot of Power Spectral Density from $88 \quad \textrm{to} \quad 2686 \quad \textrm{MHz}$
measured in Massachusetts. The picture is taken from \cite{bookcognitiveradio2009}.}}
\label{fig:PSD_WS}
\end{figure}

In \cite{HaykinBrainEmpowered}, a spectrum hole\index{spectrum hole|textbf} or a spectrum whitespace is defined as \textit{"A spectrum hole is a band of frequencies assigned to a primary user, but, at a particular time and specific geographic location, the band is not being utilized by that user"}. An example of spectrum holes in depicted in Figure \reffig{fig:PSD_WS}. The concept of cognitive radio is introduced in Chapter \ref{chapter:intro} as an SDR-based device that, using DSA, allows to efficiently exploit the spectrum holes to improve the spectrum utilization\index{spectrum utilization|textbf}. Therefore, cognitive radio\index{cognitive radio|textbf} is an intelligent wireless system that \emph{learns} from the surrounding environment, and adapts, on the fly, its parameters to the variations of the incoming Radio Frequency (RF) stimuli by dynamically changing its operating parameters i.e transmission power, carrier frequency, modulation, etc.

\begin{figure}[h]
\centering
\includegraphics[scale=0.3]{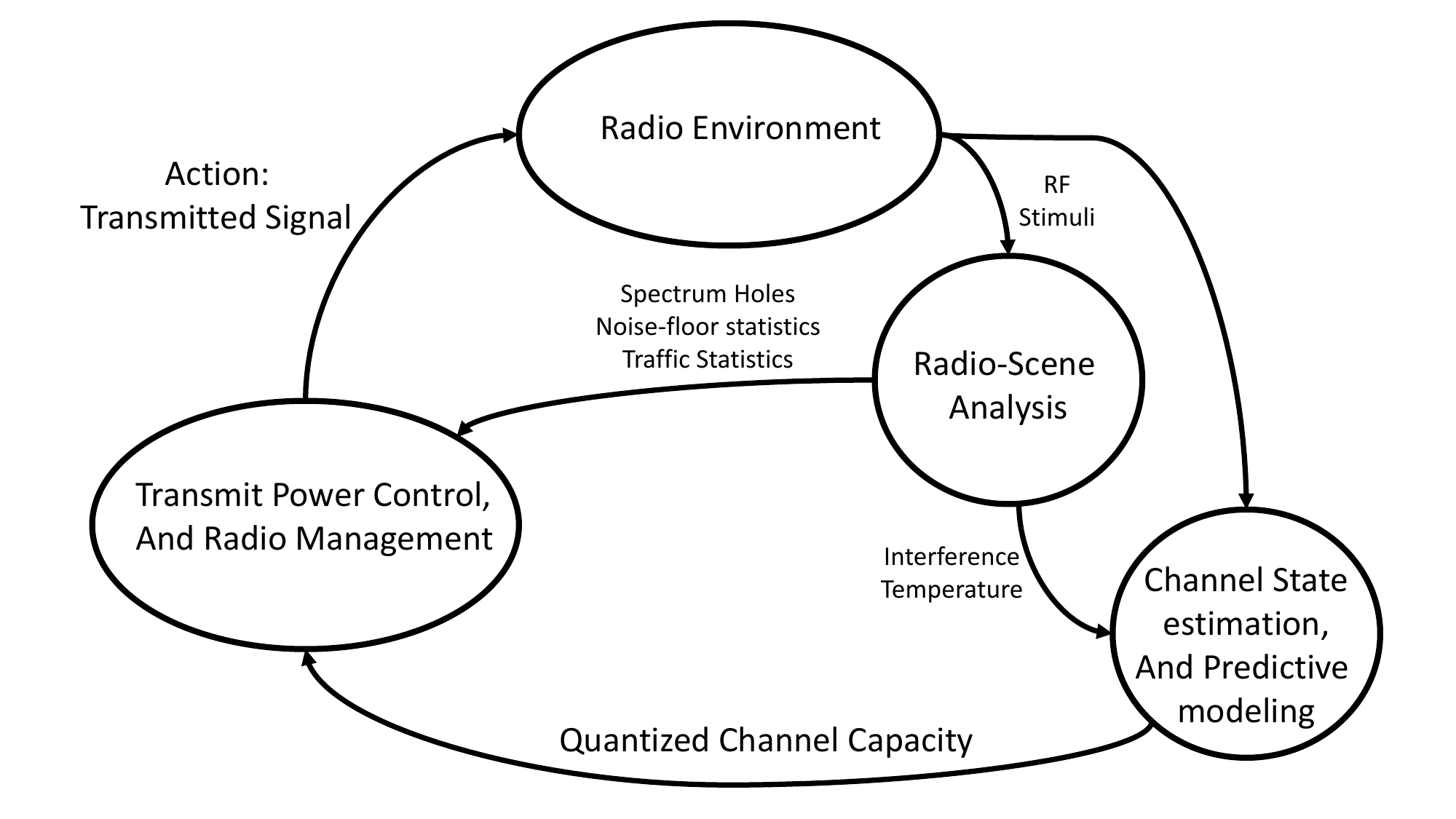}
\caption{\textit{Cognitive Cycle. The picture source is: \cite{HaykinBrainEmpowered}.}}
\label{fig:CognitiveCycle}
\end{figure}
Cognitive capability is obtained by interacting with the environment using what is called a cognitive cycle\index{cognitive cycle|textbf} illustrated in Figure \reffig{fig:CognitiveCycle}. Cognitive cycle contains three main actions: 
\begin{itemize}
\item The radio-scene analysis is used to estimate the interference temperature of the radio environment and detect the spectrum holes.
\item The channel identification is used to estimate the channel-state information and predict the channel capacity for transmission.
\item The transmit-power control and dynamic spectrum management.
\end{itemize}
A CR may construct its cognitive cycle around a specific narrow spectrum hole, or around a wideband hole or even around a set of narrow band spectrum holes in order to provide the best performance in spectrum management, efficiency, and transmission power control.

The most fundamental key enabling task and first step for learning the environment and performing other cognitive actions is the radio-scene analysis, namely, the ability to \emph{sense} the spectrum. \textit{Spectrum sensing\index{spectrum sensing|textbf} aims at of obtaining awareness about the spectrum usage and existence of primary users in a geographical area. This awareness can be obtained by using geolocation and database, by using beacons, or by local spectrum sensing at cognitive radios}\cite{HaykinBrainEmpowered}, \cite{spectrumsensingsurvey} . Spectrum sensing can be carried out by locally measuring the spectral content, or the radio frequency energy over the spectrum; or by obtaining several spectrum features together such as time, space, frequency, code, signal type, modulation, waveform, bandwidth, carrier frequency, etc. However, using several features together requires more powerful signal processing techniques and higher computational complexity. In what follows, we will analyze the energy detector, as this is the most common technique to sense the spectrum. Then, we show the benefit from cooperation between CRs to improve the accuracy of the spectrum sensing result and hence, improve spectrum utilization.

\subsection{Energy Detector}
Energy Detector\index{Energy Detector|textbf} (ED) is the most common approach to spectrum sensing for its low computational and implementation complexities\cite{EnergyDetector1}. In addition, it does not require any a-priori information about the PU signal. It is a threshold based signal detection method that, in order to decide about the spectrum occupancy, compares the output of the detector to a threshold \cite{EnergyDetector2}. This is an example of the binary detection problem as PU activity represents the system state, SU is the sensor performing the detection, and $H_0$ represents the absence of PU signal and $H_1$ its presence. Some challenges for the implementation of ED includes the threshold selection, the inability to distinguish between the PU signal and channel noise, its bad performance under low (SNR) \cite{EnergyDetectorLowSNR}, and its inefficiency in detecting spread spectrum signals \cite{EnergyDetectorwithSpreadSpectrum1},\cite{EnergyDetectorwithSpreadSpectrum2}. The signal received by the ED is written in the following form,
\begin{equation}
y(m) = s(m) + w(m),
\end{equation}
where, $s(m)$ is the PU signal, $w(m)$ is the Additive White Gaussian Noise (AWGN), and $m$ is the sample index. The information gathered by the ED as decision metric is given by the square sum of all the samples as, namely:
\begin{equation}
\Lambda = \sum\limits_{m=1}^M |y(m)|^2.
\label{eq:ED_metric}
\end{equation}
The detection and false alarm probabilities $P_D$, $P_{FA}$ of the ED are given as follows
\begin{equation}
P_D = P(\Lambda > \lambda_E|H_1),
\label{eq:ED_PD}
\end{equation}
\begin{equation}
P_{FA} = P(\Lambda > \lambda_E|H_0)
\label{eq:ED_PD}
\end{equation}
The decision threshold $\lambda_E$ can be selected by using Bayesian or Neyman-Pearson techniques as discussed in Section \ref{Sec:Detection_Theory}.

The noise can be modeled as a Gaussian random variable with zero-mean and variance $\sigma_w^2$, i.e. $w(m) = N (0, \sigma_w^2
)$. In the simplified case, the same model can be adopted for the PU signal as well, i.e. $s(m) = N (0, \sigma_s^2)$. Under these assumptions, the decision metric in Equation (\refeq{eq:ED_metric}) which is the square sum of $M$ Guassian samples, follows a chi-square distribution with $2M$ degrees of freedom $\mathcal{X}_{2M}^2$ and it is formalized as:
\begin{equation}
  \Lambda =\begin{cases}
   \frac{\sigma_w^2}{2} \mathcal{X}_{2M}^2, & \text{ under }  H_0\\
     \frac{\sigma_w^2 + \sigma_s^2}{2} \mathcal{X}_{2M}^2, & \text{ under }  H_1.
  \end{cases}
\end{equation}

The probabilities $P_{FA}$ and $P_D$ are computed as, follows \cite{EnergyDetectorPDPFAAWGN}
\begin{equation}
P_{FA} =\frac{\Gamma(\frac{M}{2},\frac{\lambda_E}{2})}{\Gamma(\frac{M}{2})},
\end{equation}
\begin{equation}
P_{D} = Q_{\frac{M}{2}}(\sqrt{2\gamma_S},\sqrt{\lambda_E}),
\end{equation}
where $\Gamma(.)$ is the incomplete gamma function \cite{BookIntegralsGammaFunction}, $Q_{\frac{M}{2}}$ is the generalized Marcum \emph{Q-function} \cite{IntegralsQfunc}, and $\gamma_S = \frac{E_S}{N_0}$ is the SNR experienced by the detector. 

More advanced detection techniques for cognitive radio offer better performance than ED, but require more a-priori information about the PU signal and has higher complexity and implementation cost. Example of these detectors are cyclo-stationary detector, matched filter detector, radio identification detector, wave-based form detector and others. A complete survey about spectrum sensing techniques can be found in \cite{spectrumsensingsurvey}, \cite{ALB11}.

\subsection{Cooperative Spectrum Sensing}
Cooperation among SUs performing spectrum sensing has been proposed to solve problems like fading, shadowing, hidden PUs and the delay in sensing time \cite{agilityCRhiddenterminal},\cite{ganesan2005cooperative}, \cite{cabric2006spectrum}. Cooperative spectrum sensing\index{Cooperative spectrum sensing|textbf} decreases the probabilities of missed detection and false alarm considerably and hence, increases the spectrum utilization and decreases the interference to PU. In addition, as it is shown analytically and numerically in \cite{lehtomaki2005analysis}, cooperative sensing provides higher capacity gains. Cooperation between SUs can be implemented in both centralized and decentralized fashion \cite{nextgenerationDSA}.                           
\subsubsection{Centralized Cooperative Spectrum Sensing}
In centralized spectrum sensing, a FC collects information from cognitive radios and then computes the global decision to identify the availability of the spectrum. Then, it broadcasts the global decision back to the cognitive network. To do so, simple and advanced detection techniques like those explained in Section \ref{Sec:Detection_Theory} can be implemented by the cognitive radios. In turn, the FC can employ a fusion rule to come out with the final decision about the PU activity in the spectrum. For instance, in \cite{CollaborativeTVdetectionDSA}, sensing results considered in both soft and hard versions are fused at the FC referred to as master node in order to detect TV channels. In \cite{lunden2007spectrum}, SUs send a quantized version of the local information to the FC which, in turn, applies a likelihood ratio test over the received likelihood ratio information to compute the final decision. The case of centralized fusion with one bit local decision is considered in \cite{HardDecisionCRNCSS}. 

\subsubsection{Decentralized Cooperative Spectrum Sensing}
In the decentralized spectrum sensing, which is often referred to as distributed sensing, cognitive radios share their information among each other but they make their own decisions as to which part of the spectrum they can use. Decentralized
sensing is more advantageous over the centralized version, in that there is no need for a backbone infrastructure and then, it has lower network implementation cost. Many decentralized fusion techniques has been considered in a cognitive radio setup for information exchange and coordination among the nodes. Examples are: consensus algorithm \cite{consensusCSSCRN1}, \cite{consensusCSSCRN2}, diffusion adaptation algorithm \cite{diffusionCSSCRN}, belief propagation algorithm \cite{beliefpropagationCSSCRN} and many others.

\section{Conclusion}
We introduced some basic notions of detection theory and outlined some detection techniques used locally at the sensors as well as the corresponding decision strategy. Specifically, we discussed the Bayesian detector, the Neyman-Pearson detector and the SPRT detector used locally at the sensors and their extension to be deployed globally at the FC. Then, we listed some common information fusion techniques that can be employed by the FC in the centralized setup. Specifically, we discussed the AND, OR and $k$-out-of-$n$ rules when sensors send hard decisions, and the SLC, MRC and SC combination rules when sensors send soft decisions. Regarding the decentralized case, we explained decentralized fusion by means of consensus algorithm as an application for the emerging field of signal processing over graphs. Finally, we gave an overview of cognitive radio networks and cooperative spectrum sensing, as an example of distributed detection and information fusion, and we specifically discussed the spectrum sensing by means of energy detector.


\chapter{Basics of Game Theory}
\label{chapter:GoT}
\emph{"Thus the expert in battle moves the enemy, and is not moved by him." }
\\
Sun Tzu, "The Art of War"
\section{Introduction}

\PARstart{\textcolor{red}G}ame{} theory is a mathematical discipline that studies the situations of competition and/or cooperation, between decision makers\index{decision makers|textbf} known as players\index{players|textbf}. Game theoretic concepts apply whenever the actions of several decision-makers are mutually dependent, that is their choices mutually affect each other. For this reason, game theory is sometimes referred to as \emph{interactive decision theory}.

Although examples of games occurred long before, the birth of modern Game Theory as a unique field was in 1944, with the book "Theory of Games and Economic Behavior" by John von Neumann and Oskar Morgenstern \cite{von2007gametheory}. Game Theory\index{Game Theory|textbf} provides tools to formulate, model and study strategic scenarios in a wide variety of application fields ranging from economics and political science to computer science. A fundamental assumption in almost all variants of Game Theory is that each decision maker is rational and intelligent. A rational player\index{rational player|textbf} is one who has certain specific preferences over the outcomes of the game. A player intelligence is its ability to always select the action that gives him the most preferable outcome, given his expectation about his opponents action. The objective of game-theory is to predict how rational players will play the game, or, to give advice on the strategies to be followed when playing the game against rational opponents.

Game-theoretic models are highly abstract representations of classes of real life situations for which satisfying solutions for players are recommended with desirable properties. Game Theory encompasses a great variety of situations depending on the number of players\index{players|textbf}, the way the players interact, the knowledge that a player has on the strategies adopted by the opponents, the deterministic or probabilistic nature of the game, etc. In all the models, the basic entity is the player, which should be interpreted as an individual or as a group of individuals, making a decision or following a strategy. A distinction can be made between situations in which the players have common goals and hence play a cooperative game\index{cooperative game|textbf} and situations in which the players have different and possibly conflicting goals. In the latter case we say that the game is non-cooperative\index{non-cooperative game|textbf}. Hybrid games\index{Hybrid game|textbf} contain cooperative and non-cooperative players. For instance, coalitions of players are formed in a cooperative game, but they play in a non-cooperative manner. Another possible classification between game-theoretical models concerns the amount of information available to the players about each other, leading to Games with Perfect or Imperfect Information. A more common classification is made between simultaneous and sequential games. Simultaneous games\index{Simultaneous game|textbf} are games where both players move simultaneously, or if they do not move simultaneously, they are unaware of the earlier players' actions so that their action are effectively simultaneous. On the contrary, sequential or dynamic games are games where players have some knowledge about earlier actions. This knowledge does not need to be complete i.e., a player may know that an earlier player did not perform one particular action, while he does not know which of the other available actions the first player actually chose.

Game representations are used to differentiate between simultaneous and sequential games\index{sequential game|textbf}. To distinguish between the two type of games, here, we introduce briefly games in normal and extensive forms. A normal form game\index{normal form game|textbf} is represented by a matrix where, for the 2-players case, one player is considered as the {\em row} player, and the other as the {\em column} player. Each row or column represents a strategy (which is the action selected by the player) and each entry in the matrix represents the payoff\index{payoff|textbf}, that is the final outcome of the game for each player for every combination of strategies. An example of a 2-player game in normal form is shown in Table \reftab{Table.ExampleNormalForm}.

\begin{table}[H]
\centering
\begin{tabular}{lll}
& Strategy 1                    & Strategy 2 \\ \cline{2-3} 
\multicolumn{1}{l|}{Strategy 1}        & \multicolumn{1}{l|}{$\quad(a,b)$} & \multicolumn{1}{l|}{$\quad(c,d)$} \\ \cline{2-3} 
\multicolumn{1}{l|}{Strategy 2}           & \multicolumn{1}{l|}{$\quad(e,f)$} & \multicolumn{1}{l|}{$\quad(j,h)$} \\ \cline{2-3} 
\end{tabular}
\caption{\textit{Example of game representation in normal form. The row player is player 1 and the column player is player 2. The entries of the table are the payoffs of the game for each pair of strategies.}\label{Table.ExampleNormalForm}}
\end{table}

\begin{figure}[h]
\centering
\includegraphics[scale=0.35]{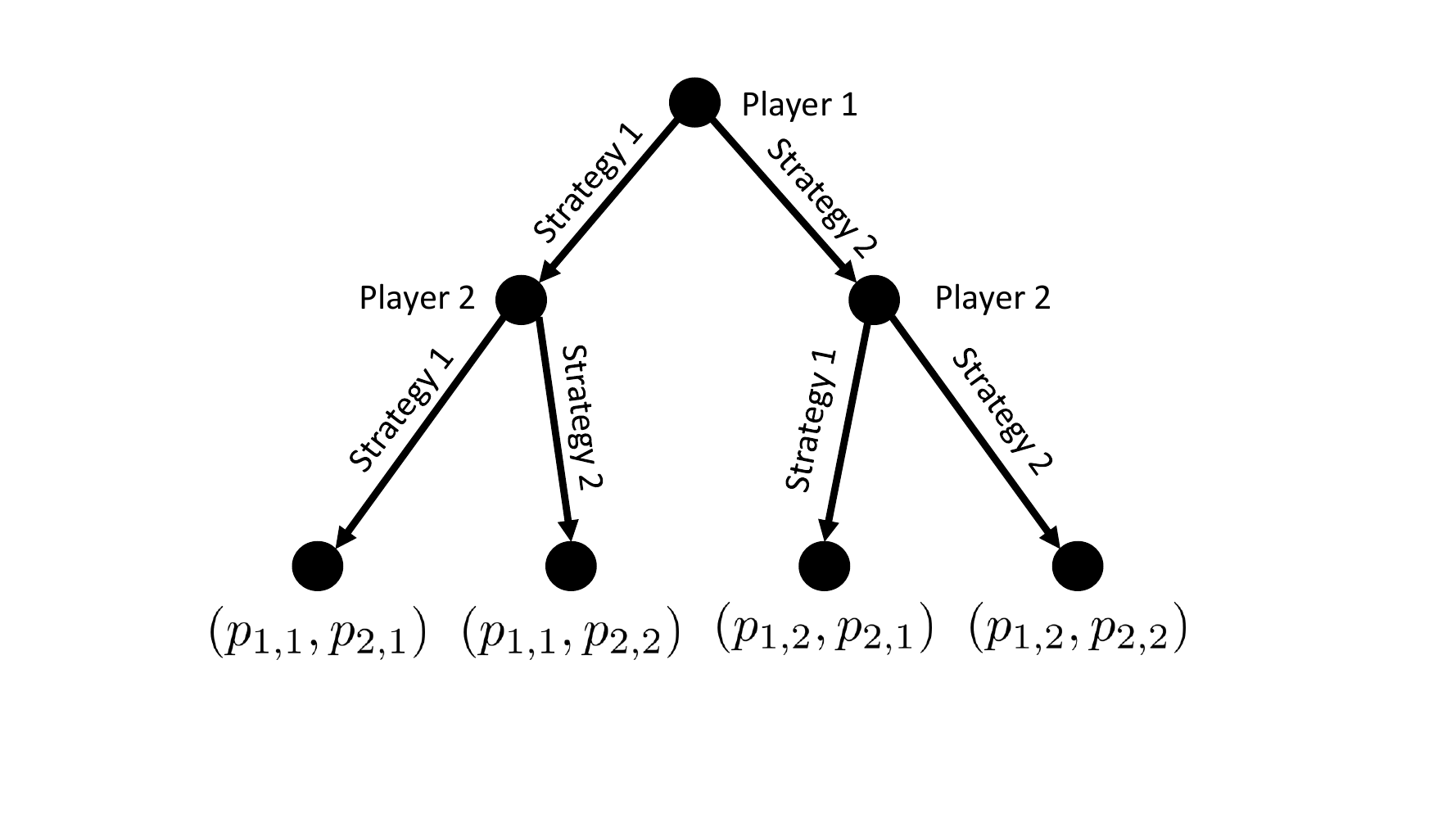}
\caption{\textit{\textit{Example of a game described in extensive form.}}}
\label{fig:Game_Extensive_Form}
\end{figure}

A game in extensive form\index{extensive form game|textbf} can be described using a game tree\index{game tree|textbf}, that is, a diagram that shows the choices (strategies) made by players at different points in time. The nodes of the tree represent the players positions in the game (in time) and the edges are the strategies. The payoffs are represented at the end of each branch of the tree. A complete tree is a tree that starts at the initial position, that is the beginning of the game, and contains all possible moves from each player. The complete tree is the extensive form game representation. An example of 2-player game in extensive form is shown in Figure \reffig{fig:Game_Extensive_Form}, where, the payoff\index{payoff|textbf} of $p_{i,j}$ refers to player $i \in \{1,2\}$ selecting a strategy $j \in \{1,2\}$.

For non-cooperative games, the normal form is used when the players choose their action or set of actions once and for all at the beginning, that is when all the players' decisions are made simultaneously. By contrast, the extensive form is used for sequential games, when each player needs to reconsider his actions when it is his turn to play\cite{osborne2004gametheory}.

\section{Normal Form Games}
The normal form (also called strategic form\index{strategic game|textbf}) is the basic game model studied in non-cooperative game theory. A game in normal form\index{game in normal form|textbf} lists each player strategies, and the
outcomes that result from each possible combination of choices. A two-player normal form game is defined by the four-tuple $G(\mathcal{S}_1,\mathcal{S}_2, \\ v_1, v_2)$, where $\mathcal{S}_1 = \{s_{1,1} \dots s_{1,n_1}\}$ and $\mathcal{S}_2 = \{s_{2,1}
\dots s_{2,n_2}\}$ are the sets of strategies available to the first and second player, and $v_l(s_{1,i}, s_{2,j}), l= 1,2$ is the payoff\index{payoff|textbf} (also called utility\index{utility|textbf}) of the game for the $l^{th}$ player, when the first player chooses the strategy $s_{1,i}$ and the second chooses $s_{2,j}$. A profile\index{profile|textbf} is a pair of strategies $(s_{1,i},s_{2,j})$. Games in normal form are compactly represented by matrices, namely payoff matrices. By considering $N$ players, a more general form of the normal form game is represented by the 3-tuple $G(N,\mathcal{S}, \mathbf{v})$ where, $N = \{1,2,\dots,n\}$ is the players set, $\mathcal{S} = \{ \mathcal{S}_1, \mathcal{S}_2, \dots ,\mathcal{S}_n \}$ are the sets of strategies, $\mathcal{S}_i = \{s_{i,1} \dots s_{i,n_i}\}$ represents the strategies set available to the players, and the vector $\mathbf{v} = (v_1, \dots,v_n)$ is the set of the game payoffs with $v_i$ corresponding to the payoff of player $i$. A vector $\mathbf{ss} = (s_{1,i_1}, \dots,s_{n,i_n}) \in \mathcal{S}$ is a strategy profile for the game with $N$ players. 

\subsection{Game Analysis}

\subsubsection{Nash Equilibirium}

Given a game, determining the best strategy that each player should follow to maximize its payoff is not easy. Even more, a profile which is optimum for both players may not exist. A common goal in Game Theory is to determine the existence of equilibrium points\index{equilibrium point|textbf}, i.e., profiles that, to a certain extent, represent a satisfactory choice for both players. While there are many definitions of equilibrium, the most famous and commonly adopted is the one by John Nash\index{John Nash|textbf} \cite{Nash50,Osb94}. In a 2-player game, a profile $(s_{1,i^*}, s_{2,j^*})$ is a Nash equilibrium if:

\begin{equation}
\begin{array}{ll}
    v_1(s_{1,i^*}, s_{2,j^*}) \ge v_1(s_{1,i}, s_{2,j^*}) & \forall s_{1,i} \in \mathcal{S}_1\\
    v_2(s_{1,i^*}, s_{2,j^*}) \ge v_2(s_{1,i^*}, s_{2,j}) & \forall s_{2,j} \in \mathcal{S}_2,
\end{array}
\label{eq.Nash}
\end{equation}

where for a zero-sum game\index{zero-sum game|textbf} $v_2 = -v_1$. In practice, a profile is a Nash equilibrium if none of the players can improve its payoff by changing its strategy unilaterally. This profile is known as the {\em equilibrium point}\index{equilibrium point|textbf} or {\em saddle point}\index{saddle point|textbf} of the game. Two types of Nash equilibria exist: pure strategy Nash equilibrium\index{Pure Strategy Nash Equilibrium|textbf} and mixed strategy Nash equilibrium\index{Mixed Strategy Nash Equilibrium|textbf}. \\

\textit{Strategies in Normal Form Games}
\begin{itemize}
\item Pure Strategy Nash Equilibirum\index{Pure Strategy Nash Equilibirum|textbf}: \\
A pure strategy Nash equilibrium is a Nash equilibrium in which each player selects a single strategy and plays it. Then, a pure strategy profile $(s_{1,i^*}, s_{2,j^*})$ is a Nash equilibrium for the game with  $s_{1,i^*}$ and $s_{2,j^*}$ are the {\em pure} strategies for player 1 and player 2, respectively. This means that none of the player will improve his payoff by changing his strategy unilaterally.
\item Mixed Strategy Nash Equilibirum\index{Mixed Strategy Nash Equilibrium|textbf}: \\
Players could also follow another, more complicated strategy. They can randomize their choice over the set of available strategies according to a certain probability distribution. Such a strategy is called a {\em mixed strategy}. Given a normal form game $G(N,\mathcal{S}, \mathbf{v})$, let $\Pi(\mathcal{Z})$ to be the set of all the probability distributions over the set $\mathcal{Z} =\{z_1,\dots,z_n\}$. Then, the {\em set of mixed strategies} for a player $i$ are all probability distributions over its strategy set $\mathcal{S}_i$, namely, $\mathbf{ss}_i = \Pi(\mathcal{S}_i)$. The set of {\em mixed strategy profiles} is the cartesian product of single mixed strategy sets $ \mathbf{ss}=\mathbf{ss}_1 \times \mathbf{ss}_2 \times \dots \times \mathbf{ss}_n$. $ss_i(s_{n,i_n})$ is the probability that a strategy $s_{n,i_n}$ will be played under mixed strategy $ss_i$. The expected payoff $v_i$ for player $i$ of the mixed strategy profile $ss = (ss_1, ss_2,\dots,ss_n)$ is defined as:
\begin{equation}
v_i(ss) = \sum\limits_{\mathbf{ss} \in \mathcal{S}} v_i(\mathbf{ss}) \prod\limits_{j=1}^n ss_j(s_{j,i})
\label{eq.payoff_mixed_str}
\end{equation}
An example of a game in normal form in which players will follow mixed strategies is the "Rock-Paper-Scissors"\index{Rock-Paper-Scissors|textbf} game with an example of its payoff matrix shown in Table \reftab{Table.RockPaperScissors}.

\begin{table}[H]
\centering
\begin{tabular}{lccc}
                              & \multicolumn{1}{l}{Rock}    & \multicolumn{1}{l}{Paper}   & \multicolumn{1}{l}{Scissors} \\ \cline{2-4} 
\multicolumn{1}{l|}{Rock}     & \multicolumn{1}{c|}{(0,0)}  & \multicolumn{1}{c|}{(-1,1)} & \multicolumn{1}{c|}{(1,-1)}  \\ \cline{2-4} 
\multicolumn{1}{l|}{Paper}    & \multicolumn{1}{c|}{(1,-1)} & \multicolumn{1}{c|}{(0,0)}  & \multicolumn{1}{c|}{(-1,1)}  \\ \cline{2-4} 
\multicolumn{1}{l|}{Scissors} & \multicolumn{1}{c|}{(-1,1)} & \multicolumn{1}{c|}{(1,-1)} & \multicolumn{1}{c|}{(0,0)}   \\ \cline{2-4} 
\end{tabular}
\caption{\textit{"Rock-Paper-Scissors" game example. The row player is player 1 and the column player is player 2.}\label{Table.RockPaperScissors}}
\end{table}
Let $r_1,p_1,s_1$ be the probabilities that player 1 plays rock, paper, and scissors, respectively. Then, the expected payoff of player 1 is: $v_1 = p_1-s_1$ if player 2 rocks, $v_1 = s_1-r_1$ if player 2 plays paper, and $v_1 = r_1-p_1$ if player 2 plays scissors. Player 2 tries to minimize player 1's payoff, so we have: $v_1 \leq p_1-s_1$,   $v_1 \leq s_1-r_1$, and $v_1 \leq r_1-p_1$. Then, player 1 must find $r_1,p_1,$ and $s_1$ that maximize his payoff subject to $r_1+p_1+s_1 = 1$. By solving this optimization problem\index{optimization problem|textbf}, the mixed strategy for player 1 turns out to be $(r_1,p_1,s_1)=(1/3,1/3,1/3)$. Optimization problems of this kind are solved by means of  {\em Linear Programming} which will discuss by the end of this subsection.
\end{itemize}

It is known that every normal form game with a finite sets of actions has at least one Nash equilibrium in mixed strategies \cite{Nash50}.

For strictly competitive games\index{strictly-competitive game|textbf}, Nash equilibrium\index{Nash equilibrium|textbf} has interesting properties. Let $G$ be a zero-sum game\index{zero-sum game|textbf} and $(s_{1,i^*}, s_{2,j^*})$ be a Nash equilibrium; then,  $s_{1,i^*}$ maximizes the first player payoff in the worst case scenario, i.e., assuming that second player selects his most profitable strategy corresponding to the most harmful action for the first player. Similarly, $s_{2,j^*}$ maximizes the second player payoff. We also have that \cite{Osb94} 

\begin{equation}
    \max_{\mathcal{S}_1} \min_{\mathcal{S}_2} v_1(s_{1,i}, s_{2,j}) =  \min_{\mathcal{S}_2} \max_{\mathcal{S}_1} v_1(s_{1,i}, s_{2,j}) = v_1(s_{1,i^*}, s_{2,j^*})
\label{eq.minimax}
\end{equation}

As a consequence of relation \eqref{eq.minimax}, if many equilibrium points exist, they all yield the same payoff. In a 2-player game, a player minmax value is always equal to its maxmin value, and both are equal to the Nash equilibirum value as shown by Von Neumann's Minimax Theorem\index{Von Neumann's Minimax Theorem|textbf} \cite{Neumann1928}.
A known result asserts that, if the two players are allowed to adopt mixed strategies over their set of actions, finding the Nash equilibrium\index{Nash equilibrium|textbf} for the game corresponds to solving one of the two Linear Programming\index{Linear Programming|textbf} (LP) problems in \eqref{eq.minimax}. \\

\textit{Linear Programming}\\

\begin{figure}[h]
\centering
\includegraphics[scale=0.35]{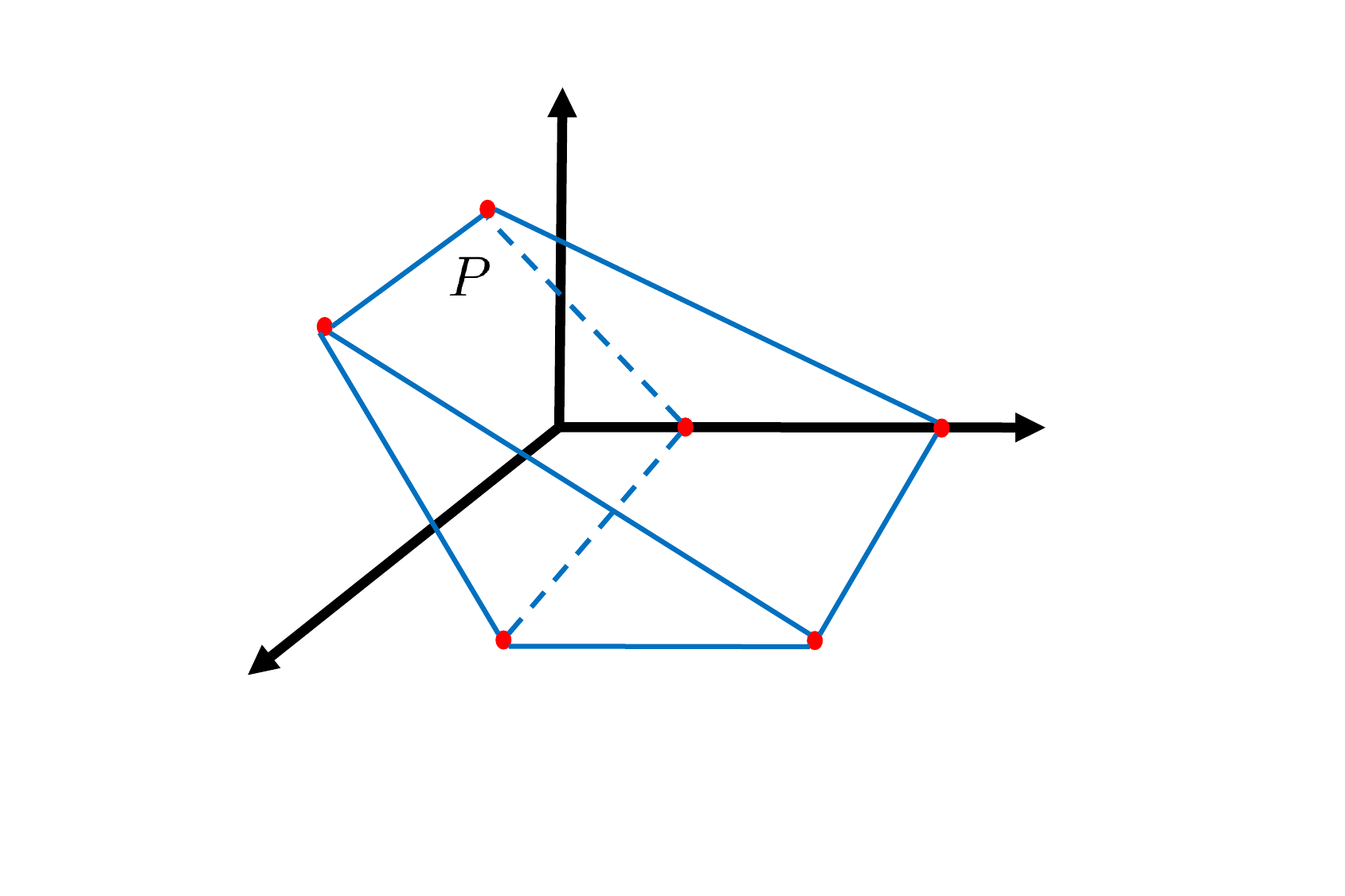}
\caption{\textit{Convex polytope example.}}
\label{fig:convex_polytope}
\end{figure}

Linear Programming\index{Linear Programming|textbf} (LP) \cite{chvatal1983linearprogramming}, or linear optimization\index{linear optimization|textbf}, is a method to evaluate the best outcome of a mathematical linear function subject to linear relationships or constraints. More formally, LP is an optimization technique to maximize or minimize a linear objective function, subject to linear equality and linear inequality constraints.  Its feasible region is a convex polytope\index{convex polytope|textbf} as in Figure \reffig{fig:convex_polytope}, which is a set defined as the intersection of finitely many half spaces, each of which is defined by a linear inequality or constraint. Its objective function is a real-valued linear function defined on this polyhedron (polytope). A linear programming algorithm finds a point in the polyhedron\index{polyhedron|textbf} (on its solid shape surface) where this function has the smallest (or largest) value if such a point exists.

A linear program can be expressed as:
\begin{equation}
\begin{aligned}
&{\text{maximize}}
& & \mathbf{c}^T x \\
& \text{subject to}
& & \mathbf{Ax} \leq \mathbf{b}, \\
& \text{and}
&& \mathbf{x} \geq 0.
\end{aligned}
\label{eq.LP_Form}
\end{equation}
where, $\mathbf{x}$ represents the vector of variables to be determined, $\mathbf{b}$ and $\mathbf{c}$ are vectors of known coefficients, $\mathbf{A}$ is a matrix of known coefficients, and $(.)^T$ is the matrix transpose. Here, $\mathbf{c}^T x$ is the objective function and the constraints of the problem are the inequalities $\mathbf{Ax} \leq \mathbf{b}$ and $\mathbf{x} \geq 0$. These constraints specify the convex polytope over which the objective function is to be optimized.

Back to the 2-player zero-sum game, Equation (\refeq{eq.minimax}) can be seen as solving two separated LP problems, namely, minimizing the payoff (the function) subject to $\mathcal{S}_2$ and then, maximizing it subject to $\mathcal{S}_1$.

\subsubsection{Dominance Solvable Games}
Despite its popularity, the practical meaning of Nash equilibrium\index{Nash equilibrium|textbf} is often unclear, since it cannot be guaranteed that the players will end up playing at the Nash equilibrium. A particular kind of normal form games for which stronger forms of equilibrium exist are called dominance solvable games\index{dominance solvable game|textbf} \cite{Osb94}. This concept is directly related to the notion of dominant and dominated strategies. A strategy is said to be {\em strictly dominant}\index{strict dominance|textbf} for one player if it is the best strategy for the player no matter how the other player decides to play.
Obviously, if such a strategy exists for one of the players, he will surely adopt it. Similarly, a strategy $s_{l,i}$ is {\em strictly dominated} by strategy $s_{l,j}$, if the payoff achieved by player $l$ choosing $s_{l,i}$ is always lower than that obtained by playing $s_{l,j}$ regardless of the choice made by the other player. Formally, in the 2-players case, a strategy $s_{1,i}$ is {\em strictly dominated} by strategy $s_{1,k}$ for a player, for instance, player 1, if

\begin{equation}
    v_1(s_{1,k}, s_{2,j}) > v_1(s_{1,i}, s_{2,j}) \quad \forall s_{2,j} \in \mathcal{S}_2.
\label{eq.dominatedDEF}
\end{equation}

Following this definition, a {\em strictly dominant} strategy is a strategy which strictly dominates all the other strategies in the strategy set.

A possible technique to solve a game is by recursive elimination\index{recursive elimination|textbf} of the dominated strategies since, all the strategies that a player definitely should not adopt can be removed from the game. In recursive elimination, first, all the dominated strategies are removed from the set of available strategies, since no rational player would ever play them. In this way, a new, possibly smaller game is obtained. Then, at this point, some strategies, that were not dominated before, may be dominated in the remaining game, and hence are eliminated. The process is repeated  until no dominated strategy exists for any player. A {\em rationalizable equilibrium} is any profile which remains after the recursive elimination of dominated strategies \cite{ChenGames,Bern84}. If at the end of the process only one profile is left, the remaining profile is said to be the {\em only rationalizable equilibrium}\index{rationalizable equilibrium|textbf} of the game, which is also the only Nash equilibrium\index{Nash equilibrium|textbf} point. A dominance solvable game\index{dominance solvable game|textbf} is a game that can be solved according to the procedure described above.

It goes without saying that the concept of rationalizable equilibrium is a stronger notion than that of Nash equilibrium \cite{weirich2007equilibrium}. In fact, under the assumption of rational and intelligent players, it can be seen that the players will choose the strategies corresponding to the unique rationalizable equilibrium since it will maximize their payoffs. An interesting, related notion of equilibrium is that of dominant equilibrium. A {\em dominant equilibrium} is a profile that corresponds to dominant strategies for both players and is the strongest kind of equilibrium that a game in normal form may have.

\subsection{Examples of Normal Form Games}
\subsubsection{Prisoner's Dilemma}
A Prisoner's dilemma\index{Prisoner's dilemma|textbf} \cite{prisonersdilemma1993} is a famous example of a 2-player normal form game in which the two players are prisoners suspected of a crime. Each player is taken separately to a room and asked to confess the crime or deny it. Based on the choice of the two players, each of them will stay a number of years in jail. The Prisoner's dilemma game can be formalized as:
\begin{itemize}
\item Players are Prisoner 1 and Prisoner 2.
\item $\mathcal{S}_1=\mathcal{S}_2 =\{\textrm{Confess}, \textrm{Deny}\}$.
\item The payoff matrix is given in Table \reftab{Table.PrisonersDilemma}. \\
\begin{table}[H]
\centering
\begin{tabular}{lll}
& Confess                    & Deny                       \\ \cline{2-3} 
\multicolumn{1}{l|}{Confess}        & \multicolumn{1}{l|}{$\quad(1,1)$} & \multicolumn{1}{l|}{$\quad(3,0)$} \\ \cline{2-3} 
\multicolumn{1}{l|}{Deny}           & \multicolumn{1}{l|}{$\quad(0,3)$} & \multicolumn{1}{l|}{$\quad(2,2)$} \\ \cline{2-3} 
\end{tabular}
\caption{\textit{Prisoner's Dilemma payoff matrix example. The row player is Prisoner 1 and the column player is Prisoner 2.}\label{Table.PrisonersDilemma}}
\end{table}
\end{itemize}
In this game, regardless of whether a prisoner decision is to confess or deny, each prisoner gets less punishment by denying the crime and hence accusing the other prisoner. The reason behind the dilemma is that prisoner 2 can either confess or deny. In the first case, if prisoner 2 confesses, prisoner 1 should deny, because going free is better than staying 1 year in jail. In the second case, if prisoner 2 denies, prisoner 1 should denies as well, because getting jailed for 2 years is better than 3. Therefore, in both cases prisoner 1 should deny the crime as well as prisoner 2 and hence the strategy profile (Deny, Deny) with payoffs $(2,2)$ is pure strategy Nash equilibrium for the game.

\subsubsection{Battle of Sexes}
The battle of sexes\index{battle of sexes|textbf} is a 2-player coordination game in normal form. In this game, a man and a woman want to go out together to watch one of two movies $F1$ and $F2$ at two different places. The players are currently in two different places and did not agree before where to go. They have to decide each on his own where to go, knowing that they cannot communicate with each other. Their main concern is to be together, however the man has a preference for $F1$ and the woman for $F2$.
The payoff for each strategy of the game in Table \reftab{Table.BattleofSexes} accounts also for the harm that the couple receives if they do not go to the same place.
\begin{table}[H]
\centering
\begin{tabular}{lll}
& $F1$                    & $F2$                       \\ \cline{2-3} 
\multicolumn{1}{l|}{$\quad F1$}        & \multicolumn{1}{l|}{$\quad(2,1)$} & \multicolumn{1}{l|}{$\quad(0,0)$} \\ \cline{2-3} 
\multicolumn{1}{l|}{$\quad F2$}           & \multicolumn{1}{l|}{$\quad(0,0)$} & \multicolumn{1}{l|}{$\quad(1,2)$} \\ \cline{2-3} 
\end{tabular}
\caption{\textit{Battle of sexes game example. Row player is the man and column player is the woman.}\label{Table.BattleofSexes}}
\end{table}

If the players go separated to watch different movies, they will receive no payoff. Instead, if they go together, only one of them will enjoy the movie and the other will receive only the payoff related to the pleasure of staying with his/her partner. Hence, there will be two Nash equilibria which are the profiles: $(2,1)$ and $(1,2)$. This battle can be solved by the use of mixed strategies. 

Suppose that the woman is likely to choose $F1$ with probability $w$ and $F2$ with probability $1-w$. Likewise, the man is likely to choose $F1$ with a probability $m$ and $F2$ with probability $1-m$. In this case, the probabilities become: $w\times m$ that both go to $F1$, $(1-m) \times w$ that the man goes to $F2$ and the woman to $F1$, $m \times (1-w)$ the man goes to $F1$ and the woman to $F2$, and $(1-m) \times (1-w)$ that both go to the $F2$. The man will receive an expected payoff of $2w$ if he goes to $F1$ and $1-w$ if he goes to $F2$. If the man mixes the two strategies of going to $F1$ and $F2$, they must have the same expected payoff so to make the other player uncertain, otherwise, the best response would be to always use the action whose expected payoff is higher. In this way, we have $2w = 1-w$ and hence, $w= 1/3$. Likewise, conditioned on the man's strategy, on the woman side we have $m=2(1-m)$ and hence, $m=2/3$. 

In the above setting in Table \reftab{Table.BattleofSexes}, we have: the probability that the two players will choose either to go together to $F1$ (or $F2$) is $2/9$, the probability that the man goes to $F2$ and the woman to $F1$ is $1/9$, and the probability that the man goes to $F1$ and the woman to $F2$ is $4/9$. Then, the expected payoff for the man becomes $2\times(2/9) + 1\times(2/9) + 0\times(4/9) = 2/3$. Similarly, the expected payoff for the woman is also $2/3$. Based on the expected payoffs, the values $(m,w) = (2/3,1/3)$ is a mixed strategy Nash equilibrium\index{Mixed Strategy Nash Equilibrium|textbf} for the game and a fair choice for both players since it returns the same payoff for both. 

\subsubsection{Common-Payoff Games}
Common-Payoff games\index{Common-Payoff games|textbf} are games in which, for every different {\em strategy profile} $\mathbf{ss}$, the players have the same payoff. These games are also known as pure coordination games or team games since all players need to coordinate on a strategy to maximize their payoff. An example of Common-Payoff game is shown in the following payoff matrix:
\begin{table}[H]
\centering
\begin{tabular}{lll}
& Left                    & Right \\ \cline{2-3} 
\multicolumn{1}{l|}{Left}        & \multicolumn{1}{l|}{$\quad(1,1)$} & \multicolumn{1}{l|}{$\quad(0,0)$} \\ \cline{2-3} 
\multicolumn{1}{l|}{Right}           & \multicolumn{1}{l|}{$\quad(0,0)$} & \multicolumn{1}{l|}{$\quad(1,1)$} \\ \cline{2-3} 
\end{tabular}
\caption{\textit{Common-payoff game example. The row player is driver 1 and the column is driver 2.}\label{Table.Commonpayoff}}
\end{table}
In this typical example two drivers drive toward each other and meet on a narrow road and they have to select the road side upon which to drive. Both have to deviate from each other in order to avoid collision. If both follow the same deviation they will manage to pass each other, but if they choose different deviations they will collide. In the payoff matrix in Table \reftab{Table.Commonpayoff}, successful passing is represented by a payoff of 1, and a collision by a payoff of 0. In this case there are two pure Nash equilibria: either both deviate to the left, or both deviate to the right. Therefore, it doesn't matter which side both drivers select, as long as they both select the same.

\subsubsection{Zero-Sum Games}
In Zero-Sum games\index{zero-sum game|textbf}, also known as strictly-competitive games\index{strictly-competitive game|textbf}, the two players have opposite goals. In this case, the two payoffs are strictly
related to each other, since for every profile we have $v_1(s_{1,i}, s_{2,j}) + v_2(s_{1,i}, s_{2,j}) = 0$. In other words, the win of one player is equal to the loss of the other, then, only one payoff must be defined. The payoff $v$ of the game can be defined by adopting the perspective of only one player, e.g., $v_1 = v$, with the understanding that the payoff of the second player is equal to $-v$. In the most common formulation of zero-sum games with perfect information, the sets $\mathcal{S}_1$, $\mathcal{S}_2$ and the payoff functions are assumed to be known to both players. In addition, it is assumed that the players choose their strategies before starting the game so that they have no idea about the choice of the other player. An example of zero-sum games is presented in the following example, namely, the game of matching pennies .\\

\textit{Matching Pennies} \\
In the game of matching pennies\index{matching pennies|textbf}, each of the two players have a coin. They both flip their coins and simultaneously show their result. If the coins match, player 1 wins both coins; otherwise, both coins go to player 2. The payoff matrix of matching pennies game is shown in the Table \reftab{Table.MatchingPennies}.
\begin{table}[H]
\centering
\begin{tabular}{lll}
& Heads                    & Tails \\ \cline{2-3} 
\multicolumn{1}{l|}{Heads}        & \multicolumn{1}{l|}{$ (1,-1)$} & \multicolumn{1}{l|}{$ (-1,1)$} \\ \cline{2-3} 
\multicolumn{1}{l|}{Tails}           & \multicolumn{1}{l|}{$ (-1,1)$} & \multicolumn{1}{l|}{$ (1,-1)$} \\ \cline{2-3} 
\end{tabular}
\caption{\textit{Matching Pennies game example. The row player is player 1 and the column is for player 2.}\label{Table.MatchingPennies}}
\end{table}

This game has no pure strategy Nash equilibrium since there is no pure strategy for any of the players and there is no "best response" by any of the players. Instead, the unique Nash equilibrium of this game is in mixed strategies wherein each player selects a head or a tail with a probability of $0.5$.

\section{Conclusion}
In this chapter we gave a brief introduction to game theory. First, we introduced games in normal form and we explained some solution concepts to these games, namely, Nash equilibrium and dominance solvability. Then, we discussed some examples of games in normal form to clarify the mechanics of the games and their solutions. As game theory may be used to model competitions, in this thesis, it will play a fundamental role in modeling the competition between the players, namely, the adversaries as the attackers and the sensor network. Throughout the thesis we will focus on 2-player games in normal form in which the first player (the attackers) and the second player (the defender) aim at acheiving opposite objectives. At one hand, the attackers want to introduce decision errors in the distributed sensor network in order to achieve some selfish or malicious objectives. On the other hand, the defender aims at protecting the sensor network against attackers and provide the most possible robust detection and decision performance. By adopting such a model, we are aiming at finding possible equilibria describing the interplay between the attackers and the defender and then, try to find out who will win the game.



\chapter{Security Attacks and Defenses in Distributed Sensor Networks}
\label{chapter:SecurityThreats}
\emph{"If you know the enemy and know yourself, you need not fear the result of a hundred battles. If you know yourself but not the enemy, for every victory gained you will also suffer a defeat. If you know neither the enemy nor yourself, you will succumb in every battle."}
\\
Sun Tzu, "The Art of War"

\emph{"Power resides where men believe it resides. It's a trick, a shadow on the wall. And a very small man can cast a very large shadow."}
\\
Lord Varys, "Game of Thrones"

\bigskip
\section{Introduction}
\PARstart{\textcolor{red}I}nformation{} fusion in distributed sensor networks\index{distributed sensor networks|textbf} may suffer from various threats and attacks. An incentive for the adversary\index{adversary|textbf} could be the critical nature of the phenomenon\index{phenomenon|textbf} such as troops passage in a battlefield, monitoring traffic flow using sensors \cite{chan2003security}, \cite{shi2004designing}, \cite{wang2006surveyWSN}, image authenticity in front of a court for crime witness \cite{BarniTondiSourceIdentificationGame},\cite{barnitondifontani2012universal}, and many others.
In addition, deceiving the detection and decision of the network about the phenomenon could be beneficial for the attacker e.g. to gain exclusive access to the spectrum in cognitive radio networks \cite{fragkiadakis2013survey}, change an item reputation in online reputation systems \cite{YSKY09} and others.
 
A fundamental and key enabler factor to ensure proper sensor networks functionality is securing them against attacks\index{attacks|textbf} and threats. A fundamental step for the protection of sensor networks is securing the information fusion process in order to ensure a {\em trusted and accurate} detection and decision about the observed phenomenon. To do so, the information fusion process\index{information fusion|textbf} should take into account the possible presence of sensors under the control of adversary\index{adversary|textbf} in the network knowing that they can have various malicious objectives.

In this chapter, we outline the most common attacks in centralized distributed sensor networks as well as in consensus-based decentralized networks. In addition, we present the most common countermeasures to protect the network from these attacks.

\section{Attacks to Distributed Sensor Networks}
\begin{figure}[h]
\centering
\includegraphics[scale=0.28]{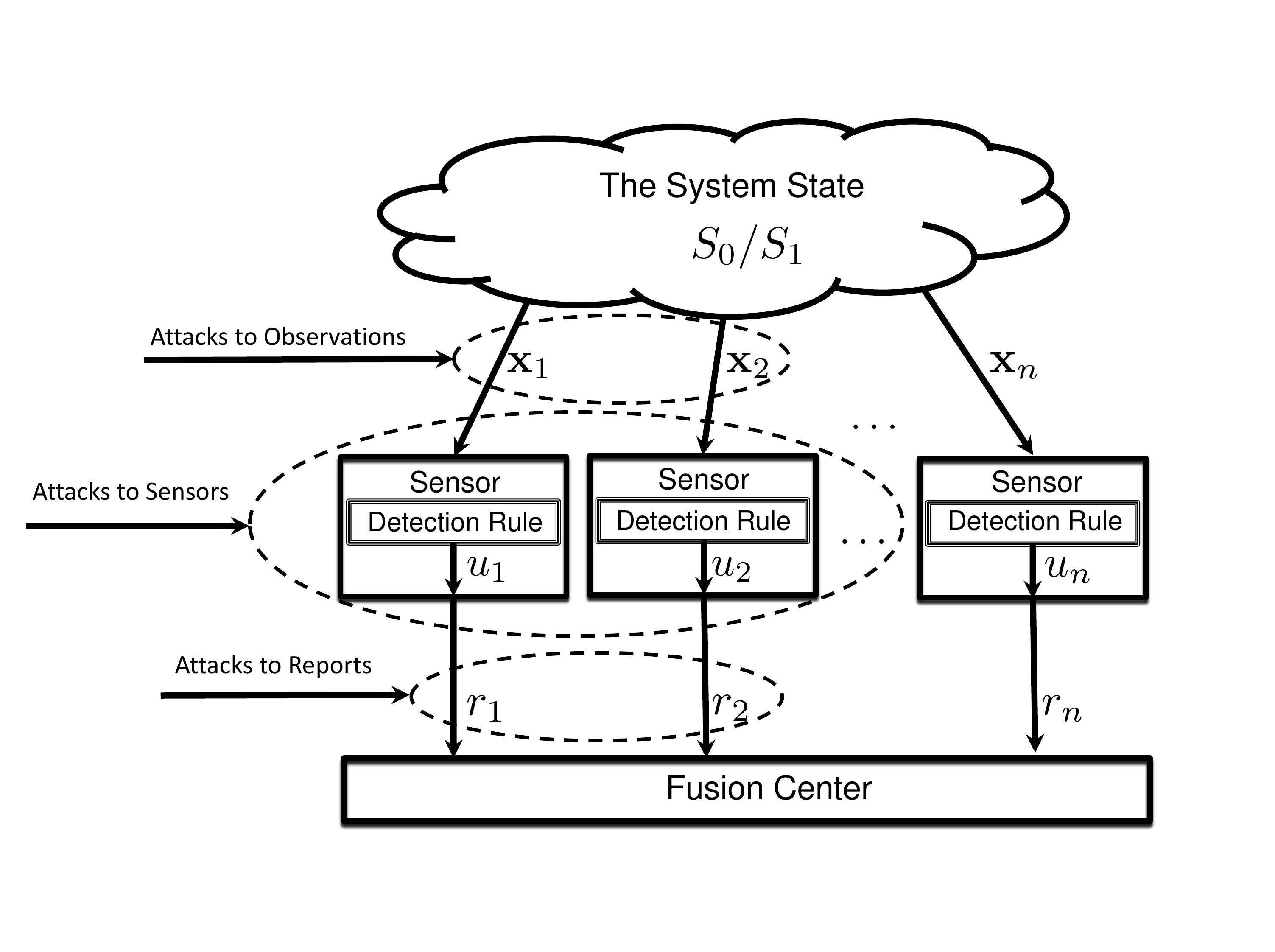}
\caption{\textit{Classification of attacks to distributed sensor networks.}}
\label{fig:Attacks_Net}
\end{figure}

We start by considering the centralized network setup illustrated in Figure \reffig{fig:Attacks_Net}. The illustration shows several possible adversarial setups. In these setups, the information fusion process will carried out at the level of the binary decisions $r_1,\dots,r_n$ provided by the sensors. The adversary can carry out its attacks in three positions, which are:
\begin{itemize}
\item The observations about the phenomenon used by the sensors to make the local decision. In this attack, the adversary\index{adversary|textbf} can access and eavesdrop the observations and modify them since they have control over the observation channel between the system and the sensor network. In this way, they can control what the sensors will observe about the phenomenon and consequently, deceive the local decision and hence, indirectly, corrupt the information fusion process performed later at the FC.
\item The sensors themselves. In this attack, a fraction (or all) of the sensors are under the control of the adversary\index{adversary|textbf}. Then, the adversary can modify the detection and decision rules, the decision thresholds or the data computed locally to be sent to the FC. Altering the information computed locally by the sensors can maliciously affect the result of the fusion process since a part of the information fused is unliterary altered by the adversary.
\item The information sent by the sensors. In this case, the attackers does not have control over the sensors\index{sensors|textbf} but instead, they have access to the links between the sensors and the FC. Alike the case of attacking the sensors, this attack\index{attack|textbf} can make the information fusion fail. The difference between the two types of attacks is that in the latter, the adversary does not have access to the observations.
\end{itemize}
 
\subsection{Attacks to the Observations}
The $n$ sensors in the network observe the phenomenon through the vectors $\mathbf{x}_1, \mathbf{x}_2, \allowbreak \dots,\mathbf{x}_n$. In this scenario, a fraction of the links, let us say $\alpha = b/n$, is under the control of the adversary\index{adversary|textbf}. The variable $\alpha$, could be either the exact fraction of the links under the control of the adversary, or the probability that a link is under the control of the adversary. The adversary can modify the vectors from $\mathbf{x}_1, \mathbf{x}_2, \dots,\mathbf{x}_b$ to $\mathbf{\tilde{x}}_1, \mathbf{\tilde{x}}_2, \dots,\mathbf{\tilde{x}}_b$ so to induce a wrong local decision\index{local decision|textbf} at the sensors $1,\dots,b$. The objective of the adversary could vary from pushing the $b$ sensors to change their decision from $H_0$ to $H_1$ and viceversa or only corrupting the functionality of the network by inducing random errors. Consequently, a fraction of the information sent to the FC will be wrong and this will affect the global decision at the FC. This setup is very general and can be used to model a variety of situations. Here, we present two examples of such an adversarial setup: the jammer attacks in wireless communication networks \cite{xu2005jammingattack} and the Primary User Emulation Attack (PUEA) in cognitive radio networks \cite{chen2008defensePUEA}.
\subsubsection{Jammer Attack}
Due to the open and shared nature of the wireless medium, together with the advancement of wireless technologies and software, wireless networks can be easily monitored, accessed and broadcast on by a transmitter. The adversary can observe the communication links and the information between wireless entities, and launch Denial of Service\index{Denial of Service|textbf} (DoS) attacks by injecting wrong information messages. Severe types of DoS attacks can be launched in wireless networks which can block the wireless medium and prevent other wireless devices from even communicating with each other. These attackers are known as {\em jammers} and continuously transmit radio frequency signals to occupy the channel and then block the information flow in the network or between the network and the system \cite{shi2004designing}. Therefore, a {\em jammer} is an adversary who intentionally trying to interfere with the transmissions and receptions of wireless communications.

The objective of a jammer\index{jammer|textbf} is to interfere with legitimate wireless communications. It can achieve this goal by either preventing the transmission of the information from the source, preventing the reception of the information by the receiver, or modifying the information exchanged between them. Many types of jammers exist depending on their objective and behavior, we may have:
\begin{itemize}
\item A Constant jammer continuously transmits signals to block the communication between some selected network entities.
\item A Deceptive jammer constantly injects regular
information into the links with no separation between subsequent transmissions. In this way, the receiver will be deceived into believing that there is
an information coming and will remain in receiving mode.
\item A Random jammer randomly alternates between
sleeping and attacking modes in order to save energy.
\item Reactive jammer: the jammer stays
quiet when the channel is idle and there is no information exchange and starts transmitting as soon as it senses any activity on the channel. Therefore, a reactive jammer targets the reception of a message. The advantage of this kind of jammer is that it is harder to detect.
\end{itemize}
A complete survey of jammer attacks and their feasibility in wireless networks can be found in \cite{xu2005jammingattack}. 

The connection between jammer attacks and attacking the observations is straightforward as the jammer can:
\begin{itemize}
\item Block the system state information\index{information|textbf} observed by the sensors\index{sensors|textbf} to let them believe that the activity does not exist which means that the system is in state $S_0$ while it can be in $S_1$.
\item Inject a specific information value on the link to change the sensor local decision from $H_0$ to $H_1$ or viceversa, like the case of a {\em deceptive jammer}.
\end{itemize}

\subsubsection{Primary User Emulation Attack}
In Cognitive Radio Networks\index{Cognitive Radio Networks|textbf}, guaranteeing a trustworthy spectrum sensing is a particularly important problem as spectrum sensing\index{spectrum sensing|textbf} is a key enabler for this technology. The main concern in spectrum sensing is the ability to distinguish between PU and SU signals. To do so, an SU should continuously scan the spectrum for the presence of PU signals in the candidate bands. If an SU detects a PU signal in the current band, it must immediately switch to another band. On the other hand, if the SU detects the presence of another SU, it runs a coexistence mechanism \cite{nextgenerationDSA,wyglinski2009cognitive} to share spectrum bands.
Distinguishing the two types of signals is not trivial, especially in hostile environments. Many techniques are proposed to increase the accuracy of PU activity detection including: Energy detectors, Cyclostationary detectors, Wave-based detectors, Matched filter detectors and so on \cite{spectrumsensingsurvey,ALB11}. The most common approach to improve spectrum sensing accuracy is {\em collaborative spectrum sensing} \cite{CollaborativeTVdetectionDSA,ghasemicollaborativeSS}, in which, the SUs collaboratively scan the spectrum and send their results, to a FC\index{Fusion Center|textbf} that makes a final decision about spectrum occupancy. This collaboration among SUs can be also implemented in a decentralized fashion \cite{nextgenerationDSA}. 

A Primary User Emulation Attacker\index{Primary User Emulation Attacker|textbf} (PUEA) \cite{fragkiadakis2013survey, clancy2008security, chen2009modeling} is an adversary that modifies the air interface of a CR to mimic the characteristics of a PU signal, thereby causing the SUs to mistakenly detect the adversary signal as a PU signal. The high reconfigurability of software-defined CR devices makes PUEAs possible and realistic \cite{jin2009mitigatingPUEA}.

When the attacker detects no PU activity, it sends
{\em jamming signals} emulating PU's activity, so
to let the SUs believe that a PU is active and hence, prevent them from using the available spectrum.
This attacker\index{attacker|textbf} can be seen as a new type of DoS jamming attack specific to cognitive radio networks scenario\cite{chen2009modeling}.

Based on the objective of the adversary, PUEAs can be classified into two classes: Selfish PUEA\index{Primary User Emulation Attacker|textbf} and Malicious PUEA. A selfish PUEA aims at increasing the usage of the spectrum by the attacker. By finding an available spectrum band and preventing other SUs from accessing that band, the adversary can gain alone the access to the spectrum band. On the other hand, a malicious PUEA aims at impeding spectrum sharing by deceiving the spectrum sensing proces. In this way, SUs always detect the presence of a PU and move to another band \cite{sharma2016PUEA}.

\subsection{Attacks to the Sensors}
In the case of attacks to the sensors\index{sensors|textbf}, the FC has to tackle with the presence of a number of malevolent sensors, which deliberately alter their information reports to induce a global decision error. According to a consolidated literature, such nodes are referred to as {\em byzantine nodes}\index{byzantine nodes|textbf} or simply {\em Byzantines}\index{Byzantines|textbf} \cite{ByzantineGeneralProblem,Vemp13}. Note that a byzantine sensor can decide to alter its report by relying on its observations of the system state $S_i, i\in \{0,1\}$, but usually it does not have access to the observations available to the other sensors and their information reports. In this setup, a fraction $\alpha$ of the $n$ sensors is under the control of the attacker which, in order to make the fusion process fail, alter the local information or decision prior to sending them to the FC\index{Fusion Center|textbf}. From the perspective of the FC, the problem can be viewed as one of robust distributed information fusion\index{information fusion|textbf} problems as the information from the sensors is a mixture of good and adversarial data. 

\subsubsection{Spectrum Sensing Data Falsification Attacks}

In a cognitive radio network setup, Spectrum Sensing Data Falsification\index{Spectrum Sensing Data Falsification Attack|textbf} (SSDF) \cite{ALB11}, \cite{yu2009defense,min2009attacktolerant,qin2009towardsawareCR, kaligineedi2008securetechniquesinCRN,SecureCSSinCRN2008,attar2012surveysecurityCRN} refers to SU that sends altered local spectrum sensing results, which will possibly result in erroneous decisions by other SUs or by the FC. The SSDF attack is an example of a {\em Byzantine} attack targeting the spectrum sensing process. In cognitive radio networks, spectrum sensing failure problem can be caused by malfunctioning SUs or SSDF attacks. A malfunctioning SU is unable to produce reliable local information and may send wrong sensing information to the FC. On the other hand, in SSDF attacks, a malicious SU intentionally sends falsified reports to the FC in the attempt to cause a failure in the information fusion process. It is shown in \cite{wang2009attackproof} that, under certain assumptions, even a single byzantine sensor\index{byzantine nodes|textbf} can make the information fusion process\index{information fusion|textbf} fail.

Depending on the attack objective and behavior, SSFD attacks can be classified into the following categories: 

\begin{itemize}
\item Malicious SUs \cite{mishra2006cooperativesensing} send false sensing results so to confuse other SUs or the FC about spectrum occupancy. The objective of malicious SSDF attack is to lead the FC or the rest of the SUs to decide the absence of a PU signal when it is present, or make them believe that there is a PU signal when there is not. Consequently, in the first case, the SUs will refrain from using the specific band, while in the second case they will cause harmful interference to PU.
\item Greedy SUs \cite{sodagari2010DoSinCRN} continuously report that a specific
spectrum band is occupied by a PU. This can be seen as a selfish attack aiming at occupying the available band alone by forcing the other SUs to evacuate it.
\item Unintentionally misbehaving SUs \cite{fragkiadakis2013survey} send wrong information reports about PU activity in the spectrum, not because they are attackers, but because of a problem in their software or hardware such as random faults or virus \cite{fitton2002securitySDR,li2009architectureSDR,xiao2009tamper}.
\end{itemize}

\subsection{Attacks to the Reports}
In this case, the adversary\index{adversary|textbf} can access the links between the sensors and the FC. This may correspond to a situation in which the adversary does not control the nodes but only the communication link between the nodes and the fusion center, or to the case of byzantine sensors which, for some reasons, cannot observe the information data at the input of the sensor, or decide not exploit such a knowledge. For the FC, both types of attacks force it to consider that a part of the received information is malevolently altered and consider this fact when fusing the information. The difference between the two cases is that: in case of the attacks to reports, the adversary does not have access to the observations about the system and could be also that it does not know the local information computed at the sensors.  

\subsection{Attacks to Consensus Algorithms for decentralized distributed sensor networks}
In a decentralized consensus algorithm\index{consensus algorithm|textbf}, a {\em Byzantine attack} can target the initial phase or the state update phase of the algorithm in order to mislead the network decision about the system state \cite{varshney2015consensuswithByzantines}. The first case is referred to as {\em data/measurement falsification attack}, while the second case is known as {\em consensus disruption attack}. Data falsification\index{data falsification attack|textbf} attackers are more capable and can disguise themselves while degrading the network detection performance using falsified data or measurements. On the other hand, a consensus disruption attack\index{consensus disruption attack|textbf} aims at corrupting the consensus operation but, it is easier to detect because of its nature \cite{varshney2015consensuswithByzantines}.
\begin{itemize}
\item Data/Measurement Falsification Attack \\
In a data falsification attack, sensors falsify their initial data or the false data could be injected from the outside in order to degrade the detection performance of the network. By doing so, the attacker tries to change the final test information which, in weighted average consensus algorithms, is the weighted average of all the initial measurements $\bar{x} = \frac{1}{n} \sum_{i \in \mathcal{N}} w_i x_i(0)$ where, $\mathcal{N}$ is the set of all the sensors, $x_i(0)$ is the initial measurement at sensor $i$, and $w_i$ is the assigned weight to the measurement $x_i(0)$. Formally, by considering the attack at a sensor $i$, we have:
\begin{equation}
\tilde{x}_i(0) = x_i(0) + \Delta_i \quad \text{or} \quad w_i \rightarrow \tilde{w}_i,
\label{eq.InitialPhaseAttack}
\end{equation}

where, $\tilde{x}_i(0)$ is falsified initial data, $\Delta_i$ is the attack power which can take any real value, and $\tilde{w}_i$ is the modified weight.

\item Consensus Disruption Attack \\
This attack aims at degrading the detection performance by disregarding the state update rule of the consensus algorithm as follows:

\begin{equation}
\tilde{x}_i(k+1) = x_i(k) + \frac{\epsilon}{w_i} \sum_{j \in \mathcal{N}_i} (x_j(k)-x_i(k)) + u_i(k)
\label{eq.StateUpdateAttackConsensus}
\end{equation}

where, $u_i(k)$ is the injected value by the byzantine sensor $i$ in the state update $x_i(k+1)$ at iteration $k+1$. Then, the attacked state update value $\tilde{x}_i(k+1)$ is sent by the adversary to all its neighbor sensors in the set $\mathcal{N}_i$ and the error propagates through the network.
\end{itemize}

Similar to the attacks in the centralized setup, attacks on consensus algorithm can have the objective of pushing the network into believing that the system state is $S_0$ while it is $S_1$ by using low values of $\Delta_i$ or $u_i(k)$ or the opposite case by making the network believe that $S_1$ is correct while it is not by using high values of the attack. In addition, the attack can be constant or probabilistic in nature, which means that the attack can inject the falsified value statically, or randomly with a certain probability $P_i$.

\section{Defenses Against Attacks to Distributed Sensor Networks}
Having presented the most common adversarial setups and attacks in distributed sensor networks, we now describe the most common countermeasures\index{countermeasures|textbf} and show how they contribute to protect the network. In general, these countermeasures can directly modify the information fusion process or introduce a pre-step before the fusion process so to filter out the adversary effect before performing the fusion.

\subsection{Defenses against Attacks to the Observations}
An asymptotic version - as a function of the network size $n$ - of the problem of attacks to the observations has been studied in \cite{Bar13} in a binary hypothesis testing framework. In this setup, by knowing the true system state\index{system state|textbf}, the attacker can corrupt a part of or all the observations. On his side, the FC runs a binary hypothesis test based on the Neyman-Pearson criterion, in which $H_0$ is the hypothesis that the system state is in a safe or normal condition and $H_1$ that it is not. The authors propose a general framework based on {\em game-theory} that encompasses a wide variety of situations including distributed detection\index{distributed detection|textbf}, data fusion, multimedia forensics\index{multimedia forensics|textbf}, and sensor networks.
In this framework, the interplay between the adversary\index{adversary|textbf} and the defender\index{defender|textbf} (FC) is modeled as a 2-player game in which the adversary tries to induce a {\em false negative} error while the FC must ensure that {\em false positive} error probability stays below a threshold. The set of strategies of the defender consists of the acceptance regions for $H_0$ ensuring a given false positive error probability. On the other hand, the strategies of the attacker are all the possible modifications of the observation sequence subject to a maximum distortion. Following this theoretical model, the authors derive the equilibrium point\index{equilibrium point|textbf} of the game\index{Game Theory|textbf}, showing that a dominant strategy exists for the defender, which means that the defender can choose its strategy without caring about what the adversary is doing. An interesting result of this work states that the defender would get no advantage from the knowledge of the attacked sensors. The reason of this behavior is due to the adoption of a NP setup at the FC, and by the assumption that the attacker acts only when $H_0$ does not hold while the FC is asked to satisfy a constraint on false positive error.

The study in \cite{Bar13} addresses the problem from the most general and theoretical point of view. Now we move to present some more practical defenses for the two attacker types we have presented in the previous section: the jammer and the PUEA.

\subsubsection{Defenses Against Jammer Attack}
For the jammer\index{jammer|textbf} attack, the typical defenses involve the usage of spread-spectrum communication such as frequency hopping or code spreading \cite{wang2006surveyWSN}. Frequency-hopping spread spectrum\index{Frequency-hopping spread spectrum|textbf} (FHSS) \cite{rappaport1996wirelesscommunicationsBook} is a method of transmitting signals by rapidly changing the carrier frequency among many available channels using a pseudo-random sequence known at both the transmitter and the receiver. The lack of knowledge of the frequency selection by the attacker makes jamming the frequency being used not possible. However, since the range of possible frequencies is limited, a jammer may instead jam a wide set of the frequencies increasing its possibility to succeed in the attack.

Code spreading\index{Code spreading|textbf} \cite{rappaport1996wirelesscommunicationsBook} is another way to defend against jamming attacks. A pseudo-random spreading sequence is used to multiplex the signals for transmission. This sequence is known to both the transmitter and the receiver and without it the attacker cannot jam the communication channel. This method is widely used in mobile networks \cite{dinan1998spreadingcodesmobilenetworks}.
However, code spreading  has high design complexity
and energy consumption, thus limiting its usage in energy limited scenarios like wireless sensor networks.

\subsubsection{Defenses Against PUEA}
In its report, the FCC \cite{FCC2002} states that: "No modification to the incumbent signal should be required to accommodate opportunistic use of the spectrum by SUs". This restriction should be followed when designing security mechanisms to defend against PUEA or any other attack specific to the cognitive radio setup. 

For PUEA\index{Primary User Emulation Attacker|textbf}, few defense mechanisms assume that the
location of the PU is known. Those mechanisms are called location-based defense mechanisms. We follow this classification of the defense mechanisms which has been introduced in \cite{fragkiadakis2013survey}.\\

\textit{Location-based defense mechanisms against PUEA} \\

In \cite{chen2008defensePUEA}, the authors utilize two pieces of information to develop their defense protocol: the location of the PU transmitter and the Received Signal Strength (RSS). The RSS information is collected by a separate wireless sensor network. The defense scheme\index{defense scheme|textbf} consists of three phases: first, it verifies if the signal characteristics are similar to those of the PU or not, then it tests the received signal energy based on the location information, and last, it tests the localization of the signal transmitter. A transmitter who does not pass any of these three phases will be considered as PUEA and will be omitted. The drawbacks of this method are: first, the location information about the PU is not always available, especially in small networks, and second, is the RSS may have large fluctuations even within small area networks.

In \cite{jin2009detectingPUEA}, Fenton's approximation and Sequential Probability Ratio Test\index{sequential probability ratio test|textbf} (SPRT) are used to analytically model the received power at the SUs. The SUs compare the received power in a band of interest to a threshold. If the power is below the threshold the band is considered to be free. On the other hand, if the band is tagged as occupied the SUs test whether the detected signal source was a legitimate PU or a PUEA. Based on the assumption that the attackers and the SUs are uniformly distributed, two statistical formulations are proposed to model the Probability Density Function (PDF) of the received power at the SU from a legitimate PU, and the PDF of the received power at the SU from malicious users. Then, the defense mechanism at the SUs tests the two PDFs using an SPRT by performing a binary hypothesis test between $H_0$ which means that the signal comes from legitimate PU, and $H_1$ according to which the signal comes from a PUEA. In this setup, several malicious users can be present in the network and the authors show that when the attackers are too close to the SUs, the false alarm and missed detection probabilities are maximized because the total received power from all the attackers is larger than the received power from the legitimate PU. A drawback of this work is the possibility of an endless loop of the SPRT that leads to very long sensing times. This work is extended in \cite{jin2009mitigatingPUEA} where the authors use Neyman Pearson composite hypothesis testing to solve the endless loop problem of SPRT. \\

Other defense proposals based on localization algorithms use the Time of Arrival (TOA), Time Difference of Arrival (TDOA), and Angle of Arrival (AOA) in order to distinguish between a  PU legitimate signal and PUEA \cite{wyglinski2009cognitive}. 
In TOA, SUs receive signals from satellites that contain their location and time information. Based on
this information, the node can calculate its own position and estimates the PU position.
TDOA \cite{dogancay2005closedTDOA} is a passive localization technique that uses the difference between the arrival times of signals transmitted by a PU but does not know the signal transmission time. TDOA measures the time differences at several receivers with known locations and computes a location estimate of the PU that permits to distinguish between a legitimate and a malicious behavior.
In the AOA technique, an SU measures the angle of arrival of the signal from two or more other SUs. If the locations of the other SUs are known, the receiver can compute its own location using triangulation \cite{niculescu2001AOS}. By using the same method, the AOA information at multiple SUs is used to determine the PU location and hence, help to distinguish between the legitimate PU and PUEA.
All of these techniques fail when the PUEA is too close to the PU when knowing the PU location gives no benefit. \\

\textit{Defense mechanisms not based on location} \\

In \cite{liu2010authenticatingPUsignal}, the authors use the channel impulse response, referred to as "link signature", to determine whether a PU transmitter changes its location or not. They propose the use of a "helper node" located in a fixed position very close to the PU. This node communicates with SUs to help them to verify the PU signals. The SUs do so by verifying the cryptographic link signatures\index{cryptographic link signatures|textbf} carried out by the helper node which communicates with SUs only when there is no PU transmission.
For this reason, the helper node has to sense the PU transmissions and also to differentiate PU signals from PUEA signals. The helper node authenticates the legitimate PU using the first and the second multipath components of the received PU signal at its interface. Then, the helper node\index{helper node|textbf} compares the ratio of the multipath components to a threshold, and if the ratio is above the threshold, it decides that the signal belongs to a legitimate PU, else that it is a PUEA.
Now, the SUs verify the PU transmission by computing the distance between the link signatures of the received signals and those sent by the helper node. If the distance is lower than a threshold, the received signal belongs to a legitimate PU, otherwise, it is a PUEA and it will be discarded.

Other proposals to defend against PUEA contradict with the FCC requirement since they try to modify the PU signal. Part of these proposals modify the PU signal to integrate into it a cryptographic signatures that permits the SU to verify the PU from the PUEA, like the work in \cite{mathur2007digitalsignaturesPUEA}, and other proposed authentication mechanisms between the PU transmitter and the SUs \cite{liu2010authenticatingPUsignal}.

\subsection{Defenses against Attacks to Sensors}
In this section, we present the proposed countermeasures\index{countermeasures|textbf} to defend against the {\em Byzantine} attacks and the Spectrum Sensing Data Falsification\index{Spectrum Sensing Data Falsification Attack|textbf} (SSDF) attack in cognitive radio networks.

As mentioned earlier in Chapter \ref{chapter:DF}, various information types can be provided by the sensors at different levels of abstraction. In this section, we consider simplest case in which the sensors send binary decisions about the phenomenon, namely, a bit $0$ under $H_0$ and $1$ otherwise. We consider this case since it is the most relevant for the rest of the thesis.  

In the absence of Byzantines\index{Byzantines|textbf}, the Bayesian optimal fusion rule has been derived in \cite{OptFusion,Var97} and it is known as Chair-Varshney rule\index{Chair-Varshney rule|textbf}. If the local error probabilities $(P_{MD}, P_{FA})$ are symmetric and equal across the sensor network, Chair-Varshney rule boils down to simple majority-based decision. 

In the presence of Byzantines, Chair-Varshney rule requires the knowledge of Byzantines' positions in the binary vector submitted to the FC along with the flipping probability $P_{mal}$ \footnote{The flipping probability is the probability that the attacker flips its binary local decision about the system state before sending it to the FC.}. Since this information is rarely available, the FC may resort to a suboptimal fusion strategy. 

An overview of the literature about distributed detection\index{distributed detection|textbf} and estimation in the presence of the Byzantines\index{Byzantines|textbf} is given in \cite{Vemp13}. The authors introduce the concept of {\em critical power} of Byzantines ($\alpha_{blind}$) defined as the fraction of Byzantines that makes the decision at the FC no better than flipping a coin. In \cite{WSNMatta}, by adopting a Neyman-Pearson\index{Neyman-Pearson|textbf} setup and assuming that the byzantine nodes know the true state of the system, the asymptotic performance - as a function of the network size $n$ - obtainable by the FC is analyzed as a function of the percentage of Byzantines in the network. By formalizing the attack problem as the minimization of the Kullback-Leibler Distance\index{Kullback-Leibler Distance|textbf} (KLD) \cite{CandT} between the information reports received by the FC under the two hypotheses $H_0$ and $H_1$, the blinding percentage, that is, the percentage of Byzantines in the network that makes the FC blind, is determined and shown to be - at least asymptotically - always equal to $50 \%$. This means that unless more than half of the sensors are {\em Byzantines}, asymptotically, the FC can provide reliable detection and decision.

By observing the system state over a longer observation window, the FC improves the estimation of the sequence of system states by gathering a number of reports provided by the sensors before making a global decision. In cooperative spectrum sensing,
for instance, this corresponds to collectively decide about the vacant spectrum bands over a time window, or, more realistically, at different frequency slots. The advantage of deciding over a sequence of states rather than on each single state separately, is that in such a way it is possible for the FC to understand which are the byzantine nodes and discard the corresponding observations (such an operation
is usually referred to as {\em Byzantine isolation})\index{Byzantine isolation|textbf}. Such a strategy is adopted in \cite{Raw11}, where the analysis of \cite{WSNMatta} is extended to a situation in which the Byzantines do not know the true state of the system. Byzantine isolation is achieved by counting the mismatches between the reports received
from each sensor and the global decision made by the FC. 
In order to cope with the lack of knowledge about the strategy adopted by the Byzantines, the decision fusion problem is casted into a game-theoretic formulation,
where each party makes the best choice without knowing the strategy adopted by the other party.

A slightly different approach is adopted in \cite{LearnByzantines}. By assuming that the FC is
able to derive the statistics of the reports\index{reports|textbf} submitted by honest sensors\index{sensors|textbf}, Byzantine isolation is carried out whenever the reports received from a node deviate from the expected statistics. In this way, a correct decision can be made also when the percentage of Byzantines exceeds $50 \%$. The limit of this approach is that it does not work when the reports sent by the Byzantines have the same statistics of those transmitted by the honest nodes. This is the case, for instance, in a perfectly symmetric setup with equiprobable system states, symmetric local error probabilities, and an attack strategy consisting of simple decision flipping.

\subsubsection{Defenses to SSDF in Cognitive Radio Networks}
The Byzantine attack in a cognitive radio context is known as SSDF attack. In this part, we present the most common countermeasures\index{countermeasures|textbf} against these attacks. In the  scenario considered here, the group of SUs send {\em binary} decisions about PU activity to the FC which, in turn, decides about the spectrum occupancy by fusing the decisions using an information fusion rule. In some of these works, when the SUs are not trusted a priori, a trust or reputation metric is assigned to each SU in the network depending on its behavior. 

In \cite{wang2009attackproof} the proposed scheme calculates {\em trust values} for SUs based on their past reports. This metric can become unstable if no attackers are present in the network or there are not enough reports. For this reason, the authors also compute a {\em consistency value} for each SU. If the consistency value and the trust value fall below certain thresholds, the SU is identified as a Byzantine and its reports are not considered in the fusion rule. The authors evaluate the proposed scheme using two fusion rules, namely, the OR rule and the $2$-out-of-$n$ rule. A drawback of this work is that only {\em one adversary} is considered in the evaluation.

The authors in \cite{RawatConf} use a {\em reputation metric} to detect and isolate attackers from honest SUs. This metric is computed by comparing the report of each SU to the final decision made at the FC. The metric increments by one if the report and the final decision mismatch (more reliable SUs have low metric values). If the reputation metric of an SU exceeds a predefined threshold, its reports are isolated and not used in the fusion rule. By adopting the majority voting as the fusion rule, the authors show that when the percentage of attackers in the network is below $40 \%$, the probability of isolating the attackers can exceed $95 \%$, while the isolation probability of the honest SUs is very near to zero. This defense scheme is similar to \cite{wang2009attackproof} with the difference that here the authors did not restore the reputation metric if an SU is temporary misbehaving and thus, \cite{wang2009attackproof} is considered to be a more fair approach.

Weighted Sequential Probability Ratio Test\index{Weighted Sequential Probability Ratio Test|textbf} (WSPRT) as a modified version of the SPRT test is proposed in \cite{tolerant_scheme} to assign a weight $w_i$ to each SU in the network as follows:
\begin{equation}
\Lambda(\mathbf{u}) = \prod_{i=1}^n \Big(\frac{P(u_i|H_1)}{P(u_i|H_0)}\Big)^{w_i} .
\label{eq.WSPRT_FC}
\end{equation}
In WSPRT shown in Equation (\refeq{eq.WSPRT_FC}), the FC computes the product of the likelihood ratios for each decision provided by the SUs. Based on the likelihood value for each SU report, the FC assigns to it a weight $w_i$ which value its contribution in the final decision made at the FC. If the report of the SU matches with the final decision at the FC, its reputation metric $w_i$ is increased by one, otherwise it is decreased. In this work, the reputation metric of a misbehaving SU can be restored to zero just after a few instants if it starts behaving correctly again. Each SU implements an SPRT and decides between two hypothesis $H_1$ for PU presence and $H_0$ for its absence, by comparing the local spectrum sensing measurement with two predefined thresholds. If the output falls between these thresholds, no decision is made and the SU takes a new sample. For the simulation, the authors assume two constant strategies for the adversary, namely, the "always true" SSDF attackers, which always reports a vacant spectrum, and the "always false" SSDF attackers, that reports always the spectrum as occupied. Furthermore, eight different information fusion rules are considered at the FC: AND, OR, Majority, SPRT, WSPRT and LRT with three different thresholds. For the "always-false" attack the simulation results show that for all fusion rules, except the OR and AND rules, the correct detection ratio decreases as the number of the attackers increases. For the other two rules, the correct sensing ratio does not significantly change, but it is lower than the other rules. For the "always-true" case, the results show that the performance of the majority rule decreases significantly as the number of attackers increase, which means that this rule is more vulnerable to this type of attack than the other rules.

In \cite{noon2010defendinghitrun}, the authors propose an intelligent attack called the "hit-and-run" attack. By knowing the fusion technique used by the FC, this attacker alternates between honest and lying modes. The attacker estimates its own {\em suspicious level} and as long as it is below a threshold $h$, it reports falsified decisions. If the suspicious level falls below a threshold, it switches to honest mode. On the defender side, when the suspicious level of an SU becomes larger than $h$, a point is assigned to this user. When the cumulative points exceed a predefined threshold, the reports of this SU are ignored permanently. Simulations show that the scheme achieves good performance for up to three attackers.
A drawback of this method is that a user is permanently
removed from a CRN if it collects enough points then, some honest but temporarily misbehaving SUs can be permanently removed from the network. In addition, the authors assume a non-realistic scenario in which the adversary knows the reports of the other SUs.

In \cite{SS2010}, a Double-Sided Neighbor Distance (DSND) algorithm is used for the detection of the attackers. An SU is characterized as an adversary if its reports to the FC are too far or too close to the reports sent by other SUs. Two attack models are considered: the independent attack, where an adversary does not know the reports of the honest SUs, and the dependent attack, where the adversary is aware the reports of the others. The results show that, in the case of the independent attack, the adversary can always be detected when the number of spectrum sensing iterations tends to infinity. For the dependent attack, the adversary can avoid been detected if it has accurate information about the missed detection and false alarm probabilities and then follows them in his attacking strategy.

\subsection{Defenses Against Attacks to Consensus Algorithm}
In this section, we present some attempts to address the security threats to consensus algorithms\index{consensus algorithm|textbf} in distributed sensor networks.

In \cite{yu2009defense}, a defense scheme against data falsification attack\index{data falsification attack|textbf} is proposed. By assuming that the attacker behavior is static and injecting constant falsified measurements, the scheme eliminates the state update with the largest deviation from the local mean among all the received state updates from the neighboring nodes. The drawback of this scheme is that it can only deal with the situation in which only one Byzantine node\index{byzantine nodes|textbf} exists and it excludes one state value even if there are no Byzantines in the network. Another drawback is that the scheme can cause unidirectional information exchange in the network. This work is extended in \cite{tang2012consensusSec} to enhance the security of the consensus algorithm by adding an authentication technique for the nodes prior to joining them to the consensus mechanism. This technique uses ID-based cryptography with threshold secret sharing and it is implemented prior to filtering the state with maximum deviation from the mean.

In \cite{yan2012vulnerabilityconsensus}, the vulnerability of distributed consensus-based spectrum sensing is analyzed and an adversary detection algorithm with an adaptive threshold is proposed. The authors propose a novel type of attack called "Covert Adaptive Data Injection Attack". By "covert" they mean that the attacker is willing to inject false data without being detected. On the other hand, "adaptive" means that the attacker uses the knowledge of the detection algorithm, and adapts its strategy based on neighbors' state update information. The defensed method developed is based on a specific model for power propagation and hence, it restricts its application.  

The authors in \cite{liu2012adaptivedeviation} propose a Byzantine mitigation technique based on adaptive local thresholds. This threshold is updated at each consensus iteration, and is used to classify the state updates received by the neighboring nodes between honest nodes and Byzantines. Based on this classification, the scheme modifies the consensus algorithm in such a way to introduce a reducing factor to the state update phase of the algorithm. The reducing factor assigned to the nodes classified as Byzantines mitigates their contributions to the state update phase and in the same time, tolerates the occasional large state update deviations of honest users. The reducing factor will eventually isolates the Byzantines state update from the rest of the network.

Consensus disruption attacks\index{consensus disruption attacks|textbf} are easier to detect
because of their nature. The identification of consensus disruption attackers has been studied in the literature of "resilient-consensus" and control theoretic techniques were developed to identify disruption attackers in a single consensus iteration \cite{pasqualetti2012consensus,sundaram2011consensus}.

\section{Conclusion}
In this chapter we have presented the most common attacks and countermeasures in different adversarial setups. We started by considering the centralized version of the problem and reviewed the attacks and defenses in several adversarial versions. In addition, we provided real-life examples of attacks and some specific mitigations for each adversarial setup. Then, we presented the attacks and defenses in decentralized consensus algorithm. 

In the rest of the thesis, we will focus on information fusion in distributed sensor network in which some of the sensors presented in the network are {\em Byzantines}. In addition to that, we will specifically consider the case in which the sensors report binary decisions to the FC. For the consensus algorithm, we will focus on the case of data falsification attack where the sensors falsify their measurements prior to any information exchange. 

\part{Adversarial Decision Fusion in The Presence of Byzantines}
\chapter{Soft Isolation Defense Mechanism Against Byzantines for Adversarial Decision Fusion}
\label{chapter:CDC}
\emph{"I'm not upset that you lied to me, I'm upset that from now on I can't believe you."}
\\
Friedrich Nietzsche

\emph{"Chaos isn't a pit. Chaos is a ladder. Many who try to climb it fail, and never get to try again — the fall breaks them. And some are given a chance to climb, but they refuse. They cling to the realm, or the gods, or love ... illusions. Only the ladder is real, the climb is all there is."}
\\
Lord Petyr Baelish, "Game of Thrones"
\section{Introduction}

\PARstart{\textcolor{red}I}n{} this chapter we address the problem of decision fusion in centralized distributed sensor networks\index{distributed sensor networks|textbf} in the presence of Byzantines. In the problem that we have considered, the fusion center is required to make a decision about the status of an observed system by relying on the information provided by the nodes. Decision fusion must be carried out in an adversarial setting, by taking into account the possibility that some of the nodes are Byzantines and they malevolently alter their reports to induce a decision error. Despite being the simplest kind of attack, this case contains all the ingredients of more complex situations, hence its analysis is very instructive and already provides interesting insights into the achievable performance of decision fusion\index{decision fusion|textbf} in distributed sensor networks under adversarial conditions.

\begin{figure}[t!]
\centering\includegraphics[width=\columnwidth]{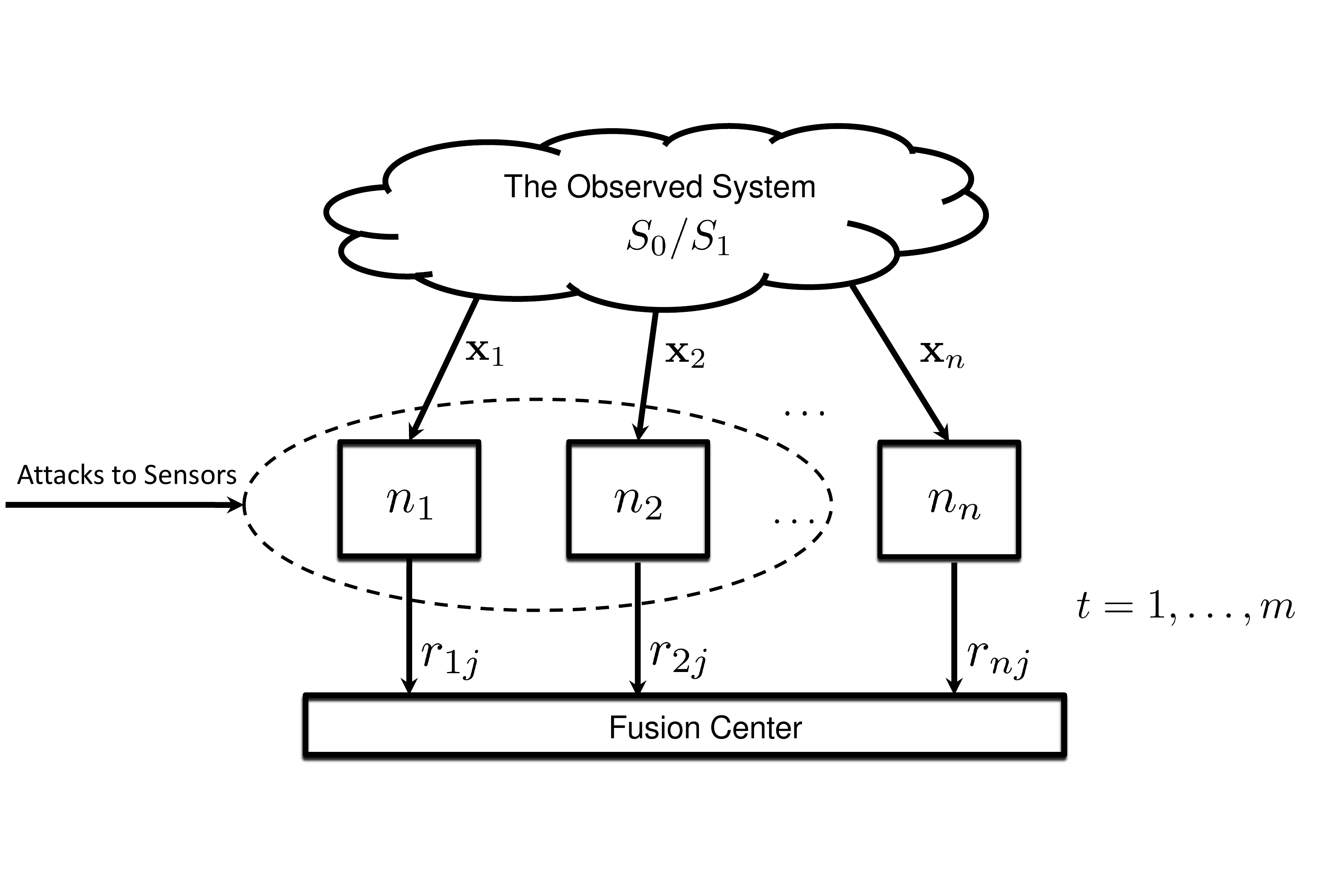}
\caption{\textit{Decision fusion under adversarial conditions.}}
\label{fig.ADVscheme}
\end{figure}

A graphical representation of the problem studied in this chapter is given in Figure \reffig{fig.ADVscheme}. The $n$ nodes of the distributed sensor network observe a system through the vectors ${\bf x}_1, {\bf x}_2 \dots {\bf x}_n$. Based on such vectors, the nodes compute $n$ reports, say ${\bf r}_1, {\bf r}_2 \dots {\bf r}_n$ and send them to a fusion center. The fusion center gathers all the reports and makes a final decision about the state of the observed system. We assume that the system can be only in two states $S_0$ and $S_1$. Additionally, the reports correspond to local decisions on the system status made by the nodes, i.e. the reports are binary values and $r_i \in \{0,1\}$ for all $i$.%

In the adopted setup, the byzantine nodes\index{byzantine nodes|textbf} do not know the true state of the system and act by flipping the local decisions with a certain probability $P_{mal}$. The fusion center first tries to understand which are the byzantine nodes and then makes a decision by discarding the suspect nodes. 

With the above ideas in mind, the goal of this chapter is twofold. First of all, it introduces a soft identification strategy whereby the fusion center\index{Fusion Center|textbf} can isolate the byzantine nodes\index{byzantine nodes|textbf} from the honest ones. Then, we introduce a game-theoretic formulation of the decision fusion problem with attacked nodes, thus providing a rigorous framework to evaluate the performance achievable by the fusion center and the Byzantines, when both of them play at the equilibrium\index{equilibrium|textbf}. The game-theoretic approach is used to compare the fusion strategy described in this chapter with the one presented in \cite{Raw11}. Finally, we demonstrate the superior performance of the soft identification scheme by means of numerical simulations.

\section{Decision Fusion with Isolation of Byzantines}
\label{sec.isolation_scheme}

\subsection{Problem formulation}

In the scenario described above, the Fusion Center (FC) uses {\em Byzantine isolation}\index{Byzantine isolation|textbf} strategy described in chapter \ref{chapter:SecurityThreats}. In the following we give an exact formulation of such an approach.

As we said, we are considering the case of binary reports. Specifically, each node makes a local decision\index{local decision|textbf} about the state of the observed system and forwards its one-bit decision to FC, which must decide  between hypothesis $H_0$ and hypothesis $H_1$. We assume that a fixed fraction $\alpha$ of the $n$ nodes (or links between the nodes and the FC) is under the control of byzantine attackers which, in order to make the information fusion process fail, corrupt the reports by flipping the one-bit local decisions with probability $P_{mal}$ (as in \cite{Raw11,KBKV13}, we assume a symmetric attacking strategy). Under this assumption, the probability that a node is Byzantine weakly depends on the state of the other nodes when the network size is large enough. By referring to Figure \reffig{fig.ADVscheme}, the above attack corresponds to the insertion of a binary symmetric channel with crossover probability $P_{mal}$ in the attacked links.

The strategy adopted by the fusion center consists in trying to identify the attacked nodes and remove the corresponding reports from the fusion process. To do so, the FC observes the decisions taken by the nodes over a time period $m$, and makes the final decision on the state of the system at each instant $j$ only at the end of $m$. To elaborate, for each instant $j$, we indicate the reports received from the nodes as $r_j^n = (r_{1j}, r_{2j},...,r_{nj})$ where $r_{ij} \in \{0,1\}$. The fusion center applies an $l$-out-of-$n$ fusion rule\footnote{In other words, the fusion center decides in favor of $H_1$ if $l$ out $n$ nodes decided for such a hypothesis.} to $r_j^n$ to make an intermediate decision on the status of the system at time $j$. Let us indicate such a decision as $d_{int}(j)$. The local decisions made by the $i$-th node over the time window $m$, are denoted as $\textbf{u}_i = \left(u_{i1},u_{i2},\dots,u_{im}\right)$ with $u_{ij}$ as the local decision\index{local decision|textbf} at instant $j \in \{1,\dots,m\}$. The relationship between $u_{ij}$ and the status of the system at time $j$ is ruled by the following equations, which take into account the probability of a decision error by the local node:
\begin{align}
\label{eq1}
    P\left(u_{ij} = 1 | H_1 \right) & = P_{d_{i}}\\
    P\left(u_{ij} = 1 | H_0 \right) & = P_{fa_{i}},
\end{align}
where $P_{d_{i}}$ and $P_{fa_{i}}$ are, respectively, the probability of correct detection and false alarm for node $i$. In the following, we assume that the states assumed by the system over subsequent instants are independent of each other. Errors at different nodes and different times are also assumed to be independent.

By assuming that transmission takes place over error-free channels, for honest nodes we have $r_{ij} = u_{ij}$, while for the byzantine nodes\index{byzantine nodes|textbf} we have $r_{ij} \neq u_{ij}$ with probability $P_{mal}$. Then, for the Byzantine reports we have:
\begin{align}
P(r_{ij} = 1| H_1) & = P_{mal} (1 - P_{d_i}) + (1-P_{mal})P_{d_i} \label{P_d^B},\\
P(r_{ij} = 1| H_0) & = P_{mal}(1 - P_{{fa}_i}) + (1 - P_{mal})P_{{fa}_i} \label{P_fa^B}.
\end{align}
Given the observation vector $r_j^n$ for each $j$ ($j=1,..,m$), in order to remove the fake reports from the data fusion process, the FC proceeds as follows: it associates to each node $i$ a {\em reputation score} $\Gamma_i$, based on the consistency of the reports received from that node with the intermediate decisions $d_{int}(j)$ over the entire time window $m$. Then, the FC isolates the nodes whose reputation is lower than a threshold $\eta$ and decides about the system state\index{system state|textbf} by fusing only the remaining reports.

\subsection{Byzantine Identification: hard reputation measure}
\label{sec.GI_hard}

As described in chapter \ref{chapter:SecurityThreats}, in the identification scheme proposed in \cite{Raw11}, the FC computes for each node $i$ a reputation score by counting the number of times that the reports received from that node are different from
the intermediate decisions $d_{int}(j)$ during the observation window $m$. The {\em reputation score} $\Gamma_{H,i}$ is hence defined as $\Gamma_{H,i} = \sum_{i=1}^m \mathcal{I}(r_{ij} = d_{int}(j))$ where $\mathcal{I}(x)$ (indicator function) is equal to 1 when its argument its true and 0 otherwise. Accordingly, the nodes whose reputation is lower than a threshold $\eta$ are removed from the fusion process. For each $j$, the final decision is taken by relying on an $l'$-out-of-$n'$ rule, where $l'$ is the final decision threshold and $n'$ is the number of nodes remaining after that the thought-to-be byzantine nodes\index{byzantine nodes|textbf} have been discarded.

In \cite{Raw11}, the above scheme is shown to be able to mitigate the effect of byzantine attacks when $\alpha < 0.5$, a situation in which the Byzantines\index{Byzantines|textbf} are not able to blind the FC by attacking the network independently (referred to as Independent Malicious Byzantine Attacks (IMBA) in \cite{Raw11}), which is the only case considered in this chapter.

\section{Decision Fusion with Soft Identification of Malicious Nodes}
\label{sec.soft_isolation}

In this section, we propose an isolation\index{isolation|textbf} strategy which removes the Byzantines\index{Byzantines|textbf} from the network according to a soft \footnote{We point out that our method is soft with regards to the identification of the Byzantines, but is not used in the final decision step.} reliability measure.
For any instant $j$ and given the vector $r_j^n$ with the reports, the new isolation strategy relies on the estimation of the following probabilities:
\begin{align}
\label{eqmauro}
    P\left(u_i(t) = 1, r_j^n\right),\\ \nonumber
    P\left(u_i(t) = 0, r_j^n\right).
\end{align}
For a honest node, in fact, such probabilities are very different from each other, since the expression for which $r_{ij} = u_{ij}$ is close to 1, while the other is very close to 0. On the contrary, for a byzantine node, the above probabilities tend to be closer. For this reason, we propose to measure the reputation score of a node as follows. For each $j$ we first compute:
\begin{eqnarray}
\label{eq10}
    R_{ij} = \left|\log\left[\frac{P\left(u_{ij} = 0, r_j^n\right)}{P\left(u_{ij} = 1, r_j^n\right)}\right]\right|,
\end{eqnarray}
that is the absolute value of the log-ratios of the two probabilities. Then we set:
\begin{eqnarray}
\label{eq10.1}
    \Gamma_{S,i} = \sum\limits_{i = 1}^{m} R_{ij}.
\end{eqnarray}
To evaluate (\ref{eq10}), we start rewriting the joint probabilities within the log as follows (for notation simplicity, we omit the index $j$):
\begin{equation}
\label{eq5}
P\left(u_i , r^n\right) = P\left(r^n | u_i, H_0\right)P\left(u_i, H_0\right) + P\left(r^n | u_i, H_1\right) P\left(u_i, H_1\right).
\end{equation}
To proceed, we make the simplifying assumptions that the reports received by the FC from different nodes are conditionally independent\footnote{That is they are independent when conditioning to $H_0$ or $H_1$.}. This is only approximately true since in our scenario we operate under a fixed number of Byzantines, and then the probability that a node is Byzantine depends (weakly) on the state of the other nodes when their number is large enough. Such dependence decreases when the number of nodes increases and disappears asymptotically due to the law of large numbers.

Let us now consider the quantity $P(r_{j'} | u_i, H_0)$. When $i=j'$, we can omit the conditioning to $H_0$ since $r_i$ depends on the system status only through $u_i$. On the other hand, when $i \ne j'$, we can omit conditioning to $u_i$, due to the conditional independence of node reports. A similar observation holds under $H_1$. Then we can write:
\begin{align}
\label{eq6}
    P\left(u_i , r^n\right) & =  P\left(r_i| u_i\right) \bigg\{P\left(u_i|H_0\right)P\left(H_0\right)\prod\limits_{j' \ne i}P\left(r_{j'} | H_0\right) \nonumber \\
    & + P\left(u_i|H_1\right)P\left(H_1\right)\prod\limits_{j' \ne i}P\left(r_{j'} | H_1\right)\bigg\},
\end{align}

where $P\left(r_i | u_i\right) = (1-\alpha P_{mal})$ if $r_i = u_i$, and $\alpha P_{mal}$, otherwise. Moreover, we have $P(u_{j'} = 1 | H_1) = P_{d_{j'}}$ and $P(u_{j'} = 1 | H_0) = P_{fa_{j'}}$. In addition:
\begin{equation}
  P\left(r_{j'} | H_0\right) = (1-\alpha P_{mal}) P\left(u_{j'} = r_{j'} | H_0 \right) 
+ \alpha P_{mal} P\left(u_{j'} \ne r_j | H_0 \right),
\end{equation}
\begin{equation}
P\left(r_{j'} | H_1\right) = (1-\alpha P_{mal}) P\left(u_{j'} = r_{j'} | H_1 \right) + \alpha P_{mal} P\left(u_{j'} \ne r_{j'} | H_1 \right).
\end{equation}
By inserting the above expressions in (\ref{eq5}) and (\ref{eq10}), we can compute the soft reputation score $\Gamma_{S,i}$.
Then, the FC relies on $\Gamma_{S,i}$ to distinguish honest nodes from byzantine ones. Specifically, the distinction is made by isolating those nodes whose reputation score $\Gamma_{S,i}$ is lower than a threshold $\eta$  (hereafter, we will set $P_{{fa}_i} = P_{fa}$ and $P_{d_i}=P_d$ $\forall i$).

We conclude this section by observing that, strictly speaking, FC is required to know $\alpha$ and the flipping probability $P_{mal}$. With regard to $\alpha$, we assume that FC knows it. As to $P_{mal}$, in the next sections, we will see that choosing $P_{mal} = 1$ is always the optimum strategy for the attackers, and hence FC can assume that $P_{mal} = 1$.

\enlargethispage{\baselineskip}

\section{A Game-Theoretical Approach to the Decision Fusion Problem}

In this section, we evaluate the performance achieved by using the \index{soft Byzantine isolation|textbf} strategy introduced in the previous section and compare it with the hard identification strategy described in \cite{Raw11}. To do so, we use a game-theoretic approach in such a way to analyze the interplay between the choices made by the attackers and the fusion center.

\subsection{The Decision Fusion Game: definition}

In the scenario presented in this chapter, the FC is given the possibility of setting the local sensor threshold for the hypothesis testing problem at the nodes and the fusion rule\index{fusion rule|textbf}, while the Byzantines\index{Byzantines|textbf} can choose the flipping probability $P_{mal}$.

With respect to \cite{KBKV13}, we study a more general version of the decision fusion game\index{decision fusion game|textbf} which includes the isolation scheme described in Section \ref{sec.isolation_scheme}. To this purpose, the FC is endowed with the possibility of setting the isolation\index{isolation|textbf} threshold $\eta$, as well as the final fusion rule after removal of byzantine nodes. Finally, the performance are evaluated in terms of overall error probability after the removal step. We suppose that the FC does not act strategically on the local sensor threshold; then $P_d$ and $P_{fa}$ are fixed and known to FC. With regard to the Byzantines\index{Byzantines|textbf} (B), they are free to decide the flipping probability $P_{mal}$.

With the above ideas in mind, we define the general decision fusion game\index{decision fusion game|textbf} as follows:
\begin{definition}
The $DF (\mathcal{S}_{FC}, \mathcal{S}_B, v)$ game is a zero-sum strategic game\index{zero-sum game|textbf}, played by the FC and B, defined by the following strategies and payoff.
\begin{itemize}
   \item The set of strategies available to the FC is given by all the possible isolation thresholds $\eta$, and the values of $l$ and $l'$ in the $l$-out-of-$n$ intermediate and final decision rules:
   \begin{equation}
   \mathcal{S}_{FC} = ~ \{(l, \eta, l') \text{;  }  l, l' = 1,..,k \text{,  } \eta_{\min} \le \eta \le \eta_{\max}\},
   \label{S_FC}
   \end{equation}
   where $\eta_{\min}$ and $\eta_{\max}$ depend on the adopted isolation scheme.
   \item The set of strategies for B are all the possible flipping probabilities:
   \begin{equation}
   \mathcal{S}_{B} = \{P_{mal} \text{, } 0 \le P_{mal} \le 1\}.
   \label{S_B}
   \end{equation}
   \item The payoff\index{payoff|textbf} $v$ is the final error probability after malicious node removal, namely $P_{e,ar}$. Of course, the FC wants to minimize $P_{e,ar}$, while B tries to maximize it.
 \end{itemize}
\end{definition}

Applying the above definition to the identification schemes introduced so far, we see that for the case of hard reputation measure ($DF_H$ game),
the values of the isolation threshold $\eta$ range in the set of integers from $0$ to $m$, while for the scheme based on the soft removal of the malicious nodes ($DF_S$ game) $\eta$ may take all the continuous values between $\eta_{\min} = \min_{i=1,..,n} R_{ij}$ and $\eta_{\max} = \max_{i=1,..,n} R_{ij}$.

\subsection{The Decision Fusion Game: equilibrium point}

With regard to the optimum choice for the Byzantines, previous works have either conjectured or demonstrated (in particular cases) that $P_{mal} = 1$ is a dominant strategy \cite{Raw11,KBKV13}. Even in our case, the simulations we carried out, some of which are described in the next section, confirms that $P_{mal}=1$ is indeed a dominant strategy for both the hard and the soft identification schemes. This means that, notwithstanding the introduction of an identification scheme for discarding the reports of malicious nodes from the fusion process, the optimum for the Byzantines is (still) always flipping the local decisions before transmitting them to the FC.
This means that for the Byzantines\index{Byzantines|textbf} it is better to use all their power ($P_{mal} = 1$) in order to make the intermediate decision fail than to use a lower $P_{mal}$ to avoid being identified.
As a consequence of the existence of a dominant strategy for $B$, the optimum strategy for FC is the triple ($l^*, \eta^*$, ${l'}^{*}$) which minimizes $P_{e,ar}$ when $P_{mal} = 1$. By exploiting a result derived in \cite{Var97} for the classical decision fusion problem and later adopted in \cite{KBKV13} in presence of Byzantines, the optimal value\index{optimal value|textbf} $l^*$ determining the intermediate fusion rule is given by
\begin{equation}
\label{optimum l}
l^* = \frac{\ln\left[(P(H_0)/P(H_1))\{(1 - p_{10})/(1 - p_{11})\}^n\right]}{\ln\left[\{p_{11}(1 - p_{10})\}/\{p_{10}(1 - p_{11})\}\right]},
\end{equation}
where $P(H_0)$ and $P(H_1)$ are the a-priori probabilities of $H_0$ and $H_1$, while $p_{10} = p(r=1|H_0)$ and $p_{11} = p(r=1|H_1)$, evaluated for $P_{mal} = 1$. With regard to $\eta$ and $l'$, we have:
\begin{equation}
\label{optimum_eta_l}
(\eta^*, l'^*) = \arg\min_{(\eta, l')} P_{e,ar}((l^*,\eta, l'),P_{mal}=1).
\end{equation}
Depending on the adopted isolation scheme, we have a different expression for $P_{e,ar}$ and then different $\eta^*$'s and $l'^*$'s as well. The minimization problem in (\ref{optimum_eta_l}) is solved numerically for both hard and soft isolation in the next section.
According to the previous analysis, ($(l^*,\eta^*, l'^*),P_{mal}^*$) is the only {\em rationalizable equilibrium} for the $DF$ game, thus ensuring that any rational player will surely choose these strategies. The value of $P_{e,ar}$ at the equilibrium represents the achievable performance for FC and is used to compare the effectiveness of data fusion based on soft and hard Byzantine isolation\index{Byzantine isolation|textbf}.

\section{Performance Analysis}
\label{sec. Analysis_Performance}

We now evaluate the performance at the equilibrium for the two games $DF_H$ and $DF_S$, showing that the soft strategy outperforms the one proposed in \cite{Raw11}, in terms of $P_{e,ar}$. We also give a comparison of the two schemes in terms of isolation error probability.

In all our simulations, we consider a sensor network with $n=100$ nodes. We assume that the probability of the two states $S_0$ and $S_1$ are the same. We run the experiments with the following settings: $P_d = 1 - P_{fa}$ takes values in the set $\{0.8, 0.9\}$ and $\alpha \in [0.4, 0.49]$, corresponding to a number of honest nodes ranging from 51 to 60. The observation window $m$ is set to 4. For each setting, the probability of error $P_{e,ar}$ of the two schemes is estimated over 50000 simulations.

Due to the symmetry of the experimental setup with respect the two states, we have that $p_{10} = p_{01} = 1- p_{11}$. Accordingly, from (\ref{optimum l}) we get that $l^* = n/2$ and then the majority rule is optimal for any $P_{mal}$ (not only at the equilibrium). Besides, still as a consequence of the symmetric setup, the optimality of the majority rule is experimentally proved also for the final fusion rule, regardless of the values of $\eta$ and $P_{mal}$. Then, in order to ease the graphical representation of the game in normal form, we fix $l^* = 50$ and $l'^* = n'/2$ and remove these parameters from the strategies available to the FC.

Tables \reftab{tab.VasrhGame} and \reftab{tab.LLRGame} show the payoff matrix for the $DF_H$ and $DF_S$ games when $\alpha = 0.46$ and $P_d = 0.8$ (very similar results are obtained for different values of these parameters). For the $DF_S$ game, the threshold values are obtained from the reliability interval $[\eta_{S,\min}, \eta_{S,\max}]$. Since the reliability measures take different values for different $P_{mal}$, a large number of thresholds have been considered, however for sake of brevity, we show the results obtained with a rather coarse quantization interval, especially far from the equilibrium point.

\begin{table}
\centering
\small
\begin{tabular}{|c|c c c c c|}
\hline
$\eta_H$ / $P_{mal}$ & 0.6 & 0.7 & 0.8 & 0.9 & 1
\\ [0.5ex]
\hline
 4 & 0.0016	& 0.0087 &	0.0354	& 0.1109	& 0.2746\\
 3 &  0.0015 & 0.0078&	0.0262	& 0.06628	& {\bf 0.1982}\\
 2 & 0.0016 & 0.0080& 	0.0281 &	0.0726 &	0.1998\\
 1 & 0.0016 &	0.0087 &	0.0354 &	0.1109 &	0.2746\\
 0 & 0.0016	& 0.0087 &	0.0354 &	0.1109 &	0.2746\\
\hline
\end{tabular}
\caption{\textit{Payoff of the $DF_H$ game for $\alpha = 0.46$ and $P_{d} = 0.8$, $P_{fa} = 0.2$.}}
\normalsize
\label{tab.VasrhGame}
\vspace{-0.2cm}
\end{table}

\begin{table}
\centering
\small
\begin{tabular}{|c|c c c c c |}
\hline
$\eta_S$ / $P_{mal}$ & 0.6 & 0.7 & 0.8 & 0.9 & 1
\\ [0.5ex]
\hline
$\eta_{S,\min}$ &   0.0009  &  0.0035  &  0.0131&    0.0596    & 0.2253\\
$\cdot$  &    0.0009  &  0.0035  &  0.0131&    0.0596 &  0.1889\\
$\cdot$  &    0.0009  &  0.0035  &  0.0131&    0.0596    & 0.1589\\
$\cdot$  &   0.0009  &  0.0035  &  0.0131&    0.0596   &  0.1401\\
$\cdot$  &   0.0009  &  0.0035  &  0.0131&    0.0596    & 0.1405\\
$\cdot$  &   0.0009  &  0.0035  &  0.0131&    0.0596   &  {\bf 0.1375}\\
$\cdot$  &  0.0009  &  0.0035  &  0.0131&    0.0596     &0.1528\\
$\cdot$  &   0.0009  &  0.0035  &  0.0131&    0.0596    & 0.1801\\
$\cdot$  &  0.0009  &  0.0035  &  0.0131&    0.0596    & 0.2192\\
$\cdot$  &  0.0009  &  0.0035  &  0.0131&    0.0596   &  0.2742\\
$\cdot$  &    0.0009  &  0.0035  &  0.0131&    0.0361  &  0.2742\\
$\cdot$  &     0.0009  &  0.0035  &  0.0131&    0.0209  &  0.2742\\
$\cdot$  &    0.0009  &  0.0035  &  0.0131&    0.0586   &  0.2742\\
$\cdot$  &      0.0009  &  0.0035  &  0.0131&    0.1108  &  0.2742\\
$\cdot$  &     0.0009 &   0.0035  &  0.0088&   0.1108  &  0.2742\\
$\cdot$  &     0.0009 &   0.0035  &  0.0054&   0.1108  &  0.2742\\
$\cdot$  &     0.0008 &   0.0021  &  0.0355&   0.1108  &  0.2742\\
$\eta_{S,\max}$ &     0.0006 &   0.0011  &  0.0355&   0.1108  &  0.2742\\
\hline
\end{tabular}
\caption{\textit{Payoff of the $DF_S$ game for $\alpha = 0.46$ and $P_{d} = 0.8$, $P_{fa} = 0.2$.}}
\label{tab.LLRGame}
\vspace{-0.5cm}
\end{table}

\begin{figure}[h!]
\begin{center}
\subfloat[]{\includegraphics[width=\columnwidth]{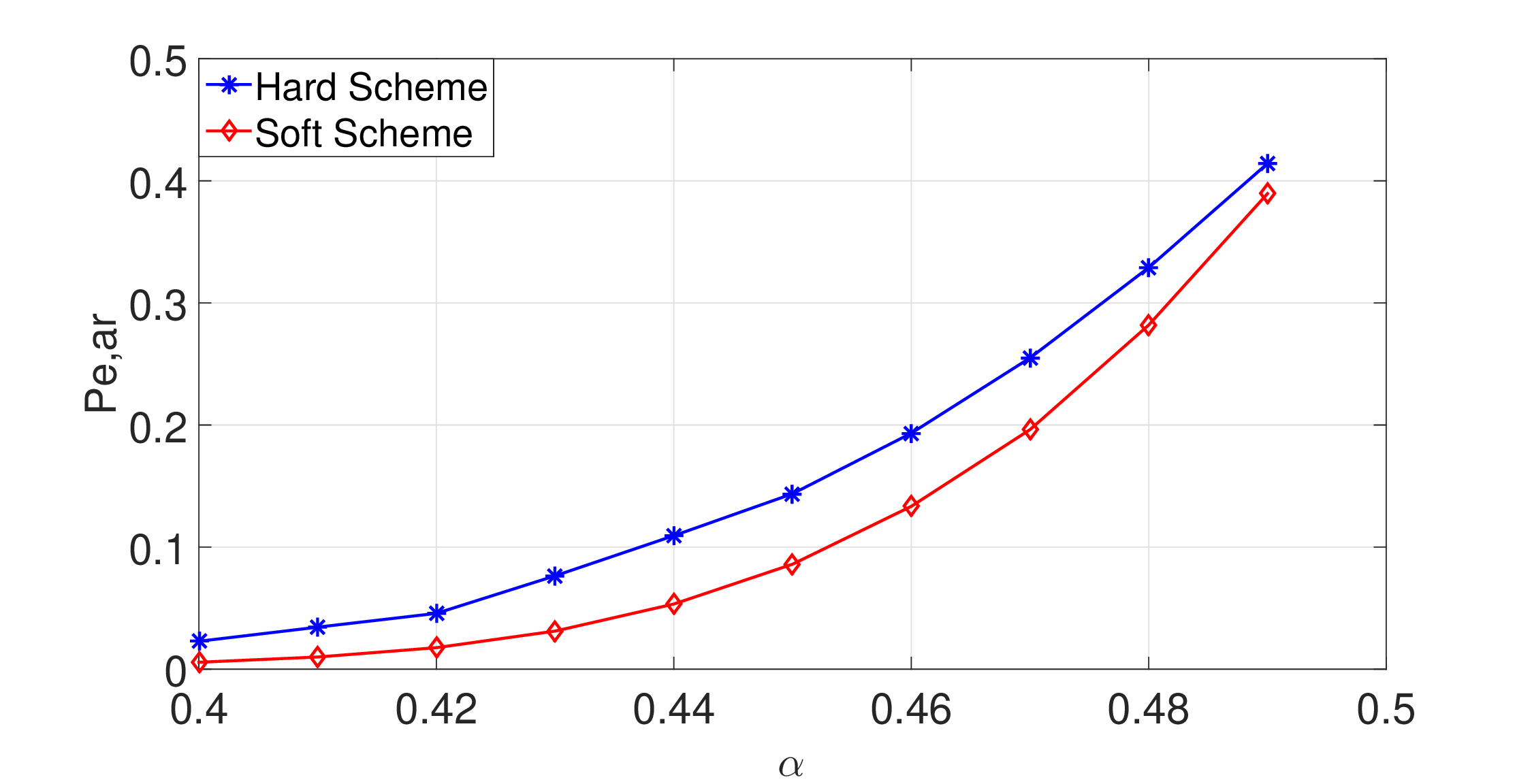}}
\hfill
\subfloat[]{\includegraphics[width=\columnwidth]{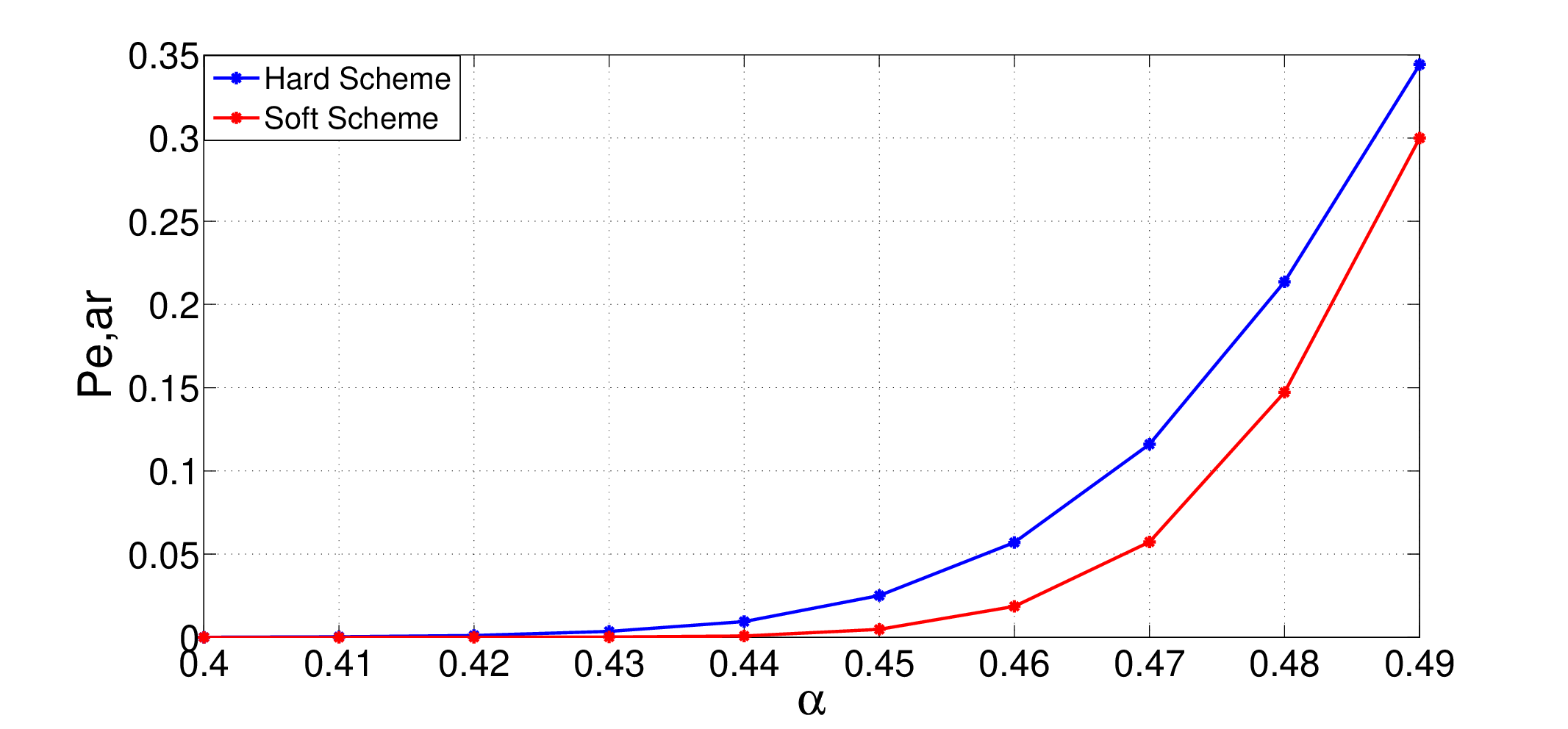}}
\vspace{-0.5cm}
\caption{\textit{Error probability $P_{e,ar}$ at the equilibrium for $P_d = 0.8$ (a) and $P_d = 0.9$ (b).}}
\label{fig:perr}
\end{center}
\end{figure}

\begin{figure}[H]
\begin{center}
\includegraphics[width=\columnwidth]{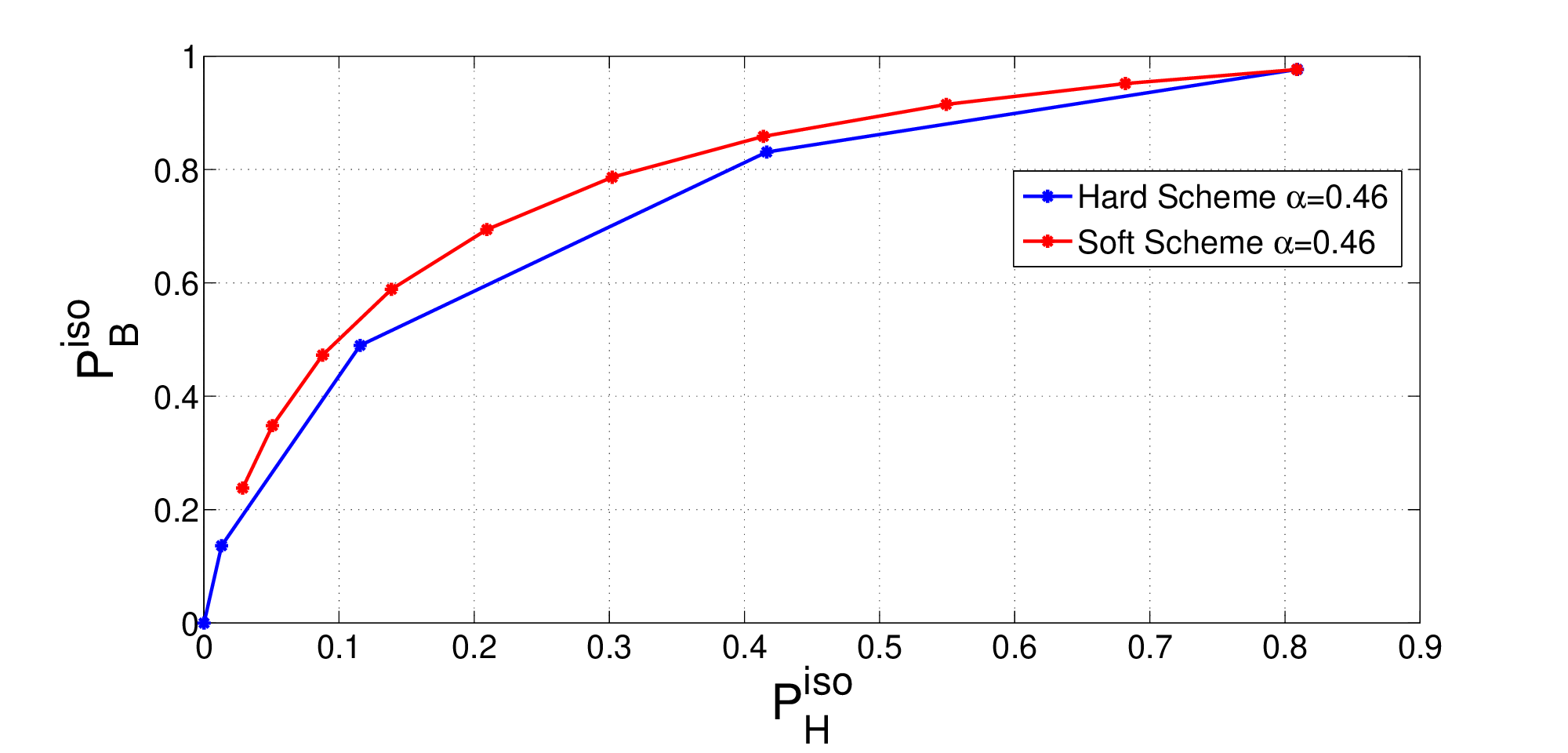}
\vspace{-0.5cm}
\caption{\textit{$P^H_{iso}$ vs. $P^B_{iso}$ at $P_{mal}=1.0$, for $\alpha=0.46$ and $P_d = 0.8$. For the soft scheme, 10 thresholds are taken.}}
\label{fig.isolation}
\end{center}
\vspace{-0.2cm}
\end{figure}

As to the strategy of the Byzantines\index{Byzantines|textbf}, the simulation results confirm the dominance of $P_{mal} = 1$ for both games.
Looking at the performance at the equilibrium, we see that the $DF_S$ game is more favorable to the FC, with a $P_{e,ar}$ at the equilibrium equal to 0.1375 against 0.1982 for the $DF_H$ game.
In Figure \reffig{fig:perr}, the two games are compared by plotting the corresponding payoffs at the equilibrium for various values of $\alpha$ in the interval $[0.4,0.49]$. Upon inspection of the figure, the superiority of the soft isolation\index{soft identification and isolation|textbf} scheme is confirmed.
Finally, we compared the two schemes in terms of capability of isolation of the byzantine nodes. The ROC\index{Receiver Operating Characteristics|textbf} curve with the probability of correct isolation ($P_{ISO}^B$) versus the erroneous isolation of honest nodes ($P_{ISO}^H$), obtained by varying $\eta$, is depicted in Figure \reffig{fig.isolation} for both schemes. The curves correspond to the case of $\alpha = 0.46$ and $P_d = 0.8$. As we can see, soft isolation allows to obtain a slight improvement of the isolation performance with respect to isolation based on a hard reputation score.

\enlargethispage{\baselineskip}

\section{Conclusions}
In this chapter, we presented a new defense scheme for decision fusion in the presence of Byzantine nodes, relying on a soft reputation measure for the identification of nodes. In order to evaluate the performance of the new scheme and compare it against prior art based on a hard reputation measure, we have used a game theoretic framework which is particularly suited to analyze the interplay between the fusion center and the Byzantines. We evaluated the equilibrium point of the game by means of simulations and used the payoff at the equilibrium to assess the validity of the soft reputation metric.


\chapter{A Game-Theoretic Framework for Optimum Decision Fusion in the Presence of Byzantines}
\label{chapter:TIFS_SPL}
\emph{"Never interrupt your enemy when he is making a mistake."}
\\
Napoleon Bonaparte
\\
\emph{"Knowledge is a Weapon, Jon. Arm yourself well before you ride forth to Battle."}
\\
George R.R. Martin, "A Dance with Dragons"
\section{Introduction}
\label{sec.intro}
\PARstart{\textcolor{red}T}his{} chapter starts from the observation that the knowledge of $P_{mal}$ and the probability distribution of Byzantines across the network would allow the derivation of the optimum decision fusion rule, thus permitting to the FC to obtain the best achievable performance. We also argue that in the presence of such an information discarding the reports received from suspect nodes is not necessarily the optimum strategy\index{optimum strategy|textbf}, since such reports may still convey some useful information about the status of the system. This is the case, for instance, when $P_{mal} = 1$. If the FC knows the identity of byzantine nodes, in fact, it only needs to flip the reports received from such nodes to cancel the Byzantines' attack. In this sense, the methods proposed in previous works, as well as the one presented in chapter \ref{chapter:CDC}, are highly suboptimal, since they do not fully exploit the knowledge of Byzantines distribution and their attacking strategy.

\begin{figure}[t!]
\centering
    \includegraphics[width= \textwidth]{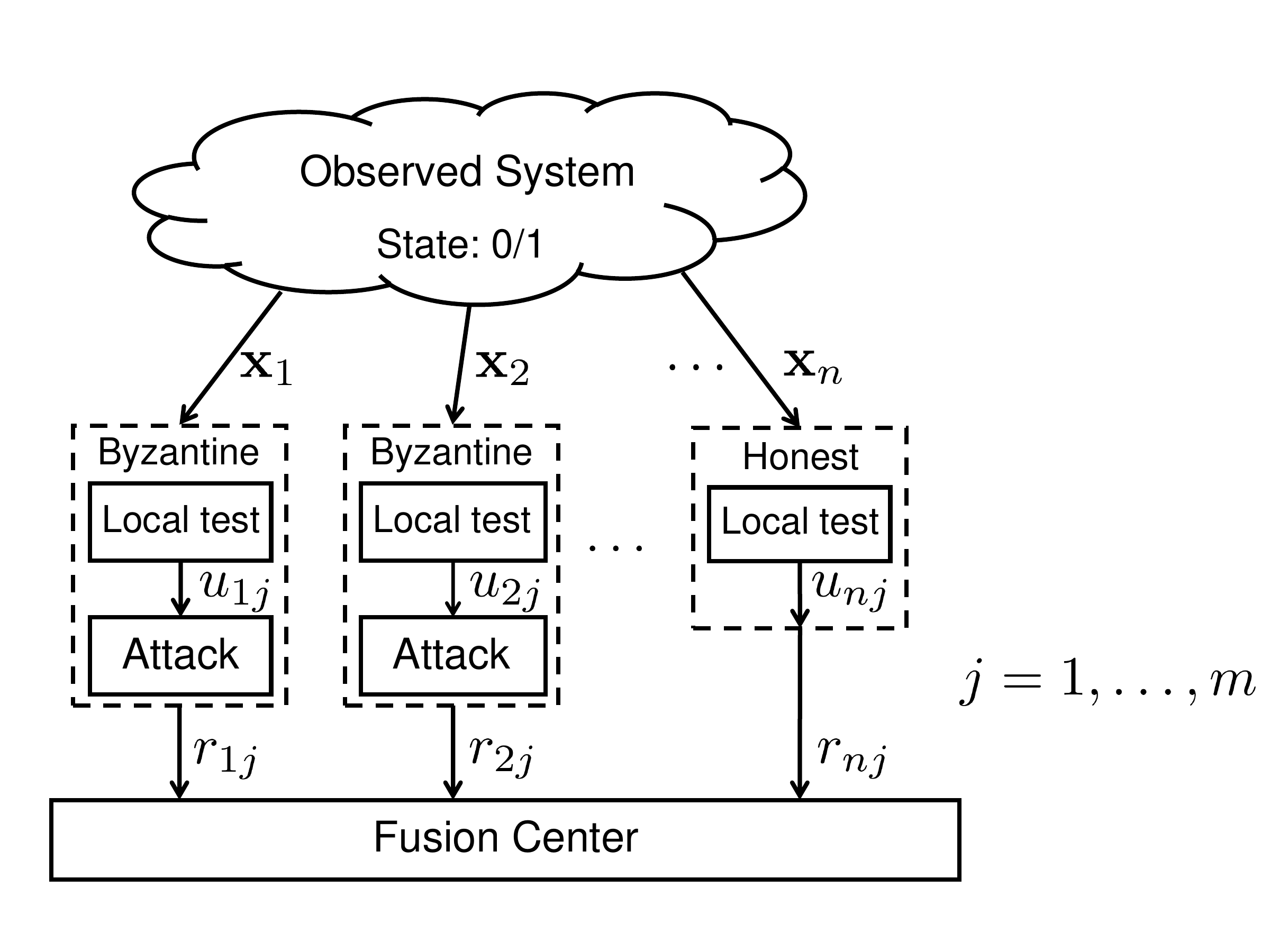}
    \caption{\textit{Sketch of the adversarial decision fusion scheme.}}
    \label{fig.setup}
\end{figure}

With the above ideas in mind, and by adopting the setup illustrated in Figure \reffig{fig.setup}, we first derive the optimum decision fusion rule when the FC knows both the probability distribution of Byzantines\index{Byzantines|textbf} and $P_{mal}$. Our analysis goes along a line which is similar to that used in \cite{OptFusion} to derive the Chair-Varshney optimal fusion rule\index{Chair-Varshney rule|textbf}. As a matter of fact, by knowing $P_{mal}$ and assuming that the probability that a node is Byzantine is fixed and independent on the other nodes, the Chair-Varshney rule can be easily extended to take into account the presence of Byzantines. In contrast to \cite{OptFusion}, however, the optimal fusion rule we derive in this chapter, makes a joint decision on the whole sequence of states hence permitting to improve the decision accuracy. Furthermore, the analysis is not limited to the case of independently distributed Byzantines. We also describe an efficient implementation of the optimum fusion strategy based on dynamic programming\index{Dynamic Programming|textbf}.

In order to cope with the lack of knowledge regarding $P_{mal}$, we introduce a game-theoretic approach according to which the FC arbitrarily sets the value of $P_{mal}$ to a guessed value $P_{mal}^{FC}$ and uses such a value within the optimum fusion rule\index{optimum fusion rule|textbf}. At the same time, the Byzantines\index{Byzantines|textbf} choose the value of $P_{mal}$ so to maximize the error probability, without knowing the value of $P_{mal}^{FC}$ used by the fusion center. The payoff is defined as the overall error probability\index{error probability|textbf}, with the FC aiming at minimizing it, while the goal of the Byzantines is to maximize it. With regard to the knowledge that the FC has about the distribution of Byzantines, we consider several cases, ranging from a maximum entropy\index{maximum entropy|textbf} scenario in which the uncertainty about the distribution of Byzantines is maximum, through a more favorable situation in which the FC knows the exact number of Byzantines present in the network. Having defined the game, we use numerical simulations to derive the existence of equilibrium points, which identify the optimum behavior for both the FC and the Byzantines in a game-theoretic sense. 

We use numerical simulations also to get more insights into the optimum strategies at the equilibrium and the achievable performance under various settings. The simulations show that in all the analyzed cases, the performance at the equilibrium outperforms those obtained in previous works (specifically in \cite{Raw11} and chapter \ref{chapter:CDC}). Simulation results also confirm the intuition that, in some instances, it is preferable for the Byzantines to minimize the mutual information\index{mutual information|textbf} between the status of the observed system and the reports submitted to the FC, rather than always flipping the decision made by the local nodes as it is often assumed in previous works\index{dual behavior|textbf}. This is especially true when the length of the observed sequence and the available information about the Byzantine distribution allow a good identification of byzantine nodes.

\section{Optimum fusion rule}
\label{sec.OptFus}

In the rest of the chapter, we will use capital letters to denote random variables and lowercase letters for their instantiations. Given a random variable $X$, we indicate with $P_X(x)$ its probability mass function (pmf). Whenever the random variable the pmf refers to is clear from the context, we will use the notation $P(x)$ as a shorthand for $P_X(x)$.

With the above notation in mind, we let $S^m = (S_1, S_2 \dots S_m)$ indicate a sequence of independent and identically distributed (i.i.d.) random variables indicating the state of the system. The independence of the different components of the state vector is a reasonable assumption in several scenarios, e.g. when they represent the status of the frequency spectrum of a cognitive radio system at different frequencies \cite{Raw11}. We assume that all states are equiprobable, that is $P_{S_j}(0) = P_{S_j}(1) = 0.5$.
We denote by $U_{ij} \in \{0,1\}$ the local decision made by node $i$ about $S_j$. We exclude any interaction between the nodes and assume that $U_{ij}$'s are conditionally independent for a fixed status of the system. This is equivalent to assuming that the local decision errors are i.i.d.

With regard to the position of the Byzantines, let $A^n = (A_1 \dots A_n)$ be a binary random sequence in which $A_i = 0$ (res. $A_i = 1$) if node $i$ is honest (res. byzantine). The probability that the distribution of Byzantines across the nodes is $a^n$ is indicated by $P_{A^n}(a^n)$ or  simply $P(a^n)$.

Finally, we let ${\bf R} = \{R_{ij}\}, ~ i = 1 \dots n, j = 1 \dots m$ be a random matrix with all the reports received by the fusion center, accordingly, we denote by ${\bf r} = \{r_{ij}\}$ a specific instantiation of ${\bf R}$. As stated before, $R_{ij} = U_{ij}$ for honest nodes, while $P(R_{ij} \ne U_{ij}) = P_{mal}$ for byzantine nodes. Byzantine nodes flip the local decisions $U_{ij}$ independently of each other with equal probabilities, so that their action can be modeled as a number of independent binary symmetric channels with crossover probability $P_{mal}$.

We are now ready to derive the optimum decision rule on the sequence of states at the FC. We stress that, while considering a joint decision on the sequence of states does not give any advantage in the non-adversarial scenario with i.i.d. states, such an approach permits to improve the accuracy of the decision in the presence of byzantine nodes. Given the received reports ${\bf r}$ and by adopting a maximum a posteriori probability criterion, the optimum decision rule minimizing the error probability can be written as:

\begin{equation}
s^{m,*} = \arg\max_{s^m} P(s^m | {\bf r}).
\label{eq.map}
\end{equation}

By applying Bayes rule and exploiting the fact that all state sequences are equiprobable, we obtain:

\begin{equation}
s^{m,*} =  \arg\max_{s^m} P({\bf r} | s^m ).
\label{eq.ML}
\end{equation}
In order to go on, we condition $P({\bf r} | s^m)$ to the knowledge of $a^n$ and then average over all possible $a^n$:

\begin{align}
s^{m,*}  
= & \arg\max_{s^m} \sum_{a^n} P({\bf r} | a^n, s^m) P(a^n) \label{eq.pseudoML}\\
= & \arg\max_{s^m} \sum_{a^n} \bigg(\prod_{i=1}^n P({\bf r}_i | a_i, s^m )\bigg) P(a^n)\label{eq.pseudoML_2}\\
= & \arg\max_{s^m} \sum_{a^n} \bigg(\prod_{i=1}^n \prod_{j=1}^m P(r_{ij} | a_i, s_j )\bigg) P(a^n),\label{eq.pseudoML_3}
\end{align}

where ${\bf r}_i$ indicates the $i$-th row of ${\bf r}$. In \eqref{eq.pseudoML_2} we exploited the fact that, given $a^n$ and $s^m$, the reports sent by the nodes are independent of each other, while \eqref{eq.pseudoML_3} derives from the observation that each report depends only on the corresponding element of the state sequence. It goes without saying that in the non-adversarial case ($P(a^n) = 1$ for $a^n = (0, \cdots,0)$ and 0 otherwise) the maximization in \eqref{eq.pseudoML_3} is equivalent to the following component-wise maximization
\begin{align}
s_j^{*}  = & \arg\max_{s_j}  \prod_{i=1}^n P(r_{ij} | s_j ), \quad \forall j = 1,\cdots, m,
\end{align}
which corresponds to the Chair-Varshney rule\index{Chair-Varshney rule|textbf}.

We now consider the case in which the probability of a local decision error, say $\varepsilon$, is the same regardless of the system status, that is $\varepsilon = Pr(U_{ij} \neq S_j|S_j = s_j)$, $s_j = 0,1$. For a honest node, such a probability is equal to the probability that the report received by the FC does not correspond to the system status. This is not the case for byzantine nodes, for which the probability $\delta$ that the FC receives a wrong report is
\begin{align}
\delta = \varepsilon (1 - P_{mal}) + (1 - \varepsilon)P_{mal}.
\end{align}
According to the above setting, the nodes can be modeled as binary symmetric channels, whose input corresponds to the system status and for which the crossover probability is equal to $\varepsilon$ for the honest nodes and $\delta$ for the Byzantines. With regard to $\varepsilon$, it is reasonable to assume that such a value is known to the fusion center, since it depends on the characteristics of the channel through which the nodes observe the system and the local decision rule adopted by the nodes. The value of $\delta$ depends on the value of $P_{mal}$ which is chosen by the Byzantines and then is not generally known to the FC. We will first derive the optimum fusion rule assuming that $P_{mal}$ is known and then relax this assumption by modeling the problem in a game-theoretic framework\index{Game Theory|textbf}.

From \eqref{eq.pseudoML_3}, the optimum decision rule can be written:
\begin{align}
\label{eq.pseudoML_symm}
s^{m,*} =  \arg\max_{s^m} & \sum_{a^n} \bigg(\prod_{i:a_i = 0}  (1-\varepsilon)^{m_{eq}(i)} \varepsilon^{m-m_{eq}(i)} \bigg. \\ & \hspace{0.5cm} \bigg. \prod_{i:a_i = 1} (1-\delta)^{m_{eq}(i)} \delta^{m-m_{eq}(i)} \bigg) P(a^n),\nonumber
\end{align}

where $m_{eq}(i)$ is the number of $j$'s for which $r_{ij} = s_j$.

As a notice, when there are no Byzantines in the network, the optimum decision in \eqref{eq.pseudoML_symm} boils down to the majority rule.

To go on with the study of the adversarial setup we need to make some assumptions on $P(a^n)$.

\subsection{Unconstrained maximum entropy distribution}
\label{sec.MaxEnt}

As a worst case scenario, we could assume that the FC has no a-priori information about the distribution of Byzantines\index{unconstrained maximum entropy|textbf}. This corresponds to maximizing the entropy\index{entropy|textbf} of $A^n$, i.e. to assuming that all sequences $a^n$ are equiprobable, $P(a^n) = 1/2^n$. In this case, the random variables $A_i$ are independent of each other and we have $P_{A_i}(0) = P_{A_i}(1) = 1/2$. It is easy to argue that in this case the Byzantines may impede any meaningful decision at the FC. To see why, let us assume that the Byzantines decide to use $P_{mal}=1$. With this choice, the mutual information\index{mutual information|textbf} between the vector state $S^m$ and ${\bf R}$ is zero and so any decision made by the FC center would be equivalent to {\em guessing} the state of the system by flipping a coin. The above observation is consistent with previous works in which it is usually assumed that the probability that a node is Byzantine or the overall fraction of Byzantines is lower than 0.5, since otherwise the Byzantines would always succeed to blind the FC \cite{Vemp13}.

\subsection{Constrained maximum entropy distributions}
\label{sec.ConstrMaxEnt}

A second possibility consists in maximizing the entropy of $A^n$ subject to a constraint which corresponds to the a-priori information available to the fusions center. We consider two cases. In the first one the FC knows the expected value of the number of Byzantines present in the network, in the second case, the FC knows only an upper bound of the number of Byzantines. In the following, we let $N_B$ indicate the number of Byzantines present in the network.

\subsubsection{Maximum entropy with given $E[N_B]$}
\label{subsec.fixed_mean}

Let $\alpha = E[N_B]/n$ indicate the expected fraction of Byzantine nodes in the network. In order to determine the distribution $P{(a^n})$ which maximizes $H(A^n)$ subject to $\alpha$, we observe that $E[N_B] = E[\sum_i A_i] = \sum_i E[A_i] = \sum_i \mu_{A_i}$, where $\mu_{A_i}$ indicates the expected value of $A_i$. In order to determine the maximum entropy distribution\index{maximum entropy|textbf} constrained to $E[N_B] = \alpha n$, we need to solve the following problem:
\begin{equation}
\max_{P(a^n): \sum_i \mu_{A_i} = n\alpha} H(A^n).
\label{eq.MaxEntConstr}
\end{equation}
We now show that the solution to the above maximization problem is obtained by letting the $A_i$'s to be i.i.d. random variables with $\mu_{A_i} = \alpha$. We have:
\begin{equation}
    H(A^n) \le \sum_i H(A_i) = \sum_i h(\mu_{A_i}),
\label{eq.Crule}
\end{equation}
where $h(\mu_{A_i})$ denotes the binary entropy\index{entropy|textbf} function\footnote{For any $p \le 1$ we have: $h(p) = p \log_2 p + (1-p) \log_2(1-p)$.} and where the last equality derives from the observation that for a binary random variable $A$, $\mu_A = P_A(1)$. We also observe that equality holds if and only if the random variables $A_i$'s are independent. To maximize the rightmost term in Equation (\refeq{eq.Crule}) subject to $\sum_i \mu_{A_i} = n \alpha$, we observe that the binary entropy is a concave function \cite{CandT}, and hence the maximum of the sum is obtained when all $\mu_{A_i}$'s are equal, that is when $\mu_{A_i} = \alpha$.

In summary, the maximum entropy\index{maximum entropy|textbf} case with known average number of Byzantines, corresponds to assuming i.i.d. node states for which the probability of being malicious is constant and known to the FC\footnote{Sometimes this scenario is referred to as Clairvoyant case \cite{Raw11}.}. We also observe that when $\alpha = 0.5$, we go back to the unconstrained maximum entropy\index{unconstrained maximum entropy|textbf} case discussed in the previous section.

Let us assume, then, that $A_i$'s are Bernoulli random variables with parameter $\alpha$, i.e., $P_{A_i}(1) = \alpha$, $\forall i$. In this way, the number of Byzantines in the network is a random variable with a binomial distribution. In particular, we have $P(a^n) = \prod_i P(a_i)$, and hence \eqref{eq.pseudoML_2} can be rewritten as:
\begin{equation}
s^{m,*} = \arg\max_{s^m} \sum_{a^n} \bigg(\prod_{i=1}^n P({\bf r}_i | a_i, s^m )P(a_i) \bigg).
\label{eq.factorization}
\end{equation}

The expression in round brackets corresponds to a factorization of $P({\bf r},\allowbreak  a^n | s^m )$.
If we look at that expression as a function of $a^n$, it is a product of marginal functions.
By exploiting the distributivity of the product with respect to the sum
we can rewrite \eqref{eq.factorization} as follows
\begin{equation}
s^{m,*} =  \arg\max_{s^m} \prod_{i=1}^{n} \bigg(\sum_{a_i \in \{0,1\}} P({\bf r}_i| a_i, s^m) P(a_i) \bigg),
\label{eq.pseudoML_Random}
\end{equation}
which can be computed more efficiently, especially for large $n$.
The expression in \eqref{eq.pseudoML_Random} can also be derived directly from \eqref{eq.ML} by exploiting first the independence of the reports and then applying the law of total probability.
By reasoning as we did to derive \eqref{eq.pseudoML_symm}, the to-be-maximized expression for the case of symmetric error probabilities at the nodes becomes

\begin{equation}
s^{m,*} =  \arg\max_{s^m} \prod_{i=1}^{n} \left[(1-\alpha)(1-\varepsilon)^{m_{eq}(i)} \varepsilon^{m-m_{eq}(i)} + \alpha(1-\delta)^{m_{eq}(i)} \delta^{m-m_{eq}(i)}\right].
 \label{eq.pseudoML_Random_2}
\end{equation}
Due to the independence of node states, the complexity\index{computational complexity|textbf} of the above maximization problem grows only linearly with $n$, while it is exponential with respect to $m$, since it requires the evaluation of the to-be-minimized function for all possible sequence $s^m$. For this reason, the optimal fusion strategy can be adopted only when the length of the observed sequence is limited.

\subsubsection{Maximum entropy with $N_B < h$}
\label{subsec.less_half}

As a second possibility, we assume that the FC knows only that the number of Byzantines\index{Byzantines distribution|textbf} $N_B$ is lower than a certain value $h$ ($h \le n$). For instance, as already observed in previous works \cite{Vemp13,WSNMatta,Raw11}, when the number of Byzantines exceeds the number of honest nodes no meaningful decision can be made. Then, as a worst case assumption, it makes sense for the FC to assume that $N_B < n/2$ (i.e., $h = n/2$), since if this is not the case, no correct decision can be made anyhow.  Under this assumption, the maximum entropy distribution\index{maximum entropy|textbf} is the one which assigns exactly the same probability to all the sequences $a^n$ for which $\sum_i a_i < n/2$. More in general, the FC might have some a priori knowledge on the maximum number of corrupted (or corruptible) links in the network, and then he can constraint $N_B$ to be lower than $h$ with $h < n/2$. To derive the optimum fusion strategy\index{optimum fusion rule|textbf} in this setting, let $\mathcal{I}$ be the indexing set $\{1,2,...,n\}$.  We denote with $\mathcal{I}_k$ the set of all the possible $k$-subsets of $\mathcal{I}$. Let $I \in \mathcal{I}_k$ be a random variable with the indexes of the byzantine nodes\index{byzantine nodes|textbf}, a node $i$ being byzantine if $i \in I$, honest otherwise. We this notation, we can rewrite \eqref{eq.pseudoML} as 

\begin{align}
s^{m,*} = \arg\max_{s^m}  \sum_{k=0}^{h - 1}\sum_{I \in \mathcal{I}_{k}} P({\bf r} | I, s^m ) p(s^m),\label{eq.ML_less_thann2_general}
\end{align}

where we have omitted the term $P(I)$ (or equivalently $P(a^n)$) since all the sequences for which $N_B < h$ have the same probability. In the case of symmetric local error probabilities, \eqref{eq.ML_less_thann2_general}
takes the following form:
%


\begin{equation}
s^{m,*} =  \arg\max_{s^m}   \sum_{k=0}^{h - 1} \sum_{I \in \mathcal{I}_{k}} \bigg(\prod_{i \in I} (1-\delta)^{m_{eq}(i)} \delta^{m-m_{eq}(i)} \prod_{i \in \mathcal{I} \setminus I} (1-\varepsilon)^{m_{eq}(i)} \varepsilon^{m-m_{eq}(i)}\bigg).
  \label{eq.ML_less_thann2_2}
\end{equation}
Since, reasonably, $h$ is a fraction of $n$, a problem with \eqref{eq.ML_less_thann2_2} is the complexity\index{computational complexity|textbf} of the inner summation, which grows exponentially with $n$ (especially for values of $k$ close to $h$). Together with the maximization over all possible $s^m$, this results in a doubly exponential complexity\index{complexity|textbf}, making the direct implementation of \eqref{eq.ML_less_thann2_2} problematic. In Section \ref{sec.DP}, we introduce an efficient algorithm based on dynamic programming\index{Dynamic Programming|textbf} which reduces the computational complexity\index{computational complexity|textbf} of the maximization in \eqref{eq.ML_less_thann2_2}.

We conclude by stressing an important difference between the case considered in this subsection and the maximum entropy\index{maximum entropy|textbf} case with fixed $E[N_B]$, with the same average number of Byzantines. In the setting with a fixed $E[N_B]$ ($< n/2$) there is no guarantee that the number of Byzantines\index{Byzantines|textbf} is always lower than the number of honest nodes, as it is the case in the setting analyzed in this subsection  when $h \le n/2$. This observation will be crucial to explain some of the results that we will present later on in the chapter.

\subsection{Fixed number of Byzantines}
\label{sec.OF_DETstates}

The final setting we are going to analyze assumes that the fusion center\index{Fusion Center|textbf} knows the exact number of Byzantines, say $n_B$. This is a more favorable situation with respect to those addressed so far. The derivation of the optimum decision fusion rule\index{optimum fusion rule|textbf} stems from the observation that, in this case, $P(a^n) \ne 0$ only for the sequence for which $\sum_i a_i = n_B$. For such sequences, $P(a^n)$ is constant and equal to $\binom{n}{n_B}^{-1}$. By using the same notation used in the previous section, the optimum fusion rules, then, is:
\begin{align}
s^{m,*} = \arg\max_{s^m}  \sum_{I \in \mathcal{I}_{n_B}} P({\bf r} | I, s^m ) p(s^m),\label{eq.ML_Determ}
\end{align}
which reduces to

\begin{equation}
s^{m,*} =  \arg\max_{s^m}   \sum_{I \in \mathcal{I}_{n_B}} \bigg(\prod_{i \in I} (1-\delta)^{m_{eq}(i)} \delta^{m-m_{eq}(i)} \prod_{i \in \mathcal{I} \setminus I} (1-\varepsilon)^{m_{eq}(i)} \varepsilon^{m-m_{eq}(i)}\bigg),
  \label{eq.ML_Determ_2}
\end{equation}
in the case of equal local error probabilities. With regard to computational complexity\index{computational complexity|textbf}, even if the summation over all possible number of Byzantines\index{Byzantines distribution|textbf} is no more present, the direct implementation of \eqref{eq.ML_Determ_2} is still very complex due to the exponential dependence of the cardinality of $\mathcal{I}_{n_B}$ with respect to $n$.

\section{An efficient implementation based on dynamic programming}
\label{sec.DP}

The computational complexity of a direct implementation of \eqref{eq.ML_less_thann2_2} and \eqref{eq.ML_Determ_2} hinders the derivation of the optimum decision fusion rule\index{optimum fusion rule|textbf} for large size networks. Specifically, the problem with \eqref{eq.ML_less_thann2_2} and \eqref{eq.ML_Determ_2} is the exponential number of terms of the summation over $\mathcal{I}_k$ ($\mathcal{I}_{n_B}$ in \eqref{eq.ML_Determ_2}). In this section, we show that an efficient implementation of such summations is possible based on Dynamic Programming\index{Dynamic Programming|textbf} (DP) \cite{DynamicProgramming}.

Dynamic programming\index{Dynamic Programming|textbf} is an optimization strategy which allows to solve complex problems by transforming them into subproblems and by taking advantage of the subproblems overlap in order to reduce the number of operations. When facing with complex recursive problems, by using dynamic programming we solve each different subproblem only once by storing the solution for subsequent use. If during the recursion the same subproblem is encountered again, the problem is not solved twice since its solution is already available. Such a re-use of previously solved subproblems is often referred in literature as memorization algorithm \cite{DynamicProgramming}. Intuitively, DP allows to reduce the complexity of problems with a structure, such that the solutions of the same subproblems can be reused many times.

We now apply dynamic programming to reduce the complexity of our problem.
Let us focus on a fixed $k$ (and $n$) and let us define the function $f_{n,k}$ as follows:

\begin{equation}
\label{function_f_n_k}
f_{n,k} =   \sum_{I \in \mathcal{I}_{k}} \bigg(\prod_{i \in I} (1-\delta)^{m_{eq}(i)} \delta^{m-m_{eq}(i)}\prod_{i \in \mathcal{I} \setminus I} (1-\varepsilon)^{m_{eq}(i)} \varepsilon^{m-m_{eq}(i)}\bigg).
\end{equation}

\begin{figure}[h!]
\centering
    \includegraphics[width=0.8\columnwidth]{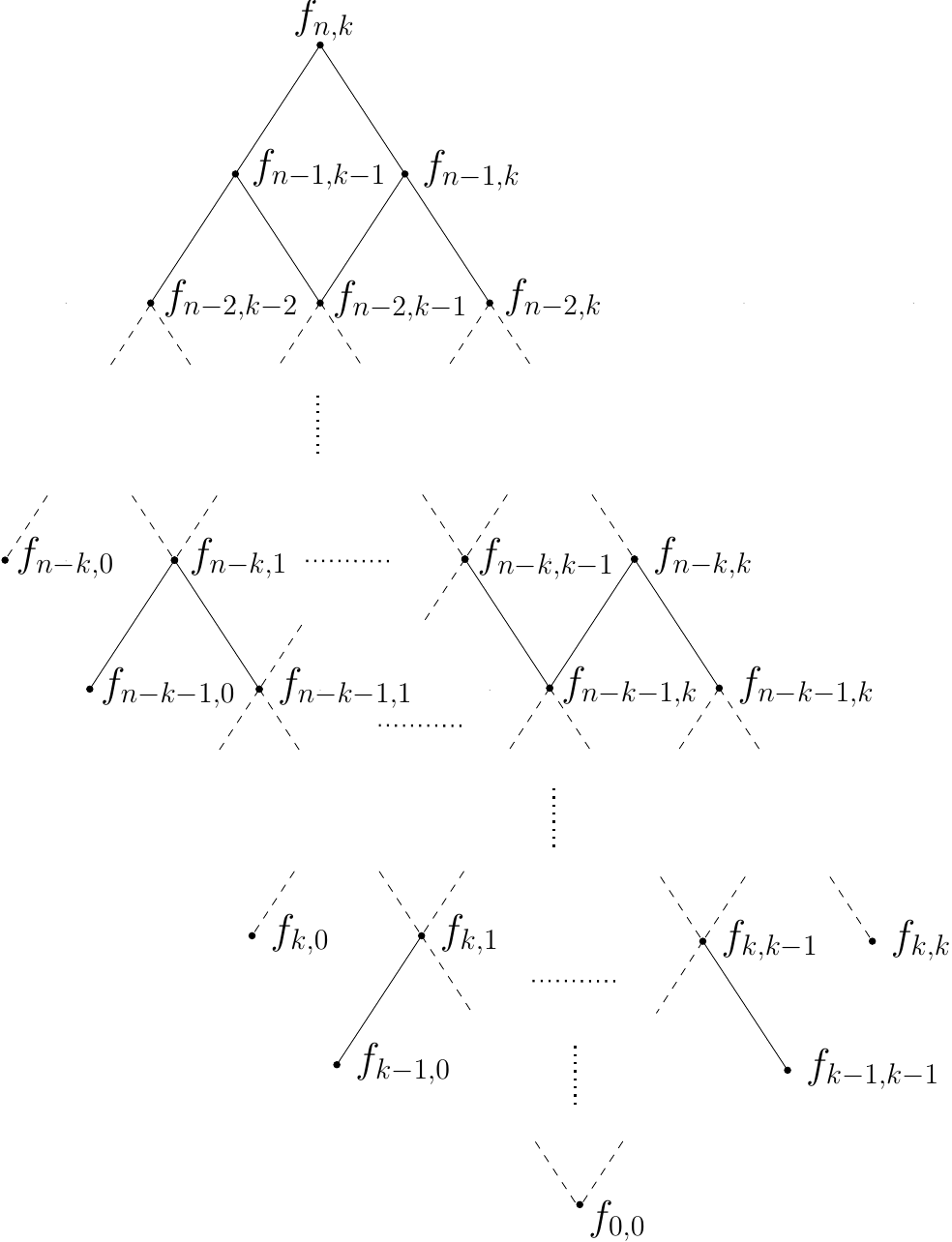}\vspace{0.4cm}
    \caption{\textit{Efficient implementation of the function in \eqref{function_f_n_k} based on dynamic programming. The figure depicts the tree with the iterations for the case $k < n-k$.}}
    \label{fig.OptRule_DPcomplexity}
\end{figure}

By focusing on node $i$, there are some configurations $I \in \mathcal{I}_{k}$ for which such a node belongs to $I$, while for others the node belongs to the complementary set $\mathcal{I} \setminus I$. Let us define $b(i) = (1-\delta)^{m_{eq}(i)} \delta^{m-m_{eq}(i)}$ and $h(i) = (1-\varepsilon)^{m_{eq}(i)} \varepsilon^{m-m_{eq}(i)}$, which are the two contributions that node $i$ can provide to each term of the sum, depending on whether it belongs to $\mathcal{I}$ or $\mathcal{I} \setminus I$. Let us focus on the first indexed node. By exploiting the distributivity of the product with respect to the sum, expression \eqref{function_f_n_k} can be rewritten in a recursive manner as:
\begin{align}
f_{n,k} =  b(1) f_{n-1,k-1} + h(1) f_{n-1,k}.
\end{align}
By focusing on the second node, we can iterate on $f_{n-1,k-1}$  and  $f_{n-1,k}$, getting:
\begin{align}
f_{n-1,k-1} =  b(2) f_{n-2,k-2} +  h(2) f_{n-2,k-1},
\label{eq.sub1}
\end{align}
and
\begin{align}
\label{eq.sub2}
f_{n-1,k}= b(2) f_{n-2,k-1} + h(2) f_{n-2,k}.
\end{align}
We notice that subfunction $f_{n-2,k-1}$ appears in both \eqref{eq.sub1} and \eqref{eq.sub2} and then it can be computed only once. The procedure can be iterated for each subfunction until we reach a subfunction whose value can be computed in closed form, that is:  $f_{r,r} = \prod_{i =n-r + 1}^n b(i)$ and $f_{r,0} = \prod_{i =n-r + 1}^n h(i)$, for some $r \le k$.
By applying the memorization strategy typical of dynamic programming, the number of required computations is given by the number of nodes in the tree depicted in Figure \reffig{fig.OptRule_DPcomplexity}, where the leaves correspond to the terms computable in closed form\footnote{The figure refers to the case $k < n-k$, which is always the case in our setup since $k < \lfloor n/2 \rfloor$.}. By observing that the number of the nodes of the tree is $k(k + 1)/2 + k(n - k - k) + k(k+1)/2 = k(n - k +1)$, we conclude that the number of operations is reduced from $\binom{n}{k}$ to $k(n - k +1)$, which corresponds to a quadratic complexity instead of an exponential one.

\section{Decision fusion with Byzantines game}
\label{sec.GT}

The optimum decision fusion rules derived in Section \ref{sec.OptFus} assume that the FC knows the attacking strategy adopted by the Byzantines, which in the simplified case studied in this chapter corresponds to knowing $P_{mal}$. By knowing $P_{mal}$, in fact, the FC can calculate the value of $\delta$ used in Equations (\refeq{eq.pseudoML_symm}), (\refeq{eq.pseudoML_Random_2}), (\refeq{eq.ML_less_thann2_2}) and (\ref{eq.ML_Determ_2}), and hence implement the optimum fusion rule\index{optimum fusion rule|textbf}. In previous works, as in \cite{RawatConf, Raw11}, it is often conjectured that $P_{mal} = 1$. In some particular settings, as the ones addressed in \cite{KBKV13} and chapter \ref{chapter:CDC}, it has been shown that this choice permits to the Byzantines to maximize the error probability\index{error probability|textbf} at the fusion center. Such an argument, however, does not necessarily hold when the fusion center can localize the byzantine nodes with good accuracy and when it knows that the byzantine nodes always flip the local decision. In such a case, in fact, the FC can revert the action of the Byzantines by simply inverting the reports received from such nodes, as it is implicitly done by the optimal fusion rules derived in the previous section. In such a situation, it is easy to argue that it is better for the Byzantines to let $P_{mal} = 0.5$ since in this way the mutual information\index{mutual information|textbf} between the system status and the reports received from the byzantine nodes is equal to zero. In general, the byzantine nodes must face the following dilemma: is it better to try to force the FC to make a wrong decision by letting $P_{mal} = 1$ and run the risk that if their location in the network is detected the FC receives some useful information from the corrupted reports, or erase the information that the FC receives from the attacked nodes by reducing to zero the mutual information between the corrupted reports and $S^m$ ?

Given the above discussion, it is clear that the FC cannot assume that the Byzantines use $P_{mal} = 1$, hence making the actual implementation of the optimum decision fusion rule impossible.

In order to exit this apparent deadlock, we model the race of arms between the Byzantines and the FC \index{decision fusion game|textbf}as a two-player, zero-sum\index{zero-sum game|textbf}, strategic game\index{strategic game|textbf}, whose equilibrium\index{equilibrium|textbf} defines the optimum choices for the FC and the Byzantines. In this model, the interplay is between the value of $P_{mal}$ adopted by the Byzantines and the value used by the FC in its attempt to implement the optimum fusion rule as game\index{decision fusion game|textbf}. For sake of clarity, in the following we indicate the flipping probability adopted by the Byzantines as $P_{mal}^B$, while we use the symbol $P_{mal}^{FC}$ to indicate the value adopted by the FC in its implementation of the optimum fusion rule. With the above ideas in mind, we introduce the following Decision Fusion Game\index{decision fusion game|textbf}.

\begin{definition}
The $DF_{Byz}(\mathcal{S}_{B}, \mathcal{S}_{FC}, v)$ game is a two player, zero-sum, strategic, game played by the FC and the Byzantines\index{players|textbf} (collectively acting as a single player), defined by the following strategies and payoff.
\begin{itemize}
\item{The sets of strategies the Byzantines and the FC can choose from are, respectively, the set of possible values of $P_{mal}^B$ and $P_{mal}^{FC}$:
\begin{eqnarray}
\mathcal{S}_B = \{ P_{mal}^B \in [0,1]\};\nonumber \\
\mathcal{S}_{FC} = \{ P_{mal}^{FC} \in [0,1]\}.
\label{eq.DFgameS}
\end{eqnarray}
}
\item{The payoff function is defined as the error probability at the FC, indicated as $P_e$
\begin{align}
v =  P_e = P(S^* \neq S).
\label{eq.Pe}
\end{align}
where $S$ is the true system state\index{system state|textbf} and $S^*$ is the decision made by FC.
}
Of course the Byzantines aim at maximizing $P_e$, while the FC aims at minimizing it.
\end{itemize}
\label{def.DFgame}
\end{definition}

Note that according to the definition of $DF_{Byz}$, the sets of strategies available to the FC and the Byzantines are continuous sets. In practice, however, continuous values can be replaced by a properly quantized version of $P_{mal}^B$ and $P_{mal}^{FC}$.

In the next section, we use numerical simulations to derive the equilibrium point\index{equilibrium point|textbf} of various versions of the game obtained by varying the probability distribution of Byzantines\index{Byzantines distribution|textbf} as detailed in Section \ref{sec.OptFus}. As we will see, while some versions of the game has a unique Nash (or even dominant) equilibrium point in pure strategies, in other cases, a Nash equilibrium\index{Nash equilibrium|textbf} exists only in mixed strategies\index{Mixed Strategy Nash Equilibrium|textbf}.

\section{Simulation results and discussion}
\label{sec.simul}

In order to investigate the behavior of the $DF_{Byz}$ game for different setups and analyze the achievable performance when the FC adopts the optimum decision strategy with parameters tuned following a game-theoretic approach, we run extensive numerical simulations. The first goal of the simulations was to study the existence of an equilibrium point in pure or mixed strategies, and analyze the expected behavior of the FC and the Byzantines at the equilibrium. The second goal was to evaluate the payoff at the equilibrium as a measure of the best achievable performance of Decision Fusion in the presence of Byzantines. We then used such a value to compare the performance of the game-theoretic approach proposed in this thesis with respect to previous works.

\subsection{Analysis of the equilibrium point of the $DF_{Byz}$ game}

As we said, the first goal of the simulations was to determine the existence of an equilibrium point for the $DF_{Byz}$ game. To do so, we quantized the set of available strategies considering the following set of values: $\mathcal{S}_B^q =\{0.5, 0.6, 0.7, 0.8, 0.9, 1\}$ and $\mathcal{S}_{FC}^q = \{0.5, 0.6, 0.7, 0.8, 0.9, 1\}$. We restricted our analysis to values larger than or equal to $0.5$ since it is easily arguable that such values always lead to better performance for the Byzantines\footnote{By using a game-theoretic terminology, this is equivalent to say that the strategies corresponding to $P_{mal}^B < 0.5$ are dominated strategies and hence can be eliminated.}. As to the choice of the quantization step, we set it to 0.1 to ease the description of the results we have got and speed up the simulations. Some exploratory test made with a smaller step gave similar results. 

Let $\textbf{V}$ denote the payoff matrix of the game, that is, the matrix of the error probabilities for each pair of strategies $(P_{mal}^{B},P_{mal}^{FC}) \in \mathcal{S}_{FC}^q \times \mathcal{S}_{B}^q$. For each setting, the payoff matrix $\textbf{v}$ is obtained by running the simulations for all the possible moves of FC and the Byzantines.

Sometimes (when the game can be solved with pure strategies), the equilibrium point easily comes out through inspection of the payoff matrix, especially when a rationalizable equilibrium exists. In the general case, we can find equilibrium point by relying on the minimax theorem \cite{Osb94}.
Let $p_B$ (res. $p_{FC}$) be a column vector with the probability distribution over the possible values of $P_{mal}^{B}$ (res. $P_{mal}^{FC}$).  The mixed strategies Nash equilibrium $(p_B^*, p_{FC}^*)$ can be found by solving separately the max-min and min-max problems:
\begin{align}
& p_B^* = {\arg\max}_{p_B(\mathcal{S}_{B}^q)} \min_{p_{FC}(\mathcal{S}_{FC}^q)}  p_B^T  \textbf{v}  p_{FC} \nonumber\\
& p_{FC}^* = {\arg\min}_{p_{FC}(\mathcal{S}_{FC}^q)}  \max_{p_B(\mathcal{S}_{B}^q)}  p_B^T \textbf{v}  p_{FC}
\label{eq.MinMax}
\end{align}
which can be reduced to two simple linear programming\index{Linear Programming|textbf} problems.

We found that among all the parameters of the game, the value of $m$ has a major impact on the equilibrium point. The value of $m$, in fact, determines the ease with which the FC can localize the byzantine nodes, and hence plays a major role in determining the optimum attacking strategy\index{optimum attacking strategy|textbf}. For this reason, we split our analysis in two parts: the former refers to small values of $m$, the latter to intermediate values of $m$. Unfortunately, the exponential growth of the complexity of the optimum decision fusion rule\index{optimum fusion rule|textbf} as a function of $m$ prevented us from running simulations with large values of $m$.

Simulations were carried out by adopting the following setup. We run 50,000 trials to compute $P_e$ at each row of the matrix. In particular, for each
$P_{mal}^B$, we used the same 50,000 states to compute $P_e$ for all $P_{mal}^{FC}$ strategies. In all the simulations, we let $P_{S_j}(0) = P_{S_j}(1) = 0.5$, $n=20$, and $\varepsilon=0.1$. We used the linear programming\index{Linear Programming|textbf} tools from Matlab Optimization Toolbox \cite{MatlabOT} to solve (\ref{eq.MinMax}).

\subsubsection{Small $m$}

For the first set of simulations, we used a rather low value of $m$, namely $m = 4$. The other parameters of the game were set as follows: $n = 20$, $\varepsilon = 0.1$. With regard to the number of Byzantines present in the network we used $\alpha = \{0.3, 0.4, 0.45\}$ for the case of independent node states studied in Section \ref{subsec.fixed_mean}, and $n_B = \{6, 8, 9\}$ for the case of known number of Byzantines (Section \ref{sec.OF_DETstates}). Such values were chosen so that in both cases we have the same average number of Byzantines, thus easing the comparing between the two settings.

Tables \reftab{tab.indip03m4} through \reftab{tab.indip045m4} report the payoff for all the profiles resulting from the quantized values of $P_{mal}^B$ and $P_{mal}^{FC}$, for the case of independent node states (constrained maximum entropy distribution\index{constrained maximum entropy|textbf}). The error probabilities in all the tables are scaled by a convenient power of 10. In all the cases $P_{mal}^B = 1$ is a dominant strategy\index{dominant strategy|textbf} for the Byzantines, and the profile $(1,1)$ is the unique rationalizable equilibrium of the game. As expected, the error probability increases with the number of Byzantines. The value of the payoff at the equilibrium ranges from $P_e = 0.0349$ with $\alpha = 0.3$ to $P_{e} = 0.3314$ with $\alpha = 0.45$. For completeness, we report the value of the error probability in the non-adversarial setup, which is $P_e = 0.34 \cdot 10^{-5}$.

\begin{table}[h]
\centering
\renewcommand{\arraystretch}{1.1}
\begin{tabular}{c| c| c| c| c| c| c|}
                       \hline
\multicolumn{1}{|c|}{\tiny{$P_{mal}^{B}$/$P_{mal}^{FC}$}} & 0.5   & 0.6   & 0.7   & 0.8   &0.9   &1.0   \\ \hline
\multicolumn{1}{|c|}{0.5}  &0.845 & 0.965 & 1.1 & 1.3 & 1.6 & 2.1\\ \hline
\multicolumn{1}{|c|}{0.6} &1.2 & 1.1 & 1.2 & 1.5e-3 & 1.8 & 2.6\\ \hline
\multicolumn{1}{|c|}{0.7}  &2.2 & 2.0 & 1.8 & 1.8e-3 & 2.1 & 3.7\\ \hline
\multicolumn{1}{|c|}{0.8} &5.4 & 5.1 & 5.0 & 5.0e-3 & 5.1 & 7.7\\ \hline
\multicolumn{1}{|c|}{0.9}  &16.2 & 16.1 & 16.5 & 16.4 & 16.0 & 19.1\\ \hline
\multicolumn{1}{|c|}{1.0} &43 & 43.1 & 46.9 & 46.8 & 41.6 & {\bf 34.9}\\ \hline
\end{tabular}

\caption{\textit{Payoff of the $DF_{Byz}$ game ($10^3 \times P_e$) with independent node states with $\alpha = 0.3$, $m= 4$, $n = 20$, $\varepsilon = 0.1$. The equilibrium point is highlighted in bold.}}
\label{tab.indip03m4}
\end{table}

\begin{table}[h]
\centering
\renewcommand{\arraystretch}{1.1}
\begin{tabular}{c| c| c| c| c| c| c|}
                       \hline
\multicolumn{1}{|c|}{\tiny{$P_{mal}^{B}$/$P_{mal}^{FC}$}} & 0.5   & 0.6   & 0.7   & 0.8   &0.9   &1.0   \\ \hline
\multicolumn{1}{|c|}{0.5}  &0.33 &0.37  &0.44  &0.58 &0.73  &0.85 \\ \hline
\multicolumn{1}{|c|}{0.6}  &0.60 &0.54  &0.59  &0.70 &0.80  &1.14 \\ \hline
\multicolumn{1}{|c|}{0.7}  &1.38 &1.20  &1.19  &1.24 &1.29  &2.40 \\ \hline
\multicolumn{1}{|c|}{0.8}  &3.88 &3.56  &3.36  &3.31 &3.35  &6.03 \\ \hline
\multicolumn{1}{|c|}{0.9}  &9.93 &9.61  &9.57  &9.55 &9.54  &11.96 \\ \hline
\multicolumn{1}{|c|}{1.0}  &20.33 &20.98  &21.70  &21.90 &21.84  & {\bf 19.19} \\ \hline
\end{tabular}

\caption{\textit{Payoff of the $DF_{Byz}$ game ($10^2 \times P_e$) with independent node states with $\alpha = 0.4$, $m= 4$, $n = 20$, $\varepsilon = 0.1$. The equilibrium point is highlighted in bold.}}
\label{tab.indip04m4}
\end{table}

\begin{table}[h]
\centering
\renewcommand{\arraystretch}{1.1}
\begin{tabular}{c| c| c| c| c| c| c|}
                       \hline
\multicolumn{1}{|c|}{\tiny{$P_{mal}^{B}$/$P_{mal}^{FC}$}} & 0.5   & 0.6   & 0.7   & 0.8   &0.9   &1.0   \\ \hline
\multicolumn{1}{|c|}{0.5}  &0.62 &0.69  &0.86  &1.34 &1.70  &1.57 \\ \hline
\multicolumn{1}{|c|}{0.6}  &1.23 &1.15  &1.26  &1.84 &2.18  &2.38 \\ \hline
\multicolumn{1}{|c|}{0.7}  &2.94 &2.64  &2.57  &3.00 &3.14  &5.33 \\ \hline
\multicolumn{1}{|c|}{0.8}  &7.89 &7.39  &7.03  &6.74 &6.81  &12.73 \\ \hline
\multicolumn{1}{|c|}{0.9}  &18.45 &17.94  &17.63  &17.08 &17.07  &22.78 \\ \hline
\multicolumn{1}{|c|}{1.0}  &34.39 &34.62  &34.84  &36.66 &36.61  & {\bf 33.14} \\ \hline
\end{tabular}

\caption{\textit{Payoff of the $DF_{Byz}$ game ($10^2 \times P_e$) with independent node states with $\alpha = 0.45$, $m= 4$, $n = 20$, $\varepsilon = 0.1$. The equilibrium point is highlighted in bold.}}
\label{tab.indip045m4}
\end{table}

Tables \reftab{tab.fixed6m4} through \reftab{tab.fixed9m4} report the payoffs for the case of fixed number of Byzantines, respectively equal to 6, 8 and 9.

\begin{table}[h!]
\centering
\renewcommand{\arraystretch}{1.1}
\begin{tabular}{c| c| c| c| c| c| c|}
                       \hline
\multicolumn{1}{|c|}{\tiny{$P_{mal}^{B}$/$P_{mal}^{FC}$}} & 0.5   & 0.6   & 0.7   & 0.8   &0.9   &1.0   \\ \hline
\multicolumn{1}{|c|}{0.5}  & {\bf 3.80} & {\bf 3.80}  &4.60  &7.60 &12.0  &29.0 \\ \hline
\multicolumn{1}{|c|}{0.6}  &3.60 &3.45  &3.90  &5.20 &8.0  &17.0 \\ \hline
\multicolumn{1}{|c|}{0.7}  &3.45 &2.80  &2.80  &3.10 &4.40  &8.75 \\ \hline
\multicolumn{1}{|c|}{0.8}  &4.10 &2.85  &2.15  &2.05 &2.25  &3.25 \\ \hline
\multicolumn{1}{|c|}{0.9}  &3.55 &2.05  &1.40  &0.95 &0.70  &0.75 \\ \hline
\multicolumn{1}{|c|}{1.0}  &2.05 &0.90  &0.35  &0.15 &0.05  &0.05  \\ \hline
\end{tabular}

\caption{\textit{Payoff of the $DF_{Byz}$ game ($10^4 \times P_e$) with $n_B = 6$, $m= 4$, $n = 20$, $\varepsilon = 0.1$. The equilibrium point is highlighted in bold.}}
\label{tab.fixed6m4}
\end{table}

\begin{table}[h!]
\centering
\renewcommand{\arraystretch}{1.1}
\begin{tabular}{c| c| c| c| c| c| c|}
                       \hline
\multicolumn{1}{|c|}{\tiny{$P_{mal}^{B}$/$P_{mal}^{FC}$}} & 0.5   & 0.6   & 0.7   & 0.8   &0.9   &1.0   \\ \hline
\multicolumn{1}{|c|}{0.5}  &1.2 &1.4  &1.9  &3.1 &6.3  &18.9 \\ \hline
\multicolumn{1}{|c|}{0.6}  &1.5 &1.4  &1.4  &2.0 &3.7  &10.0 \\ \hline
\multicolumn{1}{|c|}{0.7}  &1.4 &1.1  &0.945  &1.1 &1.7  &4.0 \\ \hline
\multicolumn{1}{|c|}{0.8}  &1.4 &0.95  &0.715  &0.58 &0.675  &1.2 \\ \hline
\multicolumn{1}{|c|}{0.9}  &2.1 &1.4  &0.995  &0.745 &0.71  &0.78 \\ \hline
\multicolumn{1}{|c|}{1.0}  &7.3 &5.7  &5.3  &3.7 &3.0  &2.9  \\ \hline
\end{tabular}

\caption{\textit{Payoff of the $DF_{Byz}$ game ($10^3 \times P_e$) with $n_B = 8$, $m= 4$, $n = 20$, $\varepsilon = 0.1$. No pure strategy equilibrium exists.}}
\label{tab.fixed8m4}
\end{table}

\begin{table}[h!]
\centering
\renewcommand{\arraystretch}{1.1}
\begin{tabular}{c| c| c| c| c| c| c|}
                       \hline
\multicolumn{1}{|c|}{\tiny{$P_{mal}^{B}$/$P_{mal}^{FC}$}} & 0.5   & 0.6   & 0.7   & 0.8   &0.9   &1.0   \\ \hline
\multicolumn{1}{|c|}{0.5}  &0.22 &0.24  &0.33  &0.63 &1.41  &4.13 \\ \hline
\multicolumn{1}{|c|}{0.6}  &0.27 &0.24  &0.27  &0.41 &0.78  &2.03 \\ \hline
\multicolumn{1}{|c|}{0.7}  &0.32 &0.24  &0.23  &0.26 &0.37  &0.82 \\ \hline
\multicolumn{1}{|c|}{0.8}  &0.54 &0.45  &0.39  &0.36 &0.41  &0.59 \\ \hline
\multicolumn{1}{|c|}{0.9}  &2.04 &1.87  &1.76  &1.58 &1.56  &1.66 \\ \hline
\multicolumn{1}{|c|}{1.0}  &9.48 &8.76  &8.37  &6.72 &5.88  & {\bf 5.51} \\ \hline
\end{tabular}

\caption{\textit{Payoff of the $DF_{Byz}$ game ($10^2 \times P_e$) with $n_B = 9$, $m= 4$, $n = 20$, $\varepsilon = 0.1$. The equilibrium point is highlighted in bold.}}
\label{tab.fixed9m4}
\end{table}

When $n_B = 6$, $P_{mal}^B = 0.5$ is a dominant strategy for the Byzantines, and the profile $(0.5, 0.5)$ is the unique rationalizable equilibrium of the game corresponding to a payoff $P_e = 3.8 \cdot 10^{-4}$. This marks a significant difference with respect to the case of independent nodes, where the optimum strategy\index{optimum strategy|textbf} for the Byzantines was to let $P_{mal}^B = 1$. The reason behind the different behavior is that in the case of fixed number of nodes, the a-priori knowledge available at the FC is larger than in the case of independent nodes with the same average number of nodes. This additional information permits to the FC to localize the byzantine nodes, which now cannot use $P_{mal}^B = 1$, since in this case they would still transmit some useful information to the FC. On the contrary, by letting $P_{mal}^B = 0.5$ the information received from the byzantine nodes is zero, hence making the task of the FC harder. When $n_B = 9$ (Table \reftab{tab.fixed9m4}), the larger number of Byzantines makes the identification of malicious nodes more difficult and $P_{mal}^B = 1$ is again a dominant strategy, with the equilibrium of the game obtained at the profile (1,1) with $P_e = 0.0551$. A somewhat intermediate situation is observed when $n_B = 8$ (Table \reftab{tab.fixed8m4}). In this case, no equilibrium point\index{equilibrium point|textbf} exists (let alone a dominant strategy) if we consider pure strategies only. On the other hand, when mixed strategies are considered, the game has a unique Nash equilibrium\index{Nash equilibrium|textbf} for the strategies reported in Table \reftab{tab.mixedNASHfixed8m4} (each row in the table gives the probability vector assigned to the quantized values of $P_{mal}$ by one of the players at the equilibrium). Interestingly the optimum strategy of the Byzantines corresponds to alternate playing $P_{mal}^{B} = 1 $ and $P_{mal}^{B} = 0.5 $, with intermediate probabilities. This confirms the necessity for the Byzantines to find a good trade-off between two alternative strategies: set to zero the information transmitted to the FC or try to push it towards a wrong decision. We also observe that the error probabilities at the equilibrium are always lower than those of the game with independent nodes. This is an expected result, since in the case of fixed nodes the FC has a better knowledge about the distribution of Byzantines.

\begin{table}[h!]
\centering
\renewcommand{\arraystretch}{1.4}
\begin{tabular}{c| c| c| c| c| c| c|}
                       \hline
\multicolumn{1}{|c|}{}& 0.5   & 0.6   & 0.7   & 0.8   &0.9   &1.0   \\ \hline
\multicolumn{1}{|c|}{$P(P_{mal}^{B})$ }  & 0.179 & 0 & 0 & 0 & 0 & 0.821 \\ \hline
\multicolumn{1}{|c|}{$P(P_{mal}^{FC})$ }  & 0 & 0  & 0  &0.846 & 0.154 &0 \\ \hline
\multicolumn{7}{|c|}{$P_e^* = 3.6e-3$} \\ \hline
\end{tabular}

\caption{\textit{Mixed strategies equilibrium for the $DF_{Byz}$ game with $n_B = 8$, $m= 4$, $n = 20$,  $\varepsilon = 0.1$. $P_e^*$ indicates the error probability at the equilibrium.}}
\label{tab.mixedNASHfixed8m4}
\end{table}

The last case we have analyzed corresponds to a situation in which the FC knows that the number of Byzantines cannot be larger than a certain value $h$ (see Sec. \ref{subsec.less_half}).

We first consider the case in which the FC knows only that the number of Byzantines\index{Byzantines distribution|textbf} is lower than $n/2$. The payoff for this instantiation of the $DF_{Byz}$ game is given in Table \reftab{tab.lessN2m4}. In order to compare the results of this case with those obtained for the case of independent nodes and that of fixed number of Byzantines, we observe that when all the sequences $a^n$ with $n_B < n/2$ have the same probability, the average number of Byzantines turns out to be 7.86. The most similar settings, then, are that of independent nodes with $\alpha = 0.4$ and that of fixed number of nodes with $n_B = 8$. With respect to the former, the error probability at the equilibrium is significantly smaller, thus confirming the case of independent nodes as the worst scenario for the FC. This is due to the fact that with $\alpha = 0.4$ it is rather likely that number of Byzantines is larger than $0.5$ thus making any reliable decision impossible. The error probability obtained with a fixed number of Byzantines equal to 8, however, is much lower. This is a reasonable result, since in that case the a-priori information available to the FC permits a better localization of the corrupted reports.

We now move to the case with $h < n/2$. Table \reftab{tab.lesshm4} reports the payoffs of the game when $N_B < n/3$. By assuming a maximum entropy distribution\index{maximum entropy|textbf} over the admissible configurations $a^n$ with $N_B < n/3$, the average number of Byzantines turns out to be 4.64. In this case, the equilibrium point shifts to $(0.5,0.5)$. This confirms the behavior discussed in the previous paragraph: since the average number of Byzantines is lower the FC is able to localize them with a batter accuracy, then it is better for the Byzantines to minimize the information delivered to the FC.
\begin{table}[h!]
\centering
\renewcommand{\arraystretch}{1.1}
\begin{tabular}{c| c| c| c| c| c| c|}
                       \hline
\multicolumn{1}{|c|}{\tiny{$P_{mal}^{B}$/$P_{mal}^{FC}$}} & 0.5   & 0.6   & 0.7   & 0.8   &0.9   &1.0   \\ \hline
\multicolumn{1}{|c|}{0.5}  &0.15 &0.17  &0.20  &0.29 &0.39  &0.51 \\ \hline
\multicolumn{1}{|c|}{0.6}  &0.17 &0.16  &0.16  &0.22 &0.29  &0.40 \\ \hline
\multicolumn{1}{|c|}{0.7}  &0.19 &0.15  &0.14  &0.16 &0.20  &0.30 \\ \hline
\multicolumn{1}{|c|}{0.8}  &0.27 &0.20  &0.17  &0.16 &0.17  &0.22 \\ \hline
\multicolumn{1}{|c|}{0.9}  &0.85 &0.76  &0.72  &0.63 &0.58  &0.63 \\ \hline
\multicolumn{1}{|c|}{1.0}  &3.81 &3.49  &3.30  &2.62 &2.24  & {\bf 2.13} \\ \hline
\end{tabular}

\caption{\textit{Payoff of the $DF_{Byz}$ game ($10^2 \times P_e$) with $N_B < n/2$. The  other parameters of the game are set as follows: $m= 4$, $n = 20$, $\varepsilon = 0.1$. The equilibrium point is highlighted in bold.}}
\label{tab.lessN2m4}
\end{table}

\begin{table}[h!]
\centering
\renewcommand{\arraystretch}{1.1}
\begin{tabular}{c| c| c| c| c| c| c|}
                       \hline
\multicolumn{1}{|c|}{\tiny{$P_{mal}^{B}$/$P_{mal}^{FC}$}} & 0.5   & 0.6   & 0.7   & 0.8   &0.9   &1.0   \\ \hline
\multicolumn{1}{|c|}{0.5} & {\bf 1.9}     &2.10 &2.30  &2.85  &3.4 &4.05   \\ \hline
\multicolumn{1}{|c|}{0.6}    &1.85  &1.75 &1.9  &2.0  &2.85 &3.80   \\ \hline
\multicolumn{1}{|c|}{0.7} & 1.3     &1.05 &0.75  &0.8  &1.30 &2.20   \\ \hline
\multicolumn{1}{|c|}{0.8} &1.7       &1.45 &1.15  &1.1  &1.15 &1.50   \\ \hline
\multicolumn{1}{|c|}{0.9} &1.25     &0.65 &0.5  &0.35  &0.35 &0.35   \\ \hline
\multicolumn{1}{|c|}{1.0}  & 0.85   &0.6 &0.4  &0.1  &0.05 &0.05   \\ \hline
\end{tabular}

\caption{\textit{Payoff of the $DF_{Byz}$ game ($10^4 \times P_e$) with $N_B < n/3$. The other parameters of the game are set as follows: $m= 4$, $n = 20$, $\varepsilon = 0.1$. The equilibrium point is highlighted in bold.}}
\label{tab.lesshm4}
\end{table}

\subsubsection{Intermediate values of $m$}

In this section we report the results that we got when the length of the observation vector increases. We expect that by comparing the reports sent by the nodes corresponding to different components of the state vector allows a better identification of the byzantine nodes, thus modifying the equilibrium of the game. Specifically, we repeated the simulations carried out in the previous section, by letting $m = 10$. Though desirable, repeating the simulations with even larger values of $m$ is not possible due to the exponential growth of the complexity of the optimum fusion rule with $m$.

Tables \reftab{tab.indip03m10} through \reftab{tab.indip045m10} report the payoffs of the game for the case of independent node states. As it can be seen, $P_{mal}^B = 1.0$ is still a dominant strategy\index{dominant strategy|textbf} for the Byzantines and the profile\index{profile|textbf} (1,1) is the unique rationalizable equilibrium of the game. Moreover, the value of $P_e$ at the equilibrium is slightly lower than for $m=4$, when $\alpha=0.3$ and $\alpha=0.4$ (see Tables \reftab{tab.indip03m4} and \reftab{tab.indip04m4}). Such an advantage disappears when $\alpha=0.45$ (see Table \reftab{tab.indip045m4}), since the number of Byzantines is so large that identifying them is difficult even with $m =10$.

\begin{table}[h!]
\centering
\renewcommand{\arraystretch}{1.1}
\begin{tabular}{c| c| c| c| c| c| c|}
                       \hline
\multicolumn{1}{|c|}{\tiny{$P_{mal}^{B}$/$P_{mal}^{FC}$}} & 0.5   & 0.6   & 0.7   & 0.8   &0.9   &1.0   \\ \hline
\multicolumn{1}{|c|}{0.5}  &0.258 &0.28  &0.39  &0.63  &1.0  &1.7 \\ \hline
\multicolumn{1}{|c|}{0.6}  &0.28 &0.226  &0.248  &0.362  &0.652  &2.0 \\ \hline
\multicolumn{1}{|c|}{0.7}  &0.346 &0.22  &0.206  &0.23  &0.314  &5.3 \\ \hline
\multicolumn{1}{|c|}{0.8}  &1.2 &0.648  &0.44  &0.428  &0.498  &13.9 \\ \hline
\multicolumn{1}{|c|}{0.9}  &8.6 &7.8  &7.6  &7.8  &7.5  &19.9 \\ \hline
\multicolumn{1}{|c|}{1.0}  &41.9 &46.7  &50.9  &59.8  &52.2  &\bf{32.9} \\ \hline
\end{tabular}

\caption{\textit{Payoff of the $DF_{Byz}$ game ($10^3 \times P_e$) with independent node states with $\alpha = 0.3$, $m= 10$, $n = 20$, $\varepsilon = 0.1$. The equilibrium point is highlighted in bold.}}
\label{tab.indip03m10}
\end{table}

\begin{table}[h!]
\centering
\renewcommand{\arraystretch}{1.1}
\begin{tabular}{c| c| c| c| c| c| c|}
                       \hline
\multicolumn{1}{|c|}{\tiny{$P_{mal}^{B}$/$P_{mal}^{FC}$}} & 0.5   & 0.6   & 0.7   & 0.8   &0.9   &1.0   \\ \hline
\multicolumn{1}{|c|}{0.5}  &0.11 &0.13  &0.19  &0.73 &2.16  &0.68 \\ \hline
\multicolumn{1}{|c|}{0.6}  &0.11 &8.32e-2  &9.96e-2  &0.26 &0.67  &1.30 \\ \hline
\multicolumn{1}{|c|}{0.7}  &0.18 &7.66e-2  &6.62e-2  &9.52e-2 &0.18  &4.87 \\ \hline
\multicolumn{1}{|c|}{0.8}  &1.10 &0.60  &0.33  &0.24 &0.28  &10.41 \\ \hline
\multicolumn{1}{|c|}{0.9}  &5.77 &4.75  &3.95  &3.53 &3.41  &13.44 \\ \hline
\multicolumn{1}{|c|}{1.0}  &20.41 &21.26  &22.65  &24.27 &26.21  &\bf{18.72} \\ \hline
\end{tabular}

\caption{\textit{Payoff of the $DF_{Byz}$ game ($10^2 \times P_e$) with independent node states with $m= 10$, $n = 20$, $\alpha = 0.4$, $\varepsilon = 0.1$. The equilibrium point is highlighted in bold.}}
\label{tab.indip04m10}
\end{table}

\begin{table}[h!]
\centering
\renewcommand{\arraystretch}{1.1}
\begin{tabular}{c| c| c| c| c| c| c|}
                       \hline
\multicolumn{1}{|c|}{\tiny{$P_{mal}^{B}$/$P_{mal}^{FC}$}} & 0.5   & 0.6   & 0.7   & 0.8   &0.9   &1.0   \\ \hline
\multicolumn{1}{|c|}{0.5}  &0.20 &0.23  &0.47  &2.88 &10.92  &1.26 \\ \hline
\multicolumn{1}{|c|}{0.6}  &0.22 &0.18  &0.24  &0.80 &2.85  &2.93 \\ \hline
\multicolumn{1}{|c|}{0.7}  &0.50 &0.19  &0.15  &0.23 &0.65  &10.64 \\ \hline
\multicolumn{1}{|c|}{0.8}  &2.61 &1.24  &0.63  &0.41 &0.59  &20.65 \\ \hline
\multicolumn{1}{|c|}{0.9}  &11.74 &9.28  &7.08  &5.65 &5.21  &25.85 \\ \hline
\multicolumn{1}{|c|}{1.0}  &34.25 &34.94  &36.01  &37.74 &39.87  &\bf{33.17}  \\ \hline
\end{tabular}

\caption{\textit{Payoff of the $DF_{Byz}$ game ($10^2 \times P_e$) with independent node states with $\alpha = 0.45$, $m= 10$, $n = 20$, $\varepsilon = 0.1$. The equilibrium point is highlighted in bold.}}
\label{tab.indip045m10}
\end{table}

The results of the simulations for the case of fixed number of nodes with $n_B = \{6,8,9\}$ are given in Tables \reftab{tab.fixed6m10} through \reftab{tab.fixed9m10}. With respect to the case of $m = 4$, the optimum strategy for the Byzantines shifts to $P_{mal}^B = 0.5$. When $n_B =6$,  $P_{mal}^B = 0.5$ is a dominant strategy, while for $n_B =8$ and $n_B =9$, no equilibrium point exists if we consider only pure strategies. The mixed strategy equilibrium\index{Mixed Strategy Nash Equilibrium|textbf} point for these cases is given in Tables \reftab{tab.mixedNASHfixed8m10} and \reftab{tab.mixedNASHfixed9m10}. By comparing those tables with those of the case $m=4$, the preference towards $P_{mal}^B = 0.5$ is evident.

\begin{table}[h!]
\centering
\renewcommand{\arraystretch}{1.1}
\begin{tabular}{c| c| c| c| c| c| c|}
                       \hline
\multicolumn{1}{|c|}{\tiny{$P_{mal}^{B}$/$P_{mal}^{FC}$}} & 0.5   & 0.6   & 0.7   & 0.8   &0.9   &1.0   \\ \hline
\multicolumn{1}{|c|}{0.5}  &\bf{1.22} &\bf{1.22}  &1.40  &2.20  &5.06  &11.0 \\ \hline
\multicolumn{1}{|c|}{0.6}  &1.12 &0.94  &1.02  &1.26  &2.56  &5.34 \\ \hline
\multicolumn{1}{|c|}{0.7}  &1.22 &0.58  &0.56  &0.64  &0.98  &2.06 \\ \hline
\multicolumn{1}{|c|}{0.8}  &1.22 &0.36  &0.32  &0.28  &0.30  &0.56 \\ \hline
\multicolumn{1}{|c|}{0.9}  &1.40 &0.20  &0.18  &0.16  &0.10  &0.18 \\ \hline
\multicolumn{1}{|c|}{1.0}  &1.52 &0.14  &0.14  &0.10  &6e-2  &4e-2 \\ \hline
\end{tabular}

\caption{\textit{Payoff of the $DF_{Byz}$ game ($10^4 \times P_e$) with $n_B = 6$, $m= 10$, $n = 20$,  $\varepsilon = 0.1$. The equilibrium point is highlighted in bold.}}
\label{tab.fixed6m10}
\end{table}

\begin{table}[h!]
\centering
\renewcommand{\arraystretch}{1.1}
\begin{tabular}{c| c| c| c| c| c| c|}
                       \hline
\multicolumn{1}{|c|}{\tiny{$P_{mal}^{B}$/$P_{mal}^{FC}$}} & 0.5   & 0.6   & 0.7   & 0.8   &0.9   &1.0   \\ \hline
\multicolumn{1}{|c|}{0.5}  &4.04 &4.44  &6.24  &10.0 &24.0  &71.0 \\ \hline
\multicolumn{1}{|c|}{0.6}  &4.02 &3.30  &3.58  &5.24 &10.0  &26.0 \\ \hline
\multicolumn{1}{|c|}{0.7}  &3.48 &2.16  &2.14  &2.16 &3.26  &7.76 \\ \hline
\multicolumn{1}{|c|}{0.8}  &3.56 &1.10  &0.88  &0.78 &0.98  &2.08 \\ \hline
\multicolumn{1}{|c|}{0.9}  &4.60 &0.68  &0.54  &0.30 &0.26  &0.44 \\ \hline
\multicolumn{1}{|c|}{1.0}  &5.20 &0.54  &0.20  &8e-2 &0  &0 \\ \hline
\end{tabular}

\caption{\textit{Payoff of the $DF_{Byz}$ game ($10^4 \times P_e$) with $n_B = 8$, $m= 10$, $n = 20$,  $\varepsilon = 0.1$. No pure strategy equilibrium exists.}}
\label{tab.fixed8m10}
\end{table}

\begin{table}[h!]
\centering
\renewcommand{\arraystretch}{1.1}
\begin{tabular}{c| c| c| c| c| c| c|}
                       \hline
\multicolumn{1}{|c|}{\tiny{$P_{mal}^{B}$/$P_{mal}^{FC}$}} & 0.5   & 0.6   & 0.7   & 0.8   &0.9   &1.0   \\ \hline
\multicolumn{1}{|c|}{0.5}  &6.74 &7.82  &12  &23 &52  &168 \\ \hline
\multicolumn{1}{|c|}{0.6}  &5.44 &4.94  &6.14  &9.40 &18  &52 \\ \hline
\multicolumn{1}{|c|}{0.7}  &4.22 &3.30  &2.78  &3.38 &5.86  &15 \\ \hline
\multicolumn{1}{|c|}{0.8}  &3.0 &2.24  &1.24  &0.78 &1.32  &3.24 \\ \hline
\multicolumn{1}{|c|}{0.9}  &5.22 &2.36  &1.34  &1.02 &0.88  &1.24 \\ \hline
\multicolumn{1}{|c|}{1.0}  &70 &40  &19  &8.90 &3.44  &2.42  \\ \hline
\end{tabular}

\caption{\textit{Payoff of the $DF_{Byz}$ game ($10^4 \times P_e$) with $n_B = 9$, $m= 10$, $n = 20$, $\varepsilon = 0.1$. No pure strategy equilibrium exists.}}
\label{tab.fixed9m10}
\end{table}

Table \reftab{tab.lessN2m10}, gives the results for the case $N_B < n/2$. As in the case of fixed number of Byzantines, the equilibrium point strategy passes from the pure strategy (1,1) to a mixed strategy (see Table \reftab{tab.mixedNASHlessm10}). Once again, the reason for such a behavior, is that when $m$ increases, the amount of information available to the FC increases, hence making the detection of corrupted reports easier. As a result, the Byzantines must find a trade-off between forcing a wrong decision and reducing the mutual information\index{mutual information|textbf} between the corrupted reports and system states\index{dual behavior|textbf}.
Eventually, Table \reftab{tab.lesshm10} reports the results of the game for the case $N_B < n/3$ and $m = 10$. As one could expect, the profile $(0.5,0.5)$ is still the equilibrium point of the game, as the optimum strategy for the Byzantines continues to be the one which minimizes the amount of information delivered to the FC.
We conclude observing that even with $m=10$, the case of independent nodes results in the worst performance.

\begin{table}[h!]
\centering
\renewcommand{\arraystretch}{1.1}
\begin{tabular}{c| c| c| c| c| c| c|}
                       \hline
\multicolumn{1}{|c|}{\tiny{$P_{mal}^{B}$/$P_{mal}^{FC}$}} & 0.5   & 0.6   & 0.7   & 0.8   &0.9   &1.0   \\ \hline
\multicolumn{1}{|c|}{0.5}  &4.46 &5.38  &6.64  &9.88 &16  &27 \\ \hline
\multicolumn{1}{|c|}{0.6}  &3.90 &3.38  &4.10  &5.90 &9.42  &19 \\ \hline
\multicolumn{1}{|c|}{0.7}  &3.04 &2.24  &1.82  &2.26 &3.68  &7.28 \\ \hline
\multicolumn{1}{|c|}{0.8}  &2.78 &1.72  &1.0  &0.72 &0.90  &1.70 \\ \hline
\multicolumn{1}{|c|}{0.9}  &3.24 &1.38  &0.62  &0.30 &0.20  &0.48 \\ \hline
\multicolumn{1}{|c|}{1.0}  &27 &15  &6.84  &4.68 &1.42  &1.04 \\ \hline
\end{tabular}

\caption{\textit{Payoff of the $DF_{Byz}$ game ($10^4 \times P_e$) with $N_B < n/2$. The other parameters of the game are set as follows: $m= 10$, $n = 20$, $\varepsilon = 0.1$. No pure strategy equilibrium exists.}}
\label{tab.lessN2m10}
\end{table}

\begin{table}[h!]
\centering
\renewcommand{\arraystretch}{1.1}
\begin{tabular}{c| c| c| c| c| c| c|}
                       \hline
\multicolumn{1}{|c|}{\tiny{$P_{mal}^{B}$/$P_{mal}^{FC}$}} & 0.5   & 0.6   & 0.7   & 0.8   &0.9   &1.0   \\ \hline
\multicolumn{1}{|c|}{0.5}  &\bf{0.5} &0.58  &0.66  &0.78 &1.1 &1.56\\ \hline
\multicolumn{1}{|c|}{0.6}  &0.44 &0.42  &0.48  &0.56 &0.88  &1.3 \\ \hline
\multicolumn{1}{|c|}{0.7}  &0.48 &0.48  &0.46  &0.48 &0.54  &0.86 \\ \hline
\multicolumn{1}{|c|}{0.8}  &0.4 &0.36  &0.3  &0.22 &0.26  &0.26 \\ \hline
\multicolumn{1}{|c|}{0.9}  &0.34 &0.3 &0.22  &0.16 &0.012  &0.016 \\ \hline
\multicolumn{1}{|c|}{1.0}  &0.34 &0.28  &0.16  &0.06 &0.02  &0.02 \\
\hline
\end{tabular}
\caption{\textit{Payoff of the $DF_{Byz}$ game ($10^4 \times P_e$) with $N_B < n/3$ in the following setup: $m= 10$, $n = 20$, $\varepsilon = 0.1$.The equilibrium point is highlighted in bold.}}
\label{tab.lesshm10}
\end{table}

\begin{table}[h!]
\centering
\renewcommand{\arraystretch}{1.4}
\begin{tabular}{c| c| c| c| c| c| c|}
                       \hline
\multicolumn{1}{|c|}{}& 0.5   & 0.6   & 0.7   & 0.8   &0.9   &1.0   \\ \hline
\multicolumn{1}{|c|}{$P(P_{mal}^{B})$ }  & 0.921 & 0 & 0 & 0 & 0 & 0.079 \\ \hline
\multicolumn{1}{|c|}{$P(P_{mal}^{FC})$ }  &0.771 &0.229  & 0  &0 & 0 &0 \\ \hline
\multicolumn{7}{|c|}{$P_e^* = 4.13e-4$} \\ \hline
\end{tabular}

\caption{\textit{Mixed strategies equilibrium for the $DF_{Byz}$ game with  $n_B = 8$, $m= 10$, $n = 20$,  $\varepsilon = 0.1$. $P_e^*$ indicates the error probability at the equilibrium.}}
\label{tab.mixedNASHfixed8m10}
\end{table}

\begin{table}[h!]
\centering
\renewcommand{\arraystretch}{1.4}
\begin{tabular}{c| c| c| c| c| c| c|}
                       \hline
\multicolumn{1}{|c|}{}& 0.5   & 0.6   & 0.7   & 0.8   &0.9   &1.0   \\ \hline
\multicolumn{1}{|c|}{$P(P_{mal}^{B})$ }  & 0.4995 & 0 & 0 & 0 & 0 & 0.5005 \\ \hline
\multicolumn{1}{|c|}{$P(P_{mal}^{FC})$ }  &0 &0  &0.66  &0.34 & 0 &0 \\ \hline
\multicolumn{7}{|c|}{$P_e^* = 1.58e-3$} \\ \hline\end{tabular}

\caption{\textit{Mixed strategies equilibrium for the $DF_{Byz}$ game with $n_B = 9$, $m= 10$, $n = 20$,  $\varepsilon = 0.1$. $P_e^*$ indicates the error probability at the equilibrium.}}
\label{tab.mixedNASHfixed9m10}
\end{table}

\begin{table}[h!]
\centering
\renewcommand{\arraystretch}{1.4}
\begin{tabular}{c| c| c| c| c| c| c|}
                       \hline
\multicolumn{1}{|c|}{}& 0.5   & 0.6   & 0.7   & 0.8   &0.9   &1.0   \\ \hline
\multicolumn{1}{|c|}{$P(P_{mal}^{B})$ }  & 0.4 & 0 & 0 & 0 & 0 & 0.6 \\ \hline
\multicolumn{1}{|c|}{$P(P_{mal}^{FC})$ }  &0 &0  &0.96  &0.04 & 0 &0 \\ \hline
\multicolumn{7}{|c|}{$P_e^* = 6.76e-4$} \\ \hline
\end{tabular}

\caption{\textit{Mixed strategies equilibrium for the $DF_{Byz}$ game with $N_B < n/2$ with $m= 10$, $n = 20$, $\varepsilon = 0.1$. $P_e^*$ indicates the error probability at the equilibrium.}}
\label{tab.mixedNASHlessm10}
\end{table}

\subsection{Performance at the equilibrium and comparison with prior works}

As a last analysis we compared the error probability obtained by the game-theoretic optimum decision fusion introduced in this chapter, with those obtained by previous works. Specifically, we compared our scheme against a simple majority-based decision fusion rule according to which the FC decides that $s_j = 1$ if and only if $\sum_i r_{ij} > n/2$ ($\mathsf{Maj}$), against the hard isolation scheme described in \cite{Raw11} ($\mathsf{HardIS}$), and the soft isolation scheme proposed in the previous chapter.

In order to carry out a fair comparison and to take into account the game-theoretic nature of the problem, the performance of all the schemes are evaluated at the equilibrium. For the $\mathsf{HardIS}$ and $\mathsf{SoftIS}$ schemes this corresponds to letting $P_{mal}^{B}$ = 1. In fact, in the previous chapter, it is shown that this is a dominant strategy for these two specific fusion schemes. As a consequence, $P_{mal}^{FC}$ is also set to 1, since the FC knows in advance that the Byzantines will play the dominant strategy. For the $\mathsf{Maj}$ fusion strategy, the FC has no degrees of freedom, so no game actually exists in this case. With regard to the Byzantines, it is easy to realize that the best strategy is to let $P_{mal}^B = 1$. When the equilibrium corresponds to a mixed strategy, the error probability\index{error probability|textbf} is averaged according to the mixed strategies at the equilibrium. Tables \reftab{tab.payoffCOMPm4} and \reftab{tab.payoffCOMPm10} show the error probability at the equilibrium for the tested systems under different setups. As it can be seen, the fusion scheme resulting for the application of the optimum fusion rule in a game-theoretic setting, consistently provides better results for all the analyzed cases. Expectedly, the improvement is more significant for the setups in which the FC has more information about the distribution of the Byzantines\index{Byzantines distribution|textbf} across the network.

\begin{table}[h]
\centering
\renewcommand{\arraystretch}{1.1}
\begin{tabular}{c| c| c| c| c|}
                       \hline
\multicolumn{1}{|l|}{} & $\mathsf{Maj}$   & $\mathsf{HardIS}$   & $\mathsf{SoftIS}$   & $\mathsf{OPT}$   \\ \hline
\multicolumn{1}{|l|}{Independent nodes, $\alpha = 0.3$}  & 0.073 & 0.048 & 0.041 & 0.035 \\ \hline
\multicolumn{1}{|l|}{Independent nodes, $\alpha = 0.4$}  & 0.239 &  0.211 & 0.201 & 0.192\\ \hline
\multicolumn{1}{|l|}{Independent nodes, $\alpha = 0.45$}   & 0.362& 0.344 & 0.338 & 0.331\\ \hline
\multicolumn{1}{|l|}{Fixed n. of nodes $n_B = 6$}  & 0.017 & 0.002 & 6.2e-4 & 3.8e-4\\ \hline
\multicolumn{1}{|l|}{Fixed n. of nodes $n_B = 8$}   & 0.125 & 0.044 & 0.016 & 0.004\\ \hline
\multicolumn{1}{|l|}{Fixed n. of nodes $n_B = 9$}  & 0.279 & 0.186  & 0.125 & 0.055\\ \hline
\multicolumn{1}{|l|}{Max entropy with $N_B  < n/2$}  & 0.154 & 0.086 & 0.052 & 0.021\\ \hline
\multicolumn{1}{|l|}{Max entropy with $N_B  < n/3$}  & 0.0041 & 5e-4 & 2.15e-4 & 1.9e-4\\ \hline
\end{tabular}

\caption{\textit{Error probability at the equilibrium for various fusion schemes. All the results have been obtained by letting $m=4$, $n = 20$, $\varepsilon = 0.1$.}}
\label{tab.payoffCOMPm4}
\end{table}

\begin{table}[h]
\centering
\renewcommand{\arraystretch}{1.1}
\begin{tabular}{c| c| c| c| c|}
                       \hline
\multicolumn{1}{|l|}{} & $\mathsf{Maj}$   & $\mathsf{HardIS}$   & $\mathsf{SoftIS}$   & $\mathsf{OPT}$   \\ \hline
\multicolumn{1}{|l|}{Independent nodes, $\alpha = 0.3$}  &0.073 &0.0364  &0.0346  &0.033  \\ \hline
\multicolumn{1}{|l|}{Independent nodes, $\alpha = 0.4$}  &0.239 &0.193  &0.19  &0.187 \\ \hline
\multicolumn{1}{|l|}{Independent nodes, $\alpha = 0.45$}   &0.363 &0.334  &0.333  &0.331 \\ \hline
\multicolumn{1}{|l|}{Fixed n. of nodes $n_B = 6$}  &0.016 &1.53e-4  &1.41e-4  &1.22e-4 \\ \hline
\multicolumn{1}{|l|}{Fixed n. of nodes $n_B = 8$}   &0.126 &0.0028  &9.68e-4  &4.13e-4 \\ \hline
\multicolumn{1}{|l|}{Fixed n. of nodes $n_B = 9$}  &0.279 &0.0703  &0.0372  &1.58e-3 \\ \hline
\multicolumn{1}{|l|}{Max entropy with $N_B  < n/2$}  &0.154 &0.0271  &0.0141  &6.8e-4 \\ \hline
\multicolumn{1}{|l|}{Max entropy with $N_B  < n/3$}  &0.0039 &9.8e-05  &7.40e-05  &5e-05 \\ \hline
\end{tabular}

\caption{\textit{Error probability at the equilibrium for various fusion schemes. All the results have been obtained by letting $m=10$, $n = 20$, $\varepsilon = 0.1$.}}
\label{tab.payoffCOMPm10}
\end{table}

\section{Conclusions}
\label{sec.conc}

We have studied the problem of decision fusion in distributed sensor networks in the presence of Byzantines. We first derived the optimum decision strategy by assuming that the statistical behavior of the Byzantines is known. Then we relaxed such an assumption by casting the problem into a game-theoretic framework in which the FC tries to guess the behavior of the Byzantines. The Byzantines, in turn, must fix their corruption strategy without knowing the guess made by the FC. We considered several versions of the game with different distributions of the Byzantines across the network. Specifically, we considered three setups: unconstrained maximum entropy distribution, constrained maximum entropy distribution and fixed number of Byzantines. In order to reduce the computational complexity of the optimum fusion rule for large network sizes, we proposed an efficient implementation based on dynamic programming. Simulation results show that increasing the observation window $m$ leads to better identification of the Byzantines at the FC. This forces the Byzantines to look for a trade-off between forcing the FC to make a wrong decision on one hand, and reducing the mutual information between the reports and the system state\index{system state|textbf} on the other hand. Simulation results confirm that, in all the analyzed cases, the performance at the equilibrium are superior to those obtained by previously proposed techniques. 

\chapter{An Efficient Nearly-Optimum Decision Fusion Technique Based on Message Passing}
\label{chapter:InfoFusion}
\emph{"Efficiency is doing things right; effectiveness is doing the right things."}
\\
Peter Drucker

\section{Introduction}

\PARstart{\textcolor{red}I}n{} the attempt to diminish the computational complexity while minimizing the loss of performance with respect to the optimum fusion rule presented in chapter \ref{chapter:TIFS_SPL}, in this chapter, we propose a nearly-optimum fusion scheme based on message passing and factor graphs\index{Factor Graph|textbf}.  Moreover, we consider a more general model for the system state\index{system state|textbf} that includes both Markovian and independent sequences. At last, we confirm the results in chapter \ref{chapter:TIFS_SPL} that the optimum strategy for the Byzantines is to follow a dual-behavior to find a trade-off between inducing global decision error at the FC and avoid being detected by trying to minimize the mutual information between the reports and the sequence of system states\index{system state|textbf}.

In chapter \ref{chapter:TIFS_SPL} we have shown that the complexity\index{computational complexity|textbf} of the optimum decision fusion\index{optimum fusion rule|textbf} algorithm grows exponentially with the length of the observation window $m$. Such a complexity prevents the adoption of the optimum decision fusion rule in many practical situations. Also the results regarding the optimum strategies of the Byzantines and the FC derived in previous chapter cannot be immediately applied to the case of large observation windows.

Message passing algorithms\index{Message Passing Algorithm|textbf}, based on the so called Generalised Distributive Law (GLD, \cite{genlaw},\cite{genlawnewlook}), have been widely applied to solve a large range of optimization problems, including decoding of Low Density Parity Check (LDPC) codes \cite{GallagerLDPC} and BCJR codes \cite{genlaw}, dynamic programming \cite{verdu1987OptwithMP}, solution of probabilistic inference problems on Bayesian networks \cite{beliefpropagationAI} (in this case message passing algorithms are known as {\em belief propagation}). Here we use message passing to introduce a near-optimal solution of the decision fusion problem with multiple observations whose complexity grows only linearly\index{linear complexity|textbf} with the size of the observation window, thus marking a dramatic improvement with respect to the exponential complexity\index{exponential complexity|textbf} of the optimal scheme proposed in chapter \ref{chapter:TIFS_SPL}.

Using numerical simulations and by first focusing on the case of small observation windows, for which the optimum solution can still be applied, we prove that the new scheme gives near-optimal performance at a much lower complexity than the optimum scheme. We then use numerical simulations to evaluate the performance of the proposed method for long observation windows. As a result, we show that, even in this case, the proposed scheme maintains the performance improvement over the simple majority rule, the hard isolation scheme in \cite{Raw11} and the soft isolation scheme described in chapter \ref{chapter:CDC}.

As opposed to chapter \ref{chapter:TIFS_SPL}, we do not limit our analysis to the case of independent system states\index{independent system states|textbf}, but we extend it to a more realistic scenario where the sequence of states obey a Markovian distribution\index{Markovian system states|textbf} \cite{HMMref} as depicted in Figure \reffig{fig.HMM}. The Markovian model is rather common in the case of cognitive radio networks \cite{CRMM1, CRMM2, CRMM3} where the primary user occupancy of the spectrum is often modelled  as a Hidden Markov Model (HMM).

The Markovian case is found to be more favourable for the FC with respect to the case of independent states, due the additional a-priori information available to the FC.

Last but not the least, we confirm that the dual optimum behavior\index{dual behavior|textbf} of the Byzantines observed in chapter \ref{chapter:TIFS_SPL} is also present in the case of large observation windows, even if in the Markovian case, the Byzantines\index{Byzantines|textbf} may continue using the maximum attack power for larger observation windows.

\section{Notation and Problem Formulation} \label{sec:Notations}
For the analysis in this chapter we adopt the same notation we used in chapter \ref{chapter:TIFS_SPL}. For the sake of clarity, here, we recapitulate such notation. Let $s^m = \left\{s_1,s_2,\ldots,s_m\right\}$ with $s_j \in \{0,1\}$ indicating the sequence of system states\index{sequence of system states|textbf} over an observation window of length $m$. The nodes collect information about the system through the vectors ${\bf x}_1, {\bf x}_2 \dots {\bf x}_n$, with ${\bf x}_i$ indicating the observations available at node $i$. Based on such observations, a node $i$ makes a local decision $u_{ij}$ about system state $s_j$. We assume that the local error probability, hereafter indicated as $\varepsilon$, does not depend on either $i$ or $j$. The state of the nodes in the network is given by the vector $a^n = \left\{a_1,a_2,\ldots,a_n\right\}$ with $a_i = 1/0$ indicating that node $i$ is honest or Byzantine, respectively. Finally, the matrix $\mathbf{R} = \left\{r_{ij}\right\}$, $i = 1,\ldots,n$, $j = 1,\ldots,m$ contains all the reports received by the FC. Specifically, $r_{ij}$ is the report sent by node $i$ relative to $s_j$. For honest nodes we have $u_{ij}= r_{ij}$ while, for Byzantines we have $p (u_{ij} \ne r_{ij} ) = P_{mal}$. The Byzantines corrupt the local decisions independently of each other.

\begin{figure}[t!]
\centering
    \includegraphics[width=0.6\textwidth]{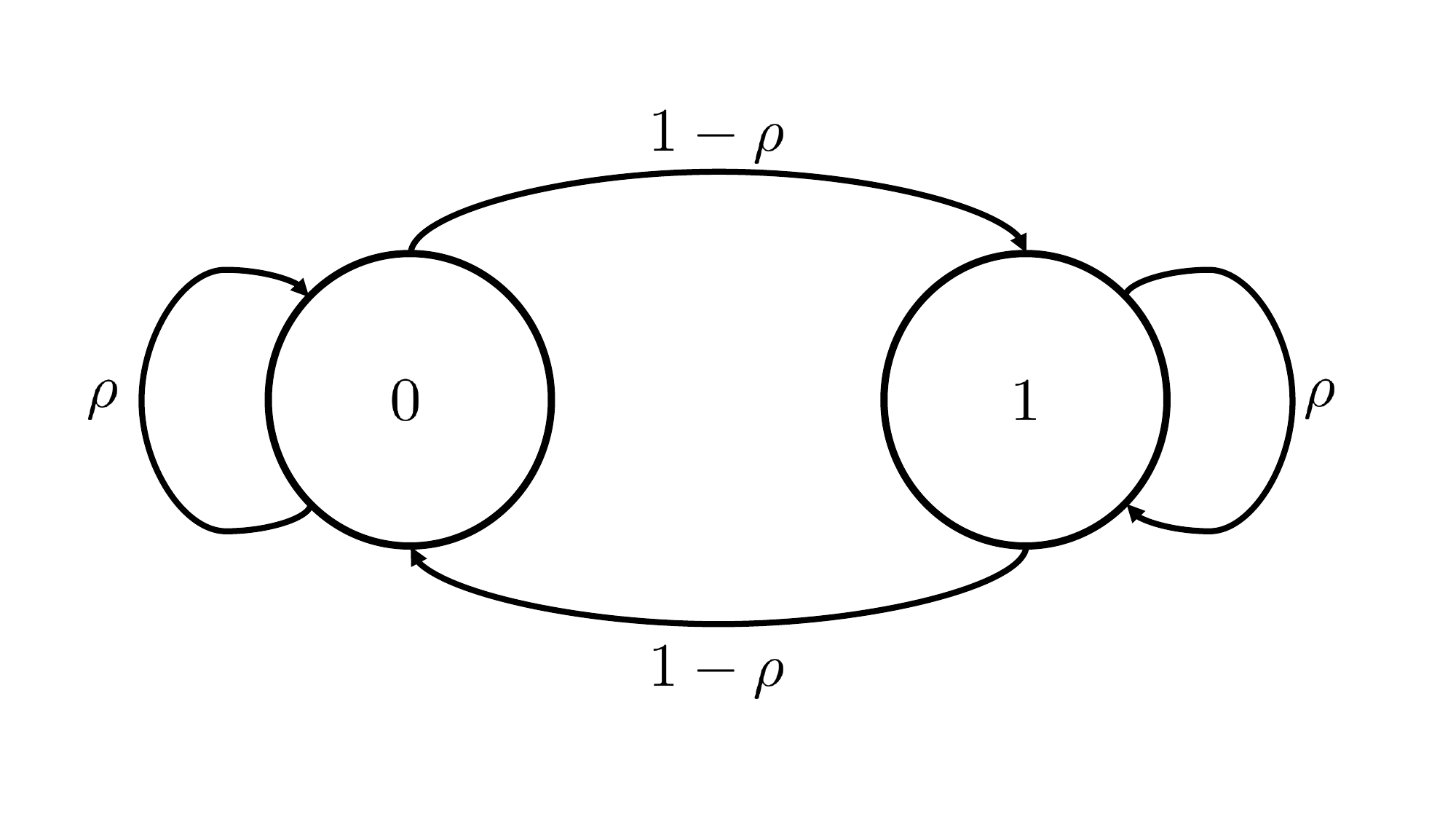}
    \caption{\textit{Markovian model for system states. When $\rho = 0.5$ subsequent states are independent.}}
    \label{fig.HMM}
\end{figure}
By assuming that the transmission between nodes and fusion center takes place over error-free channels, the report is equal to the local decision with probability 1 for honest nodes and with probability $1-P_{mal}$ for Byzantines. Hence, according to the local decision error model, we can derive the probabilities of the reports for honest nodes:

\begin{equation}
p\left(r_{ij} | s_j, a_i = 1\right) = (1-\varepsilon)\delta_{(r_{ij}-s_j)}+\varepsilon (1-\delta_{(r_{ij}-s_j)}),
\label{eqerr1}
\end{equation}

where $\delta_{(a)}$ is defined as:
\begin{equation}
    \delta_{(a)}=
    \begin{cases}
      1, & \text{if}\ a=0 \\
      0, & \text{otherwise}.
    \end{cases}
\label{delta}
\end{equation}

On the other hand, by introducing $\delta = \varepsilon(1-P_{mal}) + (1-\varepsilon)P_{mal}$, i.e., the probability that the FC receives a wrong report from a byzantine node\index{byzantine nodes|textbf}, we have:

\begin{equation}
p\left(r_{ij} | s_j, a_i = 0\right) = (1-\delta)\delta_{(r_{ij}-s_j)}+\delta (1-\delta_{(r_{ij}-s_j}))
\label{eqerr2}
\end{equation}

As for the number of Byzantines, we consider a situation in which the states of the nodes are independent of each other and the state of each node is described by a Bernoulli random variable with parameter $\alpha$, that is $p(a_i=0)=\alpha , \forall i$. In this way, the number of byzantine nodes\index{byzantine nodes|textbf} in the network is a random variable following a binomial distribution, corresponding to the constrained maximum entropy\index{maximum entropy|textbf} in chapter \ref{chapter:TIFS_SPL} with  $p\left(a^n \right) = \prod \limits_{i} p(a_i)$, where $p(a_i) = \alpha(1-a_i) + (1-\alpha)a_i$.

Regarding the sequence of states $s^m$, we assume a Markov model\index{Hidden Markov Model|textbf} as shown in Figure \reffig{fig.HMM} , i.e., $p\left(s^m\right) = \prod \limits_{j} p(s_j|s_{j-1})$. The transition probabilities are given by $p(s_j|s_{j-1}) = 1-\rho$ if $s_j = s_{j-1}$ and $p(s_j|s_{j-1}) = \rho$ when $s_j \ne s_{j-1}$, whereas for $j = 1$ we have $p(s_1|s_{0}) = p(s_1) = 0.5$.

In this chapter we look for the \emph{bitwise} Maximum A Posteriori Probability\index{Maximum A Posteriori Probability|textbf} (MAP) estimation of the system states $\left\{s_j\right\}$ which reads as follows:


\begin{flalign}
\nonumber
\hat{s}_j &= \arg \max \limits_{s_j \in \{0,1\}} ~ p\left(s_j | \mathbf{R}\right) &&\\\nonumber
&= \mathop{\arg \max}\limits_{s_j \in \{0,1\}} \sum \limits_{ \{s^m,a^n\} \backslash s_j} p\left( s^m, a^n | \mathbf{R}\right) \quad \textrm{(law of total probability)}&&\\\nonumber
&= \mathop{\arg \max}\limits_{s_j \in \{0,1\}} \sum \limits_{  \{s^m,a^n\} \backslash s_j} p\left(\mathbf{R} | s^m, a^n \right) p (s^m) p (a^n) \quad \textrm{(Bayes)} &&\\\nonumber
&= \mathop{\arg \max}\limits_{s_j \in \{0,1\}} \sum \limits_{  \{s^m,a^n\} \backslash s_j} \prod \limits_{ij}p\left(r_{ij} | s_j, a_i \right)\prod \limits_{j} p(s_j|s_{j-1}) \prod \limits_{i} p(a_i) &&\\
\label{eqNN10}
\end{flalign}

\noindent where the notation $\sum\limits_{\backslash}$ denotes a summation over all the possible combinations of values that the variables contained in the expression within the summation may assume by keeping the parameter listed after the operator $\backslash$ fixed. 
The optimization problem in \eqref{eqNN10} has been solved in the previous chapter for the case of independent system states. Even in such a simple case, however, the complexity of the optimum decision rule is exceedingly large, thus limiting the use of the optimum decision only in the case of small observation windows (typically $m$ not larger than 10). In the next section we introduce a sub-optimum solution of \eqref{eqNN10} based on message passing, which greatly reduces the computational complexity at the price of a negligible loss of accuracy.

\section{A Decision Fusion Algorithm Based on Message Passing} \label{sec:MessagePassing}
\subsection{Introduction to Sum-product message passing} \label{sec:MessagePassing Intro}
In this section we provide a brief introduction to the message passing (MP) algorithm\index{Message Passing Algorithm|textbf} for marginalization of sum-product problems\index{Sum-Product problems|textbf}. Let us start by considering $N$ binary
variables $\mathbf{z} = \{z_1,z_2,\ldots,z_N\}$, $z_i \in\{0,1\}$. Then, consider the
function $f\left(\mathbf{z}\right)$ with factorization:
\begin{equation}
f\left(\mathbf{z}\right) = \prod\limits_{k}f_k\left(\mathcal{Z}_k\right)\label{eq1}
\end{equation}
where $f_k$, $k = 1,\hdots,M$ are functions of a subset $\mathcal{Z}_k$ of the
whole set of variables.
We are interested in computing the marginal of $f$ with respect to a general variable $z_i$, defined as the sum of $f$ over all possible values of $\mathbf{z}$, i.e.:
\begin{equation}
\mu(z_i) = \sum\limits_{\mathbf{z}\backslash z_i}
\prod\limits_{k}f_k\left(\mathcal{Z}_k\right)\label{eq2}
\end{equation}
where notation $\sum\limits_{\mathbf{z}\backslash z_i}$ denotes a sum over all possible combinations of values of the variables in $\mathbf{z}$ by keeping $z_i$ fixed. We are interested in finding the value $z_i$ that optimizes Equation (\refeq{eq2}). Note that marginalization problems occur when we want to compute any arbitrary probability from joint probabilities by summing out variables we are not interested in. In this general setting, since the vector $\mathbf{z}$ has $N$ binary elements, determining the marginals by exhaustive search requires $2^N$ operations.  However, in many situations it is possible to exploit the distributive law of multiplication to get a substantial reduction in complexity.\\
To elaborate, let us associate with problem
(\ref{eq2}) a bipartite \emph{factor graph}\index{Factor Graph|textbf}, in which for each variable we draw a variable node (circle) and for each function we draw a
factor node (square). A variable node is connected to a factor node $k$ by an edge if and only if the corresponding variable belongs to $\mathcal{Z}_k$. This means that the set of vertices is partitioned into two groups (the set of nodes
corresponding to variables and the set of nodes corresponding to factors) and that an edge always connects a variable node to a factor node. \\
Let us now assume that the factor graph is a
single tree, i.e., a graph in which any two nodes are connected by exactly one path.
In this case, it is straightforward to derive an algorithm which allows to solve the marginalization problem with reduced complexity. The algorithm is the
MP algorithm, which has been broadly used in the last years in channel coding applications \cite{David}, \cite{Abr1}.

\begin{figure}[ptb]
\begin{center}
\includegraphics[width = 8cm]{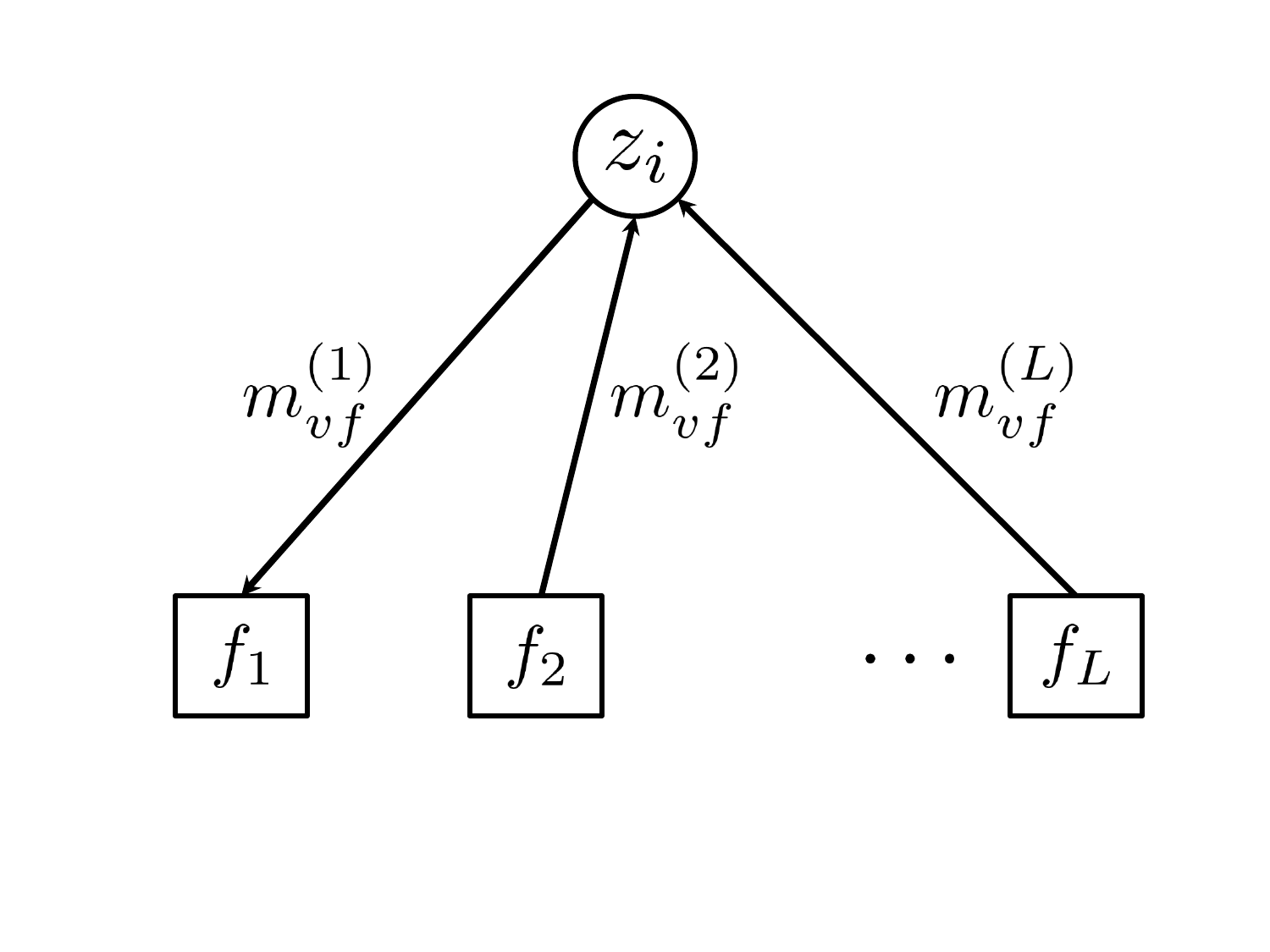}
\caption{\textit{Node-to-factor message passing.}}%
\label{Fig1}%
\end{center}
\end{figure}

\begin{figure}[ptb]
\begin{center}
\includegraphics[width = 8cm]{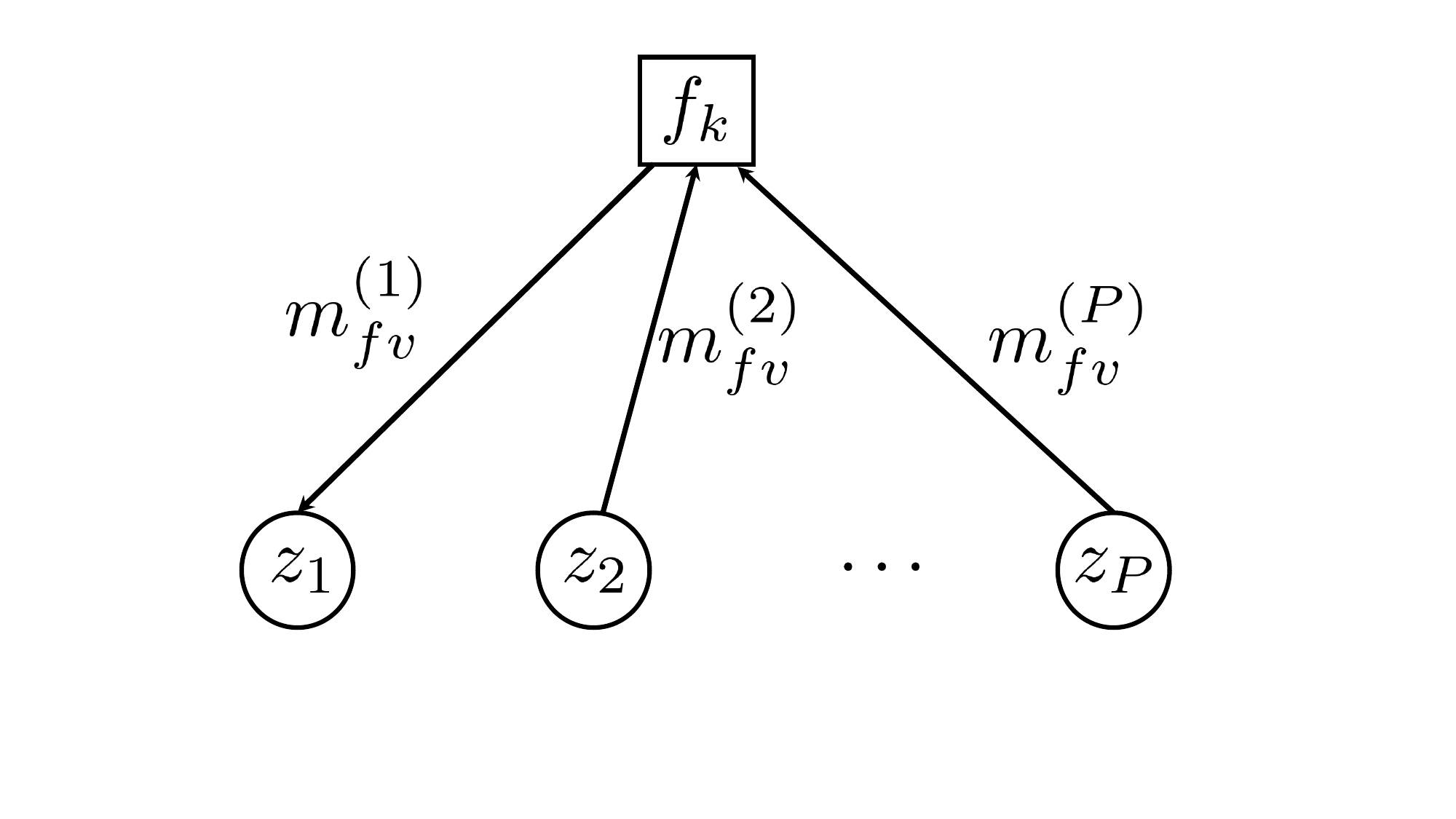}
\caption{\textit{Factor-to-node message passing.}}%
\label{Fig2}%
\end{center}
\end{figure}

To describe how the MP algorithm works, let us first define messages as $2$-dimensional vectors with binary elements, denoted by $\mathbf{m} = \left\{m(0),m(1)\right\}$. Such messages are exchanged between variable nodes\index{variable node|textbf} and function nodes\index{function node|textbf} and viceversa, according to the following rules.
Let us first consider variable-to-function messages ($\mathbf{m}_{vf}$), and take the portion of factor graph depicted in Figure \reffig{Fig1} as an illustrative example. In this graph, the variable node $z_i$ is connected to $L$ factor
nodes, namely $f_1,f_2,\ldots,f_L$. For the MP algorithm
to work properly, node $z_i$ must deliver the messages
$\mathbf{m}^{(l)}_{vf}$, $l = 1,\ldots,L$ to all its adjacent nodes.
Without loss of generality, let us focus on message
$\mathbf{m}^{(1)}_{vf}$. Such a message can be evaluated and
delivered upon receiving messages $\mathbf{m}^{(l)}_{fv}$, $l =
2,\ldots,L$, i.e., upon receiving messages from all function nodes
except $f_1$. In particular, $\mathbf{m}^{(1)}_{vf}$ may be
straightforwardly evaluated by calculating the element-wise product of
the incoming messages, i.e.:
\begin{equation}
{m}^{(1)}_{vf}(q) = \prod\limits_{j=2}^{L} {m}^{(j)}_{fv}(q) ,
\label{eq4}
\end{equation}
for $q=0,1$.
Let us now consider factor-to-variable messages, and refer to the factor graph\index{Factor Graph|textbf} of Figure \reffig{Fig2} where $P$ variable nodes are connected to the factor node $f_k$, i.e., according to the previous notation, $\mathcal{Z}_k = \{z_1,\ldots,z_P\}$. In this case, the node $f_k$
must deliver the messages $\mathbf{m}^{(l)}_{fv}$, $l = 1,\ldots,P$ to all its adjacent nodes. Let us consider again
$\mathbf{m}^{(1)}_{fv}$: upon receiving the messages
$\mathbf{m}^{(l)}_{vf}$, $l = 2,\ldots,P$, $f_k$ may evaluate the message $\mathbf{m}^{(1)}_{fv}$ as:
\begin{equation}
{m}^{(1)}_{fv}(q) = \sum\limits_{z_2,\ldots,z_P}\left[ f_k\left(q,z_2,\ldots,z_P\right)
\prod\limits_{p=2}^{P} {m}^{(p)}_{vf}(z_p)\right] \label{eq5}
\end{equation}
for $q=0,1$.

\begin{figure}[h!]
\begin{center}
\includegraphics[width = 8cm]{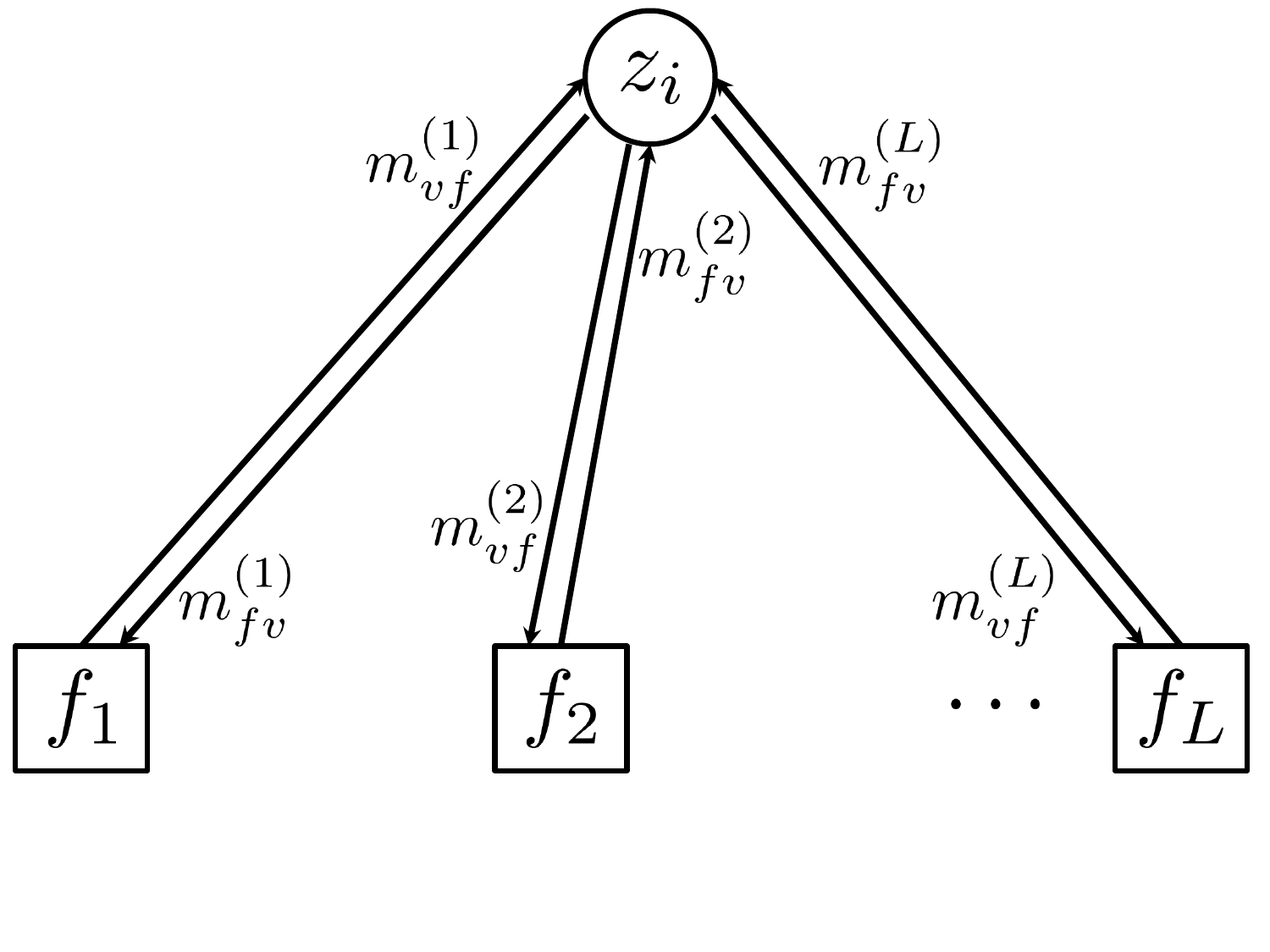}
\caption{\textit{End of message passing for node $z_i$.}}%
\label{Fig3}%
\end{center}
\end{figure}

Given the message passing rules at each node, it is possible to derive the MP algorithm which allows to compute the marginals in Equation (\refeq{eq2}).
The process starts at
the leaf nodes, i.e., those nodes which have only one connecting edge. In particular, each variable leaf node passes an all-ones message to its adjacent factor node, whilst each factor leaf node, say $f_k(z_i)$ passes the message ${m}^{(k)}_{fv}(q) = f_k(z_i = q)$ to its adjacent node $z_i$. After initialization at leaf nodes, for every edge we can compute the outgoing message as soon as all incoming messages from all other edges connected to the
same node are received (according to the message passing rules (Equation \refeq{eq4} and Equation \refeq{eq5}). When a message has been sent in both directions along every edge the algorithm stops. This situation is depicted in Figure \reffig{Fig3}: upon receiving messages from all its
adjacent factor nodes, node $z_i$ can evaluate the exact marginal as:

\begin{equation}
\mu(z_i) = \prod\limits_{k=1,\ldots,L}{m}^{(k)}_{fv}(z_i).
\label{eq6}
\end{equation}

With regard to complexity, factors to variables
message passing can be accomplished with $2^{P}$
operations, $P$ being the number of variables in $f_k$. On the other hand, variables to nodes message passing's complexity can be neglected, and, hence, the MP algorithm\index{Message Passing Algorithm|textbf} allows to noticeably reduce the complexity of the problem provided that the numerosity of $\mathcal{Z}_k$ is much lower than $N$. With regard to the optimization, Equation \refeq{eq6} evaluates the marginal for both $z_i = 0$ and $z_i = 1$, which represent the approximated computation of the sum-product for both hypotheses. Hence, the optimization is obtained by choosing the value of $z_i$ which maximizes it.

\subsection{Nearly-optimal data fusion by means of message passing}

\begin{figure}[h!]
\begin{center}
 \includegraphics[width=1.0\textwidth, height = 7cm]{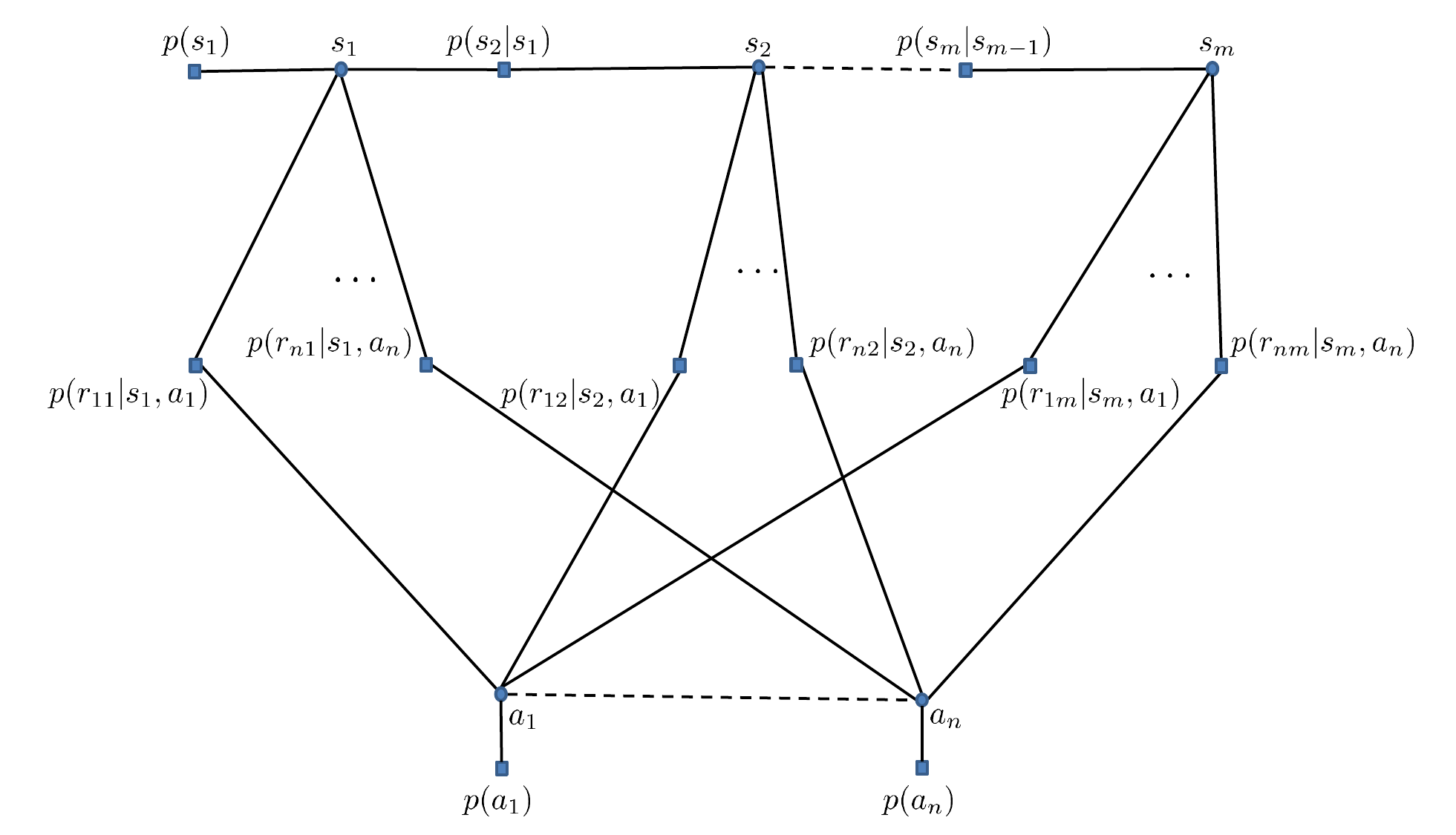} \\
\caption{\textit{Factor graph for the problem at hand.}}
\label{Factor_graph0}
\end{center}
\end{figure}

The objective function of the optimal fusion rule expressed in \eqref{eqNN10} can be seen as a marginalization of a sum product of functions of binary variables, and, as such, it falls within the MP framework described in the previous Section. More specifically, in our problem, the variables are the system states $s_j$ and the status of the nodes $a_i$, while the functions are the probabilities of the reports shown in Equations \refeq{eqerr1} and \refeq{eqerr2}, the conditional probabilities $p(s_j|s_{j-1})$, and the a-priori probabilities $p(a_i)$. The resulting bipartite graph is shown in Figure \reffig{Factor_graph0}.

It is worth noting that the graph is a loopy graph, i.e., it contains cycles, and as such it is not a tree. However, although it was originally designed for acyclic graphical models, it was found that the MP algorithm can be used for general graphs, e.g., in channel decoding problems \cite{Rich}. In general, when the marginalization problem is associated to a loopy graph, the implementation of MP requires to establish a scheduling policy to initiate the procedure, so that variable nodes may receive messages from all the connected factors and evaluate the marginals. In this case, a single run of the MP algorithm may not be sufficient to achieve a good approximation of the exact marginals, and progressive refinements must be obtained through successive iterations. However, in the presence of loopy graphs, there is no guarantee of either convergence or optimality of the final solution. In many cases, the performance of the message-passing algorithms is closely related to the structure of the graph, in general, and its cycles, in particular. Many previous works in the field of channel coding, e.g., see \cite{MaoLDPC}, reached the conclusion that, for good performance, the factor graph should not contain short cycles. In our case, it is possible to see from Figure \reffig{Factor_graph0} that the shortest cycles have order 6, i.e., a message before returning to the sender must cross at least six different nodes. We speculate that such a minimum cycles length is sufficient to provide good performance for the problem at hand. We will prove through simulations that such a conjecture is true.

\begin{sidewaysfigure}
\begin{center}
 \includegraphics[width=1.0\textwidth, height=13cm]{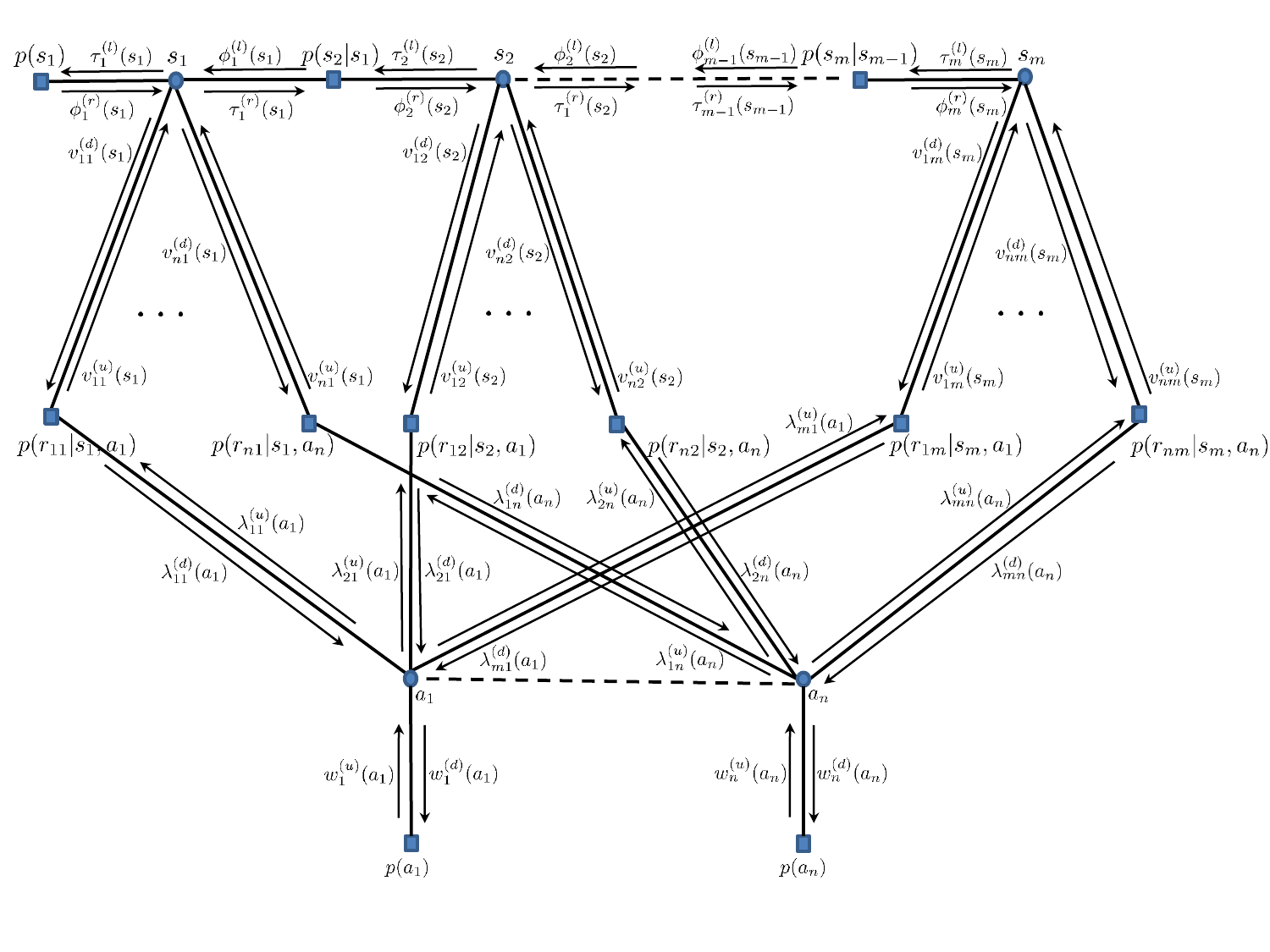} \\
\caption{Factor graph for the problem at hand with the illustration of all the exchanged messages.}
\label{Factor_graph1}
\end{center}
\end{sidewaysfigure}

To elaborate further, based on the graph\index{Factor Graph|textbf} of Figure \reffig{Factor_graph0} and on the general MP rules reported in the previous Section, we are now capable of deriving the messages for the scenario at hand. In Figure \reffig{Factor_graph1}, we display all the messages for the graph in Figure \reffig{Factor_graph0} that are exchanged to estimate in parallel each of the states $s_j, j \in \{0,1\}$ in the vector $s^m = \left\{s_1,s_2,\ldots,s_m\right\}$.
Specifically, we have:
\begin{equation}
\begin{array}{cccc}
\tau_{j}^{(l)}(s_j)  & = & \varphi_j^{(l)}(s_j) \prod\limits_{i = 1}^{n} \nu_{ij}^{(u)}(s_j)  & j = 1,\ldots,m \\
\tau_{j}^{(r)}(s_j) & = & \varphi_j^{(r)}(s_j) \prod\limits_{i = 1}^{n} \nu_{ij}^{(u)}(s_j)  & j = 1,\ldots,m \\
\varphi_j^{(l)}(s_j) & = &  \sum\limits_{s_{j+1}  = 0,1} p\left(s_{j+1} | s_j\right) \tau_{j+1}^{(l)}(s_{j+1}) &  j = 1,\ldots,m-1\\
\varphi_j^{(r)}(s_j) & = & \sum\limits_{s_{j-1} = 0,1} p\left(s_j | s_{j-1} \right)  \tau_{j-1}^{(r)}(s_{j-1}) &  j = 2,\ldots,m\\
\varphi_1^{(r)}(s_1) & = & p(s_1) & \\
\nu_{ij}^{(u)}(s_j) & = &  \sum\limits_{a_i= 0,1} p\left(r_{ij} \left| s_j,a_i\right. \right) \lambda_{ji}^{(u)}(a_i) & j = 1,\ldots,m, ~~ i = 1,\ldots,n \\
\nu_{ij}^{(d)}(s_j) & = & \varphi_j^{(r)}(s_j)  \varphi_j^{(l)}(s_j)  \prod \limits_{k = 1  \atop{k \ne i}}^{n}  \nu_{kj}^{(u)}(s_j)  & j = 1,\ldots,m-1, ~~ i = 1,\ldots,n\\
\nu_{im}^{(d)}(s_m) & = & \varphi_j^{(r)}(s_m)  \prod \limits_{k = 1  \atop{k \ne i}}^{n}  \nu_{km}^{(u)}(s_m)  &  i = 1,\ldots,n\\
\lambda_{ji}^{(d)}(a_i) & = & \sum\limits_{s_j= 0,1} p\left(r_{ij} \left| s_j,a_i\right. \right)  \nu_{ij}^{(d)}(s_j) & j = 1,\ldots,m, ~~ i = 1,\ldots,n\\
\lambda_{ji}^{(u)}(a_i) & = & \omega_i^{(u)}(a_i) \prod \limits_{q = 1  \atop{q \ne j}}^{m} \lambda_{qi}^{(d)}(a_i) & j = 1,\ldots,m, ~~ i = 1,\ldots,n\\
\omega_i^{(d)}(a_i) & = &  \prod \limits_{ j = 1 }^{m} \lambda_{ji}^{(d)}(a_i) &  i = 1,\ldots,n\\
\omega_i^{(u)}(a_i) & = & p(a_i) &  i = 1,\ldots,n\\
\end{array}
 \label{eq_all_messages}
\end{equation}

As for the scheduling policy, we initiate the MP procedure by sending the messages $\lambda_{ji}^{(u)}(a_i) = \omega_i^{(u)}(a_i)$ to all $p\left(r_{ij} \left| s_j,a_j\right. \right)$ factor nodes, and by sending the message $p(s_1)$ to the variable node $s_1$. Hence, the MP proceeds according to the general message passing rules, until all variable nodes are able to compute the respective marginals. When this happens, the first iteration is concluded. Then, successive iterations are carried out by starting from leaf nodes and by taking into account the messages received at the previous iteration for the evaluation of new messages. Hence, the algorithm is stopped upon achieving convergence of messages, or after a maximum number of iterations.

The MP scheme described above can be simplified by observing that messages can be normalized without affecting the normalized marginals. Henceforward, let us consider as normalization factors the sum of the elements of the messages, i.e., if we consider for example $\tau_{j}^{(l)}(s_j)$, the normalization factor is $\tau_{j}^{(l)}(0)+\tau_{j}^{(l)}(1)$. In this case, the normalized messages, say $\bar{\tau}_{j}^{(l)}(s_j)$ can be conveniently represented as scalar terms in the interval $(0,1)$, e.g., we can consider $\bar{\tau}_{j}^{(l)}(0)$ only since $\bar{\tau}_{j}^{(l)}(1) = 1 - \bar{\tau}_{j}^{(l)}(0)$. Accordingly, the normalized messages can be evaluated as:

\begin{flalign}
\nonumber
\bar{\tau}_{j}^{(l)} = & \frac{\bar{\varphi}_j^{(l)} \prod\limits_{i = 1}^{n} \bar{\nu}_{ij}^{(u)}}{\bar{\varphi}_j^{(l)} \prod\limits_{i = 1}^{n} \bar{\nu}_{ij}^{(u)} + (1-\bar{\varphi}_j^{(l)}) \prod\limits_{i = 1}^{n} (1-\bar{\nu}_{ij}^{(u)})} \quad j = 1,\ldots,m &&\\\nonumber
\bar{\tau}_{j}^{(r)} = & \frac{\bar{\varphi}_j^{(r)}  \prod\limits_{i = 1}^{n} \bar{\nu}_{ij}^{(u)} }{\bar{\varphi}_j^{(r)} \prod\limits_{i = 1}^{n} \bar{\nu}_{ij}^{(u)}   + (1-\bar{\varphi}_j^{(r)} ) \prod\limits_{i = 1}^{n} (1-\bar{\nu}_{ij}^{(u)} ) } \quad j = 1,\ldots,m&&\\\nonumber
\bar{\varphi}_j^{(l)} = &  \rho \bar{\tau}_{j+1}^{(l)} + (1-\rho) (1-\bar{\tau}_{j+1}^{(l)})  \quad  j = 1,\ldots,m-1 \\\nonumber
\bar{\varphi}_j^{(r)} = &  \rho \bar{\tau}_{j-1}^{(r)} + (1-\rho) (1-\bar{\tau}_{j-1}^{(r)}) \quad  j = 2,\ldots,m && \\\nonumber
\bar{\varphi}_1^{(r)} = & p(s_1 = 0) && \\\nonumber
\bar{\nu}_{ij}^{(u)} = & \frac{p\left(r_{ij} \left| 0,0\right. \right)\bar{ \lambda}_{ji}^{(u)} + p\left(r_{ij} \left| 0,1\right. \right) (1-\bar{\lambda}_{ji}^{(u)})}{\kappa_1 + \kappa_2} \\\nonumber
& \text{where,} \quad \kappa_1 = p\left(r_{ij} \left| 0,0\right. \right)\bar{ \lambda}_{ji}^{(u)} + p\left(r_{ij} \left| 0,1\right. \right) (1-\bar{\lambda}_{ji}^{(u)})  && \\\nonumber
& \text{and} \quad \kappa_2 = p\left(r_{ij} \left| 1,0\right. \right)\bar{ \lambda}_{ji}^{(u)} + p\left(r_{ij} \left| 1,1\right. \right) (1-\bar{\lambda}_{ji}^{(u)})  && \\\nonumber
& \quad j = 1,\ldots,m,  ~~ i = 1,\ldots,n && \\\nonumber
\bar{\nu}_{ij}^{(d)} = & \frac{\bar{\varphi}_j^{(r)} \bar{ \varphi}_j^{(l)}  \prod \limits_{k = 1  \atop{k \ne j}}^{n} \bar{\nu}_{ki}^{(u)}}{\bar{\varphi}_j^{(r)} \bar{ \varphi}_j^{(l)}  \prod \limits_{k = 1  \atop{k \ne i}}^{n} \bar{\nu}_{ki}^{(u)} + (1-\bar{\varphi}_j^{(r)}) (1-\bar{ \varphi}_j^{(l)})  \prod \limits_{k = 1  \atop{k \ne i}}^{n} (1-\bar{\nu}_{ki}^{(u)})}  && \\\nonumber
& \quad j = 1,\ldots,m-1, ~~ i = 1,\ldots,n &&  \\\nonumber
\bar{\nu}_{jm}^{(d)} = & \frac{\bar{\varphi}_m^{(r)} \prod \limits_{k = 1  \atop{k \ne i}}^{n} \bar{\nu}_{km}^{(u)}}{\bar{\varphi}_m^{(r)}  \prod \limits_{k = 1  \atop{k \ne i}}^{n} \bar{\nu}_{km}^{(u)} + (1-\bar{\varphi}_m^{(r)})   \prod \limits_{k = 1  \atop{k \ne i}}^{n} (1-\bar{\nu}_{km}^{(u)})}  \quad i = 1,\ldots,n && \\\nonumber
\bar{\lambda}_{ji}^{(d)}  = & \frac{p\left(r_{ij} \left| 0,0\right. \right)  \bar{\nu}_{ij}^{(d)} + p\left(r_{ij} \left| 1,0\right. \right)  (1-\bar{\nu}_{ij}^{(d)})}{\tau_1+\tau_2} &&\\\nonumber
& \text{where,}\quad \tau_1 = p\left(r_{ij} \left| 0,0\right. \right)  \bar{\nu}_{ij}^{(d)} + p\left(r_{ij} \left| 1,0\right. \right)  (1-\bar{\nu}_{ij}^{(d)})    &&\\\nonumber
& \text{and}\quad \tau_2 = p\left(r_{ij} \left| 0,1\right. \right)  \bar{\nu}_{ij}^{(d)} + p\left(r_{ij} \left| 1,1\right. \right)  (1-\bar{\nu}_{ij}^{(d)})  &&\\\nonumber
& \quad j = 1,\ldots,m,  ~~ i = 1,\ldots,n && \\\nonumber
\bar{\lambda}_{ji}^{(u)}  = & \frac{\bar{\omega}_i^{(u)} \prod \limits_{q = 1  \atop{q \ne j}}^{m} \bar{\lambda}_{qi}^{(d)} }{\bar{\omega}_i^{(u)} \prod \limits_{q = 1  \atop{q \ne j}}^{m} \bar{\lambda}_{qi}^{(d)}  + (1-\bar{\omega}_i^{(u)}) \prod \limits_{q = 1  \atop{q \ne j}}^{m} (1-\bar{\lambda}_{qi}^{(d)}) } && \\\nonumber
& j = 1,\ldots,m, ~~ i = 1,\ldots,n && \\\nonumber
\bar{\omega}_i^{(d)} = &  \frac{\prod \limits_{ j = 1 }^{m} \bar{\lambda}_{ji}^{(d)}}{\prod \limits_{ j = 1 }^{m} \bar{\lambda}_{ji}^{(d)}+ \prod \limits_{ j = 1 }^{m} (1-\bar{\lambda}_{ji}^{(d)})} \quad  i = 1,\ldots,n && \\\nonumber
\bar{\omega}_i^{(u)} = & p(a_i = 0) \quad  i = 1,\ldots,n && \\
\label{eq.eq_all_normalized_messages}
\end{flalign}

\section{Simulation Results and Discussion}
\label{sec:Simulations}

In this section, we analyze the performance of the MP decision fusion algorithm\index{Message Passing Algorithm|textbf}. We first consider the computational complexity, then we pass to evaluate the performance in terms of error probability\index{error probability|textbf}. In particular, we compare the performance of the MP-based scheme to those of the optimum fusion rule in chapter \ref{chapter:TIFS_SPL} (whenever possible), the soft isolation scheme presented in chapter \ref{chapter:CDC}, the hard isolation scheme described in \cite{Raw11} and the simple majority rule. In our comparison, we consider both independent and Markovian system states, for both small and large observation window $m$.

\subsection{Complexity Discussion}
In order to evaluate the complexity\index{computational complexity|textbf} of the message passing algorithm and compare it to that of the optimum fusion scheme\index{optimum fusion rule|textbf}, we consider both the number of operations and the running time. By number of operations we mean the number of additions, substractions, multiplications and divisions performed by the algorithm to estimate the vector of system states $s^m$.

\begin{figure}[h!]
\begin{center}
 \includegraphics[width=1.0\textwidth, height=6.5cm]{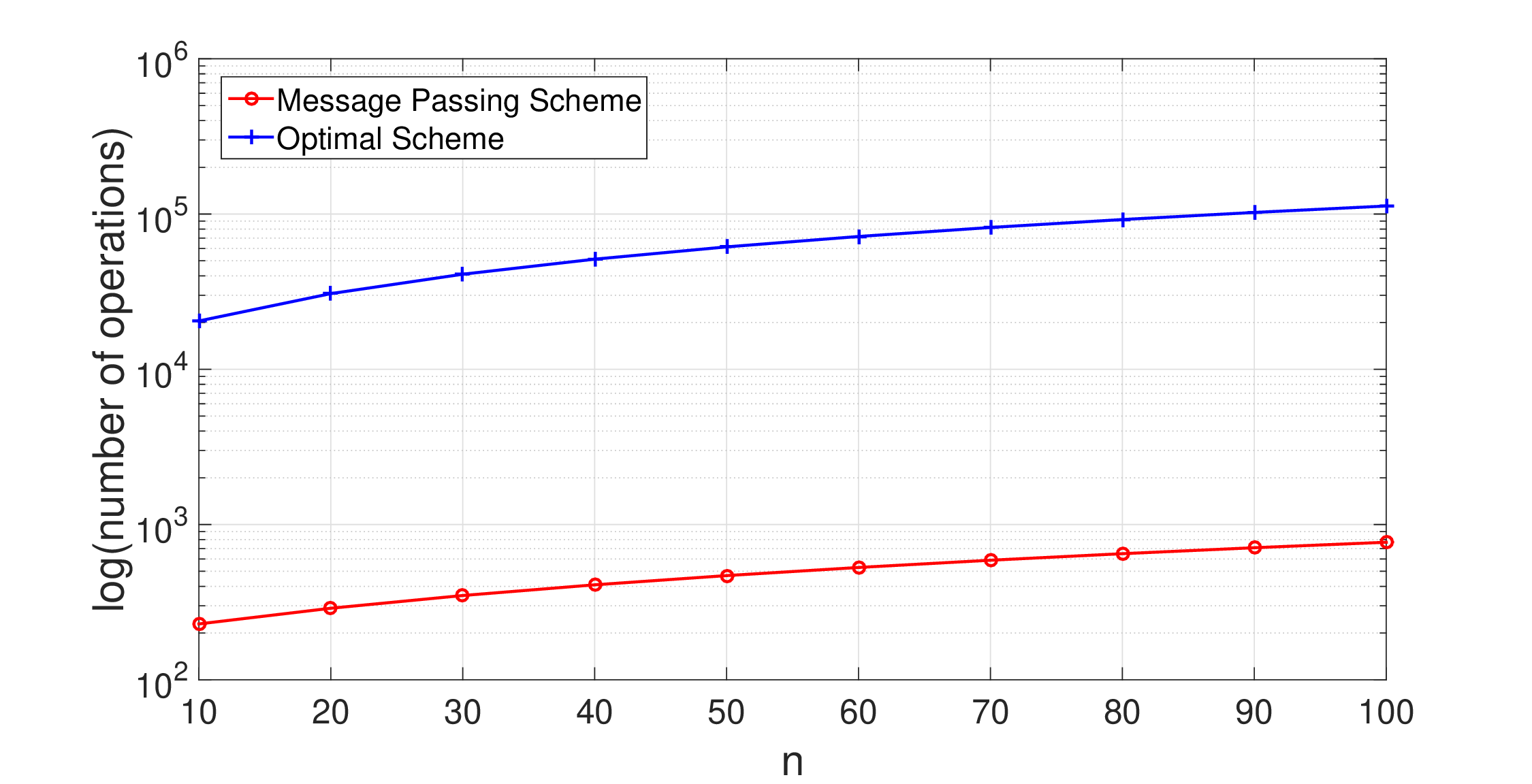} \\
\caption{\textit{Number of operations required for different $n$, $m=10$ and $5$ message passing local iterations for message passing and optimal schemes.}\label{plot_operations_m_fixed}}
\end{center}
\end{figure}

\begin{figure}[h!]
\begin{center}
 \includegraphics[width=1.0\textwidth, height=6.5cm]{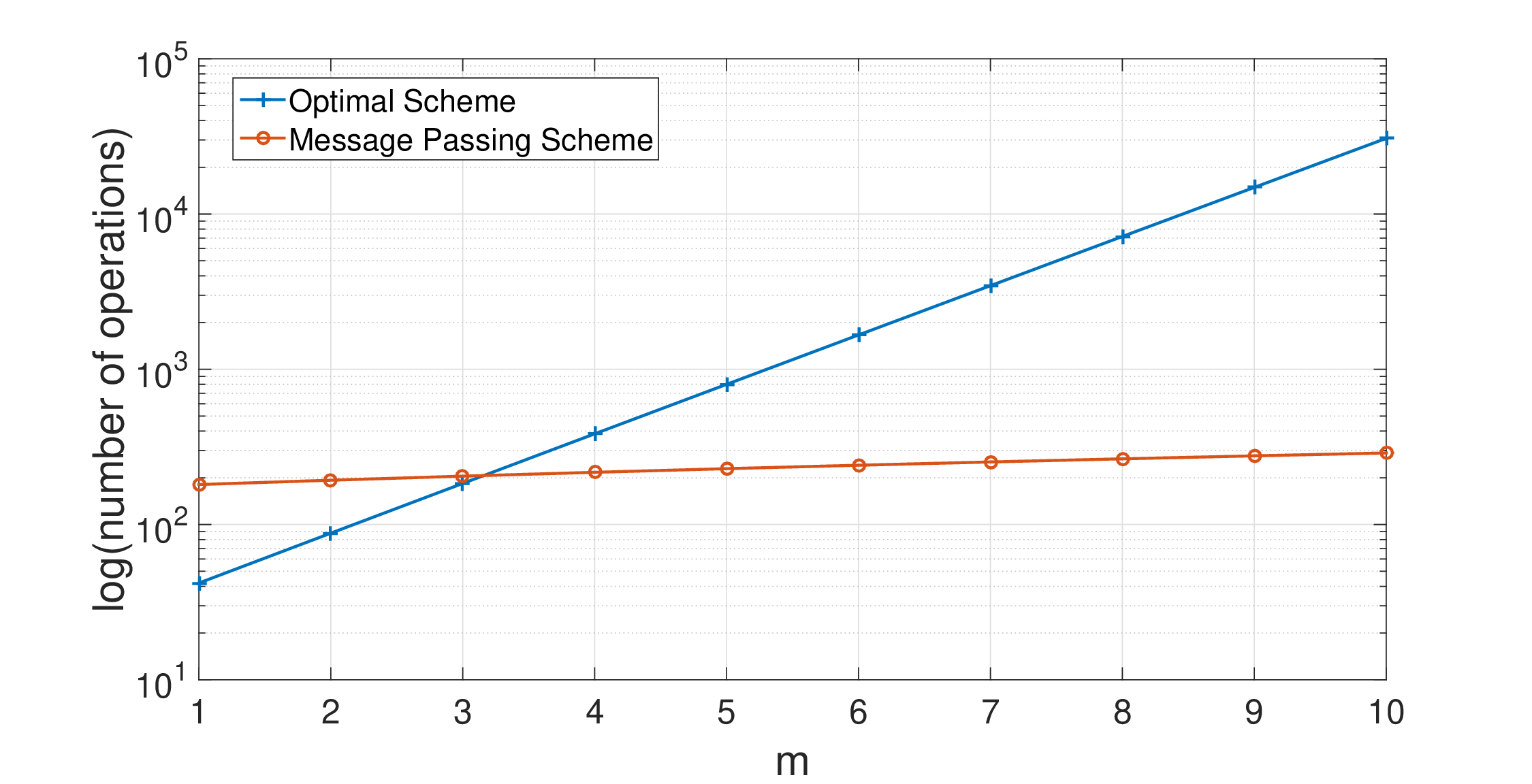} \\
\caption{\textit{Number of operations required for different $m$, $n=20$ and $5$ message passing local iterations for message passing and optimal schemes.}\label{plot_operations_n_fixed}}
\end{center}
\end{figure}

By looking at Equation \refeq{eq.eq_all_normalized_messages}, we see that running the message passing algorithm requires the following number of operations:

\begin{itemize}
\item $3n+5$ operations for each of $\bar{\tau}_{j}^{(l)}$ and $\bar{\tau}_{j}^{(r)}$.

\item $3$ operations for each of $\bar{\varphi}_j^{(l)}$ and $\bar{\varphi}_j^{(r)}$.

\item $11$ operations for $\bar{\nu}_{ij}^{(u)}$.
\item $3n+5$ operations for $\bar{\nu}_{ij}^{(d)}$.
\item $3n+2$ operations for $\bar{\nu}_{im}^{(d)}$.
\item $11$ operations for $\bar{\lambda}_{ji}^{(d)}$.
\item $3m+2$ operations for each of $\bar{\lambda}_{ji}^{(u)}$ and $\bar{\omega}_i^{(d)}$.

\end{itemize}

\noindent summing up to $12n+6m+49$ operations for each iteration over the factor graph\index{Factor Graph|textbf}. On the other hand, in the case of independent node states, the optimal scheme in chapter \ref{chapter:TIFS_SPL} requires $2^m(m+n)$ operations. Therefore, the MP algorithm is much less computationally expensive since it passes from an exponential\index{exponential complexity|textbf} to a linear complexity\index{linear complexity|textbf} in $m$. An example of the difference in computational complexity between the optimum and the MP algorithms is depicted in Figure \reffig{plot_operations_m_fixed} for fixed $m$ and in Figure \reffig{plot_operations_n_fixed} for fixed $n$.


\begin{table}[h!]
\centering
\renewcommand{\arraystretch}{1.1}
\begin{tabular}{c| c| c|}
\hline
\multicolumn{1}{|c|}{Setting/Scheme} & Message Passing   & Optimal      \\ \hline
\multicolumn{1}{|c|}{$n=20$,$\alpha=0.45$}  &943.807114 &1.6561e+04   \\ \hline
\multicolumn{1}{|c|}{$n=100$,$\alpha=0.49$}  &4888.821497 &2.0817e+04   \\ \hline
\end{tabular}
\caption{\textit{Running Time (in seconds) for the Optimal and the Message Passing algorithms for: $m=10$, $\varepsilon=0.15$, $\text{Number of Trials} = 10^5$ and $\text{Message Passing Iterations} = 5$.}\label{tab:timespent}}
\end{table}

With regard to time complexity, Table \reftab{tab:timespent} reports the running time of the MP and the optimal schemes. For $n=20$, the optimal scheme running time is $17.547$ times larger than that of the message passing algorithm\index{Message Passing Algorithm|textbf}. On the other hand, for the case of $n=100$, the optimal scheme needs around $4.258$ times more than the message passing scheme. The tests have been conducted using Matlab 2014b running on a machine with 64-bit windows 7 OS with 16,0 GB of installed RAM and Intel Core i7-2600 CPU @ 3.40GHz.

\subsection{Performance Evaluation}

In this section, we use numerical simulations to evaluate the performance of the message passing algorithm and compare them to those obtained by other schemes. The results are divided into four parts. The first two parts consider, respectively, simulations performed with small and large observation windows $m$. Then, in the third part, we investigate the optimum behaviour of the Byzantines over a range of observation windows size. Finally, in the last part, we compare the case of independent and Markovian system states.

The simulations were carried out according to the following setup. We considered a network with $n = 20$ nodes, $\varepsilon = 0.15$, $\rho = \{0.95, 0.5\}$ corresponding to Markovian and independent sequence of system states, respectively. The probability $\alpha$ that a node is Byzantine is in the range $[0,0.45]$ corresponding to a number of Byzantines between 0 and 9. As to $P_{mal}$ we set it to either  0.5 or 1\footnote{It is know from chapter \ref{chapter:TIFS_SPL} that for the Byzantines the optimum choice of $P_{mal}$ is either 0.5 or 1 depending on the considered setup.}. The number of message passing iterations is 5. For each setting, we estimated the error probability over $10^5$ trials.

\subsubsection{Small m}

\begin{figure}[h!]
\begin{center}
 \includegraphics[width=0.9\textwidth]{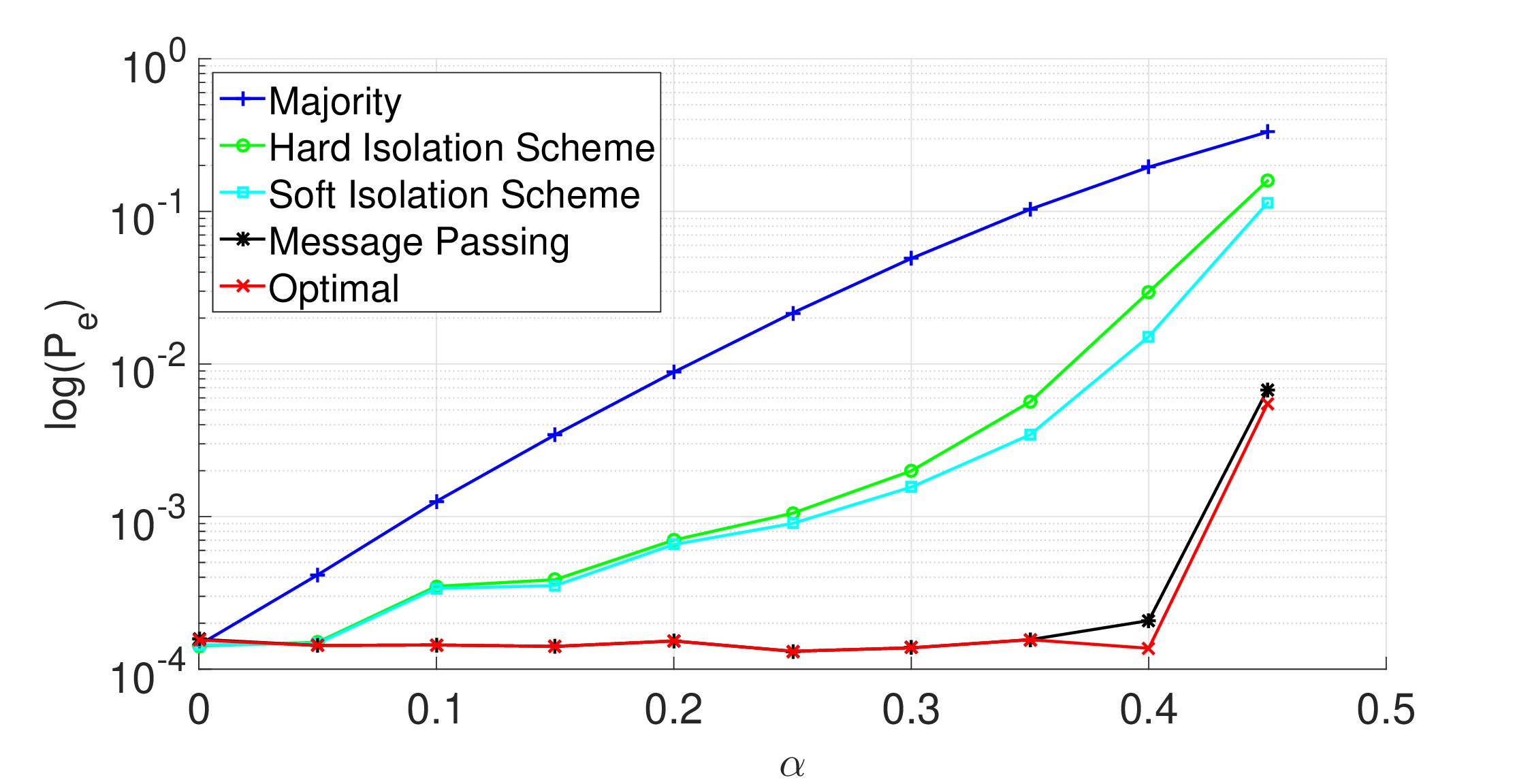} \\
\caption{\textit{Error probability as a function of $\alpha$ for the following setting: $n=20$, independent Sequence of States $\rho = 0.5$, $\varepsilon =0.15$, $m=10$ and $P_{mal}=1.0$.}}
\label{Perf_one_Independent}
\end{center}
\end{figure}

\begin{figure}[h!]
\begin{center}
 \includegraphics[width=0.9\textwidth]{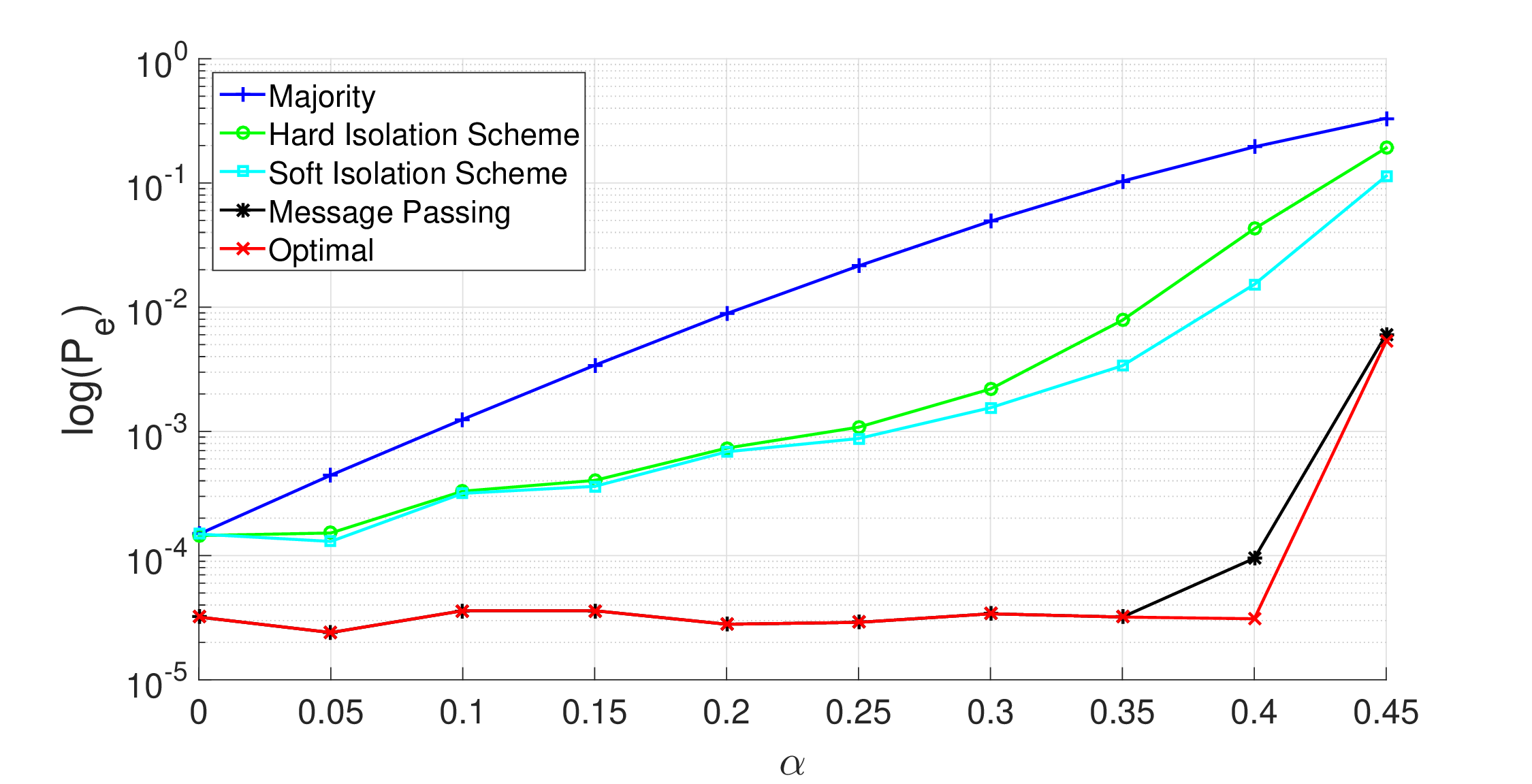} \\
\caption{\textit{Error probability as a function of $\alpha$ for the following setting: $n=20$, Markovian Sequence of States $\rho = 0.95$, $\varepsilon =0.15$, $m=10$ and $P_{mal}=1.0$}}
\label{Perf_one_markov}
\end{center}
\end{figure}

To start with, we considered a small observation window, namely  $m=10$. With such a small value of $m$, in fact, it is possible to compare the performance of the message passing algorithm to that of the optimum decision fusion rule.
The results we obtained are reported in Figure \reffig{Perf_one_Independent}. Upon inspection of the figure, the superior performance of the message passing algorithm over the Majority, Soft and Hard isolation schemes is confirmed. More interestingly, the message passing algorithm gives nearly optimal performance, with only a negligible performance loss with respect to the optimum scheme.

Figure \reffig{Perf_one_markov} confirms the results shown in Figure \reffig{Perf_one_Independent} for Markovian system states ($\rho = 0.95$).

\subsubsection{Large m}

\begin{figure}[h!]
\begin{center}
 \includegraphics[width=0.9\textwidth]{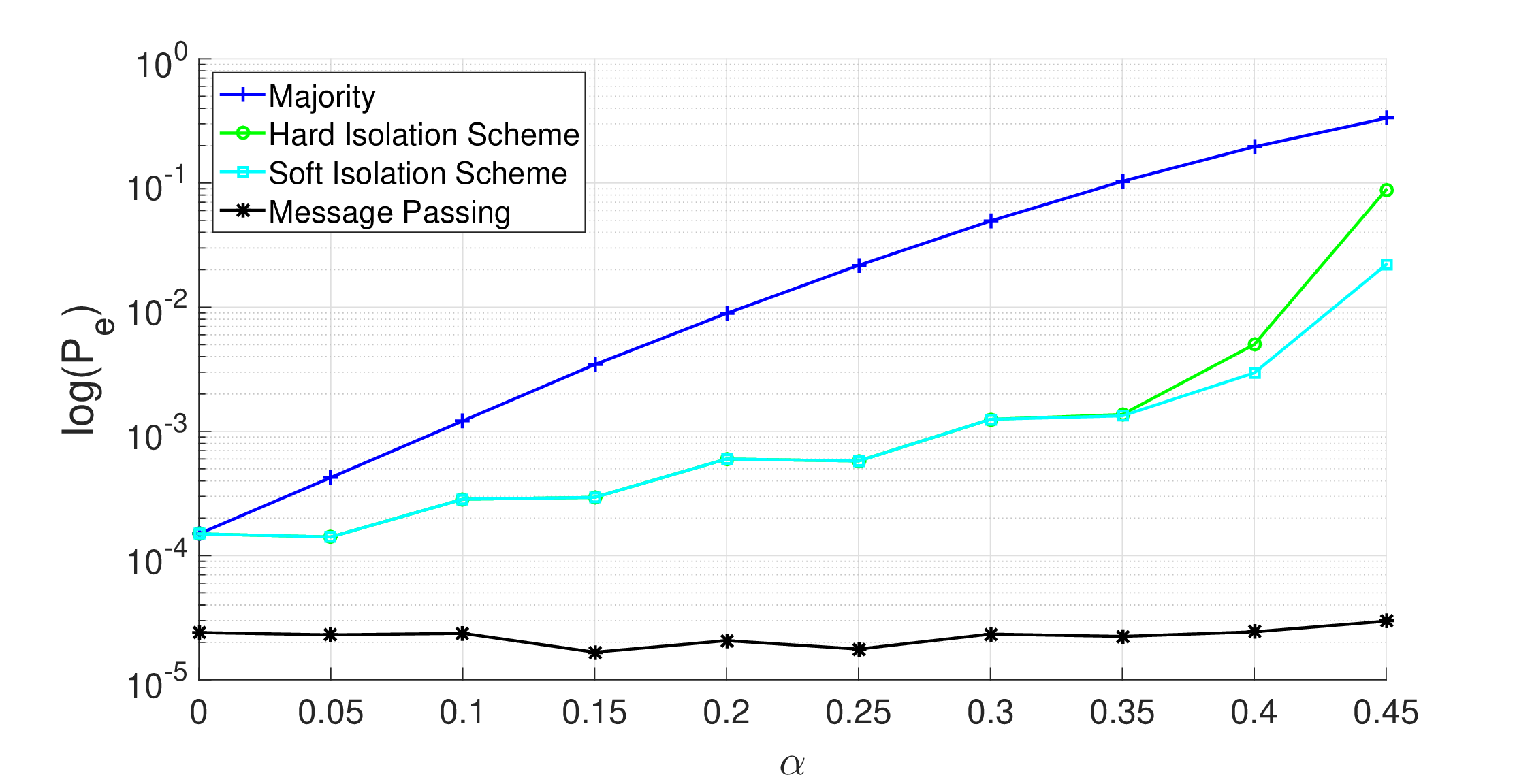} \\
\caption{\textit{Error probability as a function of $\alpha$ for the following setting: $n=20$, Markovian Sequence of States $\rho = 0.95$, $\varepsilon =0.15$, $m=30$ and $P_{mal}=1.0$.}}
\label{Perf_one_markov_m_30_Pmal1}
\end{center}
\end{figure}

\begin{figure}[h!]
\begin{center}
 \includegraphics[width=0.9\textwidth]{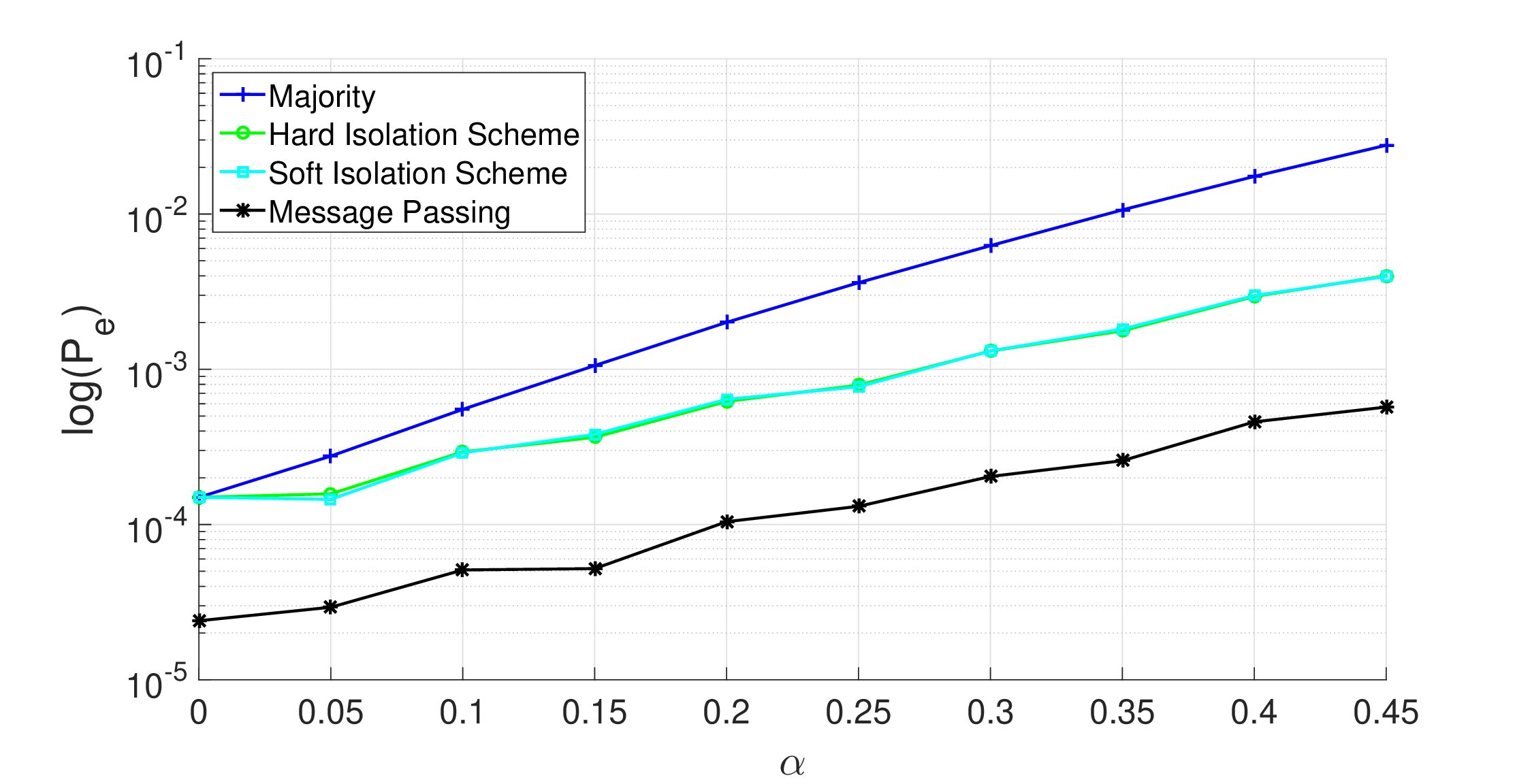} \\
\caption{\textit{Error probability as a function of $\alpha$ for the following setting: $n=20$, Markovian Sequence of States $\rho = 0.95$, $\varepsilon =0.15$, $m=30$ and $P_{mal}=0.5$.}}
\label{Perf_one_markov_m_30_Pmal05}
\end{center}
\end{figure}

Having shown the near optimality of the message passing fusion algorithm for small values of $m$; we now leverage on the small computational complexity\index{computational complexity|textbf} of such a scheme to evaluate its performance for large values of $m$ ($m = 30$). As shown in Figure \reffig{Perf_one_markov_m_30_Pmal1}, by increasing the observation window all the schemes give better performance, with the message passing algorithm always providing the best performance. Interestingly, in this case, when the attacker uses $P_{mal}=1.0$, the message passing algorithm permits to almost nullify the attack of the Byzantines for all the values of  $\alpha$. The reason is that, using $P_{mal}=1.0$ conveys more information to the FC about the Byzantines and consequently, makes their detection easier. Concerning the residual error probability, it is due to the fact that, even when there are no Byzantines in the network ($\alpha=0$), there is still an error floor caused by the local errors at the nodes $\varepsilon$. For the case of independent states, such an error floor is around $10^{-4}$. In Figure \reffig{Perf_one_markov_m_30_Pmal1} and \reffig{Perf_one_markov_m_30_Pmal05}, this error floor decreases to about $10^{-5}$ because of the additional a-priori information available in the Markovian case.

\subsubsection{Optimal choice of $P_{mal}$ for the Byzantines}

One of the main results presented in the previous chapter, is that setting $P_{mal} = 1$ is not necessarily the optimal choice for the Byzantines. In fact, when the FC manages to identify which are the malicious nodes, it can exploit the fact the malicious nodes always flip the result of the local decision to get useful information about the system state. In such cases, it is preferable for the Byzantines to use $P_{mal} = 0.5$ since in this way the reports send to the FC does not convey any information about the status of the system. However, in chapter \ref{chapter:TIFS_SPL}, it was not possible to derive exactly the limits determining the two different behaviours for the Byzantines due to the impossibility of applying the optimum algorithm in conjunction with large observation windows. By exploiting the low complexity of the message passing scheme, we are now able to overcome the limits of the analysis carried out in previous chapter.

\begin{figure}[h!]
\begin{center}
 \includegraphics[width=0.9\textwidth]{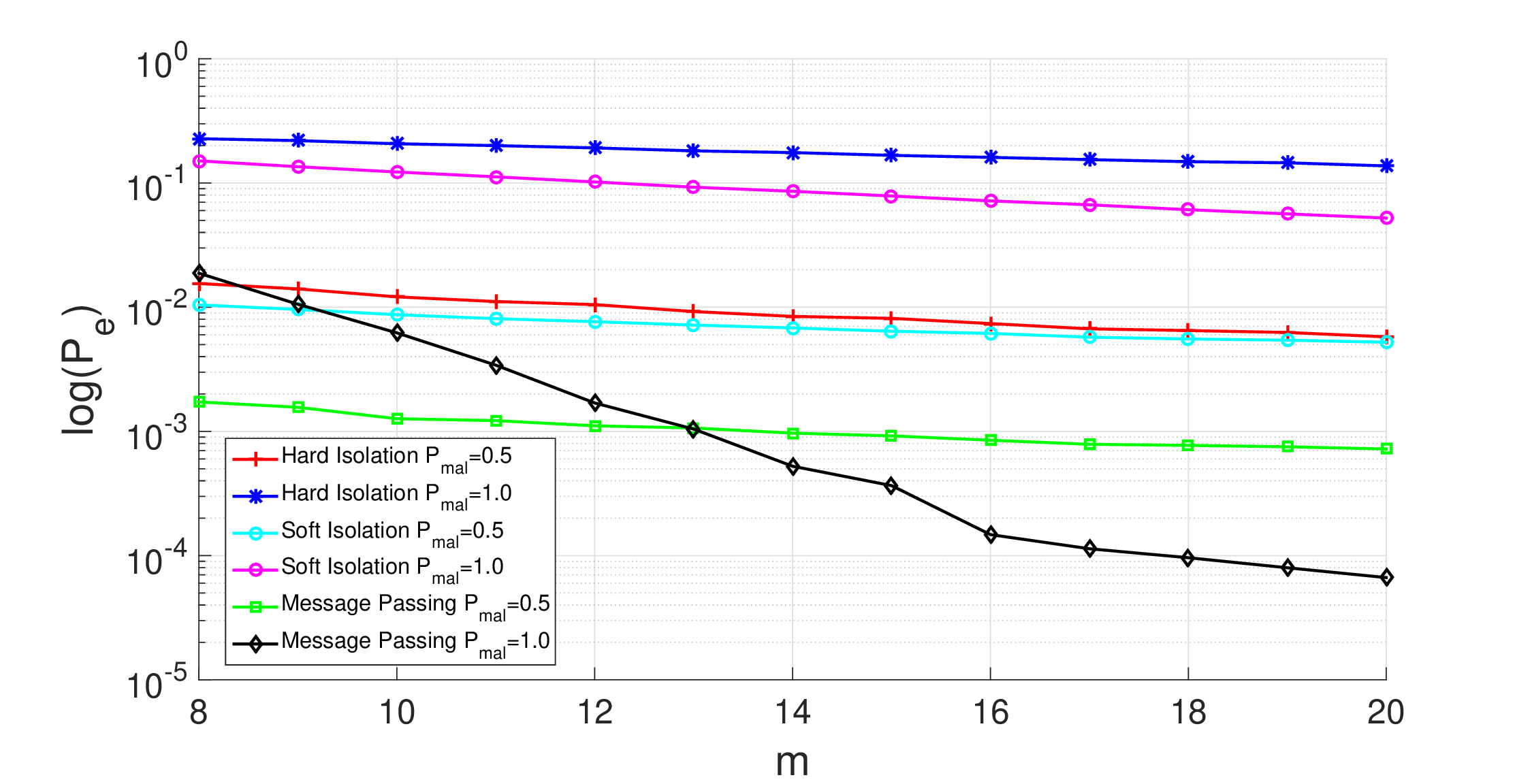} \\
\caption{\textit{Error probability as a function of $m$ for the following settings: $n=20$, Markovian Sequence of States $\rho = 0.95$, $\varepsilon =0.15$ and $\alpha=0.45$.}}
\label{Perf_dual_behavior_alpha_045_markov}
\end{center}
\end{figure}

\begin{figure}
\begin{center}
 \includegraphics[width=0.9\textwidth]{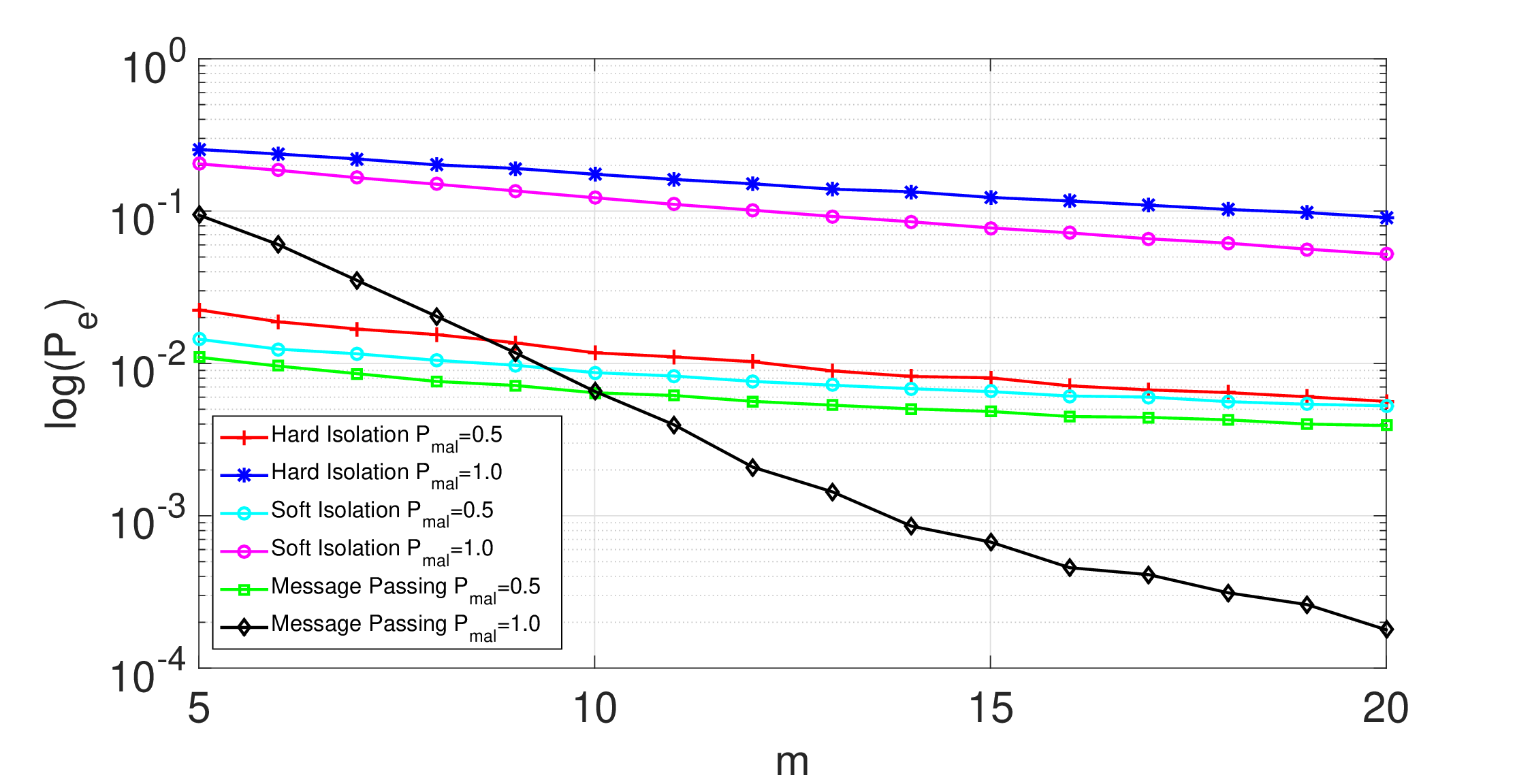} \\
\caption{\textit{Error probability as a function of $m$ for the following settings: $n=20$, independent Sequence of States $\rho = 0.5$, $\varepsilon =0.15$ and $\alpha=0.45$.}}
\label{Perf_dual_behavior_alpha_045_independent}
\end{center}
\end{figure}

Specifically, we carried out an additional set of experiments by fixing $\alpha = 0.45$ and varying the observation window in the interval [5,20]. The results we obtained confirm the general behaviour observed in chapter \ref{chapter:TIFS_SPL}. For instance, in Figure \reffig{Perf_dual_behavior_alpha_045_markov}, $P_{mal} = 1.0$ remains the Byzantines' optimal choice up to $m=13$, while for $m > 13$, it is preferable for them to use $P_{mal}=0.5$. Similar results are obtained for independent system states as shown in Figure  \reffig{Perf_dual_behavior_alpha_045_independent}.

\subsubsection{Comparison between independent and Markovian System States}

In this part, we provide a comparison between the cases of Markovian and independent system states.

By looking at Figure \reffig{Perf_dual_behavior_alpha_045_markov} and \reffig{Perf_dual_behavior_alpha_045_independent}, we see that the Byzantines  switch their strategy from $P_{mal}=1$ to $P_{mal} = 0.5$ for a smaller observation window ($m=10$) in the case of independent states\index{independent system states|textbf} (the switching value for the Markovian case is $m = 13$). We can explain this behaviour by observing that in the case of Markovian states\index{Markovian system states|textbf}, using $P_{mal} = 0.5$ results in a strong deviation from the Markovianity assumption of the reports sent to the FC thus making it easier the isolation of byzantine nodes\index{Byzantines isolation|textbf}. This is not the case with $P_{mal} = 1$, since, due to the symmetry of the adopted Markov model, such a value does not alter the expected statistics of the reports.

\begin{figure}[h!]
\begin{center}
 \includegraphics[width=0.9\textwidth]{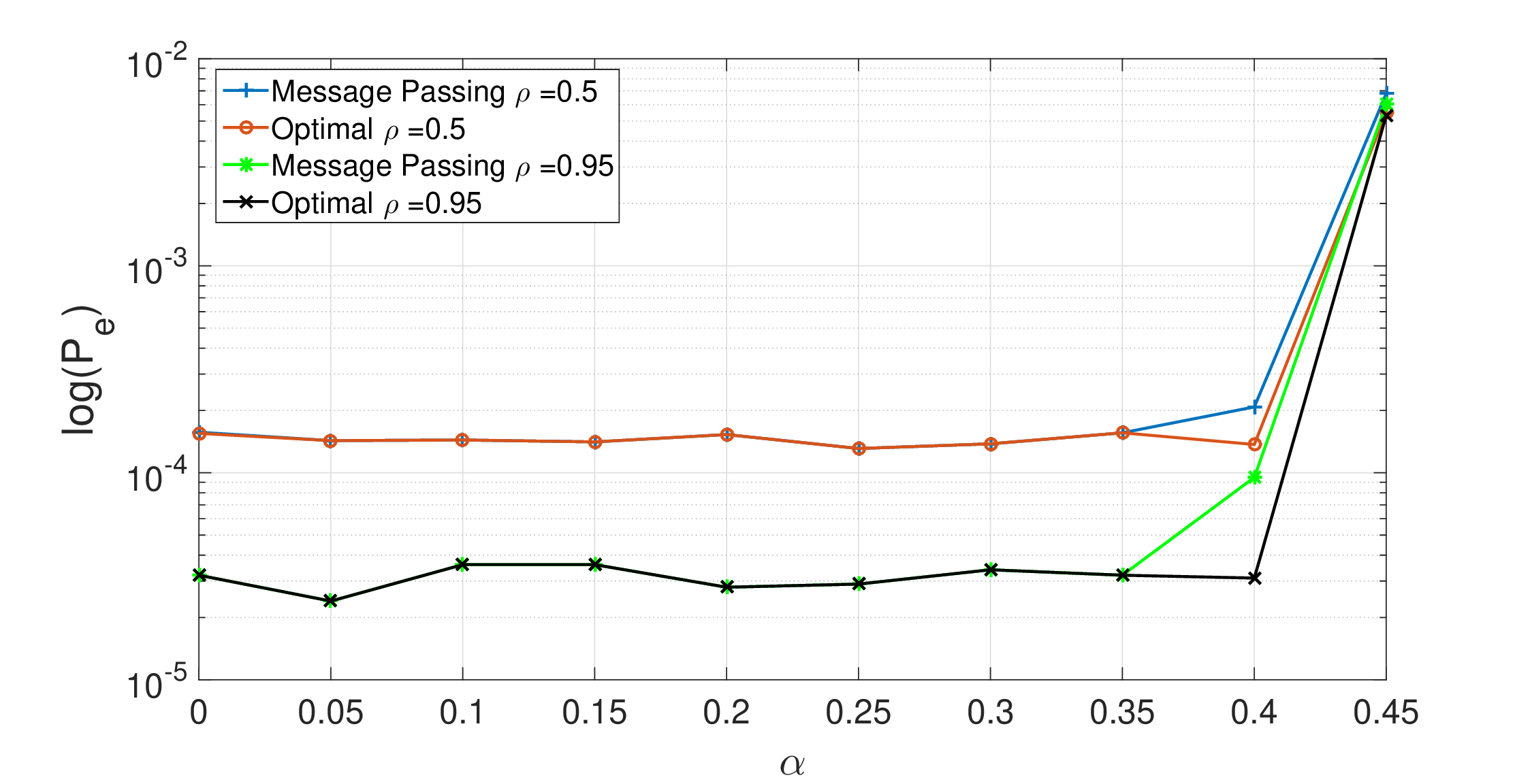} \\
\caption{\textit{Comparison between the case of independent and Markovian system states ($n=20$, $\rho = \{0.5, 0.95\}$, $\varepsilon =0.15$, $m=10$,  $P_{mal}=1.0$).}}
\label{Perf_Ind_vs_Markov}
\end{center}
\end{figure}

As a last result, in Figure \reffig{Perf_Ind_vs_Markov}, we compare the error probability for the case of independent and Markov sources. Since we are interested in comparing the achievable performance for the two cases, we consider only the performance obtained by the optimum and the message passing algorithms. Upon inspection of the figure, it turns out that the case of independent states is more favourable to the Byzantines than the Markov case. The reason is that the FC may exploit the additional a-priori information\index{a-priori information|textbf} available in the Markov case to identify the Byzantines and hence make a better decision. Such effect disappears when $\alpha$ approaches $0.5$, since in this case the Byzantines tend to dominate the network. In that case, the Byzantines' reports prevail the pool of reports at the FC and hence, the FC becomes nearly {\em blind} so that even the additional a-priori information about the Markov model does not offer a great help.

\section{Conclusions}
\label{sec:conclusion}
In this chapter, we proposed a near-optimal message passing algorithm based on factor graph for decision fusion in distributed sensor networks in the presence of Byzantines. The effectiveness of the proposed 
scheme is evaluated by means of extensive numerical simulations both for the case of independent and Markov sequence of states. 
Experiments showed that, when compared to the optimum fusion scheme, the proposed scheme permits to achieve near-optimal performance at a much lower computational cost: specifically, by adopting the new algorithm based on message passing we were able to reduce the complexity from exponential to linear. Such reduction of the complexity permits to deal with large observation windows, thus further improving the performance of the decision.
Results on large observation windows confirmed the dual behavior of the optimum attacking strategy,
looking for a trade-off between pushing the FC to make a wrong decision on one hand and reducing the mutual information between the reports and the system state on the other hand.
In addition, the experiments showed that the case of independent states is more favorable to Byzantines than the Markovian case, due to the additional a-priori information available at the FC in the Markovian case. 

\chapter{Consensus Algorithm
with Censored Data for Distributed Detection with Corrupted Measurements}
\label{chapter:GameSec}
\emph{"We cannot solve our problems with the same thinking we used when we created them."}
\\
Albert Einstein

\section{Introduction}

\PARstart{\textcolor{red}I}n{} centralized networks it is possible to adopt an optimum decision strategy based on the entire set of measurements collected by the network,. At the the same time, centralized solutions present a number of drawbacks, most of which related to the security of the network. For instance, the FC represents a single point of failure or a bottleneck for the network, and its failure may compromise the correct behavior of the whole network. In addition, due to privacy considerations or power constraints, the sensors may prefer not to share the gathered information with a remote device. For the above reasons, decentralized solutions are sometimes more attractive. 

In this chapter, we consider decentralized distributed detection based on consensus algorithm in adversarial sensor networks. By focusing on the measurement falsification attack with corruption of the physical link presented in chapter \ref{chapter:DF}, we introduce a preliminary isolation step in which each node in the network may discard its own measurement based on the available a priori knowledge of the measurements statistics under the two states of the system\index{system state|textbf}. Then, the consensus algorithm proceeds as usual, with the nodes which discarded their measurments no longer taking part in the consensus algorithm. In fact, they only observe the exchanged messages and epxloit the outcome of the consensus algorithm run by the remaining nodes (assuming that the networks is not disconnected due to the removal of the nodes with censored data). Under some assumptions on network topology, that prevents that isolation step disconnects the network, the convergence of the consensus algorithm is preserved. By following the principles of adversarial signal processing\index{Adversarial Signal Processing|textbf} \cite{AdvSP}, we assume that in turn the attacker may adjust the strength of the falsification attack to avoid that the fake measurements are discarded. We then formalize the interplay between the network and the attacker as a zero-sum competitive game and use simulations to derive the equilibrium point of the game.

\section{Distributed Detection based on consensus algorithm}
\label{sec.dd_main}
In this section, we describe the distributed detection system considered in this chapter, when no adversary is present and introduce the consensus algorithm the detection system relies on.

\subsection{The Network Model}
\label{sec.net_mod}

The network is modeled as an undirected graph $\mathcal{G}$ where the information can be exchanged in both directions between the nodes. A graph $\mathcal{G} = (\mathcal{N}, \mathcal{E})$ consists of set of nodes $\mathcal{N} = \{n_1, ...,n_n\}$  and set of edges $\mathcal{E}$  where $(n_i,n_j) \in \mathcal{E}$ if and only if there is a communication link between $n_i$ and $n_j$,  i.e., they are neighbors.
The neighborhood of a node $n_i$ is indicated as $\mathcal{N}_i = \{n_j \in \mathcal{N} : (n_i,n_j) \in \mathcal{E}$ \}. For sake of simplicity, we sometimes refer to $\mathcal{N}_i$ as the set of indexes $j$ instead than directly the nodes.  A graph $\mathcal{G}$ can be represented by its adjacency matrix $A = \{a_{ij}\}$ where $a_{ij} = 1$, if $(n_i,n_j) \in \mathcal{E}$, $0$ otherwise. The degree matrix $D$ of $\mathcal{G}$ is a diagonal matrix with $d_{ii} = a_{i1} + a_{i2}  + ... + a_{in}$, $d_{ij} = 0$, $\forall i$, $j \neq i$ \cite{godsilgraphtheory}.

\subsection{The Measurement Model}
\label{sec.dd_model}
Let $S$ be the status of the system under observation: we have $S=0$, under hypothesis $H_0$ and $S=1$ under hypothesis $H_1$. We use the capital letter $X_i$ to denote the random variable describing the measurement at node $n_i$, and the lower-case letter $x_i$ for a specific instantiation.
By adopting a Gaussian model, the probability distribution of each measurement $x_i$ under the two hypothesis is given by\footnote{We are assuming that the statistical characterization of the measurement at all the nodes is the same.}:
\begin{equation}
  P_{X}(x)=\begin{cases}
    \mathcal{N}(-\mu,\sigma), \text{ under $H_0$},\\
    \mathcal{N}(\mu,\sigma),  \text{under $H_1$},
  \end{cases}
  \label{model}
\end{equation}
where, $\mathcal{N}(\mu,\sigma)$ is the Normal Distribution with mean $\mu$ and variance $\sigma^2$.

Let us denote with $U$ the result of the final (binary) decision. An error occurs if $U \neq S$. By assuming that the measurements are conditionally independent, that is they are independent conditioned to the status of the system, the optimum decision  strategy consists in computing the mean of the measurements, $\bar{x} = \sum_{i} x_i /n $ and comparing it with a threshold $\lambda$ which is set based on the a-priori probability ($\lambda = 0$ in the case of equiprobable system states\index{system state|textbf}).
In a distributed architecture based on consensus, the value of $\bar{x}$ is computed iteratively by means of a proper message exchanging procedure between neighboring nodes, the final decision is made at each single node by comparing $\bar{x}$ with $\lambda$.

In this chapter we consider the case of equiprobable system states. It is worth observing that  our analysis, included the game formulation in Section \ref{sec.Game_theo}, can be extended to the general case in which this assumption does not hold.

\subsection{The Consensus Algorithm}
\label{subsec.cons}

In this chapter, we consider the consensus algorithm\index{consensus algorithm|textbf} as described in chapter \ref{chapter:DF}. In the symmetric setup considered in this chapter, the decision threshold of the algorithm is set to $\lambda = 0$.

\section{Measurement falsification attack against consensus-based detection}
\label{sec.Attack_Cons}
In this section, we consider the impact that one or more fake measurements have on the output of the consensus algorithm.

\subsection{Consensus Algorithm with Corrupted Measurements}
\label{subsec.cons_corr}
In the binary decision setup we are considering, the objective of the attacker\index{attacker|textbf} is inducing one of the two decision errors (or both of them): decide that $S = 0$ when $H_1$ holds (False Alarm), and decide that $S = 1$ when $H_0$ holds (Missed Detection).
We make the worst case assumption that the attacker\index{Measurement Falsification Attack|textbf} knows the true system state\index{system state|textbf}. In this case, he can try to push the network toward a wrong decision by replacing one or more measurements so to bias the average computed by the consensus algorithm.
Specifically, for any corrupted node, the attacker forces the measurement to a positive value $\Delta_0$ under $H_0$ and to a negative value $\Delta_1$ under $H_1$.
For the symmetric setup,  reasonably, $\Delta_0 =  -\Delta_1 = \Delta > 0$.
In the following we assume that the attacker corrupts a fraction $\alpha$ of the nodes, i.e the number of attacked nodes is $n_A = \alpha n$. 

Given the initial vector of measurements, the consensus value the network converges to because of the attack is:
\begin{equation}
\bar{x} = \frac{1}{n} \sum\limits_{i \in \mathcal{N}_H} x_i(0) \pm \frac{n_A \Delta}{n},
\end{equation}
where $\mathcal{N}_H$ is the set of nodes with uncorrupted measurements ($|\mathcal{N}_H| = n - n_A$).

By referring to the model described in Section \ref{sec.dd_model}, it is easy to draw a relation between $\Delta$, $\alpha$ and the probability $p$ that the attacker induces a decision error.
By exploiting the symmetry of the considered setup we can compute $p$ by considering the behavior under one hypothesis only, that is we have $p= P(U = 1| H_0) = P(\bar{X} > 0 | H_0)$.

In the following we indicate with $\bar{X}(\mathcal{N})$ the average of the measurements made by the nodes in a set $\mathcal{N}$.

The error probability $p$ for a given $n_A$ can be written as:

\begin{align}
p = & P(\bar{X} > 0 | H_0) =  P\left(\frac{n-n_A}{n}\bar{X}(\mathcal{N}_H) >  - \frac{n_A \Delta}{n} \bigg| H_0\right) \\ \nonumber
= & P\left(\bar{X}(\mathcal{N}_H) > \frac{n}{n-n_A}\bigg(  - \frac{n_A \Delta}{n}\bigg) \bigg| H_0\right) \\ \nonumber
= &  \int\limits_{- \frac{n_A \Delta}{n-n_A}}^{\infty}  \mathcal{N}( - \mu, \sigma/\sqrt{n - n_A}). \nonumber
\end{align}

\begin{figure}[t!]
\centering\includegraphics[width=\columnwidth]{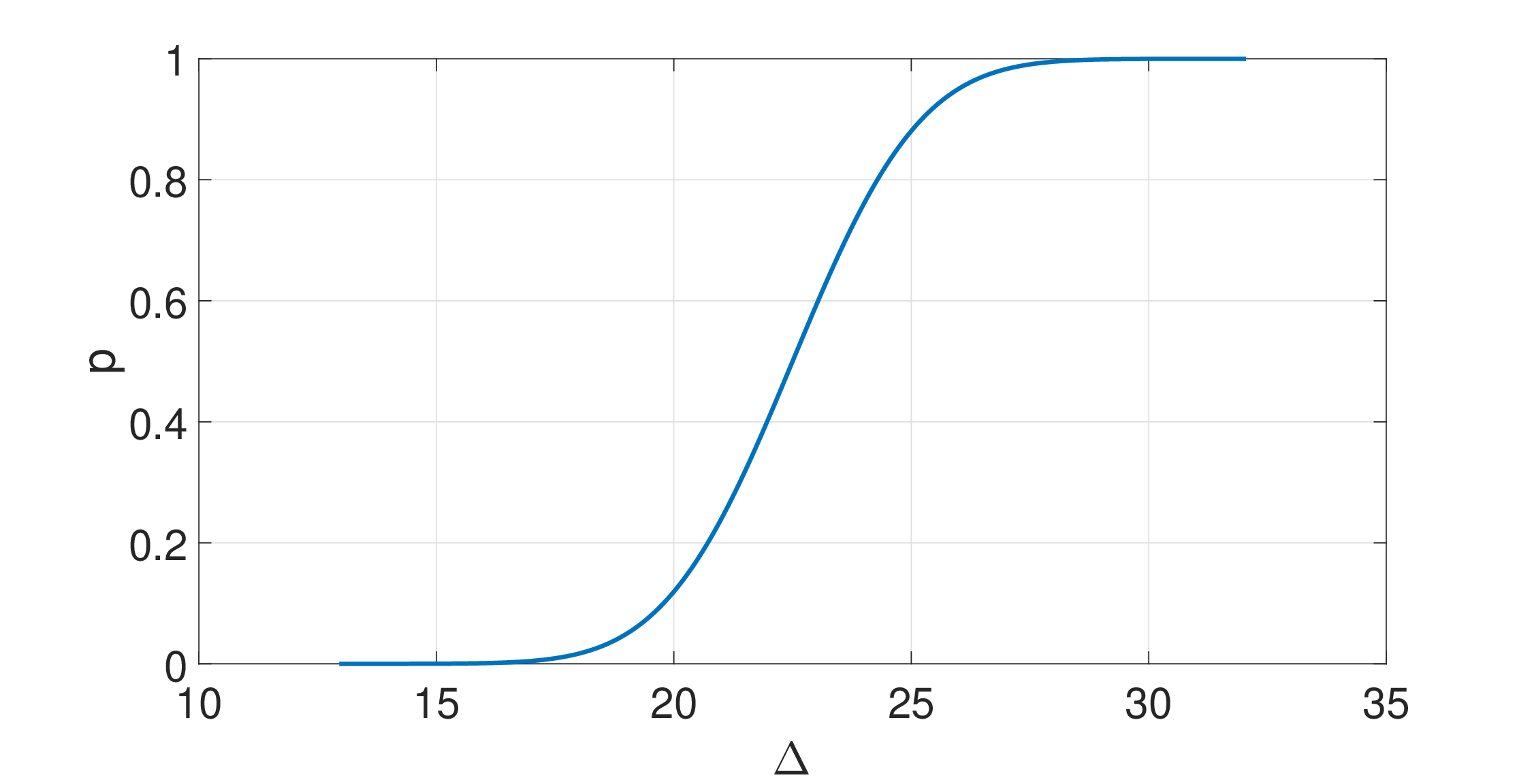}
\caption{\textit{Success probability of the attack versus $\Delta$  in the adversarial setup $n=20$, $\mu = 2.5$, $\sigma = 1$, $N_A=2 (\alpha=0.1)$}}
\vspace{-0.5cm}
\label{fig:Deltavsp}
\end{figure}

\begin{figure}[t!]
\centering\includegraphics[width=\columnwidth]{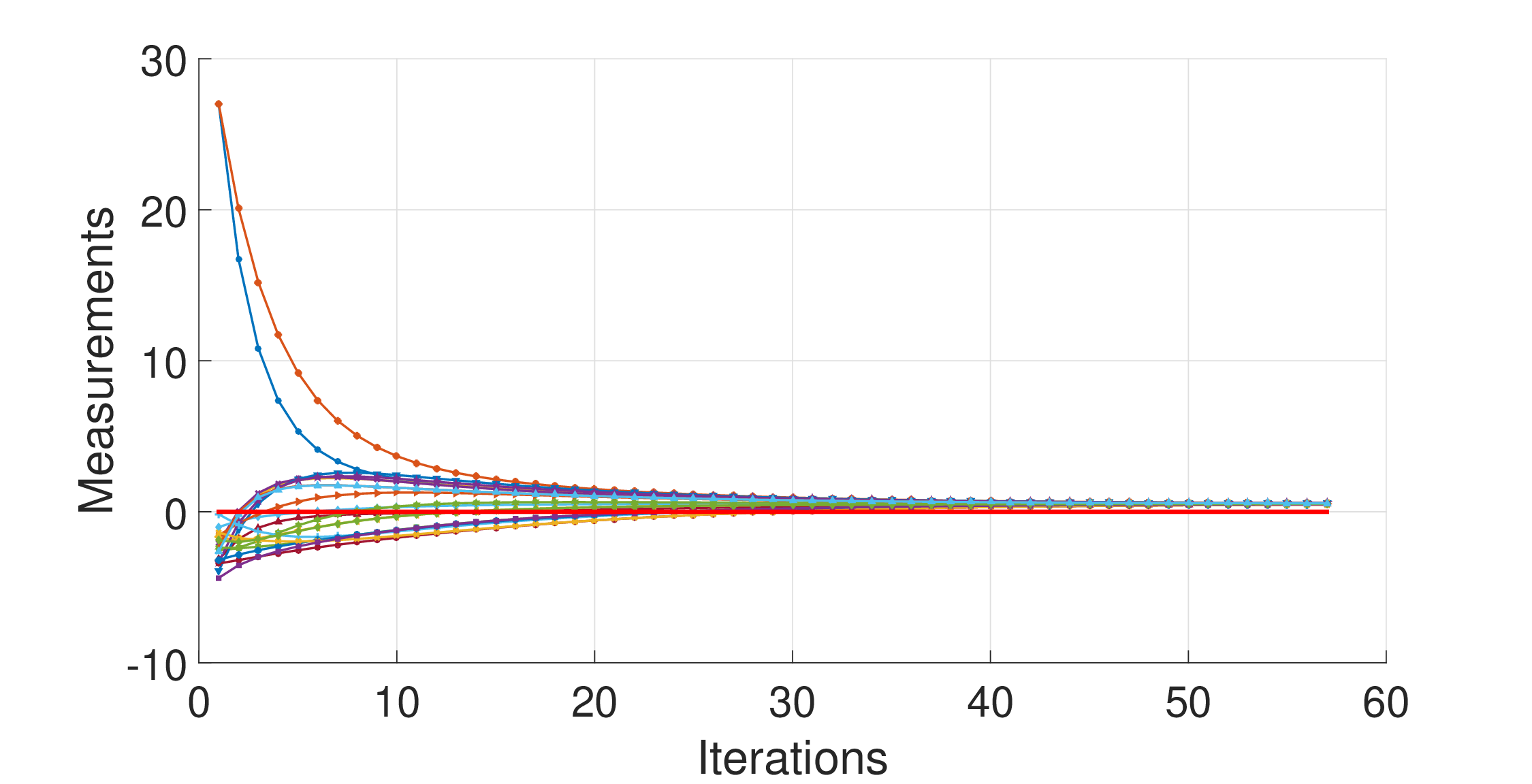}
\caption{\textit{Effect of the attack on the convergence of the consensus algorithm for $\Delta = 27$, the network decides for $H_1$ even if $H_0$ is true.}}
\vspace{-0.5cm}
\label{fig.AttackProof}
\end{figure}

Clearly, if there is no limit to the value of $\Delta$, the attacker will always succeed in inducing a wrong decision (see for example Figures \reffig{fig:Deltavsp} and \reffig{fig.AttackProof}).

This shows how harmful the attack can be against distributed detection based on consensus algorithm.

\section{Consensus Algorithm with Censored Data} 
\label{sec.Cons_removal}

With centralized fusion it is quite easy to detect false measurements, since they assume outlier values with respect to the majority of the measurements. In a distributed setting, however, this is not easy since, at least in the initial phase of the consensus algorithm (see chapter \ref{chapter:DF}), each node sees only its measurement and has no other clue about the system status.

In contrast to most previous works \cite{sundaram2011consensus,yan2012vulnerabilityconsensus}, we tackle with the problem of measurement falsifications at the initial phase of the consensus algorithm (see for instance \cite{olfati2007consensuscooperation}), by letting each node discard its measurement if it does not fall within a predefined interval containing most of the probability mass associated to both $H_0$ and $H_1$.
In the subsequent phases the remaining nodes continue exchanging messages as usual according to the algorithm, whereas the nodes which discarded their measurements only act as receivers and do not take part into the protocol.
Due to the removal, the measurements exchanged by the nodes follows a censored Gaussian distribution\index{Censored data|textbf}, i.e. the distribution which results by constraining the (initial) Gaussian variable to stay within an interval \cite{helsel2005nondetects}.
Specifically, the nodes discards all the measurements whose absolute values are large than a removal threshold $\eta$. By considering the results shown in Figure \reffig{fig:Deltavsp}, we see that, in the setup considered in the figure, if we let $\eta = 17.5$ the error probability drops to nearly zero since the attacker must confine the choice of $\Delta$ to values lower than 17.5.
\begin{figure}[t!]
\centering\includegraphics[width=\columnwidth]{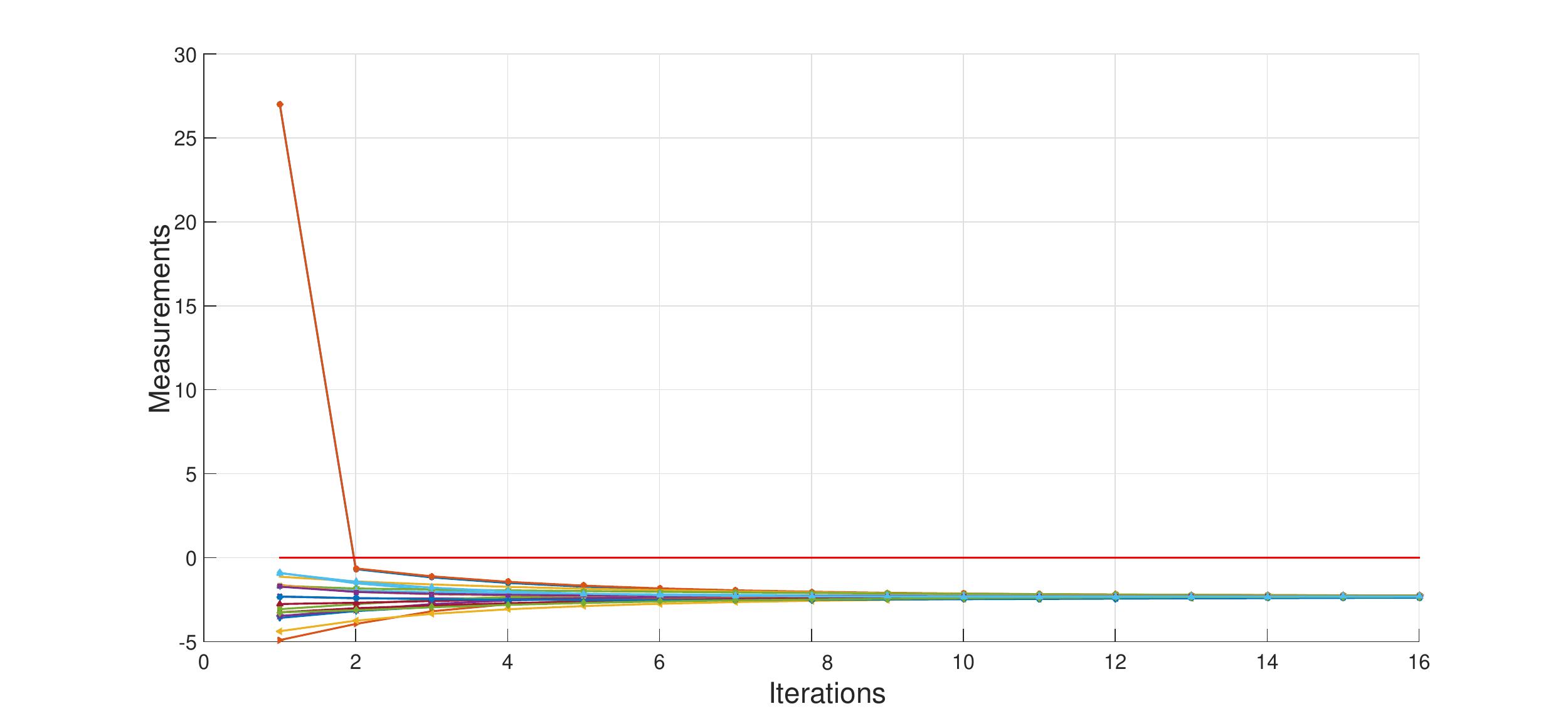}
\caption{\textit{Consensus algorithm with censored data   in the adversarial setup $n=20$, $\mu = -2.5$, $\sigma = 1$, $n_A=2 (\alpha=0.1)$, $\Delta = 27$, $\eta = 25$}}
\vspace{-0.5cm}
\label{fig:CensoredConsensus}
\end{figure}
The proposed strategy is simple, yet effective as it can be seen in the example in Figure \reffig{fig:CensoredConsensus}, and allow us to use a game theoretical approach to set the parameters (see Section \ref{sec.Game_theo}).

For our analysis, we assume that the network topology\index{network topology|textbf} is such that the connectivity of the network is preserved with high probability and then the algorithm converges to the average of the measurements that have not been discarded. For a given graph, this fact is characterized by the node connectivity, namely, the maximum number of nodes whose removal does not cause a disconnection \cite{ModernGT}. Convergence is guaranteed for instance in the following cases (see \cite{GraphRobustness} for an extensive analysis of the connectivity properties for the various topologies): Fully-connected graph\index{Fully-connected graph|textbf}; Random Graph\index{Random Graph|textbf} \cite{RandomGraphs}, when the  probability of having a connection between two nodes is large enough; Small-World Graph\index{Small-World Graph|textbf} \cite{SmallWorldGraph} when the neighbour list in ring formation is large and the rewiring probability is large as well; Scale-Free Graph\index{Scale-Free Graph|textbf} \cite{ScaleFreeGraphs} , for sufficiently large degree of the non-fully meshed nodes.

We now give a more precise formulation of the consensus algorithm based on censored data\index{Censored data|textbf}. Let us denote with $\mathcal{R}$
the set of the remaining nodes after the censorship, that is
\begin{equation}
\mathcal{R} = \{n_j \in \mathcal{N}: -\eta < x_j < \eta\},
\end{equation}
and let $\mathcal{R}_i$ be the 'active' neighborhood of node $i$ after censorship, $i \in \mathcal{R}$ (i.e. the set of the nodes in the neighborhood of $i$ which take part in the protocol). 
The update rule for node $i \in \mathcal{R}$ can be written as:
\begin{equation}
x_i(k+1) = x_i(k)  + \epsilon \sum_{j \in \mathcal{R}_i} (x_j(k)-x_i(k)),
\end{equation}
where $0 < \epsilon < ( \max\limits_i \mathcal{N}_i)^{-1}$, and the degree refers to the network after the removal of the suspect nodes, that is to the graph ($\mathcal{R}, \mathcal{E}$) (instead of  ($\mathcal{N}, \mathcal{E}$)).

Under the conditions on network topologies listed before, the consensus algorithm converges to the average value computed over the measurements made by the nodes in $\mathcal{R}$, namely  $\bar{x}(\mathcal{R})$. Otherwise, disconnection may occur and is possible that different parts of the network (connected components) converge to different values.

\enlargethispage{\baselineskip}

\section{Game-Theoretic Formulation}
\label{sec.Game_theo}
The consensus algorithm with censored data\index{Censored data|textbf} is expected to be robust in the presence of corrupted measurements\index{Measurement Falsification Attack|textbf}. On the other hand, we should assume that the attacker is aware that the network nodes remove suspect measurements in the initial phase, hence he will adjust the attack strength $\Delta$ to avoid that the false measurements are removed. We model the interplay between the attacker and the network as a two-player zero sum game\index{zero-sum game|textbf} where each player tries to maximize its own payoff\index{payoff|textbf}. Specifically, we assume that the network designer, hereafter referred as the defender\index{defender|textbf} ($D$), does not know the attack strength\index{attack strength|textbf} $\Delta$, while the attacker\index{attacker|textbf} (A) does not know the value of the removal threshold $\eta$ adopted by the defender.

With these ideas in mind, the Consensus-based Distributed Detection game $\mathcal{CDD}(\mathcal{S_A},\mathcal{S_D},v)$ is a two-player, strategic game\index{strategic game|textbf} played by the attacker and the defender, defined by the following strategies and payoff.

\begin{itemize}
\item{The space of strategies of the defender\index{defender|textbf} and the attacker\index{attacker|textbf} are respectively

\begin{align}
& \mathcal{S_D} = \{ \eta \in [0, \infty) \} \nonumber\\
& \mathcal{S_A} = \{ \Delta \in [0, \infty) \};
\label{eq.DDFGS}
\end{align}
}
{\em The reason to limit the strategies of $D$ to values larger than $0$ is to avoid removing correct measurements at the defender side and to prevent to vote for the correct hypothesis at the attacker side.}

\item{The payoff\index{payoff|textbf} function is defined as the final error probability, 

\begin{align}
v =  P_e = P(U \neq S) = P(\bar{X} > 0 / H_0), 
\label{eq.Pe}
\end{align}
where $\bar{X} = \bar{X}(\mathcal{R})$, that is the mean computed over the nodes that remain after the removal.
The attacker wishes to maximize $v$, whereas the defender wants to minimize it.
}
\end{itemize}

Note that according to the definition of the $\mathcal{CDD}$ game, the sets of strategies of the attacker and the defender are continuous sets.
We remind that, in this chapter, we consider  situations in which the network remains connected after the isolation and then the convergence of the algorithm is preserved.
Notice that, with general topologies, when disconnection may occur, the payoff function should be redefined in terms of error probability at the node level.

In the next section, we use numerical simulations to derive the equilibrium point of the game under different settings and to evaluate the payoff at the equilibrium.

\section{Simulation Results}
\label{sec.sim_res}

We run numerical simulations in order to investigate the behavior of the $\mathcal{CDD}$ game for different setups and analyze the achievable performance when the attacker and the defender adopt their best strategies with parameters tuned following a game-theoretic formalization. Specifically, the first goal of the simulations is to study the existence of an equilibrium point\index{equilibrium point|textbf} for the $\mathcal{CDD}$ game and analyze the expected behavior of the attacker and the defender at the equilibrium. The goal is to evaluate the payoff at the equilibrium as a measure of the best achievable performance of distributed detection with the consensus algorithm based on censored data.

For our experiments, we quantize the values of $\eta$ and $\Delta$ with step $0.2$ and then we consider the following sets: $\mathcal{S}_D = \{\eta \in \{0, 0.2,...\}\}$ and $\mathcal{S}_A = \{\Delta \in \{0, 0.2,...\}\}$.
Simulations were carried out according to the following setup. We considered a network with $n= \{20,50\}$ nodes where the measurement of each node is corrupted with probability $\alpha \in \{0.1, 0.2\}$. We assume that the probability that the measurement of a node is corrupted does not depend on the other nodes (independent node corruption). By following the model introduced in Section \ref{sec.dd_model}, the measurements are drawn according to a Gaussian distribution with variance $\sigma^2 = 1$ and  mean $- \mu$ and $\mu$ under $H_0$ and $H_1$ respectively. In our tests, we took $\mu = \{1,2\}$.
For each setting, we estimated the error probability of the decision based on censored data\index{Censored data|textbf} over $10^5$ trials.
Then, we determined the mixed strategies Nash equilibrium by relying on the minimax theorem\index{Von Neumann's Minimax Theorem|textbf} \cite{Osb94}.

\enlargethispage{\baselineskip}

\begin{figure}[h!]
\centering
\includegraphics[width= \textwidth,height=6cm]{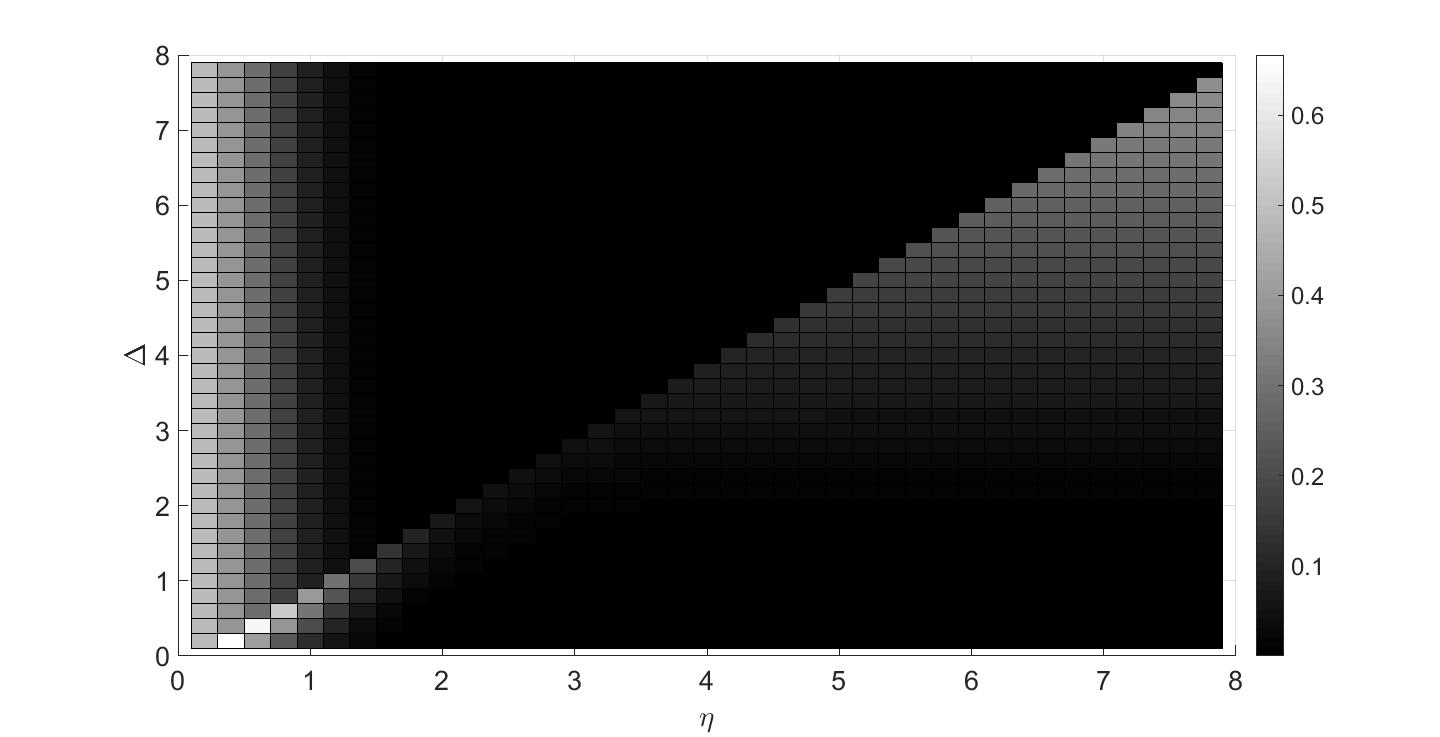}
    \caption{\textit{Payoff matrix of the game with $n=20$, $\alpha=0.1$ and $\mu = 1$ ($\text{SNR}=4$).}}
    \label{fig.GamePSNR2Alpha01}
\end{figure}

\begin{figure}[h!]
\begin{center}
\subfloat[]{\includegraphics[width=\columnwidth]{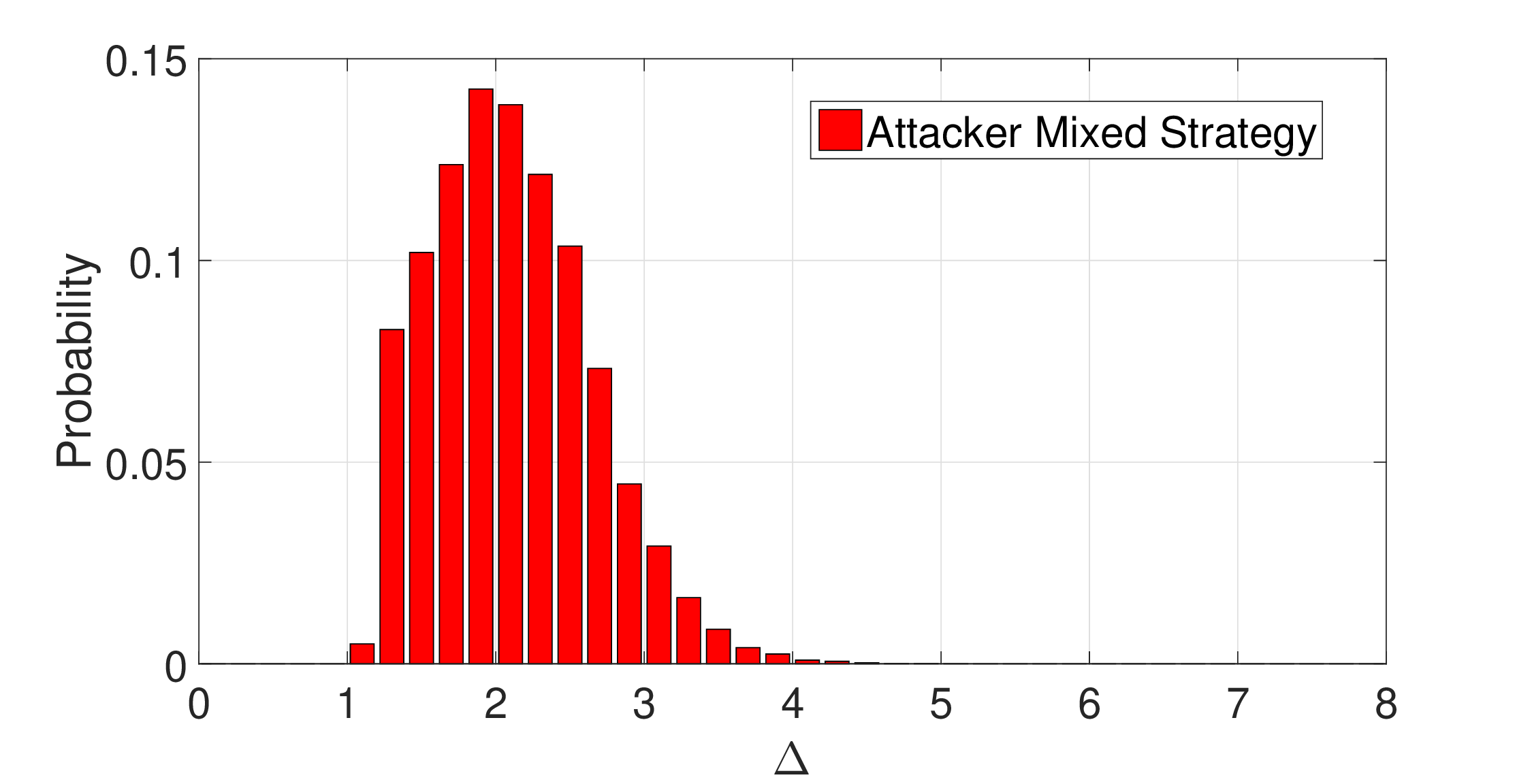}}
\label{fig:StrAPSNR2alpha01}
\subfloat[]{\includegraphics[width=\columnwidth]{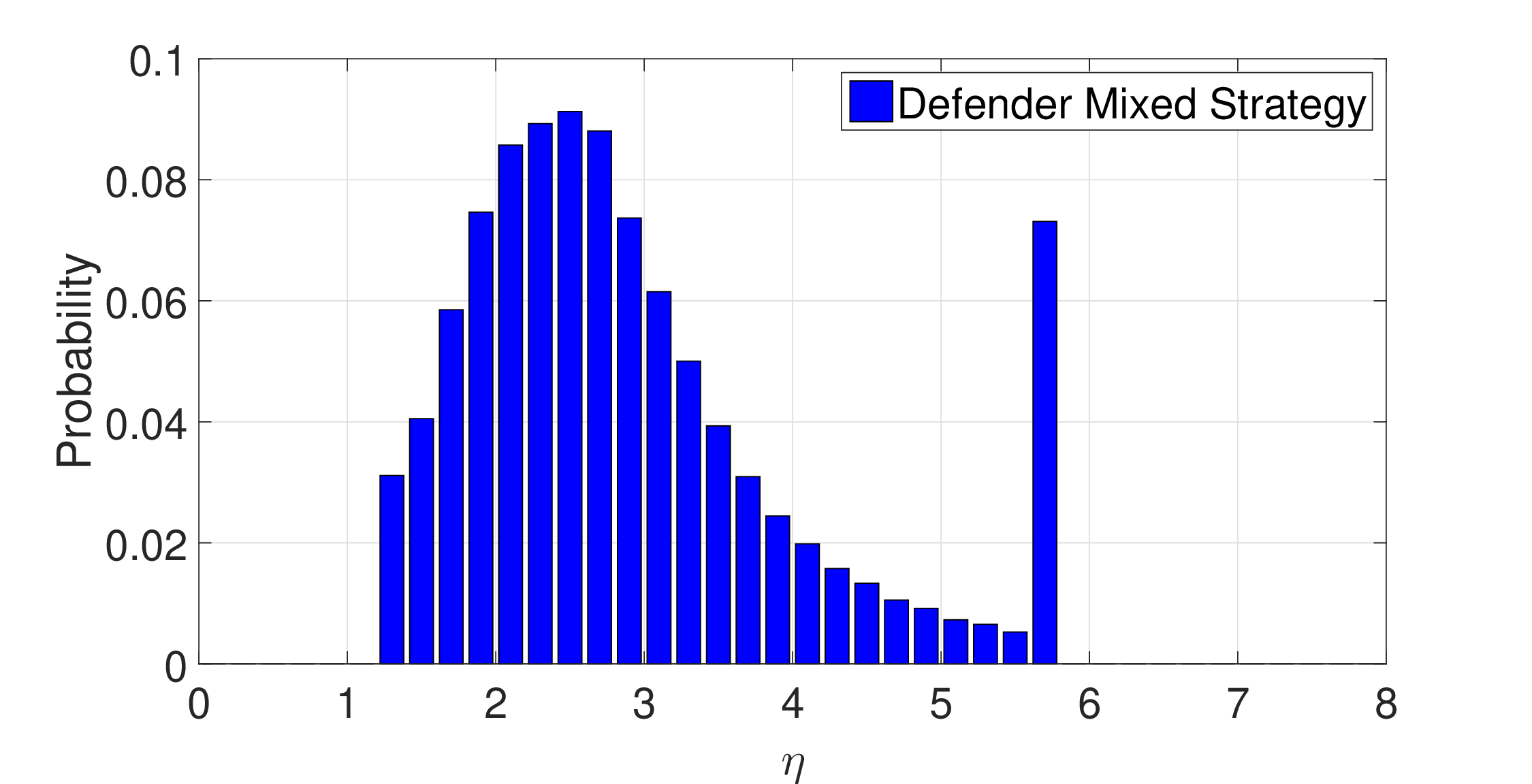}}
\label{fig:StrDPSNR2alpha01}
\caption{\textit{Equilibrium strategies in the following setup: $n=20$, $\alpha=0.1$, $\mu = 1$, ($\text{SNR}=4$). Payoff at the equilibrium: $v = 0.0176$. (a) Attacker Mixed Strategy (b) Defender Mixed Strategy.}}
\label{fig.GameStrPSNR2Alpha01}
\end{center}
\end{figure}

\begin{figure}[h!]
\centering
\includegraphics[width= \textwidth,height=6cm]{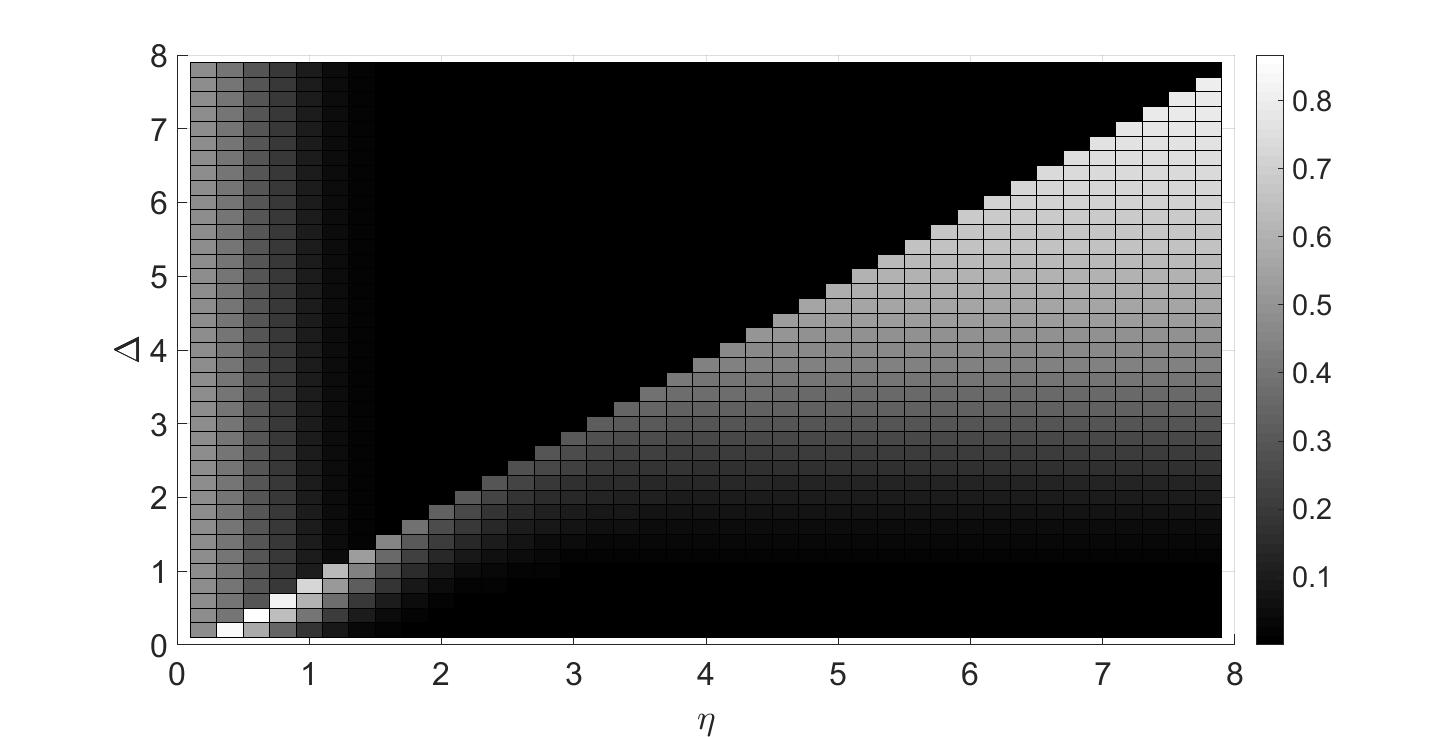}
    \caption{\textit{Payoff matrix of the game with $n=20$, $\alpha=0.2$ and $\mu = 1$ ($\text{SNR}=4$).}}
    \label{fig.GamePSNR2Alpha02}
\end{figure}

\begin{figure}[h!]
\begin{center}
\subfloat[]{\includegraphics[width=\columnwidth]{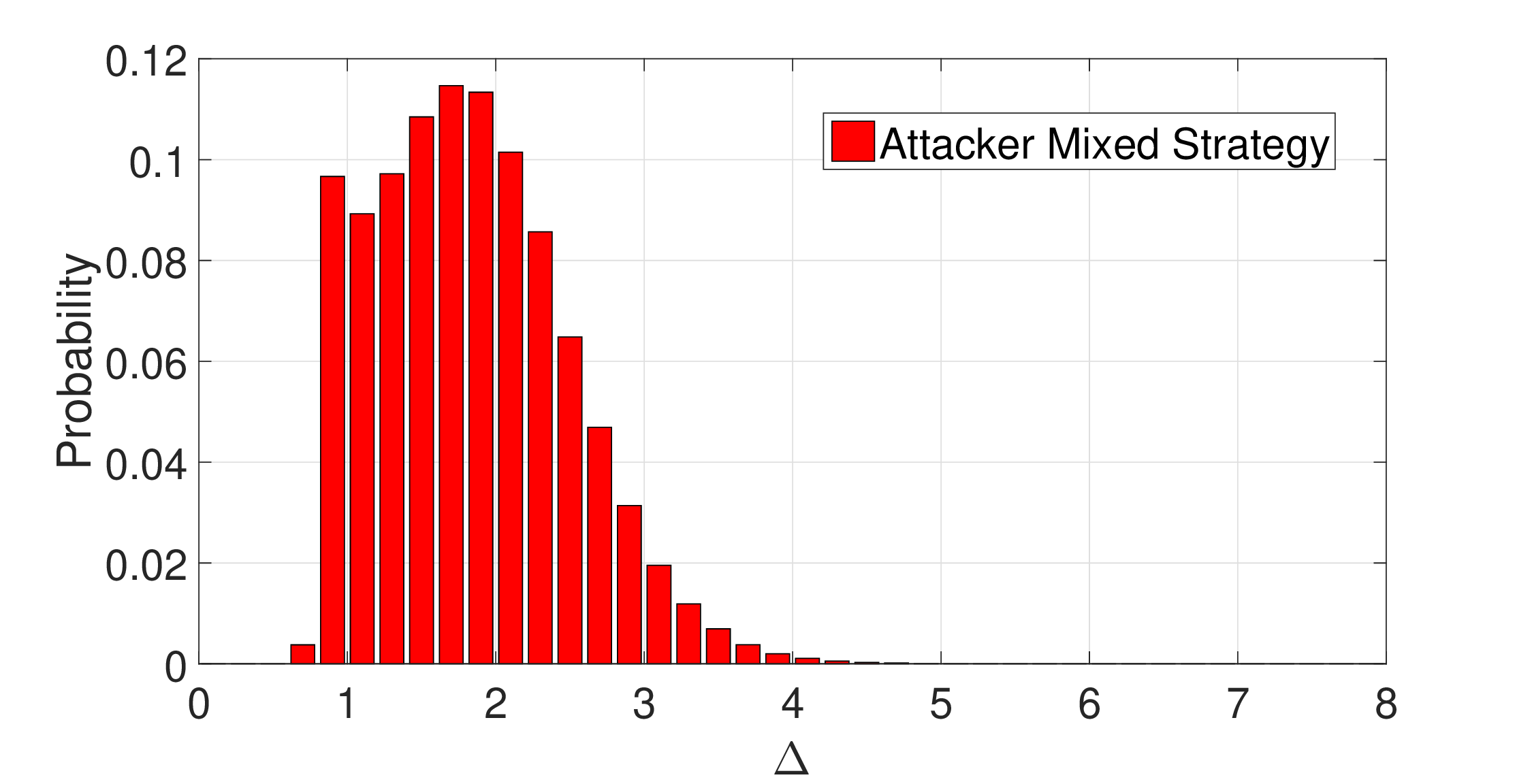}}
\label{fig:StrAPSNR2alpha02}
\subfloat[]{\includegraphics[width=\columnwidth]{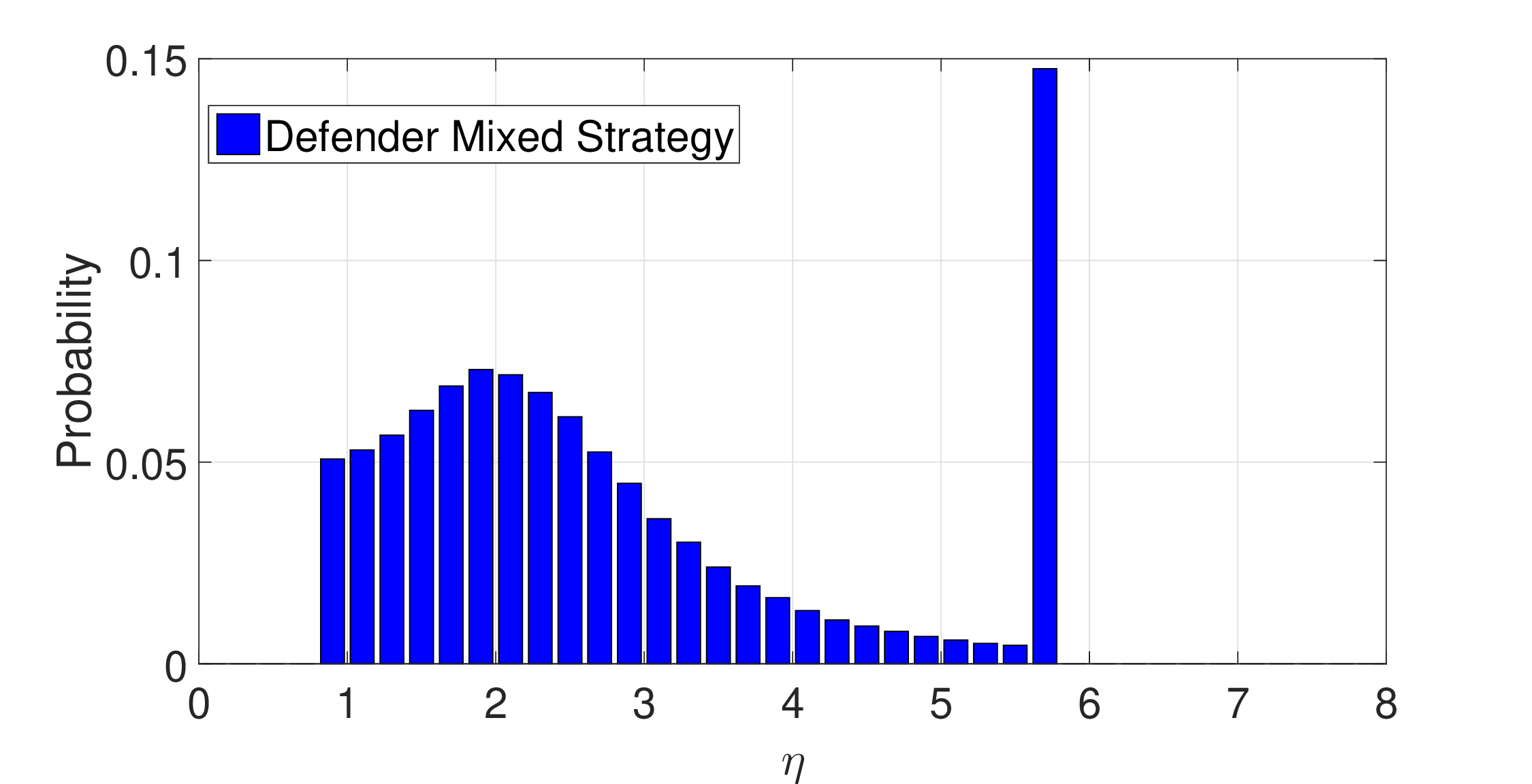}}
\label{fig:StrDPSNR2alpha02}
\caption{\textit{Equilibrium strategies in the following setup: $n=20$, $\alpha=0.2$, $\mu = 1$ ($\text{SNR}=4$). Payoff at the equilibrium: $v = 0.1097$. (a) Attacker Mixed Strategy (b) Defender Mixed Strategy.}}
\label{fig.GameStrPSNR2Alpha02}
\end{center}
\end{figure}

Figure \reffig{fig.GamePSNR2Alpha01} shows the payoff matrix in gray levels for the game with $\alpha = 0.1$ and $\mu  = 1$ (i.e., $SNR = 4$).
The stepwise behavior of the values of the payoff in correspondence of the diagonal is due to the hard isolation (for each $\Delta$, when $\eta <\Delta$ all the corrupted measurements are removed, while they are kept for $\eta \ge \Delta$). When very low values of $\eta$ are considered, the error probability increases because many 'honest' (good) measurements are removed from the network and the decision is based on very few measurements (in the limit case, when all measurements are removed, the network decides at random, leading to $P_e = 0.5$).
Figure \reffig{fig.GameStrPSNR2Alpha01} shows the mixed strategies at the equilibrium. By focusing on the distribution of the defense strategy, D seems to follow the choice of A by choosing the value $\eta$ which is one step ahead of $\Delta$, a part for the presence of a peak, that is a probability mass (of about 0.075) assigned to the value $ \eta = 5.6$, which is the last non-zero value. Interestingly, a closer inspection of the payoff matrix shows that all the strategies above this value are dominated strategies; hence, reasonably, the defender never plays them (assigning them a 0 probability). This is quite expected since for larger $\eta$ it is  unlikely that an observation falls outside the range $[-\eta,\eta]$ and then 'censoring' does not significantly affect the 'honest' measurements (i.e. $\mathcal{R} = \mathcal{N}$ with very high probability). When this is the case, it is clearly is better for D to choose small $\eta$, thus increasing the probability of removing the corrupted measurements.

\begin{figure}[h!]
\centering
\includegraphics[width= \textwidth,height=6cm]{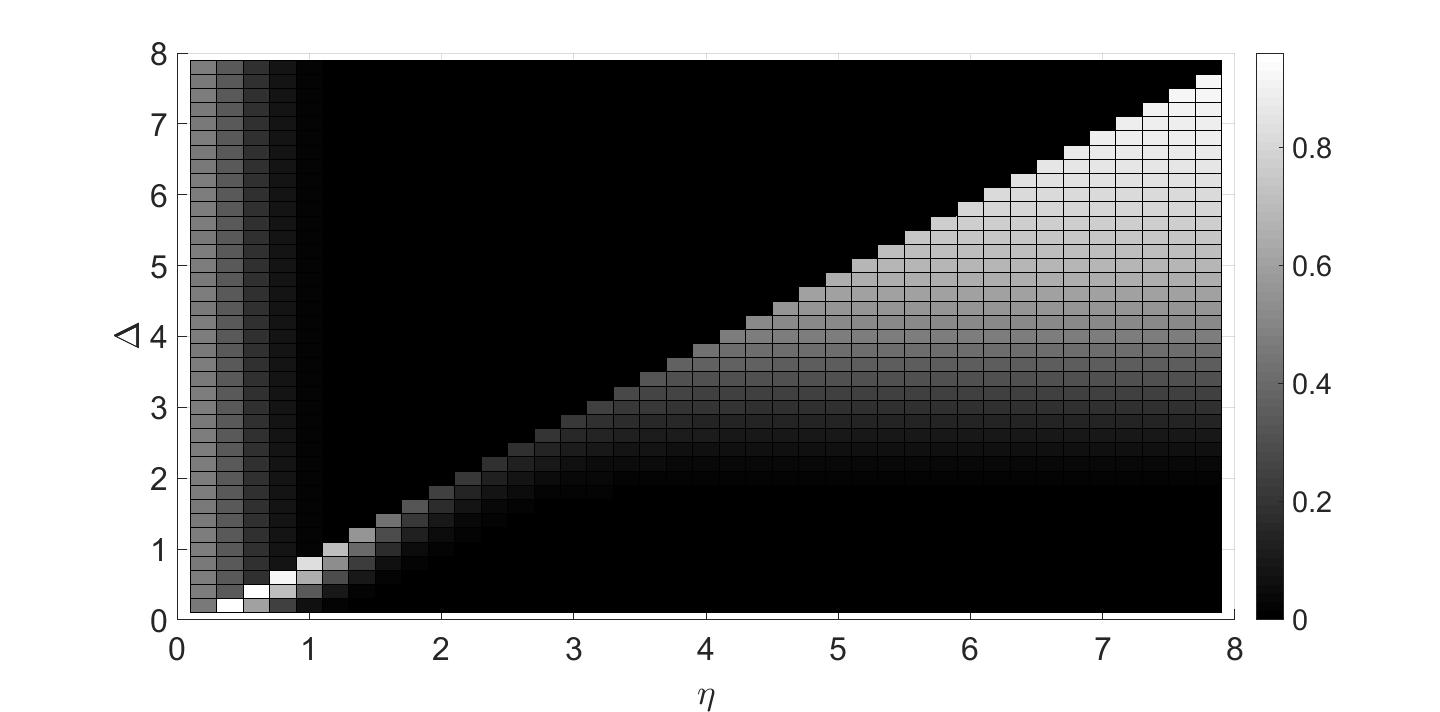}
    \caption{\textit{Payoff matrix of the game with $n=50$, $\alpha=0.2$ and $\mu = 1$ ($\text{SNR}=4$).}}
    \label{fig.GamePSNR2Alpha02N50}
\end{figure}

\begin{figure}[h!]
\begin{center}
\subfloat[]{\includegraphics[width=\columnwidth]{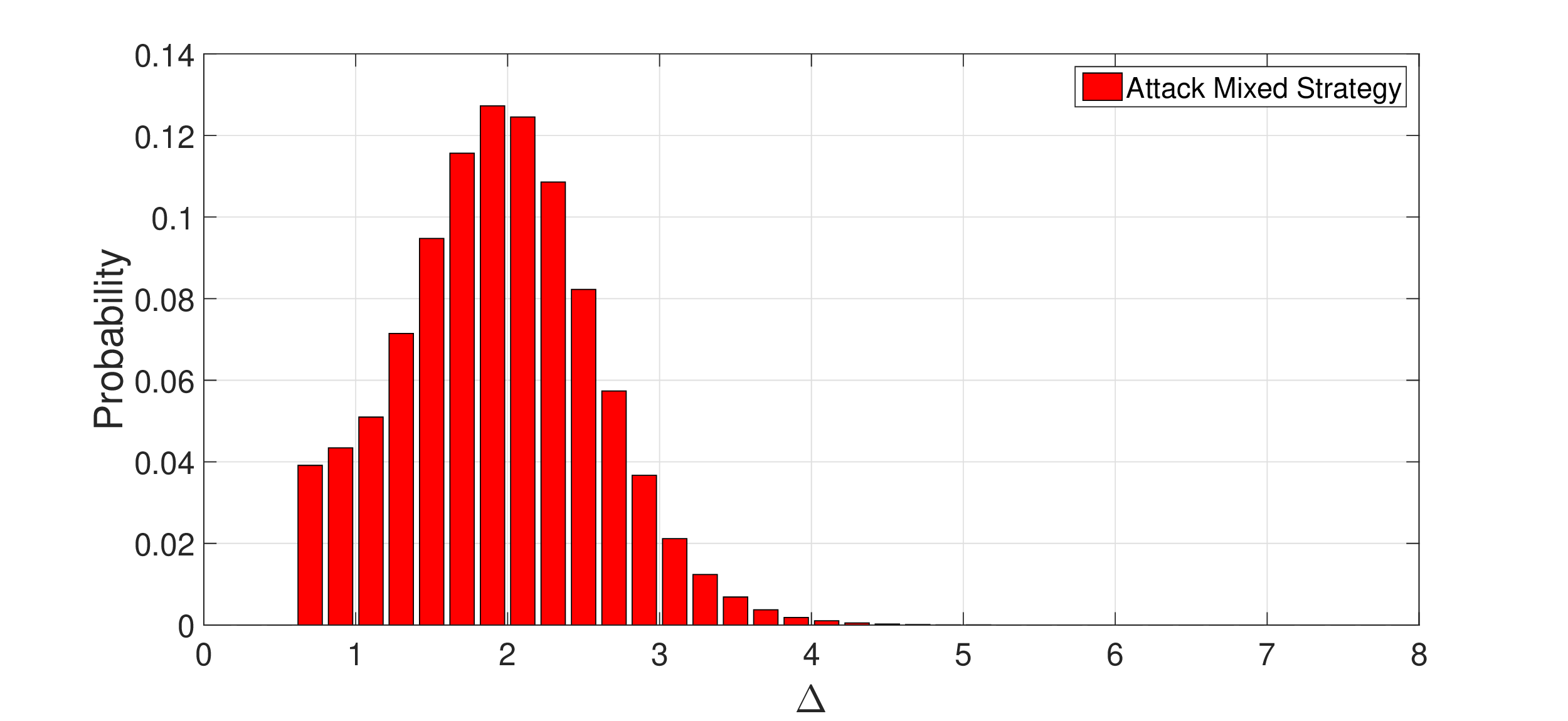}}
\label{fig:StrAPSNR2alpha02N50}
\subfloat[]{\includegraphics[width=\columnwidth]{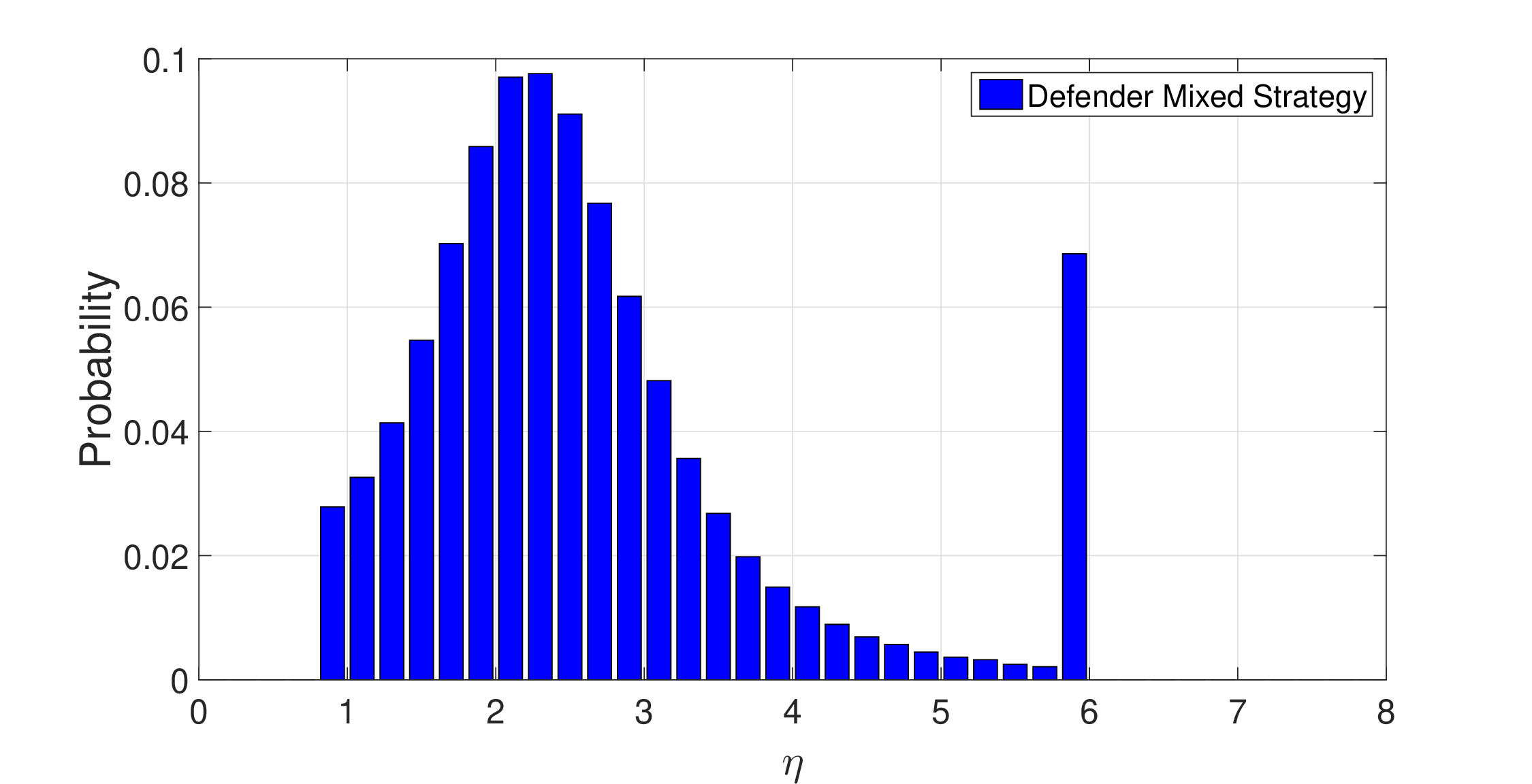}}
\label{fig:StrDPSNR2alpha02N50}
\caption{\textit{Equilibrium strategies in the following setup: $n=50$, $\alpha=0.2$, $\mu = 2$ ($\text{SNR}=4$). Payoff at the equilibrium $v = 0.0556$. (a) Attacker Mixed Strategy (b) Defender Mixed Strategy.}}
\label{fig.GameStrPSNR2Alpha02N50}
\end{center}
\end{figure}

A possible explanation for the peaked behavior is the following. When $\eta$ decreases, D starts removing good measurements which fall in the tail of the Gaussian under the corresponding hypothesis, whose values are not limited to $\Delta$, but can take arbitrarily large values. Depending also on the setup considered, it may happen that the positive contribution they give to the correct decision is more relevant than the negative contribution given by the values introduced by A.  When this is the case, it is better for the defender to use all the measurements. Therefore, the behavior of the defender at the equilibrium has a twofold purpose: trying to remove the corrupted measurements on one hand (by choosing $\eta$ one step ahead of $\Delta$) and avoiding to rule out the large good measurements on the other (by selecting the critical $\eta$).
The error probability at the equilibrium is 0.0176 thus showing that the proposed scheme allows to get correct detection with high probability despite the data corruption performed by A.

Figure \reffig{fig.GameStrPSNR2Alpha02} shows the equilibrium strategies for $\alpha = 0.2$ for the game payoff matrix in Figure \reffig{fig.GamePSNR2Alpha02}. Since the removal of the large good measurements has more impact when $\alpha$ is large, a bit higher weight is associated in this case to the peak. The error probability at the equilibrium is $v = 0.1097$.
Finally, Figure \reffig{fig.GameStrPSNR2Alpha02N50} shows the equilibrium mixed strategies for D and A for the payoff matrix in Figure \reffig{fig.GamePSNR2Alpha02N50} when $n=50$ , $\alpha = 0.2$ and $\mu = 2$.

\section{Conclusion}
\label{sec.conclusion}

We proposed a consensus algorithm based on censored data which is robust to measurement falsification attacks.
Besides, we formalized the interplay between the attacker and the network in a game-theoretic sense, and we numerically derived the optimal strategies for both players and the achievable performance in terms of error probability in different setups.
Simulation results show that, by adopting the proposed scheme, the network can still achieve correct detection through consensus, despite the presence of corrupted measurements.


\chapter{Conclusion}
\label{chapter:Conclusion}
\emph{"Begin at the beginning and go on
till you come to the end; then stop."}
\\
Lewis Carroll, "Alice in Wonderland"
\\

\section{Introduction}
\PARstart{\textcolor{red}I}n{} this thesis we provided a game-theoretic approach for adversarial information fusion in distributed sensor networks. We presented several solutions to tackle with the presence of the adversaries in such networks in both centralized and decentralized fashion. In this chapter, we summarize the main contributions of our work and outline some possible directions for future research.

\section{Summary}
Starting from the fact that information fusion in distributed networks is of great importance in many applications i.e, cognitive radio networks, multimedia forensics, wireless sensor networks and many others, 
we observed that securing these networks against attacks and threats is crucial and a key enabling factor for their proper functionality. Motivated by that, we studied these networks by considering the possible presence of adversaries that aim at corrupting their functionality. Following the concepts of adversarial signal processing, we provided a game-theoretic approach to study the security of adversarial information fusion in distributed sensor networks.

In chapter \ref{chapter:DF} and \ref{chapter:GoT} we presented a review of the basic notions of detection theory and game theory. In chapter \ref{chapter:SecurityThreats} we presented an extensive review of the literature concerning the attacks and their mitigation techniques in adversarial distributed sensor networks. Then, in chapters \ref{chapter:CDC}, \ref{chapter:TIFS_SPL}, \ref{chapter:InfoFusion} and \ref{chapter:GameSec} we provided possible solutions to mitigate the effect of the attacks in these networks and we used a game-theoretic formulation to study the ultimate performance that can be achieved by the attacker and the defender.

Specifically, we started by considering an adversarial decision fusion setup in which the nodes send to the FC a vector of binary decisions about the system state. As a first contribution, we developed a novel soft identification and isolation scheme to exclude the reports sent by the adversary, namely the Byzantines, from the decision fusion process. By adopting such a scheme, the FC can assign a reliability value to each node. As an additional contribution, we formalized the competition between the Byzantines and the FC in a game-theoretic sense and we studied the existence of an equilibrium point for the game. Then, we derived the payoff in terms of the decision error probability when the players play at the equilibrium. By using numerical simulations, we showed that the soft isolation scheme outperforms the defense mechanisms proposed in previous works.

As a second contribution, we derived the optimum decision fusion rule in the presence of Byzantines in a centralized setup. By observing the system over an observation window, we adopted the Maximum A Posteriori Probability (MAP) rule while assuming that the FC knows the attack strategy of the Byzantines and their distribution across the network. With regard to the knowledge that the FC has about the distribution of Byzantines over the network, we considered many cases. First, we examined an unconstrained maximum entropy scenario in which the uncertainty about the distribution of Byzantines is maximum Then, we considered a more favorable scenario to the FC in which the maximum entropy case is subject to a constraint. In this scenario, the FC has more a-priori information about Byzantines's distribution i.e the average or the maximum number of Byzantines in the network. Finally, we considered the most favorable situation in which the FC knows the exact number of Byzantines present in the network. Concerning the complexity of the optimal fusion rule, we developed an efficient implementation based on dynamic programming. Thereafter, we introduced a game-theoretic framework to cope with the lack of knowledge regarding the Byzantines strategy. In such a framework, the FC makes a "guess" by selecting arbitrarily a Byzantine's attacking strategy within the optimum fusion rule. By considering the decision error probability as the payoff, we studied the performance of the Byzantines as well as the FC at the equilibrium for several setups. As a main result of this chapter, we showed that the attacker should follow a mixed strategy Nash equilibrium as opposed to what was believed in previous works. This strategy reaches a trade-off between inducing decision error at the FC and avoiding being caught. Finally, by comparing the performance of the optimum fusion rule to those of previous works, we showed its superior performance over all the other schemes.

By revisiting the complexity of the optimum fusion rule, as an additional contribution, we proposed a near-optimal message passing approach based on factor graph. We considered a more general model for the observed system in which we examine both independent and Markovian sequences. Then, we showed that the message passing algorithm can give near-optimal performance while reducing the complexity from exponential to linear as a function of the observation window size. In addition, we showed that the case of independent states is more favorable to the Byzantines than the Markovian case, due to the additional a-priori information available at the FC in the Markovian case. Furthermore, based on large observation windows, we confirmed the dual behavior in the attacking strategy of the Byzantines, looking for a trade-off between pushing the FC to make a wrong decision on one hand and reducing the mutual information between the reports and the system state on the other.

In the last part of the thesis, we considered a decentralized version of the data fusion process based on consensus algorithm. We focused on case in which the adversary attacks the links between the system being monitored and the sensors. To make the network more robust, we proposed a primary isolation step to be carried at the node level to filter out the falsified information injected by the attacker. Then, we employed game-theory to model the competition between the adversary and the network. At last, we used numerical simulations to derive and study the equilibrium points of the game and the performance at the equilibrium.

\section{Open Issues}

There are some avenues for future work on the topics addressed in this thesis. As a first research direction we can improve the performance of the Byzantines by letting them exploit the knowledge of the observation vectors. By exploiting the knowledge of such information, the Byzantines can focus their attack on the most uncertain cases thus avoiding to flip the local decision when it is expected that the attack will have no effect on the FC decision. Considering a scenario where the nodes can send more extensive reports rather than one single bit \cite{varshneydatafusion} is another interesting extension of the work done in this thesis.

In decentralized networks, extending the game-theoretic approach to include graph disconnections as a part of the defender payoff and then apply our analysis to general topologies is an interesting research direction. In addition, we would like to extend the analysis to more complicated statistical models for the measurements, e.g. to the case of chi-square distribution, and to consider more complicated versions of the game, e.g. by allowing the players to adopt randomized strategies.

Even more interestingly, we could apply the concepts of adversarial signal processing to more complex and fully adaptive networks. The reason is that, these networks have been playing an increasingly important role in our daily life because of their technological, social, and economical aspects on people's life. As examples of these networks Facebook, Twitter, Amazon and many others. As a first example, on facebook, we would like to apply the concept of adversarial information fusion to combat against fake user profiles who generate forged/negative reviews or ratings against a business page and cause financial loss. Another example is about twitter, for instance, we can counter the users or profiles who, by faking information into their tweets, spread falsified information about an event or a trend that can be critical.

Although being aware of some specific limitations for some applications, we believe that studying adversarial information fusion can bring an important contribution to the security of networks with distributed sensors. Due to its wide applicability and going in line with the development of a general framework for adversarial signal processing, it is easy to foresee an increasing interest in the topic studied in this thesis in the near future.




\renewcommand{\publ}{}
\cleardoublepage \small \addcontentsline{toc}{chapter}{Bibliography}
\bibliography{Kassem}
\bibliographystyle{IEEEtran1}

\cleardoublepage 
\addcontentsline{toc}{chapter}{Index}
\printindex

\cleardoublepage
\addcontentsline{toc}{chapter}{Author's Information}

\chapter*{Author's Information}
\markboth{Author's Information}{}
\vspace{-3cm}

\begin{wrapfloat}{figure}{l}{0.5\textwidth}
\centering
\setlength{\fboxsep}{0pt}%
\setlength{\fboxrule}{0.1pt}%
\fbox{\includegraphics[width=0.4\textwidth]{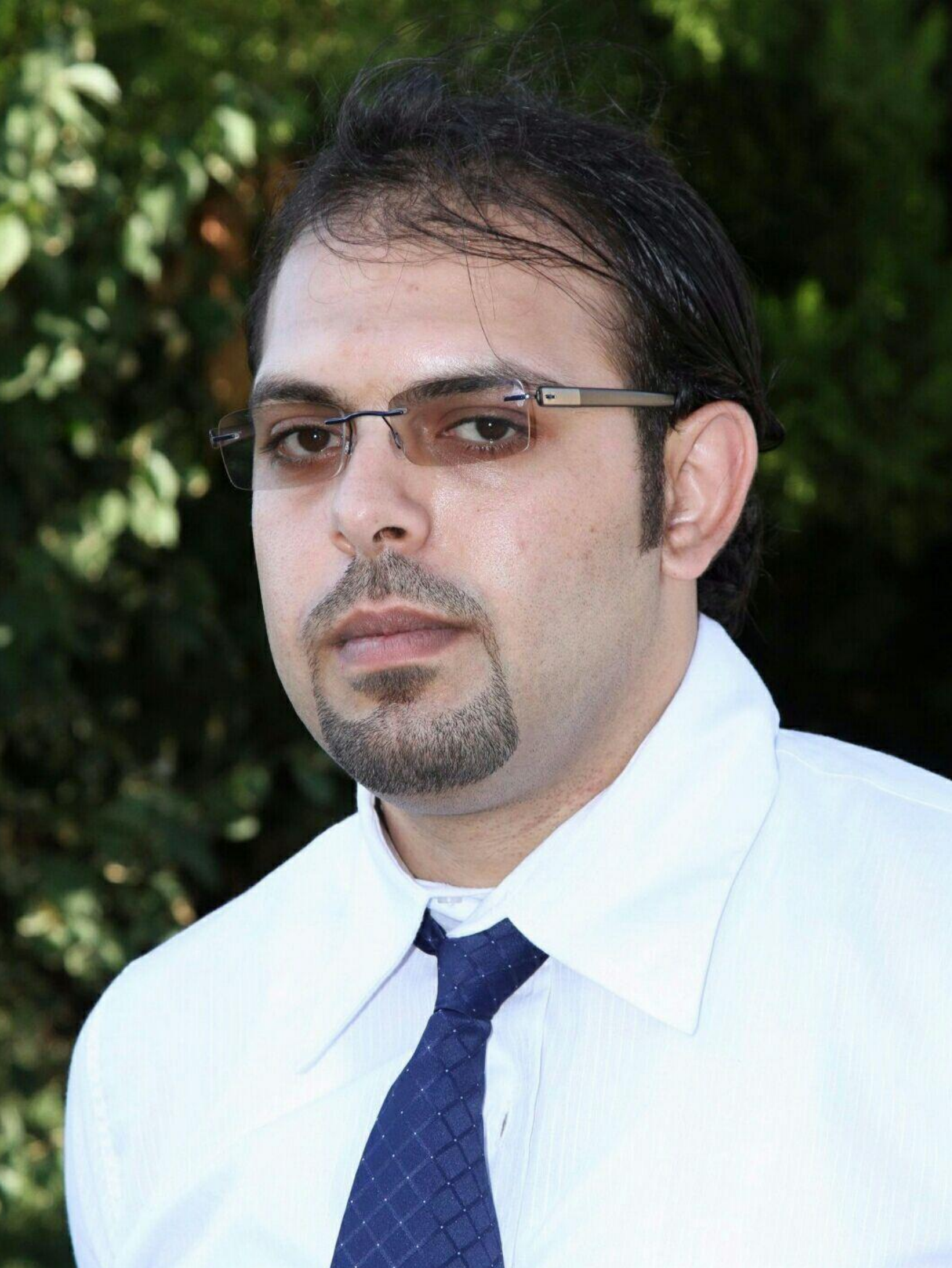}}
\end{wrapfloat}


\textcolor{red}{K}ASSEM KALLAS, the author of this thesis was born on August, $27^{th}$ $1988$ in Lebanon. He grew up in the south of his country, a land of conflicts that throughout its history, passed from one occupation to another. During his childhood, he studied at different religious schools where he encountered different religions and without being fascinated by any of them. At school, he was very good in scientific subjects i.e math, physics etc. and completely ignored other materials like "language grammar and literature" as much as they ignored him. Moreover, his student book of \textit{"behavior"} was tragic. Ironically, while his name was always found at the top of the troublemaker's list, it was also shining at the student's honor list. At high school, he was selected for the mathematics-physics category, in which he successfully passed the official governmental examination in $2006$. Just two days later, the middle east celebrated his success with the most terrible war that he ever saw. Fortunately, he survived and joined the Lebanese International University because it was his only choice since he was besieged with the family in the south and was not able to search for other universities. He got the Master of Science degree in \textit{Computer and Communications Engineering} and was expecting to find a good job. However, it was a wishful thinking. After his disappointment, in $2013$, he decided to continue his education in the second level master class in \textit{Wireless Systems and Related Technologies} at Politecnico di Torino. After graduation, he searched for a long time for Ph.D position and finally, at the last breath, he got the acceptance letter from University of Siena. Then, he chose to join the Visual Information Processing and Protection (VIPP) research group led by Professor Mauro Barni because the research field seemed the most interesting to him. During his Ph.D, he published some papers in important conferences and prestigious journals and reviewed many as well. He learned a lot from his supervisor Professor Barni who made out of him a scientific researcher. Also, Professor Barni offered him the opportunity to attend many summer schools including the annual meeting of Italian National Telecommunications and Information Theory Group. After all, he made his family's dream come true by his effort to become a "Dr.".

\includepdf[pages=-]{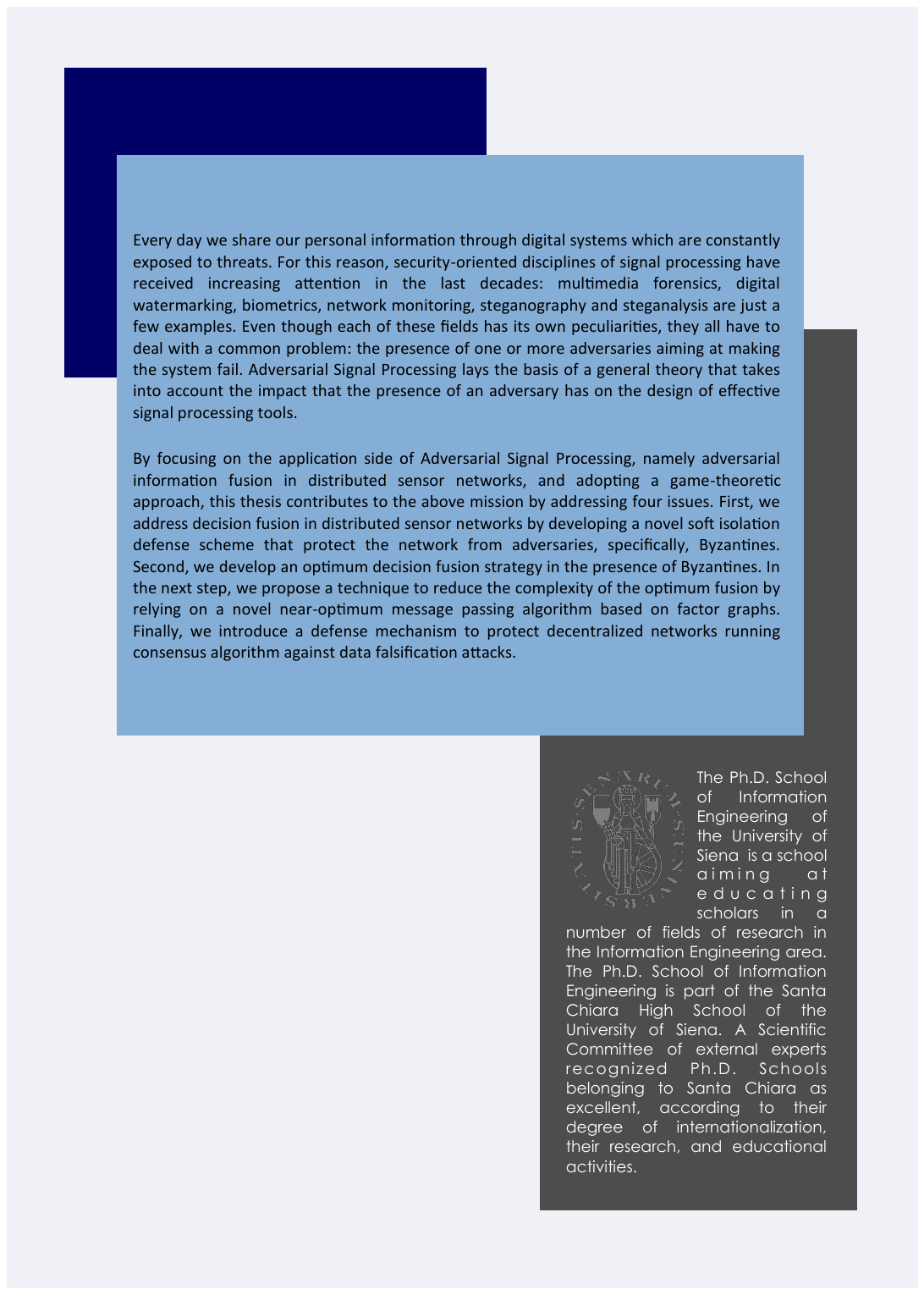}
\end{document}